\documentclass[11pt]{cernrep}
 
\usepackage{epsfig, url, psfrag, color}
\usepackage{fancyhdr, multind}
\usepackage[square,comma,numbers]{natbib}
\usepackage{dcolumn, bm, mathrsfs, xspace, graphicx,subfigure}
\usepackage{array,amsfonts,amssymb,amsmath,longtable}

\usepackage{qcd-tev4lhc}

\begin{document}



\date{\today}
\documentlabel{hep-ph/0610005 \\ FERMILAB-Conf-06-359}


\title{\bf Tevatron-for-LHC Report of the QCD Working Group}



\author{
(TeV4LHC QCD Working Group)
\mbox{M. Albrow,$^{1}$}
\mbox{M. Begel,$^{2}$}
\mbox{D. Bourilkov,$^{3}$}
\mbox{M. Campanelli,$^{4}$}
\mbox{F. Chlebana,$^{1}$}
\mbox{A. De~Roeck,$^{5}$}
\mbox{J.R. Dittmann,$^{6}$}
\mbox{S.D. Ellis,$^{7}$}
\mbox{B. Field,$^{8}$}
\mbox{R. Field,$^{3}$}
\mbox{M. Gallinaro,$^{9}$}
\mbox{W. Giele,$^{1}$}
\mbox{K. Goulianos,$^{9}$}
\mbox{R.C. Group,$^{3}$}
\mbox{K. Hatakeyama,$^{9}$}
\mbox{Z. Hubacek,$^{10}$}
\mbox{J. Huston,$^{11}$}
\mbox{W. Kilgore,$^{12}$}
\mbox{T. Kluge,$^{13}$}
\mbox{S.W. Lee,$^{14}$}
\mbox{A. Moraes,$^{15}$}
\mbox{S. Mrenna,$^{1}$}
\mbox{F. Olness,$^{16}$}
\mbox{J. Proudfoot,$^{17}$}
\mbox{K. Rabbertz,$^{18}$}
\mbox{C. Royon,$^{19,1}$}
\mbox{T. Sjostrand,$^{5,20}$}
\mbox{P. Skands,$^{1}$}
\mbox{J. Smith,$^{21}$}
\mbox{W.K. Tung,$^{7,11}$}
\mbox{M.R. Whalley,$^{22}$}
\mbox{M. Wobisch,$^{1}$}
\mbox{M. Zielinski$^{2}$}
}

\institute{
\mbox{$^{1}$ Fermilab},
\mbox{$^{2}$ Univ. of Rochester},
\mbox{$^{3}$ Univ. of Florida},
\mbox{$^{4}$ Univ. of Geneva},
\mbox{$^{5}$ CERN},
\mbox{$^{6}$ Baylor Univ.},
\mbox{$^{7}$ Univ. of Washington},
\mbox{$^{8}$ Florida State Univ.},
\mbox{$^{9}$ The Rockefeller Univ.},
\mbox{$^{10}$ Czech Technical Univ.},
\mbox{$^{11}$ Michigan State Univ.},
\mbox{$^{12}$ Brookhaven National Lab.},
\mbox{$^{13}$ DESY},
\mbox{$^{14}$ Texas Tech Univ.},
\mbox{$^{15}$ Univ. of Glasgow},
\mbox{$^{16}$ Southern Methodist Univ.},
\mbox{$^{17}$ Argonne National Lab.},
\mbox{$^{18}$ Univ. of Karlsruhe},
\mbox{$^{19}$ DAPNIA/SPP, CEA/Saclay},
\mbox{$^{20}$ Lund Univ.},
\mbox{$^{21}$ Stony Brook Univ.},
\mbox{$^{22}$ Univ. of Durham}
}

\maketitle

\begin{abstract}
The experiments at Run 2 of the Tevatron have each accumulated over 1 fb$^{-1}$ of high-transverse momentum data.
Such a dataset allows for the first precision (i.e. comparisons between theory and experiment at
the few percent level) tests of QCD at a hadron collider.  While the Large Hadron Collider has been designed
as a discovery machine, basic QCD analyses will still need to be performed to understand the working environment.
The Tevatron-for-LHC workshop was conceived as a communication link to pass on the expertise of the Tevatron
and to test new analysis ideas coming from the LHC community.  The TeV4LHC QCD Working Group focussed on
important aspects of QCD at hadron colliders:  jet definitions, extraction and use of Parton Distribution Functions,
the underlying event, Monte Carlo tunes, and diffractive physics.  
This report summarizes some of the results achieved during this workshop.

\end{abstract}

\clearpage
\tableofcontents
\clearpage

\label{chap:qcd}

\section{{Introduction and Overview}} 
\label{sec:introduction}

Quantum Chromodynamics (QCD) is the underlying theory for scattering processes, both hard and soft, at hadron-hadron colliders. At the LHC, experimental particle physics will enter a new regime of complexity. But, both the signal channels for possible new physics as well as their backgrounds, will be composed of building blocks ($W$ and $Z$ bosons, 
photons, leptons, heavy quarks, jets, {\it etc.}) which have been extensively studied at the Tevatron, 
both singly and in combination. Measurements have been carried out at the Tevatron, 
for both inclusive and exclusive final states, in regions that can be described by simple 
power-counting in factors of $\alpha_s$ and in regions where large logarithms need to be re-summed. 

In this document, we summarize some of the experience that has been gained at the Tevatron, 
with the hope that this knowledge will serve to jump-start the early analyses to be carried out at the LHC. 
The main topics covered are: (1) jet algorithms;
(2) aspects of parton distribution functions, including heavy
flavor; (3) event generator
tunings; (4) diffractive physics; and (5) an exposition of useful
measurements that can still be performed at the Tevatron into the LHC startup period.

Most physics analyses at the Tevatron or LHC involve final states with jets. Thus, jet definitions and algorithms are crucial for accurate measurements of many physics channels. 
Jet algorithms are essential to map the long distance hadronic 
degrees of freedom observed
in detectors onto the short distance colored partons of the underlying hard
scattering process most easily described in perturbation theory. 
Any mismatch between these two concepts will ultimately limit how
well we can measure cross sections including jets and how well we
can measure the masses of (possibly new) heavy particles. The report on jet algorithms reviews the history of jet algorithms at the Tevatron, their application to current Run 2 analyses, the differences that arise between comparisons to parton-level final states and real data, and some of the current controversies. Suggestions are made for improvements to the Midpoint cone algorithm that should remove some of the controversy, and which should serve as a robust algorithm for analyses at the LHC. Comparisons are made between inclusive jet analyses using the cone and $k_T$ algorithms and excellent agreement is noted. A plea is made for analyses at both the Tevatron and LHC to make use of both algorithms wherever possible. 

Parton distribution functions (PDF's) are another essential ingredient
for making predictions and
performing analyses at hadron colliders.  The contributions 
to this report concern the development
of tools for evaluating PDF's and their uncertainties and for
including more datasets into the fits, and also 
strategies and first results on extracting heavy flavor PDF's
from the Tevatron.
The {\LHAPDF} tools provides a uniform framework for including
PDF fits with uncertainties into theoretical calculations. 
At the LHC, as at the Tevatron, this will be important for estimating 
background rates 
and extracting cross section information (of possibly new objects).
{\fastNLO} is a powerful tool for performing 
very fast pQCD calculations for given observables for
arbitrary parton density functions.
This will enable future PDF fits to include data sets (such as 
multiply-differential dijet 
data in hadron-hadron collisions and the precise DIS jet data from HERA)
that have been neglected so far because the computing time
for conventional calculations was prohibitive.
Finally,
as it is expected that some aspect of physics beyond the standard model
will couple proportional to mass, heavy flavor PDF's will be
needed to calculate production cross sections of Higgs-boson-like
objects.  First results on the extraction of heavy flavor PDF's at the
Tevatron are presented, as well as a theoretical study of the
treatment of heavy flavor PDF's in Higgs boson calculations.

For good or for bad, the bulk of our understanding of the Standard Model
at hadron colliders relies on parton shower event generators.  While
these tools are based on perturbative QCD, the details of their predictions
do depend on tuneable parameters.
Our ability to estimate backgrounds to new physics searches, at least early
on in the running of the LHC, will rely on quick, accurate tunes.  
The report contains several contributions on Run II methods
for tuning parameters associated with the parton shower and the
underlying event, with comments on how these tunes apply to the LHC and what the current estimated uncertainties are.

The success of the diffractive physics program at the Tevatron
has raised the profile of such experimental exploration.  
Three contributions to the report highlight the measurements
performed at the Tevatron, and the opportunities at the LHC
of even discovering new physics through exclusive production
channels.

The final contribution was inspired by the pointed questions of 
a Fermilab review committee, and states the case for running
the Tevatron into the LHC era.


\clearpage
\section{{Jet Algorithms}} 

\newcommand\thissection[1]{\subsubsection*{#1}}

\begin{figure}[h]
\centerline{
\includegraphics[width=\textwidth]{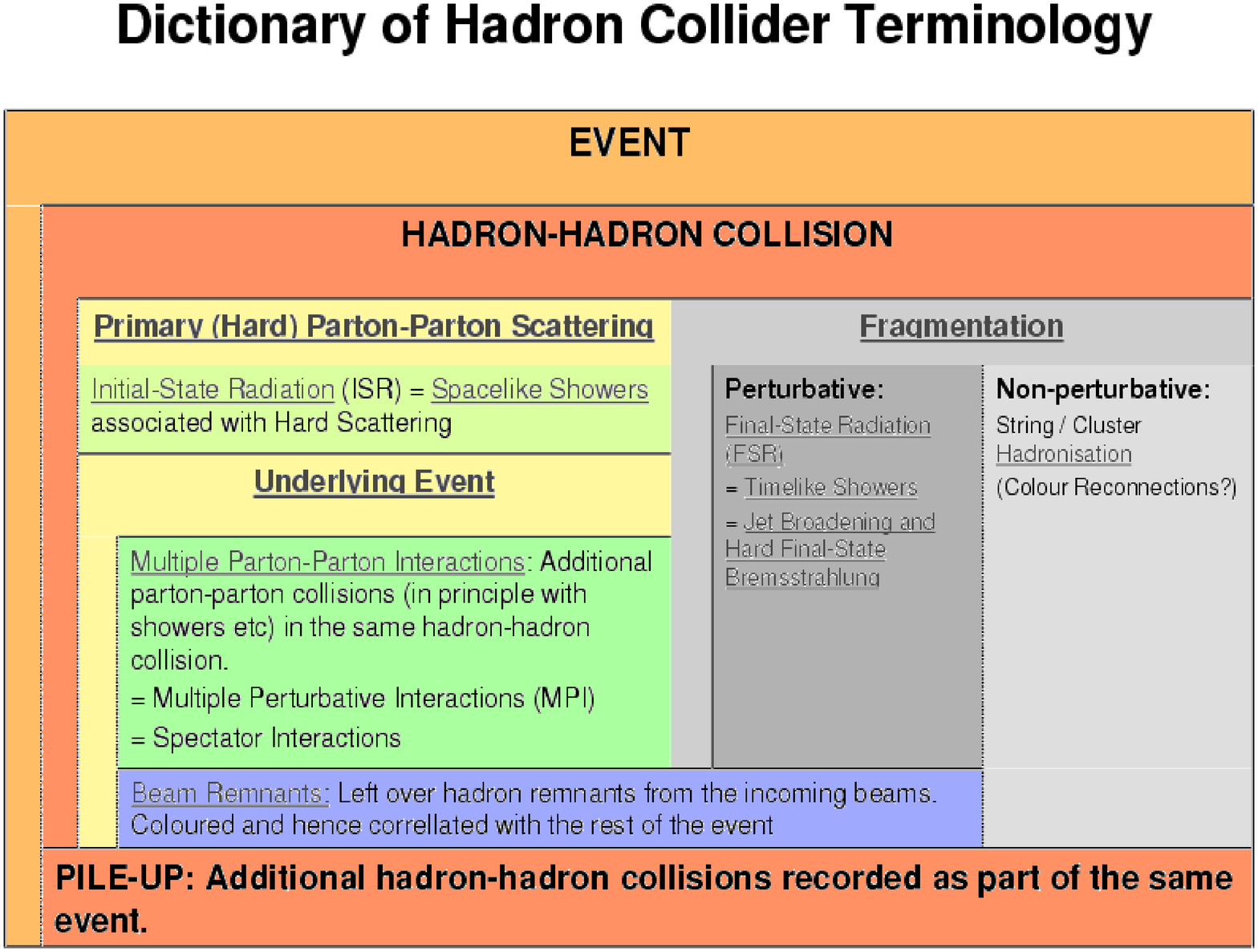}
}
\caption{Dictionary of Hadron Collider Terms}%
\label{dictionary}
\end{figure}

The fundamental challenge when trying to make theoretical predictions or
interpret experimentally observed final states at hadron colliders is that the theory of the
strong interactions (QCD) is most easily applied to the short distance 
($\ll$ 1 fermi) degrees of freedom, the color-charged quarks and gluons, 
while the
long distance degrees of freedom seen in the detectors are color singlet bound
states of these degrees of freedom. \ We can approximately picture the
evolution between the short-distance and long distance states as involving
several (crudely) distinct steps. \ First comes a color radiation step when
many new gluons and quark pairs are added to the original state, dominated by partons
that have low energy and/or are nearly collinear with the original short
distance partons.  These are described by the parton showers in Monte Carlo programs and
calculated in terms of summed perturbation theory. \ The next step involves a
non-perturbative hadronization process that organizes the colored degrees of
freedom from the showering and from the softer interactions of other initial
state partons (the underlying event simulated in terms of models for multiple
parton interactions of the spectators) into color-singlet hadrons with
physical masses. \ This hadronization step is estimated in a model dependent
fashion (\textit{i.e}., a different fashion) in different Monte Carlos. The
union of the showering and the hadronization steps is what has historically
been labeled as fragmentation, as in fragmentation functions describing the
longitudinal distribution of hadrons within final state jets. \ In practice,
both the radiation and hadronization steps tend to smear out the energy that
was originally localized in the individual short distance partons, while the
contributions from the underlying event (and any \textquotedblleft
pile-up\textquotedblright\ from multiple hadron collisions) add to the energy
originally present in the short distance scattering (a \textquotedblleft
splash-in\textquotedblright\ effect). \ Finally the hadrons, and their decay
products, are detected with finite precision in a detector. \ \ This
vocabulary (and the underlying picture) is summarized in 
Fig.~\ref{dictionary}\cite{Dict}.  It is worthwhile noting that the usual na\"ive 
picture of hard scattering events, as described in Fig.~\ref{dictionary},
includes not only the showering of the scattered short-distance partons as noted
above, typically labeled as Final State Radiation (FSR), but also showering from 
the incoming partons prior to the scattering process, labeled as Initial State 
Radiation (ISR).  This separation into two distinct processes is not strictly valid in
quantum mechanics where interference plays a role; we must sum the amplitudes 
before squaring them and not just sum the squares.  However, the numerical dominance
of collinear radiation ensures that the simple picture presented here and 
quantified by Monte Carlo generated events, 
without interference between initial and final state processes, provides a
reliable first approximation.  We will return to this issue below.

In order to interpret the detected objects in terms of the underlying short
distance physics, jet algorithms are employed to associate \textquotedblleft
nearby\textquotedblright\ objects into jets. \ The jet algorithms are intended
to cluster together the long distance particles, or energies measured in
calorimeter cells, with the expectation that the kinematics (energy and
momentum) of the resulting cluster or jet provides a useful measure of the
kinematics (energy and momentum) of the underlying, short-distance partons.
\ The goal is to characterize the short-distance physics, event-by-event, in
terms of the discrete jets found by the algorithm. \ A fundamental assumption
is that the basic mismatch between colored short-distance objects and the
colorless long-distance objects does not present an important 
limitation. \ 

As noted, jet algorithms rely on the merging of objects that are, by some
measure, nearby each other. \ This is essential in perturbation theory where
the divergent contributions from virtual\ diagrams must contribute in
exactly the same way as the divergent contributions from soft and collinear
real emissions in order that these contributions can cancel.  It is only
through this cancellation that jet algorithms serve to define 
an IR-safe (finite) quantity. The
standard measures of \textquotedblleft nearness\textquotedblright\ (see
\cite{Blazey:2000qt}) include pair-wise relative transverse momenta, as in the
$k_{T}$ algorithm, or angles relative to a jet axis, as in the cone algorithm.
\ By definition a \textquotedblleft good\textquotedblright\ algorithm yields
stable (\textit{i.e}., similar) results whether it is applied to a state with
just a few partons, as in NLO\ perturbation theory, a state with many partons
after the short distance partons shower as simulated in a Monte Carlo, a state
with hadrons as simulated in a Monte Carlo including a model for the
hadronization step and the underlying event, or applied to the observed tracks
and energy deposition in a real detector. As we will see, this constitutes
a substantial challenge.  \ Further, it is highly desirable that the identification
of jets be insensitive to the contributions from the simultaneous
uncorrelated soft collisions that occur during pile-up at high luminosity. 
Finally we want to be able to apply the \textit{same} algorithm (in detail) 
at each level in the evolution of the hadronic final state.  This implies that 
we must avoid components in the algorithm that make sense when applied to
data but not to fixed order perturbation theory, or vice versa.  This constraint
will play a role in our subsequent discussion. 

\begin{figure}[h]
\centerline{
\includegraphics[width=\textwidth]{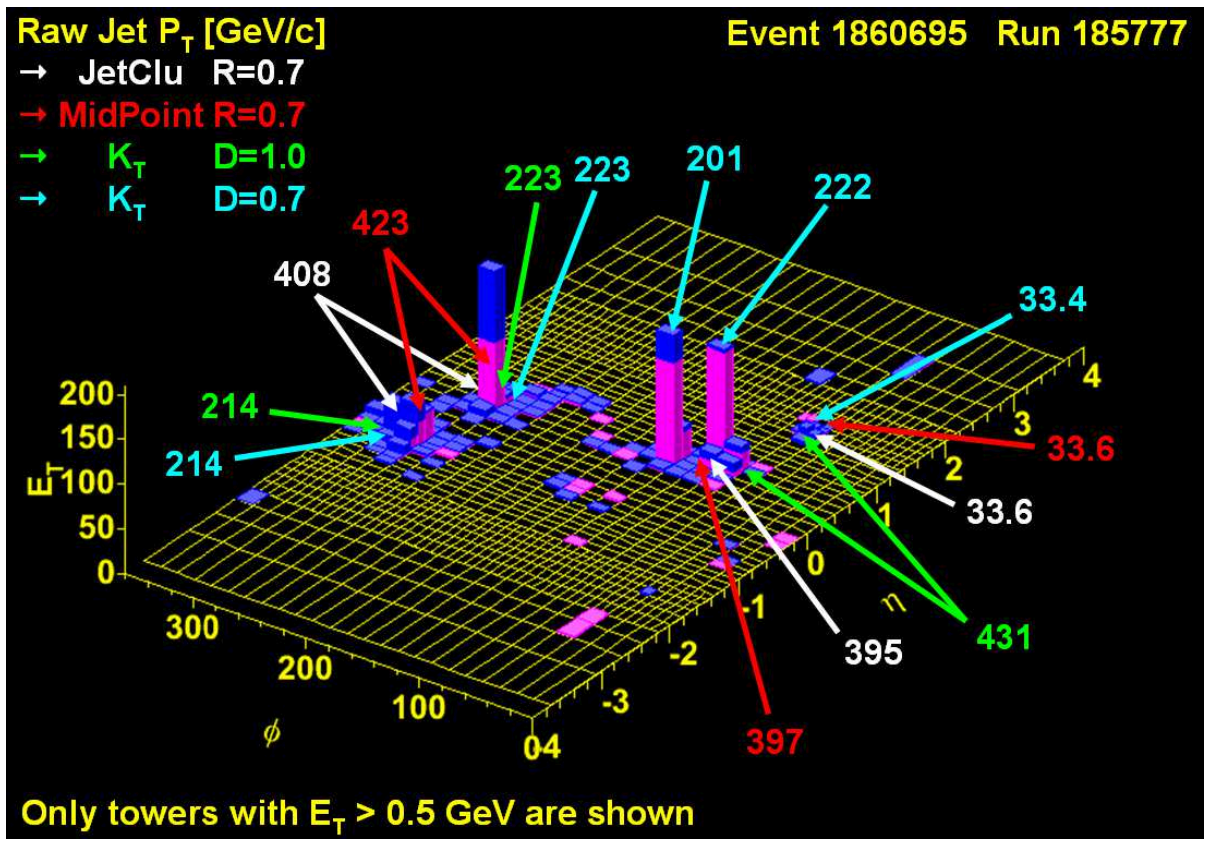}
}
\caption{Impact of different jet clustering algorithms on an interesting
event. }%
\label{lego3}%
\centerline{
}\end{figure}

For many events, the jet structure is clear and
the jets, into which the individual towers should be assigned, are fairly
unambiguous, i.e. are fairly insensitive to the particular definition of
a jet.
However, in other events such as Fig.~\ref{lego3}, the complexity
of the energy depositions means that different algorithms will result in
different assignments of towers to the various jets. This is not a problem
if a similar complexity is exhibited by the theoretical calculation, which
is to be compared to the data. \ However, the most precise and thoroughly
understood theoretical calculations arise in fixed order perturbation theory,
which can exhibit only limited complexity, \textit{e.g}., at most two partons
per jet at NLO. \ On the other hand, for events simulated with parton shower
Monte Carlos the complexity of the final state is more realistic, but the
intrinsic theoretical uncertainty is larger. \ Correspondingly the jets
identified by the algorithms vary if we compare at the perturbative, shower,
hadron and detector level. 
Thus it is essential to understand these limitations of jet algorithms
and, as much as possible, eliminate or correct for them. \ It is the goal of
the following discussion to highlight the issues that arose during Run I at
the Tevatron and outline their current understanding and possible solution
during Run II and at the LHC.

\thissection{Parton Level Vs Experiment - A Brief History of Cones}

\begin{figure}[h]
\subfigure[(An ideal) Monte Carlo generated event with 2 large energy jets
and 1 small energy jet in the LEGO\ plot.]{\includegraphics[width=.5\textwidth]{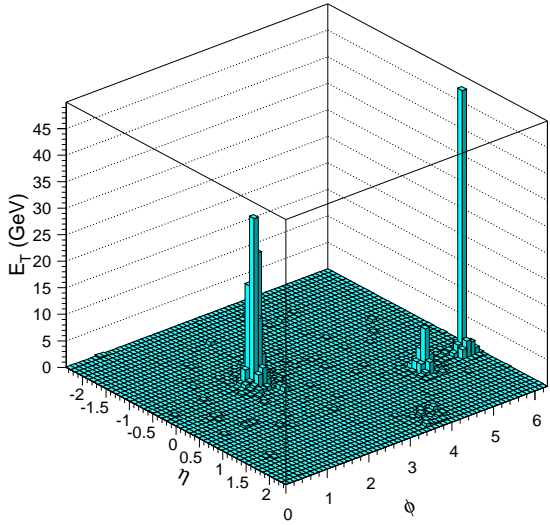}}
\subfigure[The corresponding flow structure of the trial cones.]{\includegraphics[width=.5\textwidth]{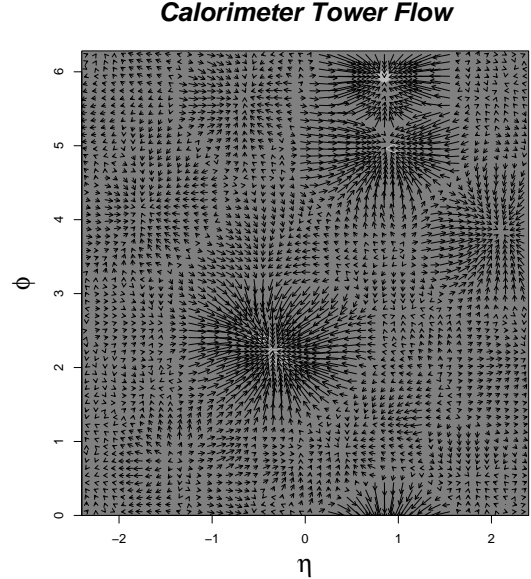}}
\caption{Illustration of the flow of trial cones to a stable jet solution.}
\label{towerflow}%
\end{figure}

The original Snowmass\cite{Snowmass} implementation of the cone algorithm can
be thought of in terms of a simple sum over all objects within a cone,
centered at rapidity (the actual original version used the pseudorapidity
$\eta$) and azimuthal angle $\left(  y_{C},\phi_{C}\right)  $ and defining a
$p_{T}$-weighted centroid via%
\begin{align*}
k  &  \subset C\text{ iff } \sqrt{\left(  y_{k}-y_{C}\right)  ^{2}+\left(
\phi_{k}-\phi_{C}\right)  ^{2}} \leq R_{cone},\\
\overline{y}_{C}  &  \equiv\frac{\sum_{k\subset C}y_{k}\ast p_{T,k}}
{\sum_{l\subset C}p_{T,l}},\text{ }\overline{\phi}_{C}\equiv\frac
{\sum_{k\subset C}\phi_{k}\ast p_{T,k}} {\sum_{l\subset C}p_{T,l}}.
\end{align*}
If the $p_{T}$-weighted centroid does not coincide with the geometric center
of the cone, $\left(  \overline{y}_{C},\overline{\phi}_{C}\right)  \neq\left(
y_{C},\phi_{C}\right)  $, a cone is placed at the $p_{T}$-weighted centroid
and the calculation is repeated. \ This simple calculation is iterated until a
\textquotedblleft stable\textquotedblright\ cone is found, $\left(
\overline{y}_{C},\overline{\phi}_{C}\right)  =\left(  y_{C},\phi_{C}\right)
$, which serves to define a jet (and the name of this algorithm as the
iterative cone algorithm). \ Thus, at least in principle, one can think in
terms of placing trial cones everywhere in $\left(  y,\phi\right)  $ and
allowing them to \textquotedblleft flow\textquotedblright\ until a stable cone
or jet is found. \ This flow idea is illustrated in Fig, \ref{towerflow},
where a) illustrates the LEGO plot for a simple (quiet) Monte Carlo generated
event with 3 apparent jets and b) shows the corresponding flows of the trial
cones. Compared to the event in Fig.~\ref{lego3} there is little ambiguity in
this event concerning the jet structure.

To facilitate the subsequent discussion and provide some mathematical
structure for this image of \textquotedblleft flow\textquotedblright\,
we define
the \textquotedblleft Snowmass potential\textquotedblright\ in terms of the
2-dimensional vector $\overrightarrow{r}=\left(  y,\phi\right)  $ via%
\[
V\left(  \overrightarrow{r}\right)  =-\frac{1}{2}\sum_{k}p_{T,k}\left(
R_{cone}^{2}-\left(  \overrightarrow{r}_{k}-\overrightarrow{r}\right)
^{2}\right)  \Theta\left(  R_{cone}^{2}-\left(  \overrightarrow{r}%
_{k}-\overrightarrow{r}\right)  ^{2}\right)  .
\]
The flow described by the iteration process is driven by the \textquotedblleft
force\textquotedblright%
\begin{align*}
\overrightarrow{F}\left(  \overrightarrow{r}\right)   &  =-\overrightarrow
{\nabla}V\left(  \overrightarrow{r}\right)  =\sum_{k}p_{T,k}\left(
\overrightarrow{r}_{k}-\overrightarrow{r}\right)  \Theta\left(  R_{cone}%
^{2}-\left(  \overrightarrow{r}_{k}-\overrightarrow{r}\right)  ^{2}\right) \\
&  =\left(  \overrightarrow{\overline{r}}_{C\left(  \overrightarrow{r}\right)
}-\overrightarrow{r}\right)  \sum_{k\subset C\left(  r\right)  }p_{T,k},
\end{align*}
where $\overrightarrow{\overline{r}}_{C\left(  \overrightarrow{r}\right)
}=\left(  \overline{y}_{C\left(  \overrightarrow{r}\right)  },\overline{\phi
}_{C\left(  \overrightarrow{r}\right)  }\right)  $ and $k\subset C\left(
\overrightarrow{r}\right)  $ is defined by $\sqrt{\left(  y_{k}-y\right)
^{2}+\left(  \phi_{k}-\phi\right)  ^{2}}$ $\leq R_{cone}$. \ As desired, this
force pushes the cone to the stable cone position.

Note that in the Run II analyses discussed below 4-vector techniques are used
and the corresponding E-scheme centroid is given instead by
\begin{align*}
k  &  \subset C\text{ iff }\sqrt{\left(  y_{k}-y_{C}\right)  ^{2}+\left(
\phi_{k}-\phi_{C}\right)  ^{2}}\leq R_{cone},\\
p_{C}  &  =\left(  E_{C},\overrightarrow{p}_{C}\right)  ={\sum_{k\subset C}%
}\left(  E_{k},\overrightarrow{p_{k}}\right)  ,\text{ }\overline{y}_{C}%
\equiv\frac{1}{2}\ln\frac{E_{C}+p_{z,C}}{E_{C}-p_{z,C}},\text{ }\overline
{\phi}_{C}\equiv\tan^{-1}\frac{p_{y,C}}{p_{x,C}}.
\end{align*}
In the NLO perturbative calculation these changes in definitions result in
only tiny numerical changes.

\begin{figure}[h]
\centerline{
\includegraphics[width=\textwidth,trim= 0 50 0 0]{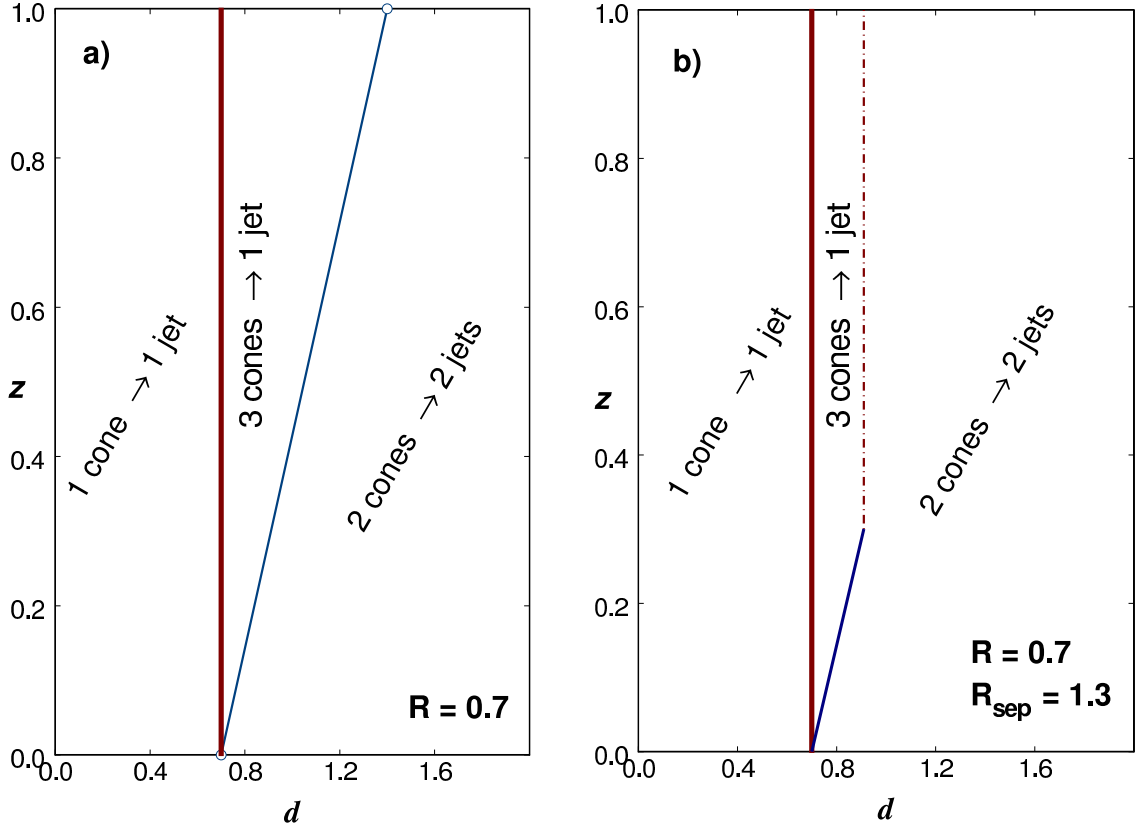}
}
\caption{Perturbative 2-parton phase space: $z=p_{T,2}/p_{T,1}\left(
p_{T,1}\geq p_{T,2}\right)  $, $d=\sqrt{\left(  y_{1}-y_{2}\right)
^{2}+\left(  \phi_{1}-\phi_{2}\right)  ^{2}}$ for a) the naive $R_{sep}=2$
case and b) for $R_{sep}=1.3$ case suggested by data. }%
\label{pertthy}%
\centerline{
}\end{figure}

\begin{figure}[!h]
\centerline{ 
\includegraphics[width=\textwidth,trim= 0 50 0 0]{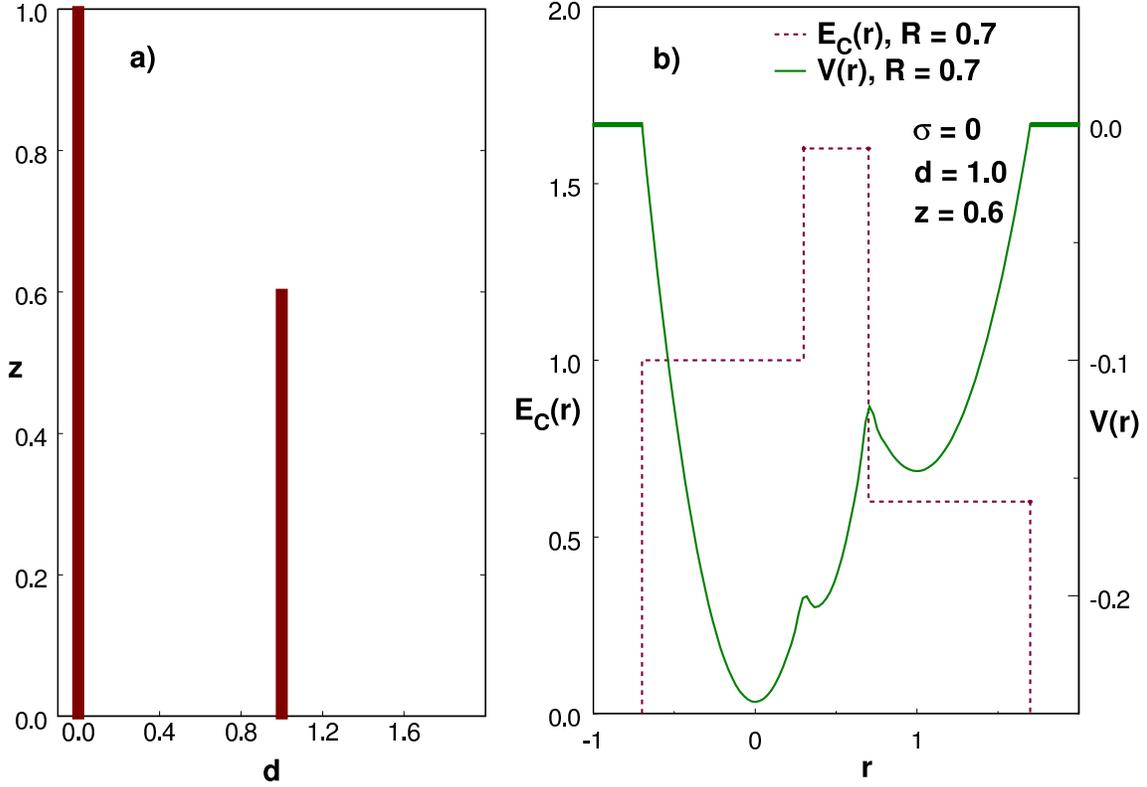}
}
\caption{2-parton distribution in $\left(  d,z\right)  $ in a) with $d=1.0$,
$z=0.6$ and the corresponding energy-in-cone, $E_{C}\left(  r\right)  $, and
potential, $V\left(  r\right)  $. }%
\label{edist}%
\centerline{
}\end{figure}

To understand how the iterative cone algorithm works consider first its
application to NLO level in perturbation theory (see, \textit{e.g.},
\cite{Ellis:1992en,Ellis:1990ek,Ellis:1989vm,Ellis:1988hv}), where there are most 2 partons in a cone. \ As defined above, the
cone algorithm specifies that two partons are included in the same jet
(\textit{i.e}., form a stable cone) if they are both within $R_{cone}$
(\textit{e.g}., 0.7 in $\left(  y,\phi\right)  $ space) of the centroid, which
they themselves define. \ This means that 2 partons of equal $p_{T}$ can
form a single jet as long as their pair-wise angular separation does not
exceed the boundaries of the cone, $\Delta R=2R_{cone}$. \ On the other hand,
as long as $\Delta R>R_{cone}$, there will also be stable cones centered
around each of the partons. \ The corresponding 2-parton phase space for
$R_{cone}=0.7$ is illustrated in Fig. \ref{pertthy} a) in terms of the ratio
$z=p_{T,2}/p_{T,1}\left(  p_{T,1}\geq p_{T,2}\right)  $ and the angular
separation variable $d=\sqrt{\left(  y_{1}-y_{2}\right)  ^{2}+\left(  \phi
_{1}-\phi_{2}\right)  ^{2}}$. \ To the left of the line $d=R_{cone}$ the two
partons always form a single stable cone and jet, while to the far right,
$d>2R_{cone}$, there are always two distinct stable cones and jets, with a
single parton in each jet. \ More interesting is the central region,
$R_{cone}<d<2R_{cone}$, which exhibits both the case of two stable cones (one
of the partons in each cone) and the case of three stable cones (the previous
two cones plus a third cone that includes both partons). \ The precise outcome
depends on the specific values of $z$ and $d$. \ (Note that the exactly
straight diagonal boundary in the figure corresponds to the $p_{T}$-weighted
definition of the Snowmass algorithm, but is only slightly distorted, $<2\%$,
when full 4-vector kinematics is used as in the Run II algorithms.) \ \ To see
the three stable cone structure in terms of the 2-parton \textquotedblleft
Snowmass potential\textquotedblright\ consider the point $z=0.6$ and $d=1.0$,
which is the 3 cones $\rightarrow$ 1 jet region. \ The corresponding potential
is illustrated in Fig. \ref{edist}. \ This potential exhibits the expected 3
minima corresponding to a stable cone at each parton and a more central stable
cone that includes both partons. \ A relevant point is that the central
minimum is not nearly as deep (\textit{i.e.}, as robust) as the other two.
As we shall see, this minimum often does not survive the smearing inherent
in the transition from the short distances of fixed order perturbation
theory to the long distances of the physical final state. 
\ As indicated by the labeling in Fig. \ref{pertthy}, in the 3 stable cone 
region the
original perturbative calculation\cite{Ellis:1992en,Ellis:1990ek,Ellis:1989vm,Ellis:1988hv} kept as the jet the 2-in-1 stable
cone, maximum $p_{T}$ configuration, \textit{i.e}., the cone that included all
of the energy in the other two cones consistent with the merging discussion
below. \ 

As we will see, much of the concern and confusion about the cone
algorithm arises from the treatment of this 3 stable cone region. \ \ It is
intuitively obvious that, as the energy in the short-distance partons is
smeared out by subsequent showering and hadronization, the structure in this
region is likely to change. \ In particular, while two individual, equal
$p_{T}$ partons separated by nearly $2R_{cone}$ may define a stable cone, this
configuration is unlikely to yield a stable cone after showering and hadronization.

\begin{figure}[!h]
\centering
\includegraphics[width=.975\textwidth,trim= 80 40 80 0]{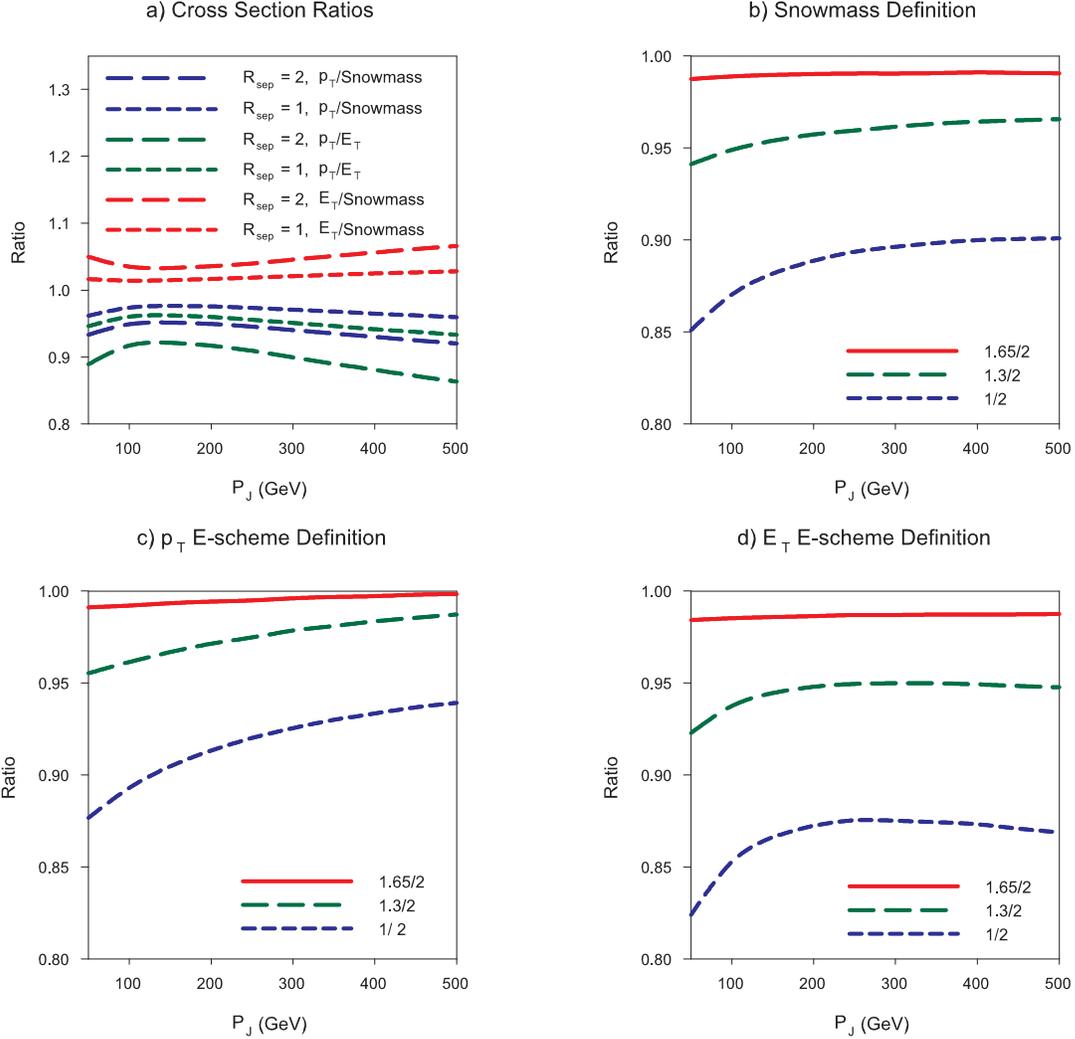}
\label{allET}
\caption{Ratios of the NLO inclusive cone jet cross section versus the 
jet momentum for 3 definitions of the kinematics for various values of
$R_{sep}$.  The Snowmass definition uses $p_T$ weighting and a jet
momentum defined by the scalar sum $P_{J} = {\sum_{k}^{}} {p_{T,k}}$.  The E-scheme
algorithms use 4-vector kinematics (as recommended for Run II) and
either $P_J = p_T =\left\vert {\vec{p}_{T}}\right\vert $ (c)
or $P_J = E_T = E \sin \theta$ (d).  The parts 
of the figure illustrate a) ratios of different choices of $P_J$ versus $P_J$
for $R_{sep} = 2$ and $R_{sep} = 1$; b) ratio to the default value $R_{sep} = 2$
for $R_{sep} = 1.65$, $R_{sep} = 1.3$ and $R_{sep} = 1$ using the Snowmass
definitions for the kinematics and for $P_J$; c) the same as b) except using 4-vector 
kinematics and $P_J = p_{T}$; d) the same as c) but with $P_J = E_T$.
}
\end{figure}

Iterative cone algorithms similar to the one just described were employed by
both the CDF and \dzero \ collaborations during Run I with considerable success.
\ There was fairly good agreement with NLO perturbative QCD (pQCD) 
for the inclusive jet cross
section over a dynamic range of order $10^{8}$. \ During Run I the data were
corrected primarily for detector effects and for the contributions of the
underlying event. \ In fact, a positive feature of the cone algorithm is that,
since the cone's geometry in $\left(  y,\phi\right)  $ space is (meant to be)
simple, the correction for the \textquotedblleft splash-in\textquotedblright%
\ contribution of the (largely uncorrelated) underlying event (and pile-up) is
straightforward. \ (As we will see below the corrections being used in Run II
are more sophisticated.) \ The uncertainties in both the data and the theory
were $10\%$ or greater, depending on the kinematic regime, and helped to
ensure agreement. \ However, as cone jets were studied in more detail, various
troubling issues arose. \ For example, it was noted long ago\cite{Abe:1991ui,Abbott:1997fc}
that, when using the experimental cone algorithms implemented at the Tevatron,
two jets of comparable energy\footnote{These studies were performed
using artificial events, constructed by overlaying jets from 2 different 
events in the data.  The fact that these
are not ``real'' events does not raise any serious limitations.} are
not merged into a single jet if they are separated by an angular distance
greater than approximately 1.3 times the cone radius, while the simple picture
of Fig. \ref{pertthy} a) suggests that merging should occur out to an angular
separation of $2R_{cone}$. \ Independently it was also noted that the
dependence of the experimental inclusive jet cross section on the cone radius
$R_{cone}$\cite{Abe:1991ea} and the shape of the energy distribution within a
jet\cite{Abe:1992wv} both differed discernibly from the NLO predictions (the
data was less $R_{cone}$ dependent and exhibited less energy near the edge of
the cone). \ All three of these issues seemed to be associated with the
contribution from the perturbative configuration of two partons with
comparable $p_{T}$ at opposite sides of the cone ($z\simeq1$, $d\simeq
2R_{cone}=1.4$ in Fig. \ref{pertthy} a)) and the data suggested a lower
contribution from this configuration than present in the perturbative result.
\ To simulate this feature in the perturbative analysis a phenomenological
parameter $R_{sep}$ was added to the NLO implementation of the cone
algorithm\cite{Ellis:1992qq}. \ In this "experiment-aware" version of the
perturbative cone algorithm two partons are not merged into a single jet if
they are separated by more than $R_{sep}\ast R_{cone}$ from each other,
independent of their individual distance from the $p_{T}$-weighted jet
centroid. \ Thus the two partons are merged into the same jet if they are
within $R_{cone}$ of the $p_{T}$-weighted jet centroid and within 
$R_{sep}\ast R_{cone}$ of each other; otherwise the two partons are identified as separate
jets. In order to describe the observed $R_{cone}$ dependence of the cross
section and the observed energy profile of jets the specific value
$R_{sep}=1.3$ was chosen (along with a \textquotedblleft
smallish\textquotedblright\ renormalization/factorization scale $\mu=p_{T}%
/4$), which was subsequently noted to be in good agreement with the
aforementioned (independent) jet separation study. \ The resulting 2 parton
phase space is indicated in Fig. \ref{pertthy} b). \ In the perturbative
calculation, this redefinition, leading to a slightly lower average $p_{T}$
for the leading jet, lowers the NLO jet cross section by about 5\% (for
$R=0.7$ and $p_{T}=100$ GeV/c). \ \ It is important to recognize that the
fractional contribution to the inclusive jet cross section of the merged 2
parton configurations in the entire wedge to the right of $d=R_{cone}$ is only
of order $10\%$ for jet $p_{T}$ of order 100 GeV/c, and, being proportional to
$\alpha_{s}\left(  p_{T}\right)  $, decreases with increasing $p_{T}$. \ Thus
it is no surprise that, although this region was apparently treated
differently by the cone algorithm implementations of CDF and \dzero \ during Run I as 
discussed below, there were
no relevant cross section disagreements above the $>10\%$ uncertainties.
\ While the parameter $R_{sep}$ is \textit{ad hoc} and thus an undesirable
complication in the perturbative jet analysis, it will serve as a useful
pedagogical tool in the following discussions.  To illustrate this
point quantitatively  Fig.~\ref{allET} shows
the dependence on $R_{sep}$ for various choices of the 
jet momentum $P_J$ at NLO in perturbation theory.  The curves labeled
Snowmass use the $p_T$ weighted kinematics described above with $P_J$ 
given by the scalar sum of the transverse momenta of the partons in the cone.
The two E-scheme algorithms use full 4-vector kinematics (as recommended for
Run II in \cite{Blazey:2000qt}) and $P_J$ equal to either the magnitude of the true 
(vector sum) 
transverse momentum (the recommended choice), 
or the ``transverse energy'' defined by 
$P_J = E_T = E \sin \theta$ (as defined by CDF in Run I).  
Thus this last variable knows about both the
momentum and the invariant mass of the set of partons in the cone,
which can be sizable for well separated parton pairs.
The differences in the various ratios
for different values of $R_{sep}$ tell us about how the 2-parton
configurations contribute.  For example, Fig.~\ref{allET} a) tells
us that, since, for a given configuration of 2 partons in a cone,
$E_T > p_{T,Snowmass} > p_T$, the cross sections at a given value of
$P_J$ will show the same ordering.  Further, as expected, the differences
are reduced if we keep only configurations with small angular separation,
$R_{sep} = 1$.  
From Fig.~\ref{allET} b) we confirm the earlier statement
that lowering $R_{sep}$ from 2 to 1.3 yields a $5\%$ 
change for the Snowmass
algorithm cross section with $P_J = 100$ GeV, 
while lowering it all the way to
$R_{sep} = 1$, \textit{i.e.}, removing all of the triangular region,
lowers the 100 GeV Snowmass jet cross section by approximately 
$12\%$.  
Figs.~\ref{allET} c) and d) confirm that 4-vector kinematics with
$P_J = p_T$ exhibits the smallest sensitivity to $R_{sep}$, \textit{i.e.},
to the 2-parton configurations in the triangle.  The choice
$P_J = E_T$, with its dependence on the mass of the pair, exhibits the
largest sensitivity to $R_{sep}$.  These are good reasons
to use the recommended E-scheme kinematics with $P_J = p_T$.

\begin{figure}[h]
\centerline{
\psfig{figure=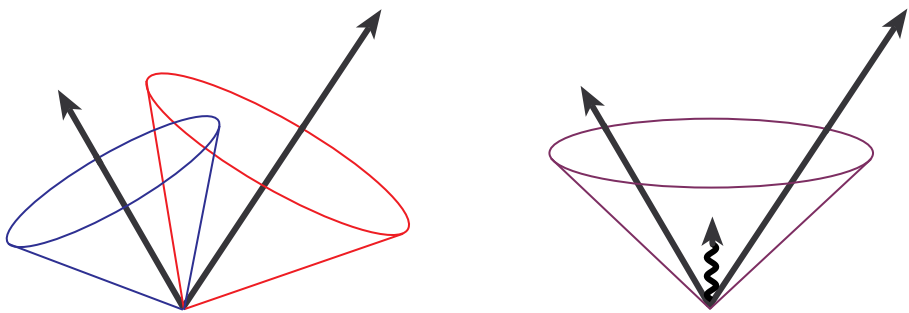,width=12cm}
}
\caption{Two partons in two cones or in one cone with a (soft) seed present. }%
\label{2inone}%
\centerline{
}\end{figure}

The difference between the perturbative implementation of the iterative cone
algorithm and the experimental implementation at the Tevatron, which is
simulated by $R_{sep}$, was thought to arise from several sources. \ While the
perturbative version (with $R_{sep}=2$) analytically includes all 2-parton
configurations that satisfy the algorithm (recall Fig. \ref{pertthy} a)), the
experiments employ the concept of \textit{seeds} to reduce the analysis time
and place trial\ cones only in regions of the detector where there are seeds,
\textit{i.e}., pre-clusters with substantial energy. \ This procedure
introduces another parameter, the lower $p_{T}$ threshold defining the seeds,
and also decreases the likelihood of finding the 2-showers-in-one-jet
configurations corresponding to the upper right-hand corner of the 3 cones
$\rightarrow$ 1 jet region of Fig. \ref{pertthy} a) and the middle minimum in
Fig. \ref{edist} b). \ Thus the use of seeds contributes to the need for
$R_{sep}<2$. \ Perhaps more importantly, the desire to match theory with
experiment means that we should include seeds in the perturbative algorithm.
\ This is accomplished by placing trial cones only at the locations of each
parton and testing to see if any other partons are inside of these cones.
\ Thus at NLO, two partons will be merged into a single jet only if they are
closer than $R_{cone}$ in $\left(  y,\phi\right)  $ space. \ This corresponds
to $R_{sep}=1.0$ in the language of Fig. \ref{pertthy} and produces a larger
numerical change in the analysis than observed, \textit{i.e.}, we wanted
$R_{sep}\simeq1.3$. \ More importantly at the next order in perturbation
theory, NNLO, there are extra partons that can play the role of low energy
seeds. \ The corresponding parton configurations are illustrated in Fig.
\ref{2inone}. \ At NLO, or in the virtual correction in NNLO, the absence of
any extra partons to serve as a seed leads to two distinct cones as on the
left, while a (soft) real emission at NNLO can lead to the configuration on
the right where the soft gluon \textquotedblleft seeds\textquotedblright\ the
middle cone that includes all of the partons. \ The resulting separation
between the NNLO virtual contribution and the NNLO soft real emission
contribution (\textit{i.e}., they contribute to different jet configurations)
leads to an undesirable logarithmic dependence on the seed $p_{T}$
threshold\cite{Seymour}. \ In the limit of an arbitrarily soft seed $p_{T}$
cutoff, the cone algorithm with seeds is no longer IR-safe. \ By introducing
seeds in the algorithm we have introduced exactly what we want to avoid in
order to be Infrared Safe, sensitivity to soft emissions. \ From the theory
perspective seeds are a very bad component in the algorithm and should be
eliminated. \ The labeling of the Run I cone algorithm with seeds
as Infrared Unsafe has
led some theorists to suggest that the corresponding analyses should be
disregarded. \ This is too strong a reaction, especially since the numerical
difference between the jet cross section found in data using seeds is expected
to be less than 2\% different from an analysis using a seedless algorithm. \ 
A more useful approach will be to either avoid the use of seeds, or to correct 
for them in the analysis of the data, which can then be compared to a 
perturbative analysis without seeds.
We will return to this issue below.

To address the issue of seeds on the experimental side and the $R_{sep}$
parameter on the phenomenological side, the Run II study\cite{Blazey:2000qt}
recommended using the MidPoint Cone Algorithm, in which, having identified 2
nearby jets, one always checks for a stable cone with its center at the
MidPoint between the 2 found cones. \ Thus, in the imagery of Fig.
\ref{2inone}, the central stable cone is now always looked for, whether there
is a actual seed there or not. \ It was hoped that this would remove the
sensitivity to the use of seeds and remove the need for the $R_{sep}$
parameter. \ While this expectation is fully justified with the localized,
short distance configuration indicated in Fig. \ref{2inone}, more recent
studies suggest that at least part of the difficulty with the missing stable
cones at the midpoint position is due to the (real) smearing effects on the
energy distribution in $\left(  y,\phi\right)  $ of showering and
hadronization, as will be discussed below. \ Also it is important to note
that, in principle, IR-safety issues due to seeds will reappear in
perturbation theory at order NNNLO, where the midpoint is not the only problem
configuration (a seed at the center of a triangular array of 3 hard and
merge-able partons can lead to IR-sensitivity).

Before proceeding, we must consider another important issue that arises when
comparing the cone algorithm applied in perturbation theory with its
application to data at the Tevatron. \ The practical definition of jets
equaling stable cones does not eliminate the possibility that the stable cones
can overlap, \textit{i.e}., share a subset (or even all) of their calorimeter
towers. \ To address this ambiguity, experimental decisions had to be made as
to when to completely merge the two cones (based on the level of overlap), or,
if not merging, how to split the shared energy. \ \ Note that there is only a
weak analogy to this overlap issue in the NLO calculation. As described in
Fig. \ref{pertthy} a), there is no overlap in either the left-hand (1 cone
$\rightarrow$ 1 jet) or right-hand (2 cones $\rightarrow$ 2 jets) regions,
while in the middle (3 cones $\rightarrow$ 1 jet) region the overlap between
the 3 cones is 100\% and the cones are always merged. \ Arguably the
phenomenological parameter $R_{sep}$ also serves to approximately simulate not
only seeds but also the role of non-complete merging in the experimental
analysis. \ In practice in Run I, CDF and \dzero \ chose to use slightly
different merging parameters. \ Thus, largely unknown to most of the theory
community, the two experiments used somewhat different cone jet algorithms in
Run I. \ (The CDF collaboration cone algorithm, JETCLU\cite{JETCLU}, also
employed another \textquotedblleft feature\textquotedblright\ called
ratcheting, that was likewise under-documented. \ Ratcheting ensured that any
calorimeter tower in an initial seed was always retained in the corresponding
final jet. \ Stability of the final cones was partially driven by this
\textquotedblleft no-tower-left-behind\textquotedblright\ feature.)

\ Presumably the two experiments reported compatible jet physics result in Run
I due to the\ substantial $\left(  \geq10\%\right)  $ uncertainties. \ Note
also that after the splitting/merging step, the resulting cone jets will not
always have a simple, symmetric shape in $\left(  y,\phi\right)  $, which
complicates the task of correcting for the underlying event and leads to
larger uncertainties. \ In any case the plan for Run II as outlined in the Run
II\ Studies\cite{Blazey:2000qt}, called for cone jet algorithms in the two
collaborations as similar as possible. Unfortunately, as described in more
detail below, events during Run II have moved the collaborations off that
track and they are currently employing somewhat different cone algorithms.
\ On the merging question, CDF in Run II merges two overlapping cones when
more than 75\% of the smaller cone's energy overlaps with the larger jet.
\ When the overlap is less, the overlap towers are assigned to the nearest
jet. \ \dzero , on the other hand, uses a criterion of a 50\% overlap in order to
merge. \ While it is not necessary that all analyses use the same jet
algorithm, for purposes of comparison or the combination of datasets
it would be very useful for the
experiments to have one truly common algorithm. \ 

\thissection{Run II Cone Algorithm Issues}

\begin{figure}[h]
\centerline{
\includegraphics[width=\textwidth,trim=0 20 0 30]{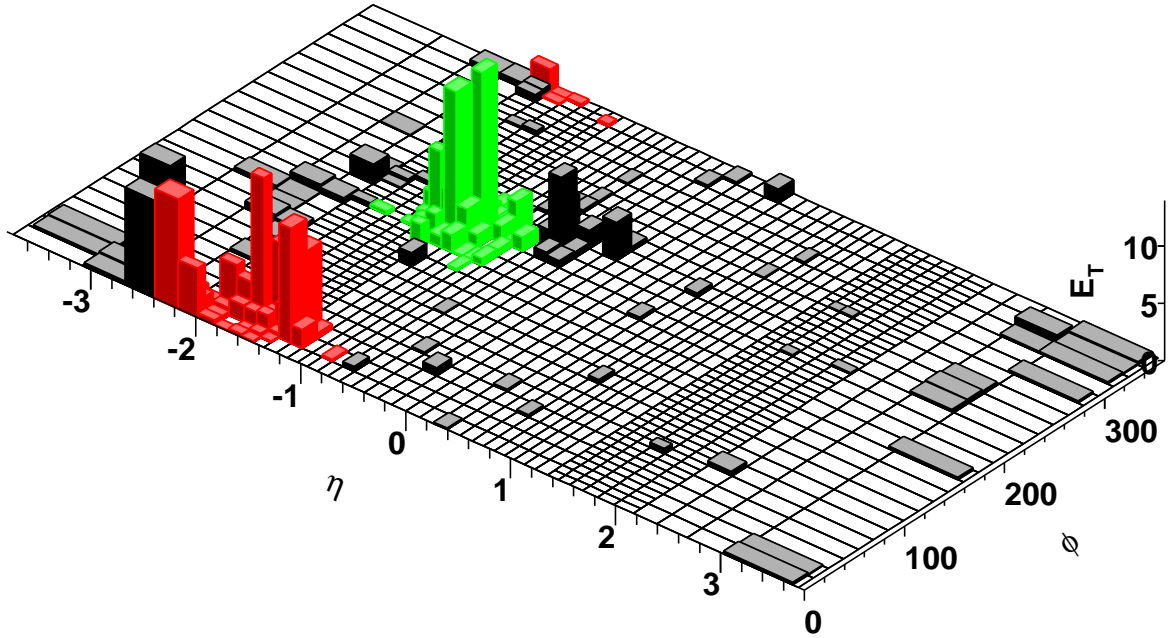}
}
\caption{An example of a Monte Carlo inclusive jet event where the midpoint
algorithm has left substantial energy unclustered. }%
\label{dark_towers}%
\centerline{
}\end{figure}

In studies of the Run II\ cone algorithms, a previously unnoticed problem has
been identified\cite{Ellis:2001aa} at the particle and calorimeter level, which is
explicitly not present at the NLO parton level. \ It is observed in a
(relatively small) fraction of the events that some 
energetic particles/calorimeter
towers remain unclustered in any jet. \ This effect is understood to arise in
configurations of two nearby (\textit{i.e}., nearby on the scale of the cone
size) showers, where one shower is of substantially larger energy. \ Any
\ trial cone at the location of the lower energy shower will include
contributions from the larger energy shower, and the resulting $p_{T}%
$-weighted centroid will migrate towards the larger energy peak. This feature
is labeled \textquotedblleft dark towers\textquotedblright\ in Ref.
\cite{Ellis:2001aa}, \textit{i.e.,} clusters that have a transverse momentum large
enough to be designated either a separate jet or to be included in an existing
nearby jet, but which are not clustered into either. \ A Monte Carlo event
with this structure is shown in Fig. \ref{dark_towers}, where the towers
unclustered into any jet are shaded black.

\begin{figure}[h]
\centerline{
\includegraphics[width=\textwidth,trim=0 50 0 0]{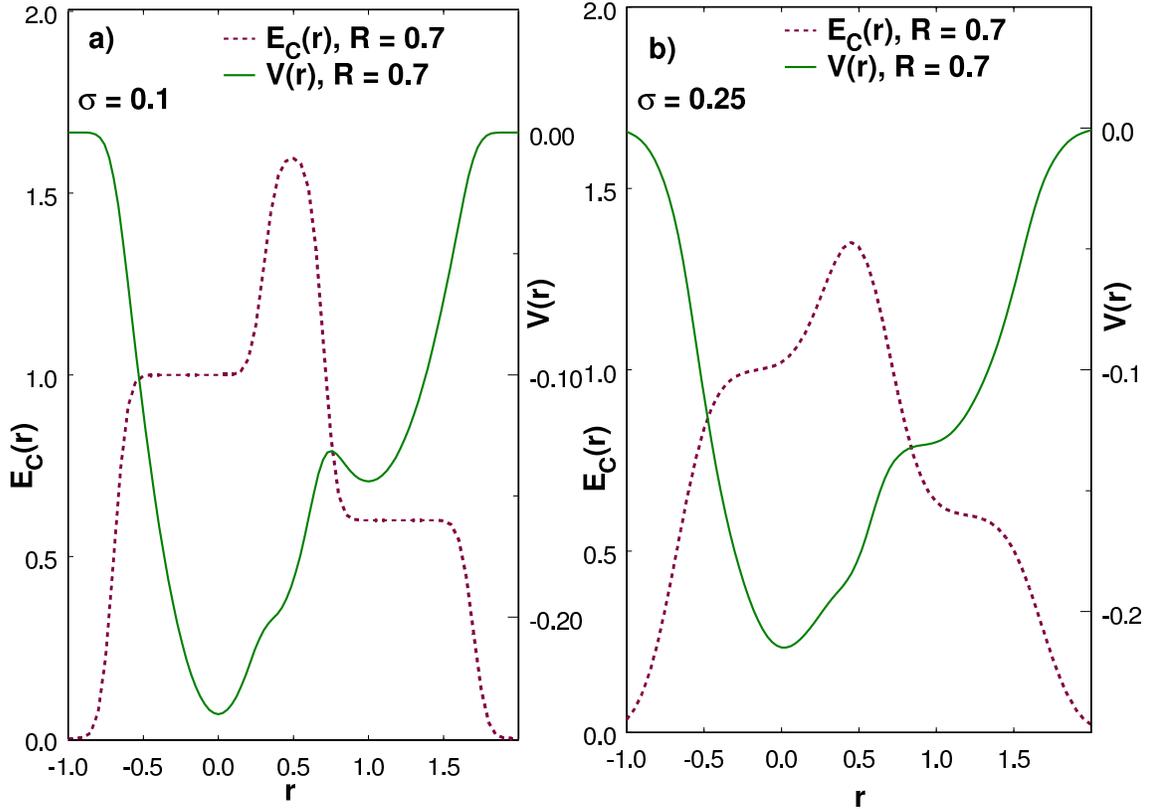}
}
\caption{Energy-in-cone and potential distributions corresponding to Gaussian
smearing with a) $\sigma=0.1$ and b) $\sigma=0.25$ for $d=1.0$ and $z=0.6$. }%
\label{smearpt}%
\centerline{
}\end{figure}

A simple way of understanding the dark towers can be motivated by returning to
Fig. \ref{edist}, where the only smearing in $\left(  y,\phi\right)  $ between
the localized energy distribution of part a) (the individual partons) and the
\textquotedblleft potential\textquotedblright\ of part b) arises from the size
of the cone itself. \ On the other hand, we know that showering and
hadronization will lead to further smearing of the energy distribution and
thus of the potential. \ Sketched in Fig. \ref{smearpt} is the potential (and
the energy-in-cone) distributions that results from Gaussian smearing with a
width of a) $\sigma=0.1$ and b) $\sigma=0.25$ (in the same angular units as
$R=0.7$). \ In both panels, as in Fig. \ref{edist}, the partons have $p_{T}$
ratio $z=0.6$ and angular separation $d=1.0$. \ Note that as the smearing
increases from zero as in panel a), we first lose the (not so deep) minimum
corresponding to the midpoint stable cone (and jet), providing another piece
of the explanation for why showers more than $1.3\ast R_{cone}$ apart are not
observed to merge by the experiments. \ \ In panel b), with even more
smearing, the minimum in the potential near the shower from the lower energy
parton also vanishes, meaning this (lower energy) shower is part of no stable
cone or jet, \textit{i.e}., leading to dark towers. \ Any attempt to place the
center of a trial cone at the position of the right parton will result in the
centroid \textquotedblleft flowing\textquotedblright\ to the position of the
left parton and the energy corresponding to the right parton remaining
unclustered in any jet. (Note that the Run I CDF algorithm, JETCLU with
Ratcheting, limited the role of dark towers by never allowing a trial cone to
leave the seed towers, the potential dark towers, behind.) \ The effective
smearing in the data is expected to lie between $\sigma$ values of 0.1 and
0.25 (with shower-to-shower fluctuations and some energy dependence,
being larger for smaller $p_{T}$ jets) making this discussion relevant, but
this question awaits further study as outlined below. \ Note that Fig.
\ref{smearpt} also suggests that the Midpoint algorithm will not entirely fix
the original issue of missing merged jets. \ Due to the presence of (real)
smearing this middle cone is almost never stable and the merged jet configuration
will not be found even though we have explicitly looked for it with the
midpoint cone. \ Thus even using the recommended Midpoint algorithm (with
seeds), as the \dzero \ collaboration is doing (with the also recommended
$f_{merge}=0.5$ value), there may remain a phenomenological need for the
parameter value $R_{sep}<2$.

A potential solution for the dark towers problem is described in Ref.
\cite{Ellis:2001aa}. \ The idea is to decouple the jet finding step from the jet
construction step. \ In particular, the stable cone finding procedure is
performed with a cone of radius half that of the final jet radius,
\textit{i.e}., the radius of the search cone, $R_{search}=R_{cone}/2$. \ This
procedure reduces the smearing in Figs. \ref{edist} and \ref{smearpt}, and
reduces the phase space for configurations that lead to dark towers (and
missing merged jets).
\begin{figure}[h]
\centerline{
\includegraphics[width=.8\textwidth,trim= 0 50 0 0]{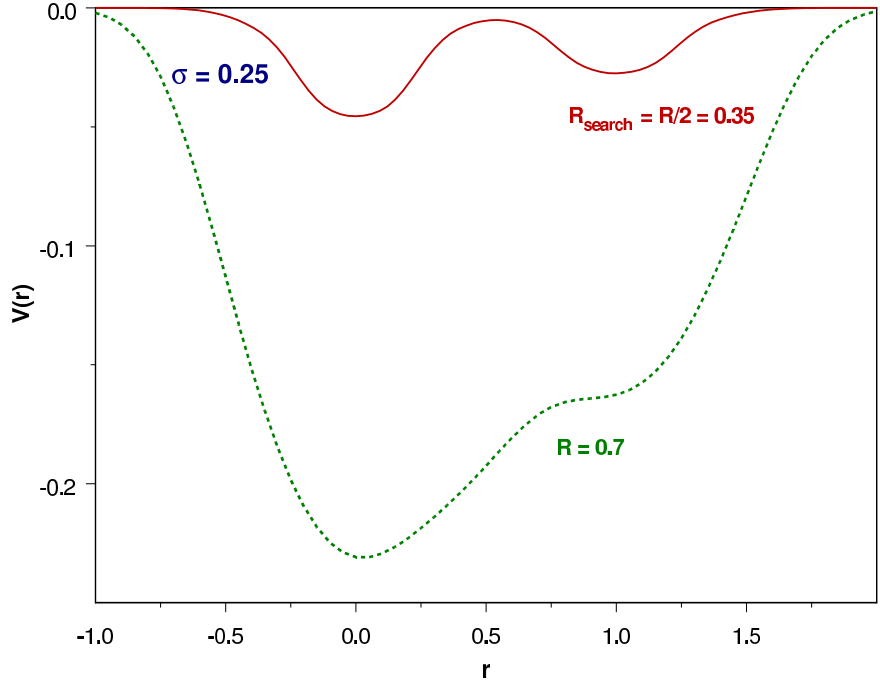}
}
\caption{The stable cone finding potential with the reduced search cone,
$R_{search}=R_{cone}/2$. \ The original potential from Fig. \ref{smearpt}, panel b)
with $R_{search}=R_{cone}$ is indicated as the dashed curve. }%
\label{smearpotfix}%
\centerline{
}\end{figure}This point is illustrated in Fig. \ref{smearpotfix}, which shows
the potential of Fig. \ref{smearpt}, panel b) corresponding to the reduced
radius search cone. \ Note, in particular, that there is again a minimum at
the location of the second parton. \ Seeds placed at each parton will yield
a stable cone at each location even after the smearing.  \ Using the smaller
search cone size means there is less influence from the (smeared) energy
of nearby partons.  
After identifying the locations of stable 
cones, the larger cone size, \textit{e.g.}, $R_{jet}=R_{cone}=0.7$, is used to
sum all objects inside and construct the energy and momentum of the jet (with
no iteration). All pairs of stable cones separated by less 
than $2 R_{cone}$ are then used to define midpoint seeds as in the usual MidPoint Cone 
Algorithm.  A trial cone of size $R_{cone}$ is placed at each such seed and iterated to
test for stability.  (Note that this midpoint cone is iterated with cone size
$R_{cone}$, not the smaller $R_{search}$, contrary to what is described
in the literature.)  Thus, just as in the MidPoint Cone Algorithm, stable midpoint 
cones will be found by the CDF Search Cone Algorithm.  However, as already discussed,
we expect that there will be no stable midpoint cone due to the smearing.  
Note that, even with the 
reduced smearing when using the smaller search cone radius, there is
still no central stable cone in the potential of Fig. \ref{smearpotfix}. 
On the other hand, as applied to NLO perturbation theory without smearing, the Search 
Cone Algorithm should act like the usual MidPoint Cone Algorithm and yield the
na\"ive result of Fig. \ref{pertthy} a).  The net impact of adding the step with the
smaller initial search cone as applied to data is an approximately 5\%
increase in the inclusive jet cross section. \ In fact,
as applied to data the Search Cone Algorithm identifies so many more stable
cones, that the CDF collaboration has decided to use the Search Cone Algorithm
with the merging parameter $f_{merge}=0.75$ (instead of $0.5$) to limit the
level of merging. \ 

Unfortunately a disturbing feature of the Search Cone
Algorithm arises when it is applied at higher orders in perturbation theory 
as was pointed out during this Workshop\cite{Search}.
At NNLO in perturbation theory the Search Cone Algorithm can act much like
the seeds discussed earlier. \ In particular, the search cone algorithm can
identify a (small, radius $R_{search}$) stable (soft) cone between two
energetic cones, exactly the soft gluon between 2 energetic partons
configuration discussed earlier. \ The soft search cone is stable exactly
because it \textquotedblleft fits\textquotedblright\ between the two energetic
partons without including either; the spacing between the two energetic
partons can be in the range $2R_{search}=R_{cone}<\Delta R<2R_{cone}$. \ Then,
when the radius of the (stable, soft) search cone is increased to $R_{cone}$,
the resulting full size cone will envelop, and serve to merge, the two
energetic partons. \ This can occur even when the two energetic partons do not
constitute a stable combined cone in the standard cone algorithm. Thus at NNLO
the search cone algorithm can exhibit an IR-sensitivity very similar to, and
just as undesirable as, the seed-induced problem discussed earlier. \ The
conclusion is that the search cone algorithm, while it does address the dark
tower issue, creates its own set of issues and is not considered to be a real
solution of the dark tower problem.

In summary, the \dzero \ collaboration is using the Midpoint Cone algorithm with
$f_{merge}=0.5$ (and seeds) to analyze Run II\ data, while the CDF
collaboration is using the Search Cone algorithm with $f_{merge}=0.75$ (with
seeds). \ CDF is encouraged to return to also using the Midpoint Cone
algorithm.  The two collaborations are encouraged to determine an optimum
value of $f_{merge}$ that is both common and appropriate to future high 
luminosity running. \ To 
compare fixed order perturbation theory
with data there must be corrections for detector effects, for the splash-in
contributions of the underlying event (and pile-up) and for the splash-out
effects of showering and hadronization. \ It is the response to these last
effects that distinguishes the various cone algorithms and drives the issues
we have just been discussing. \ The fact that the splash-in and splash-out
corrections come with opposite signs and can cancel in the uncorrected data
for the inclusive jet cross section, may help explain why Run I comparisons
with perturbation theory sometimes seemed to be better than was justified (with
hindsight). \ We will return to the question of Run II\ corrections below.
\ The conclusion from the previous discussion is that it would be very helpful
to include also a correction in the experimental analysis that accounts for
the use of seeds. \ Then these experimental results could be compared to
perturbative results without seeds avoiding the inherent infrared problems
caused by seeds in perturbative analyses.  At the same time, the analysis
described above suggests that using the MidPoint Cone Algorithm, to 
remove the impact of seeds at NLO, does not eliminate the impact of 
the smearing due to showering and hadronization, which serves to render
the midpoint cone of fixed order perturbation theory unstable.  Thus we should
still not expect to be able to compare data to NLO theory with $R_{sep} = 2$ 
(in Run II analyses \dzero \ is comparing to NLO with $R_{sep} = 2$, while CDF 
is still using $R_{sep}=1.3$).

\thissection{k$_{T}$ Algorithms}

With this mixed history of success for the cone algorithm, the (as yet) less
well studied $k_{T}$ algorithm\cite{KTAlgo,Catani:1992zp,Catani:1993hr} apparently
continues to offer the possibility of nearly identical analyses in both
experiments and in perturbation theory. \ Indeed, the $k_{T}$ algorithm, which
was first used in electron-positron colliders, appears to be conceptually
simpler at all levels. Two partons/particles/calorimeter towers are combined
if their relative transverse momentum is less than a given measure. \ To
illustrate the clustering process, consider a multi-parton final state.
\ Initially each parton is considered as a proto-jet. The quantities
$k_{T,i}^{2}=p_{T,i}^{2}$ and $k_{T,(i,j)}^{2}=min(p_{T,i}^{2},p_{T,j}%
^{2})^{.}\Delta R_{i,j}^{2}/D^{2}$ are computed for each parton $i$ and each
pair of partons $ij$, respectively. As earlier $p_{T,i}$ is the transverse
momentum of the $i^{th}$ parton, $\Delta R_{i.j}$ is the distance (in $y,\phi$
space, $\Delta R_{i.j}=\sqrt{\left(  y_{i}-y_{j}\right)  ^{2}+\left(  \phi
_{i}-\phi_{j}\right)  ^{2}}$) between each pair of partons. \ $D$ is the
parameter that controls the size of the jet (analogous to $R_{cone}$). If the
smallest of the above quantities is a $k_{T,i}^{2}$, then that parton becomes
a jet and is removed from the proto-jet list. \ If the smallest quantity is a
$k_{T,(i,j)}^{2}$, then the two partons $\left(  i,j\right)  $ are merged into
a single proto-jet by summing their four-vector components, and the two
original entries in the proto-jet list are replaced by this single merged
entry. \ This process is iterated with the corrected proto-jet list until all
the proto-jets have become jets, \textit{i.e}., at the last step the
$k_{T,(i,j)}^{2}$ for all pairs of proto-jets are larger than all $k_{T,i}%
^{2}$ for the proto-jets individually (\textit{i.e}., the remaining proto-jets
are well separated) and the latter all become jets. \ 

Note that in the pQCD NLO inclusive $k_{T}$ jet calculation, the parton pair
with the smallest $k_{T}^{2}$ may or may not be combined into a single jet,
depending on the $k_{T,i}^{2}$ of the individual partons. Thus the final state
can consist of either 2 or 3 jets, as was also the case for the cone algorithm.
\ In fact, the pQCD NLO result for the inclusive $k_{T}$ jet cross
section\cite{KTAlgo} suggests near equality with the cone jet cross section in
the case that $D\simeq1.35R_{cone}$ (with no seeds, $R_{sep}=2$). \ Thus the 
inclusive
cone jet cross section with $R_{cone}=0.7$ ($R_{sep}=2$) is comparable in
magnitude to the inclusive $k_{T}$ jet cross section with $D=0.9$, at least at
NLO. \ In the NLO language illustrated in Fig. \ref{pertthy} the condition
that the partons be merged in the $k_{T}$ algorithm is that $z^{2}\left(
d^{2}/D^{2}\right)  <z^{2}$ or $d<D$. \ Thus at NLO the $k_{T}$ algorithm
corresponds to the cone algorithm with $R_{cone}=D$, $R_{sep}=1$. \ The
earlier result, $D\simeq1.35R_{cone}$ (with $R_{sep}=2$), is just the NLO
statement that the contribution of the rectangular region $0\leq
d\leq1.35R_{cone}$, $0\leq z\leq1$ is numerically approximately equal to the
contribution of the rectangular region $0\leq d\leq R_{cone}$, $0\leq z\leq1$
plus the (3 stable cone) triangular region $R_{cone}\leq d\leq\left(
1+z\right)  R_{cone}$, $0\leq z\leq1$.

In contrast to the cone case, the $k_{T}$ algorithm has no problems with
overlapping jets and, less positively, every calorimeter tower is assigned to
some jet. \ While this last result made some sense in the $e^{+}e^{-}$
collider case, where every final state particle arose from the short-distance
process, it is less obviously desirable in the hadron collider case. \ While
the $k_{T}$ algorithm tends to automatically correct for the splash-out effect
by re-merging the energy distribution smeared by showering and hadronization
into a single jet, this same feature leads to a tendency to enhance the
splash-in effect by "vacuuming up" the contributions from the underlying event
and including them in the large $k_{T,i}^{2}$ jets.\ \ This issue is
exacerbated when the luminosity reaches the point that there is more than one
collision per beam bunch crossing and pile-up is significant. \ This is now
true at the Tevatron and will certainly be true eventually at the LHC. \ Thus
while the (splash-out) fragmentation corrections for the $k_{T}$ algorithm are
expected to be smaller than for cone algorithms, the (splash-in) underlying
event corrections will be larger. \ This point presumably provides at least a
partial explanation for the only marginal agreement between theory and
experiment in the Run I results with the $k_{T}$ algorithm from the
\dzero \ Collaboration\cite{D0ktRunI}. \ A test of our understanding of these
corrections will be provided by the comparison of the $D$ and $R_{cone}$
parameter values that yield comparable experimental jet cross sections. \ 
If we can reliably correct back to the fixed order perturbative level
for both the cone and $k_{T}$ algorithms, we should see
$D \simeq 1.35 R_{cone}$.  Note that this result assumes that the cone
jet cross section has been corrected to the value corresponding to
$R_{sep} = 2$.\ On the other hand, under-corrected splash-in contributions in
the $k_{T}$ algorithm will require $D < 1.35 R_{cone}$ for comparable jet cross
section values (still assuming that $R_{sep} = 2$ describes the cone results).
If the cone algorithm jet cross section has under-corrected splash-out
effects ($R_{sep} < 2$), we expect that an even smaller ratio of $D$ to $R_{cone}$
will required to obtain comparable jet cross sections (crudely we expect
$D < 1(1.35)R_{cone}$ for $R_{sep} = 1(2)$).

Another concern with the $k_{T}$ algorithm is the computer time needed to
perform multiple evaluations of the list of pairs of proto-jets as 1 pair is
merged with each pass, leading to a time that grows as $N^{3}$, where $N$ is
the number of initial proto-jets in the event. \ Recently\cite{hep-ph/0512210} an
improved version of the $k_{T}$ algorithm has been defined that recalculates
only an intelligently chosen sub-list with each pass and the time grows only
as $N\ln N$, for large $N$.

It should also be noted that, although it would appear that the $k_{T}$
algorithm is defined by a single parameter $D$, the suggested code for the
$k_{T}$ algorithm on its \textquotedblleft official\textquotedblright\ web
page\cite{CEDARktjet} includes 5 parameters to fully define the specific
implementation of the algorithm. \ Thus, as is the case for the cone
algorithm, the $k_{T}$ algorithm also exhibits opportunities for divergence
between the implementation of the algorithm in the various experiments, and
care should be taken to avoid this outcome.

\thissection{Run II Jet Results}

\begin{figure}[h]
\centerline{
\psfig{figure=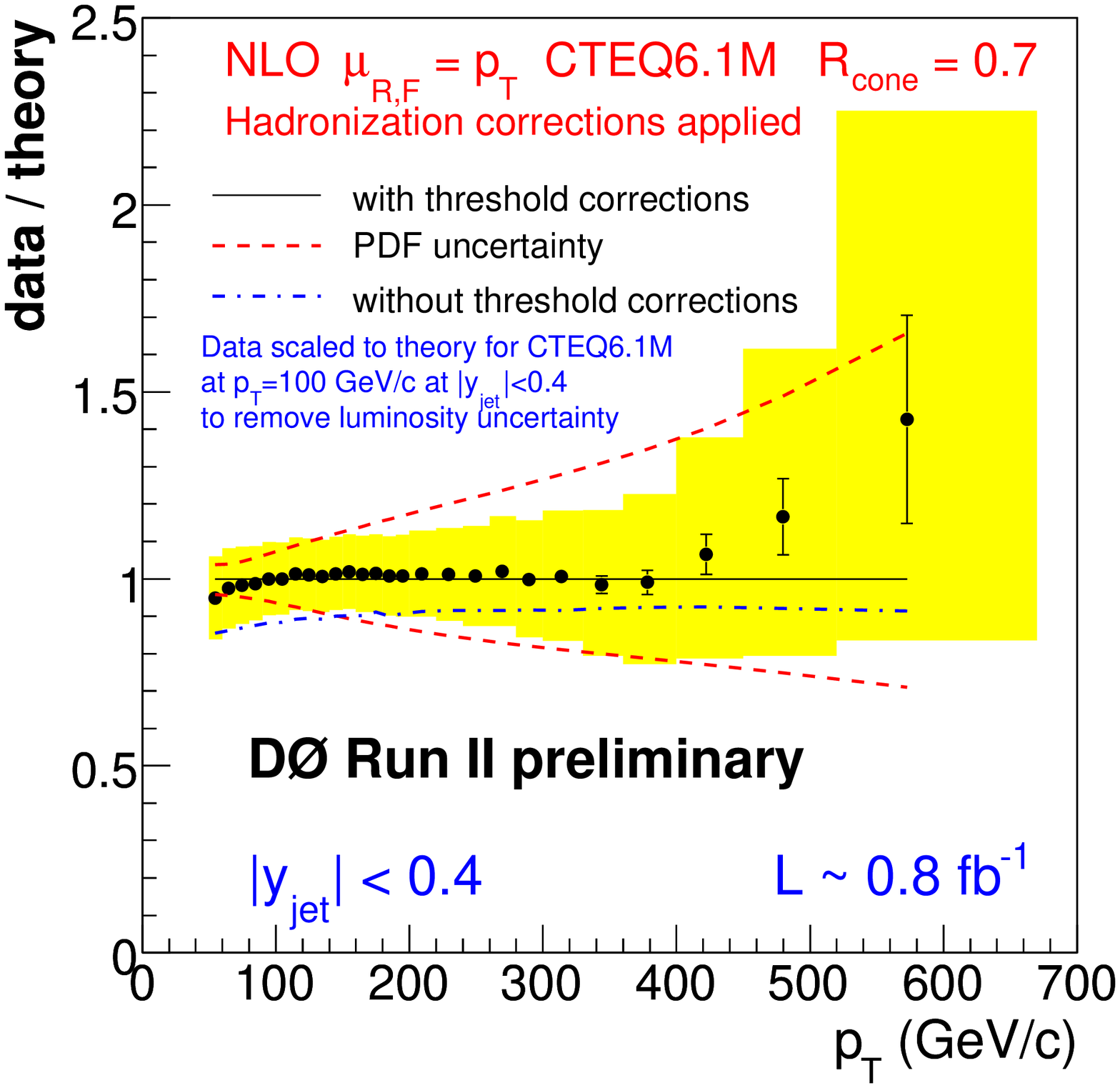,width=7cm, height=7cm}
\psfig{figure=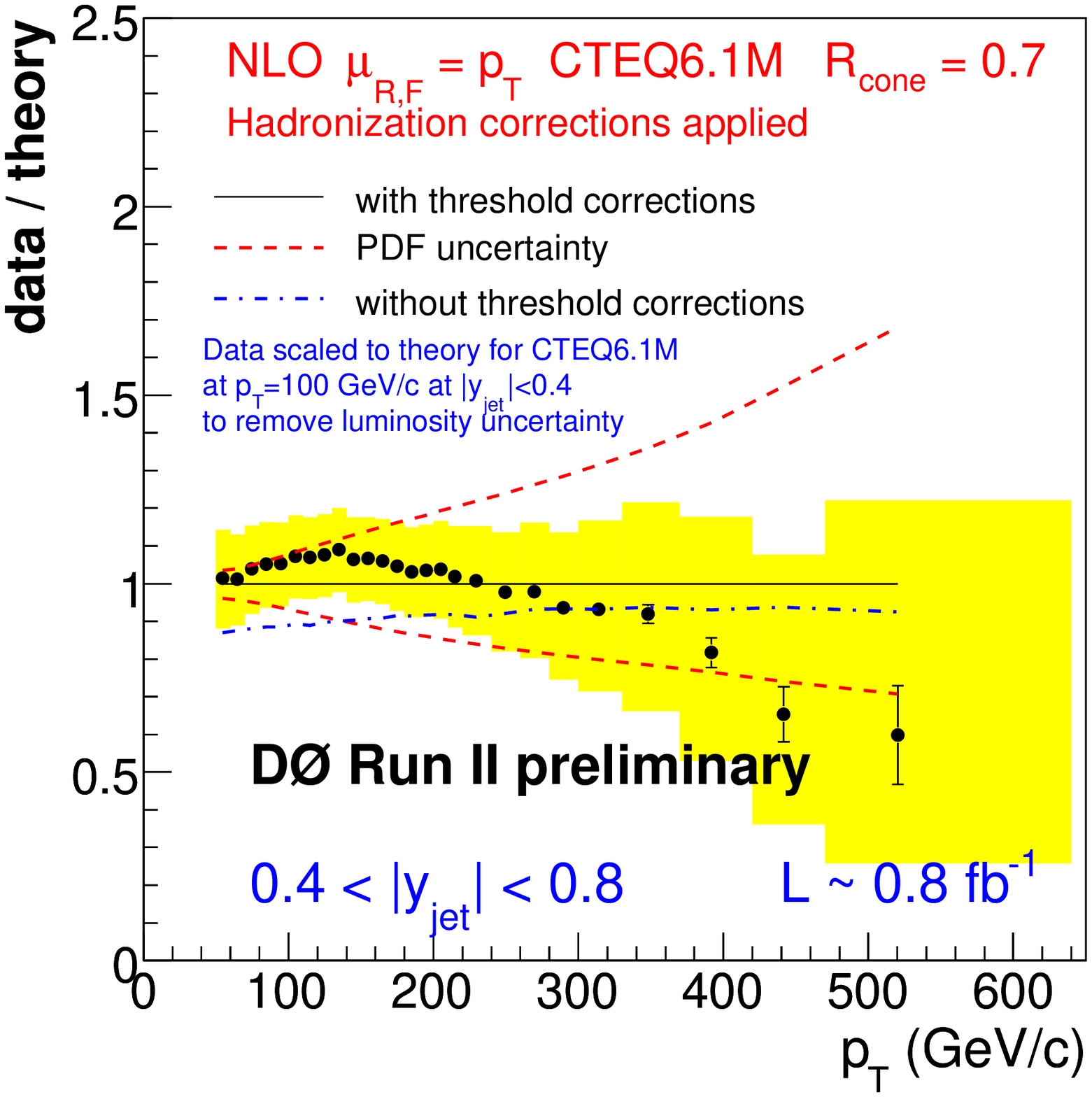,width=7cm, height=7cm}
}
\caption{The \dzero \ Run II inclusive jet cross section using the MidPoint
algorithm (\mbox{$R_{\rm cone} = 0.7$}, \mbox{$f_{\rm merge} = 0.50$}) 
compared with 
theory in two rapidity ranges $0<|y|<0.4$ (left) and $0.4<|y|<0.8$ (right).
The theory prediction includes the parton-level NLO calculation
(\mbox{$R_{\rm sep}=2$}) plus ${\cal O}(\alpha_s^4)$ threshold corrections
and hadronization corrections.}%
\label{D0cone2}%
\centerline{
}\end{figure}

\begin{figure}[h]
\centerline{
\psfig{figure=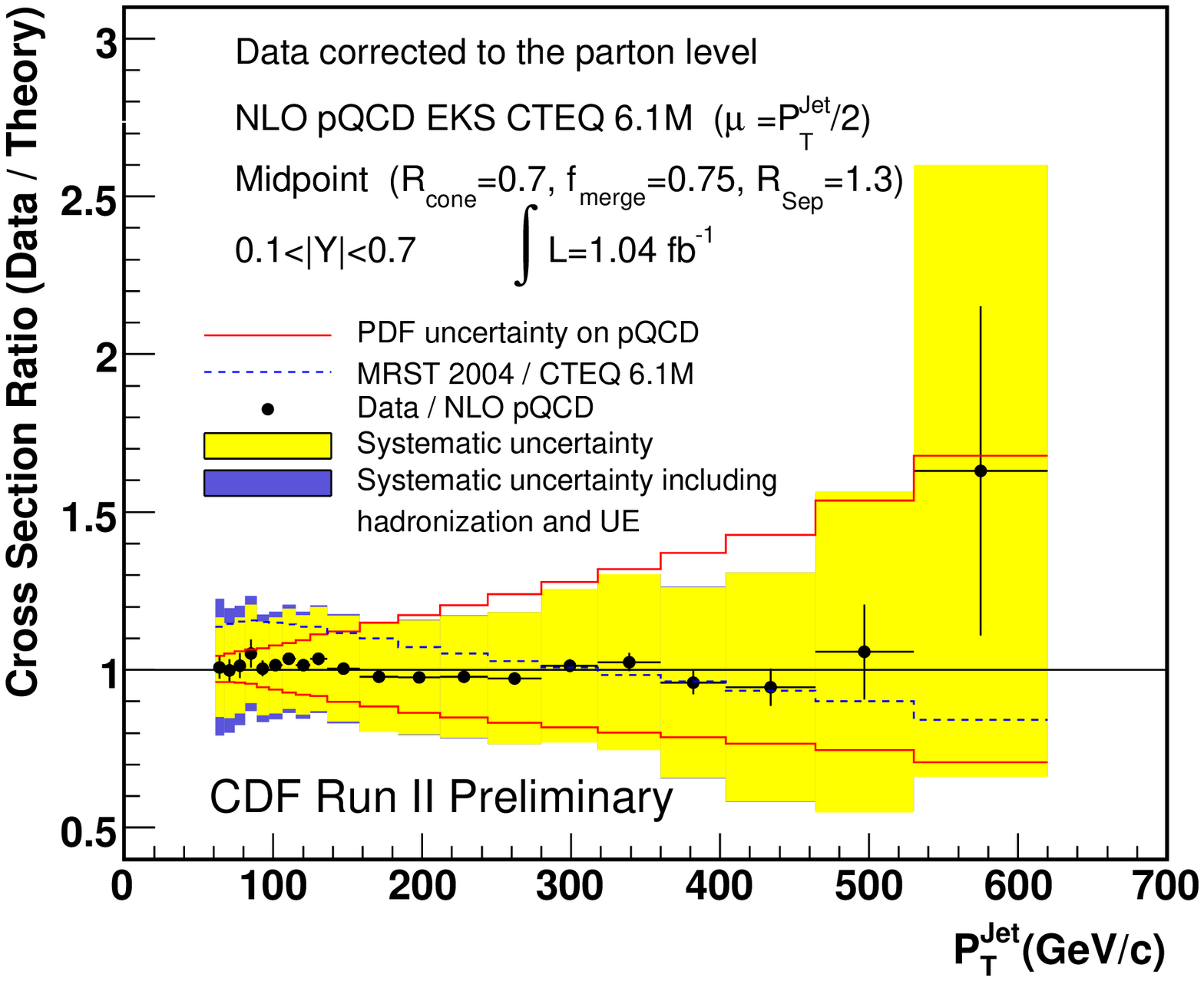,width=7cm}
\psfig{figure=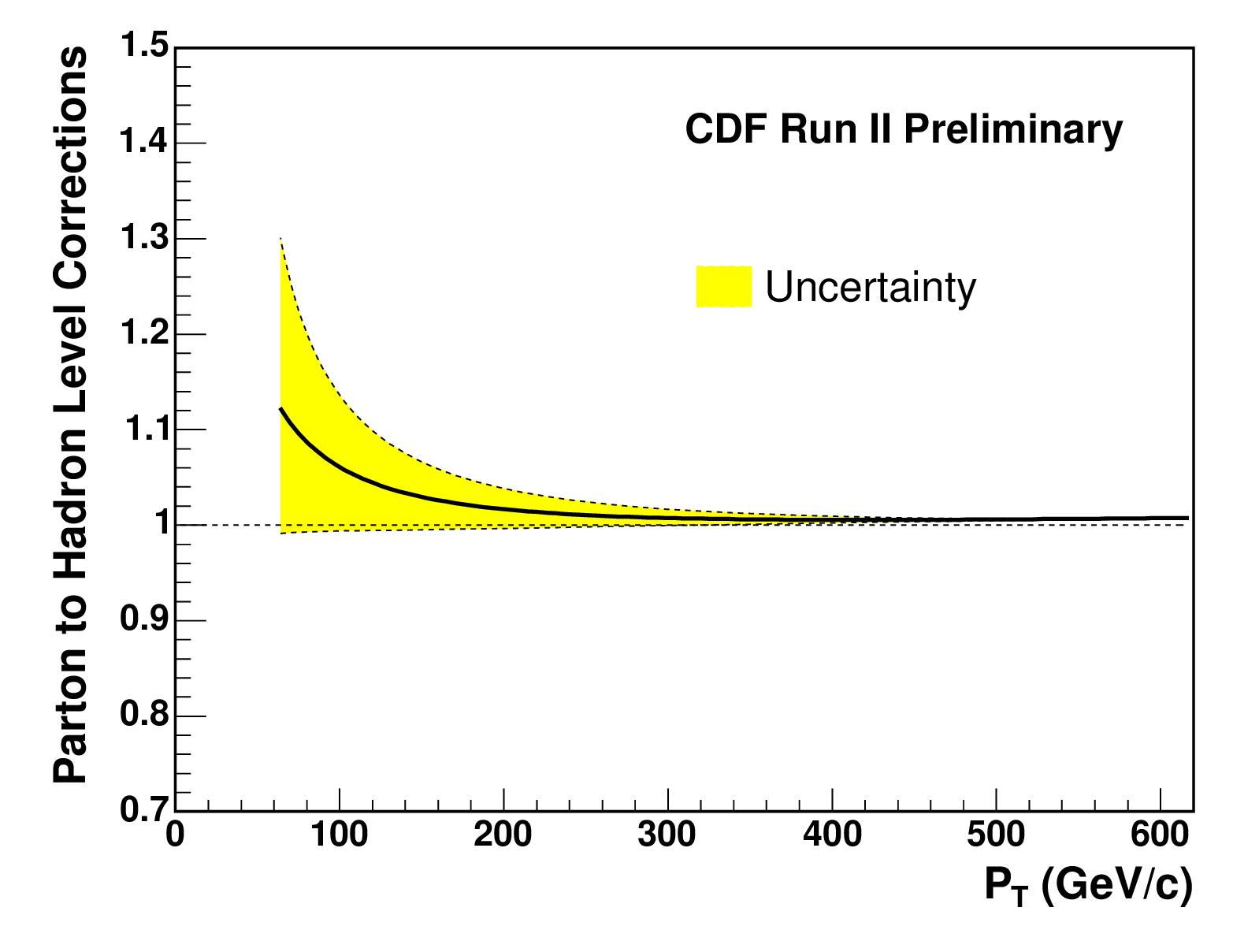,width=7cm}
}
\caption{The CDF Run II inclusive jet cross section using the Search Cone
algorithm ($R_{\rm cone}$ = 0.7, $f_{\rm merge}$ = 0.75, 
$R_{\rm search}=R_{\rm cone}/2$) 
compared with parton-level NLO QCD with $R_{\rm sep} = 1.3$ (left). 
The data have been extrapolated (\textit{i.e}., corrected) to the
parton level using the parton to hadron correction factor (right). The
hadron-level data are multiplied by the reciprocal of this factor. }%
\label{CDFcone2}%
\centerline{
}\end{figure}

Preliminary Run II inclusive cone jet results\cite{hep-ex/0512020,D0ConeRunII} 
suggest that, even with differing algorithms, the two
collaborations are in reasonable agreement as indicated in Figs. \ref{D0cone2}
and \ref{CDFcone2}. \ \ On the other hand, the challenge, as noted above, is
to continue to reduce the systematic uncertainties below the current 10\% to
50\% level, which effectively guarantees agreement, if the primary differences
are also at the 10\% level. \ Indeed, the current studies of the corrections
due to splash-in, \textit{i.e}., the underlying event (and pile-up), and the
splash-out corrections due to hadronization are much more sophisticated than
in Run I and presented in such a way that they can be applied either to the
data (corrected for the detector) or to a theoretical (perturbative)
calculation. \ The evaluation of these corrections is based on data and the
standard Monte Carlos, \PY~and~\HW, especially Tune A of \PY, which
seems to accurately simulate the underlying event in terms of multiple parton
interactions, including the observed correlations with the hard scattering
process.\cite{Field:2005yw} \ 

For CDF the multiple interaction (pile-up) correction is measured by
considering the minimum bias momentum in a cone placed randomly in $\left(
y,\phi\right)  $ with the constraint that $0.1<\left\vert y\right\vert <0.7$
(so the rapidity range matches the range for the jet cross section
measurement). The $p_{T}$ in the cone is measured as a function of the number
of vertices in the event. The slope, $A_{1}$, of the straight line fit to
$\left\langle p_{T}\right\rangle _{cone}$ versus the number of vertices is the
$p_{T}$ that needs to be removed from the raw jet for each additional vertex
seen in an event, where the number of vertices is proportional to the number
of additional interactions per crossing. \ The measurement of the correction
is therefore affected by fake vertices. \ The correction to the inclusive jet
cross section decreases as the jet $p_{T}$ increases. \ The towers that are
within the cone are summed as 4-vectors just as in the Midpoint jet algorithm.
\ The summation of the towers uses the following prescription: for each tower
construct the 4-vectors for the hadronic and electromagnetic compartments of
the calorimeter (correctly accounting for the location of the z-vertex). The
4-vectors are then summed. This method, while it does approximate the Midpoint
algorithm, makes no attempt to account for the splitting/merging that is
performed by the cone jet algorithm (resulting in jets that are not shaped
like ideal cones). \ This random cone method is a reasonable
approach when the number of additional interactions is small. At CDF, the
correction to a Midpoint jet (cone radius of 0.7) is $\sim 1$ GeV/c per
jet. The effect of this correction is significantly different if there is 1
additional vertex per event than if there are 10. \ It may be the case that
for a large number of additional vertices the systematic uncertainty
associated with the pile up correction may become comparable to the other
systematic uncertainties. \ The systematic uncertainty assigned to this
correction is determined in part by its inclusion in the generic correction
scheme used by CDF. \ The systematic uncertainty is made large enough to cover
the variation of correction as derived in different samples. \ Note that the
jet clustering has a threshold of 100 MeV, towers below this are not included
in any jet. \ Additional energy deposited in a cone can be added to a tower
below threshold and thus cause it to be included in the jet or be added to a
tower that was already in the jet. \ Following the methods used in Run I, the
correction for pile up was derived with 3 tower thresholds, 50 MeV, 100 MeV,
and 150 MeV, which provides some check of the two ways that the pile up energy
can be added to a tower. \ An alternative approach is to derive the correction
based on making the shape of the inclusive cross section independent of the
instantaneous luminosity (and this approach has been used to compare the
corrections in the cone algorithm with those in the $k_{T}$ algorithm).

The CDF hadronization correction (parton to hadron as described here) for the
inclusive jet cross section (cone or $k_{T}$) is obtained using \PY~(Tune
A), as noted above. The correction is simply the ratio of the 
hadron level inclusive jet cross
section with multiple parton
interactions (MPI) turned on over the parton level
(after showering) inclusive jet cross section with MPI turned off. \ 
This results in a $\sim12\%$ correction (for the
Search Cone algorithm) at 60 GeV/c. \ Although it is unphysical to explicitly
separate the effects of organizing the partons into hadrons and including the
contributions of the underlying event (UE), it is still a useful approximation
in the context of the Monte Carlos and produces fairly stable and intuitively
appealing results. \ Further, since only an underlying event correction was
employed in the Run I analyses, the option to apply just the underlying event
correction provides a connection between the results from the 2 runs. \ The
underlying event correction found in this way corresponds to adding
approximately 1 GeV to the perturbative jet, crudely in agreement with the Run
I numbers. \ The definitions for the separate corrections are as follows:%
\begin{align*}
C_{i}^{UE}   =\frac{\sigma_{i}^{hadron(UE)}}{\sigma_{i}^{hadron(no-UE)}}, 
C_{i}^{had}    =\frac{\sigma_{i}^{hadron(no-UE)}}{\sigma_{i}^{parton(no-UE)}
}, 
C_{i}^{p\rightarrow h}    =\frac{\sigma_{i}^{hadron(UE)}}{\sigma
_{i}^{parton(no-UE)}}.
\end{align*}
In these expressions, UE means MPI turned on, no-UE means MPI turned off. The
beam-beam remnants in \PY~tend to end up at large rapidity and their effect
on the central rapidity jet cross section is not included here. \ The
systematic uncertainty assigned to the hadronization correction comes from
comparing, in \PY~and \HW, the fragmentation and UE components. 
\PY~and~\HW~have very similar fragmentation corrections. \ However, as
expected, the UE corrections are different. The resulting systematic
uncertainty comes exclusively from the different UE correction and was found
to be $\sim30\%$ for jets near 60 GeV/c.

\begin{figure}[!h]
\centerline{
\psfig{figure=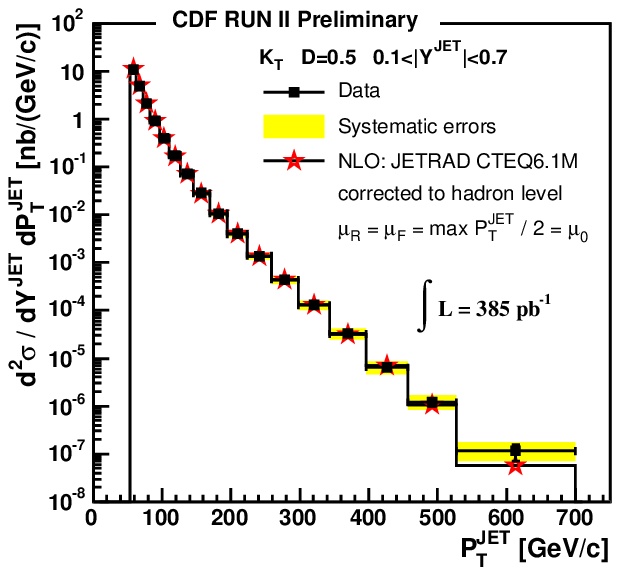,width=5cm}
\psfig{figure=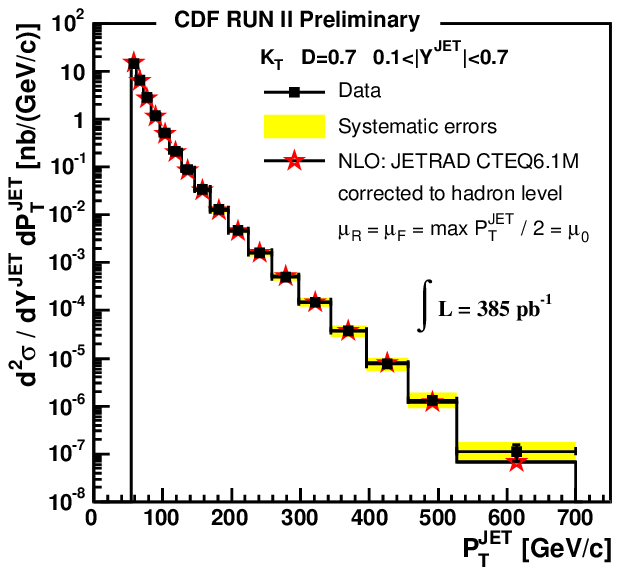,width=5cm}
\psfig{figure=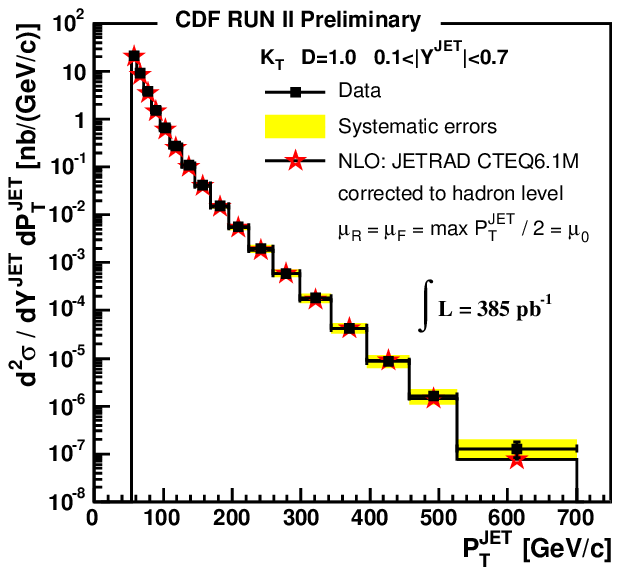,width=5cm}
}
\centerline{
\psfig{figure=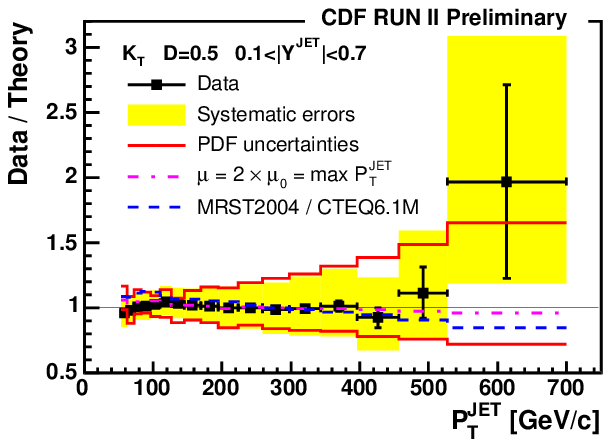,width=5cm}
\psfig{figure=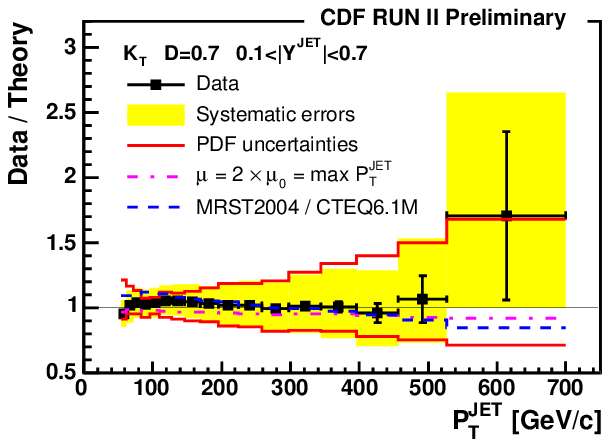,width=5cm}
\psfig{figure=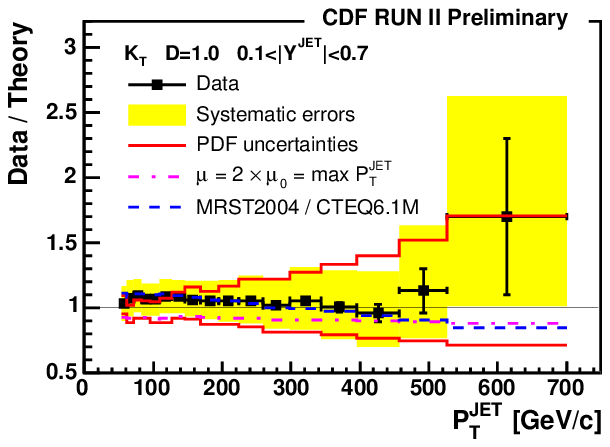,width=5cm}
}
\centerline{
\psfig{figure=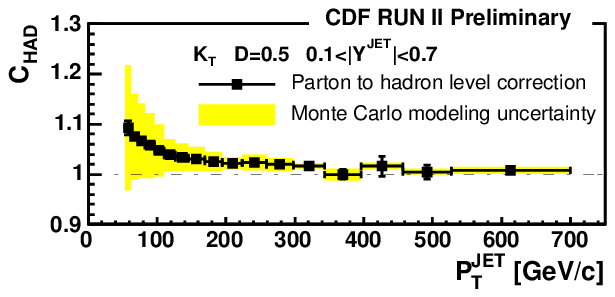,width=5cm}
\psfig{figure=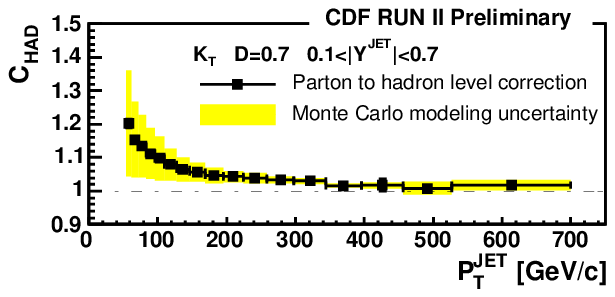,width=5cm}
\psfig{figure=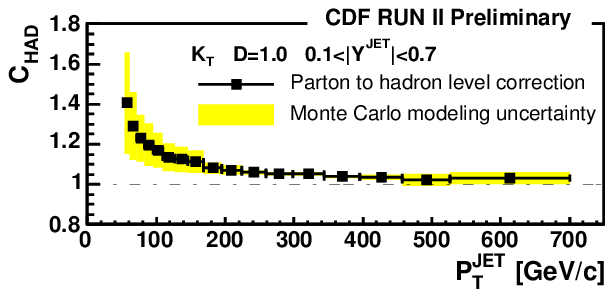,width=5cm}
}
\caption{CDF comparison of NLO theory versus data for the $k_{T}$ algorithm
(with 3 values of $D$), including the Monte Carlo determined correction for
showering and fragmentation effects as indicated in the\ bottom row of figures.
}%
\label{CDFkt2}%
\end{figure}

CDF has also undertaken a similar study of corrections for the $k_{T}$
algorithm\cite{hep-ex/0512062}. \ \ In this case, without the fixed geometry of
the cone algorithm, the multiple interaction correction is extracted from the
data by asking that the shape of the measured inclusive jet cross section be
independent of the instantaneous luminosity after the correction is applied.
\ It was also confirmed that, within systematic uncertainties, the corrections
are consistent between the cone and $k_{T}$ algorithms for $D=R_{cone}$. \ As
a further test the luminosity independent inclusive jet cross section shape
test (after correction) was applied also to the cone algorithm. \ This
approach yields only a slightly larger correction than found by the method
above using minimum bias events, and still within the expected systematic
uncertainty. \ The slightly larger correction is presumably due to the
remaining effects from merging/splitting of the cones. \ As mentioned above,
the $k_{T}$ algorithm\ hadronization correction is determined just as in the
cone algorithm case. \ The resulting corrections yield quite satisfactory
agreement between the corrected theory and experiment as indicated in Fig.
\ref{CDFkt2} for a range of values of $D$. \ \ This figure also exhibits the
quite substantial systematic uncertainty in these corrections (of order 50\%
to 80\% of the correction) and an overall systematic uncertainty (see the
middle row of graphs in Fig. \ref{CDFkt2}) of order 20\% at low $p_{T}$ to
close to 100\% at large $p_{T}$. \ Thus the current systematic uncertainty of
the $k_{T}$ algorithm results are comparable to those for the cone algorithm,
as indicated in Figs. \ref{D0cone2} and \ref{CDFcone2}.

\thissection{Non-perturbative contributions to jet measurements}

\begin{figure}[!h]
\centering
\includegraphics[width=.8\textwidth, trim= 0 50 0 100]{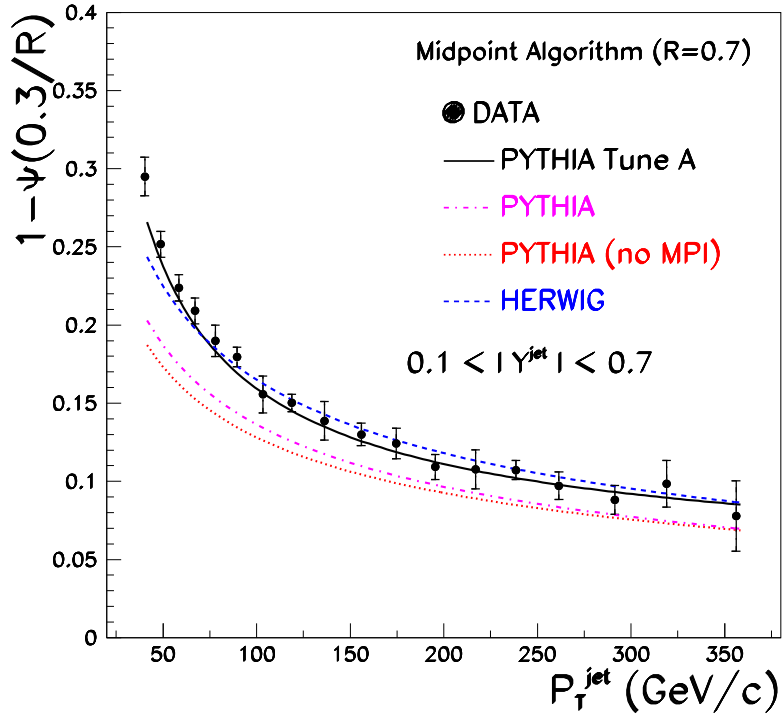}
\caption{Measured integrated jet shape compared to different Monte Carlo
models for the underlying event\cite{hep-ex/0505013}. }%
\label{shapes}%
\end{figure}

As noted above, in hadron-hadron collisions, the measured inclusive jet
production cross section at the particle level, regardless of the jet
algorithm considered, includes all-orders and non-perturbative contributions
from the underlying event and the fragmentation into hadrons that are not
present in the fixed-order parton-level calculation, which become significant
at low jet transverse momentum. A proper comparison between the data and the
theoretical prediction requires good control of such contributions.
Experimentally, they are estimated using leading-order parton-shower Monte
Carlo generators, and the variation of the predicted jet cross sections after
turning off the interaction between beam remnants and the hadronization in the
Monte Carlo. This procedure is model dependent, and strongly relies on the
Monte Carlo providing a good description of those observables in the data that
are most sensitive to non-perturbative contributions like, for example, the
internal structure of the jets. Recent precise measurements on jet
shapes\cite{hep-ex/0505013}, as indicated in Fig. \ref{shapes}, have allowed the
detailed study of the models employed to describe the underlying event in
inclusive jet production at the Tevatron (see also Ref. \cite{Field:2005yw}). Future
measurements of the underlying event in Run II, for different hadronic final
states, promise to play a major role in the early understanding of the
measured jet cross sections at the LHC.

\thissection{More details}

\begin{figure}[!h]
\centerline{
\psfig{figure=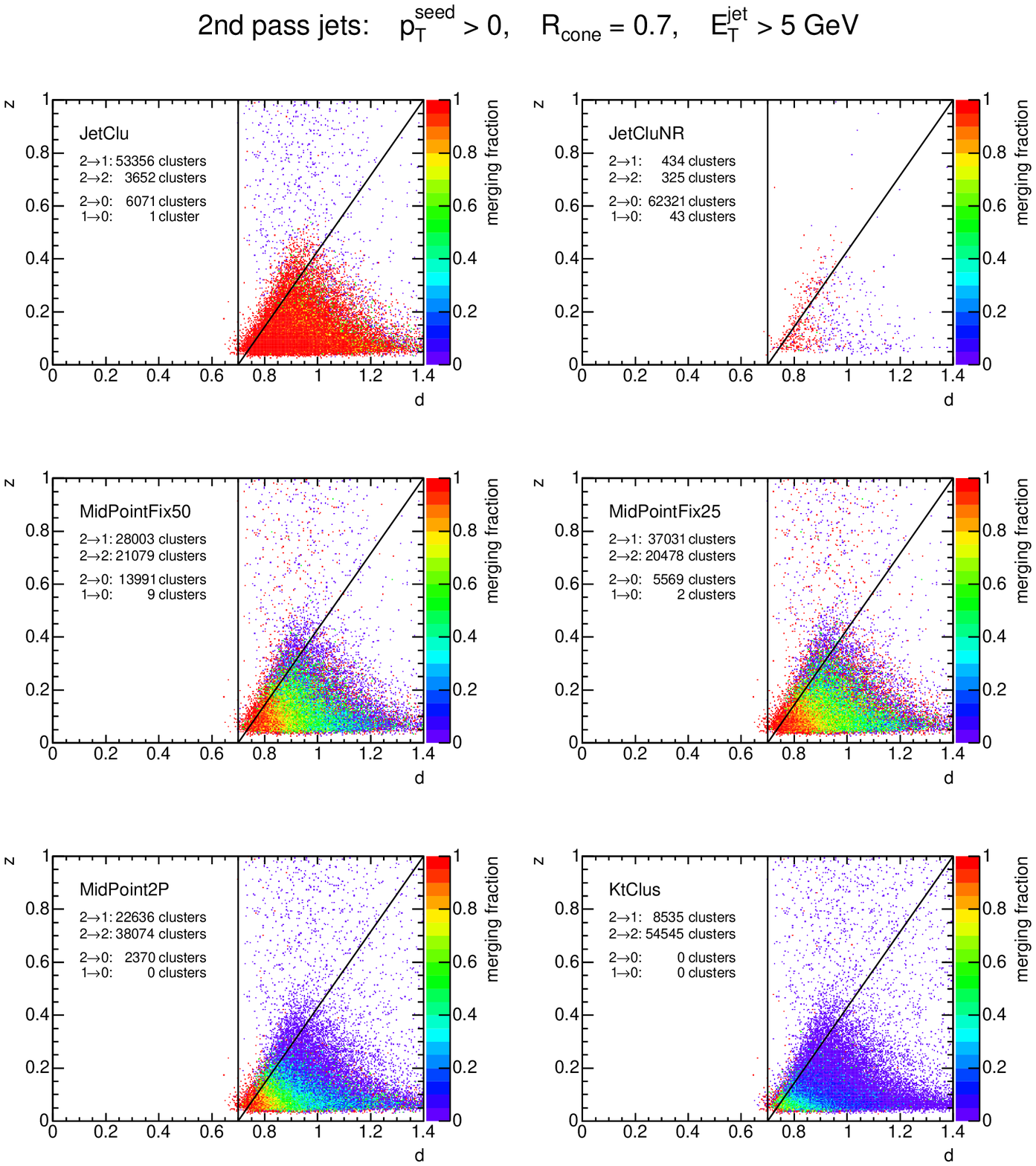,width=.95\textwidth}
}
\caption{Merging probabilities in the $z,d$ plane. }%
\label{d_vs_z_color}%
\centerline{
}\end{figure}

Several of the issues noted above are illustrated in Fig.~\ref{d_vs_z_color},
taken from a recent analysis\cite{Matt}. \ The goal of the analysis, and the
figure, is to understand how various cone jet algorithms deal with the issues
leading to the dark towers, \textit{i.e}., with configurations of nearby
showers described in the notation of Fig. \ref{pertthy}. \ The reference
algorithm is the standard MidPoint Cone Algorithm with a seed tower threshold
of 1 GeV and applied to a set of events simulated with \PY (Tune A)
using CTEQ5L parton distribution functions.
After stable cones and jets are found, defining the 1st pass jets,
the towers in these jets are removed from the analysis. \ The remaining
towers, the dark towers, are then identified into 2nd pass jets by simply
placing a cone of size 0.7 about the highest $p_{T}$ tower, calling that a 2nd
pass jet and removing the enclosed towers from the analysis as long as
the total $p_T$ in the cone is greater than 5 GeV. \ This process is
repeated to generate the list of 2nd pass jets. \ For simplicity the 2nd pass 
jets are constructed without iteration or splitting/merging. 
The 2nd pass jets can be thought of as
populating the $\left(  d,z\right)  $ plane around the closest 1st pass jet.
Recall that $d$ is the angular separation and $z$ is the $p_T$ ratio
($z=p_{T,2}/p_{T,1}$). \ \ Next apply the 6 other jet algorithms to the
same data. \ JETCLU is the CDF Run I algorithm (with ratcheting and 
$f_{merge} = 0.75$);  JETCLUNR
is the same algorithm but without ratcheting;
MidPointFix50 is a Search Cone Algorithm with $R_{search}=R_{cone}/\sqrt{2}$
and $f_{merge} = 0.5$;
MidPointFix25 is a Search Cone Algorithm with $R_{search}=R_{cone}/2$
and $f_{merge}=0.5$. \ The
MidPoint2P algorithm, which also uses $f_{merge} = 0.5$, is similar to the 
reference (standard) MidPoint algorithm
with a 2nd pass, except that in this case the 2nd pass cones (with the 1st pass
towers removed) are iterated and a midpoint stable cone is looked for. \ In
the final step of MidPoint2P the contents of the 2nd pass cones are calculated
using all towers, but no iteration, and then standard splitting/merging is
applied to both 1st and 2nd pass cones. \ The final case studied is the $k_T$
Algorithm with $D = R_{cone} = 0.7$. \ 

To produce the plot the jets found by
each of these 6 algorithms are identified with the 2nd pass and closest 1st pass
jets found by the reference algorithm by comparing the highest $p_{T}$ towers.
\ For example, if the highest $p_{T}$ towers in both a 2nd pass jet and the
closest 1st pass jet are in the same JETCLU jet, we conclude that JETCLU
merges these 2 clusters (a $2\rightarrow1$ clustering). \ If the highest
$p_{T}$ towers in both a 2nd pass jet and the closest 1st pass jet are in
different JETCLU jets, we conclude that JETCLU does not merge these 2 clusters
(a $2\rightarrow2$ clustering). \ If the highest $p_{T}$ tower in a 2nd pass
jet is not in any JETCLU jet, we conclude that it remains a dark tower also in
JETCLU (a $2\rightarrow0$ clustering). \ The final, unlikely scenario is that
the highest $p_{T}$ tower in a 1st pass jet is not in any JETCLU jet and we
conclude that JETCLU is ignoring this 1st pass jet (a $1\rightarrow0$
clustering). \ Applying the same tests to each of the 6 algorithms yields
the plot in Fig. \ref{d_vs_z_color} which compares the amount of (relative
probability of) $2\rightarrow1$ merging and $2\rightarrow2$ splitting for the
various algorithms. \ The total numbers of cluster pairs in each category are
also listed for each algorithm on the individual plots. \ An \textquotedblleft
ideal\textquotedblright\ cone jet algorithm, \textit{i.e}., one that matches
well to NLO\ perturbation theory, will exhibit a large merging fraction (red)
in the triangular region above the diagonal, a low merging fraction (blue)
below the diagonal and few remaining dark towers.  The JETCLU algorithm
exhibits considerable merging everywhere, especially below the diagonal,
presumably due to ratcheting. \ In contrast JETCLUNR without ratcheting
exhibits little merging and yields essentially the same dark tower content
as the reference MidPoint Cone Algorithm (\textit{i.e.}, 
mostly $2\rightarrow0$ clustering).
Thus, while ratcheting ensures a low level 
of $2\rightarrow0$ clustering, \textit{i.e}.,
few dark towers, the high level of merging over the entire region is undesirable.
The MidPoint Fix25 algorithm, the one currently in use at
CDF (except here $f_{merge}=0.5$ instead of 0.75), has the desirable features 
of relatively low merging below the diagonal
and relatively high merging above the diagonal with few remaining dark towers.
\ This explains the original appeal of the search cone algorithm, but, of
course, the current analysis does not speak to the IR-sensitivity issue for this
algorithm when applied to perturbation theory. \ Similar comments apply also
to the MidPointFix50 algorithm, but with somewhat less merging and more
remaining dark towers. \ The MidPoint2P algorithm has a lower level of overall
merging (a lower $2\rightarrow1$ clustering count and a higher $2\rightarrow2$
clustering count), and a lower number of remaining dark towers. \ This
suggests that using a 2nd pass to find jets can address the dark tower issue.
However, this approach does not address
the problem of the 2-in-1 stable cone solutions that disappear when the
smearing effect of fragmentation is included.  This conclusion again
emphasizes the difficulty of matching the behavior of jet algorithms at the
parton and hadron levels.  The final plot for the $k_{T}$ Algorithm illustrates
the expected result that this algorithm yields very little merging of objects 
separated by
an angular distance of more than $D = 0.7$, \textit{i.e.}, it acts like a cone
algorithm with 
$R_{sep}\simeq 1.0$.

\begin{figure}[!h]
\centerline{
\psfig{figure=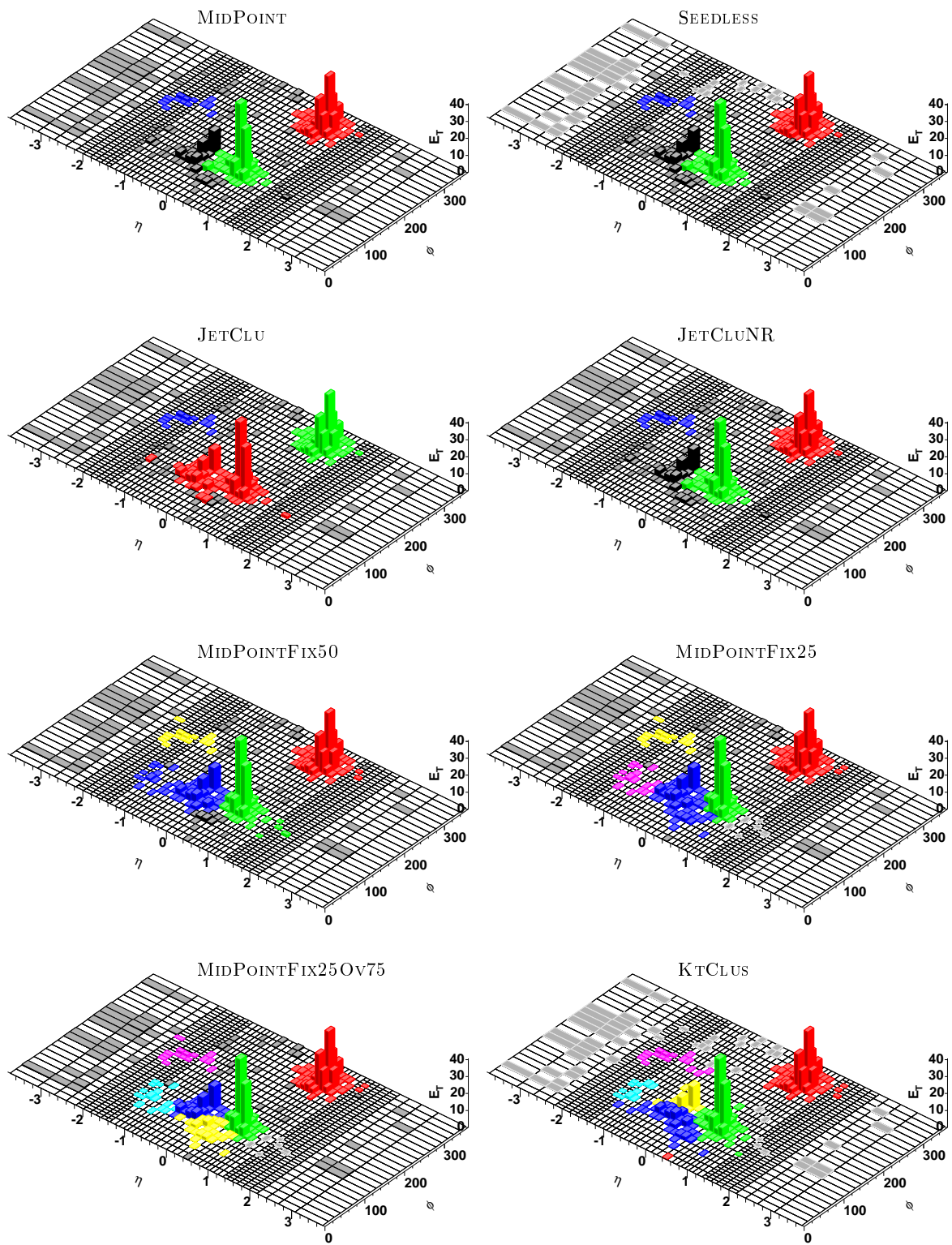,height=.9\textheight}
}
\caption{Jets and dark towers found by various algorithms in the same event. }%
\label{jets2}%
\centerline{
}\end{figure}

In our discussions above about of how well jet algorithms are working at the
Tevatron Run II, we saw that there are detectable differences ($\sim$10\%) 
between the CDF and \dzero implementations of the cone algorithm. \ These
differences arise to a large extent from how the jet algorithms handle
configurations where two energetic partons are nearby each other in $\left(
y,\phi\right)  $ on the scale of the cone size $R_{cone}$, \textit{i.e}.,
nearby but not collinear. \ As suggested by the simulated event shown in Fig.
\ref{jets2}, in general the legacy CDF algorithm JETCLU will merge the showers
from the two partons (consider in particular the two tower configuration near
rapidity $0$, azimuth $100^\circ$)
as the ratcheting feature can lead to a stable central cone, while the lower energy
shower is left as dark towers when ratcheting is turned off (JetCluNR).  The Run II
Midpoint Cone Algorithm and the seedless algorithm likewise find
only the more energetic\ jet with the secondary shower not included in any jet
(\textit{i.e.}, as dark towers).  The two Search Cone Algorithms (MidPointFix50 and 
MidPointFix25, both with $f_{merge} = 0.5$), identify the secondary shower as a second 
jet with some differences in the very low energy objects.  The Search Cone
Algorithm with $f_{merge} =0.75$ (MidPointFix25Ov75, which is essentially the current CDF 
algorithm) finds the secondary shower to be two jets, \textit{i.e.}, 
showing less merging.  Finally the $k_{T}$ Algorithm identifies all energetic towers 
into jets with less merging than the cone algorithms with $f_{merge} = 0.5$, and only
small differences from the $f_{merge} =0.75$ algorithm.

\thissection{The Future}

In order to study the jet algorithm situation more thoroughly, we desire an 
analysis tool that
provides NLO accuracy for the jets, \textit{i.e}., reduced factorization scale
dependence (suggesting small theoretical uncertainty), plus an accurate
treatment of energetic radiation at angles of order $R_{cone}=0.7$ with
respect to the core shower direction.  At the same time we must include both
showering and hadronization. \ This is just what the development of MC/NLO
tools (such as \MN\cite{Frixione:2002bd,Frixione:2003ei}) is meant to do 
for us. \ The challenge with light jet
calculations in MC/NLO is that, since every object participating in the
short-distance process is colored, they can all produce their own shower.
\ Thus the subtraction process outlined in 
Refs.\cite{Frixione:2002bd,Frixione:2003ei} must be performed
for every parton and, to some extent, tuned to minimize the occurrence of
events with negative weights. \ Note that the early work in this field focused
on processes with color only in the initial state in order to minimize the
required \textquotedblleft bookkeeping\textquotedblright. \ Only recently,
with the addition of the single top 
process\cite{Frixione:2005vw,Frixione:2006he}, where the top quark
has essentially no time to radiate before it decays, have colored
objects been included in the final state. \ It is now time to develop the full
light jet \MN code so that the optimization of the jet algorithm can be studied.
Such code would allow detailed analyses of the corrections necessary to go
from the long distance hadronic states measured in the detectors back to the
NLO short distance partonic states of perturbation theory.  Such studies could
illuminate the dependence on various experimental analysis
parameters that are currently largely hidden from view.  For example, how much
does the final jet cross section depend of the $f_{merge}$ parameter used
in the split/merge step of the cone algorithm.  As noted above, CDF and \dzero
are currently using different values for this parameter.  Similarly there
is a question about how energetic a stable cone must be ($p_T > p_{T,min}$)
in order to be 
included in the split/merge process.  CDF includes all stable cones ($p_{T,min} = 0$), while
\dzero includes only those above 3 GeV ($=p_{T,min}$).  It is important that we understand
quantitatively the impact of the different choices for both these parameters.  
Their effects are certainly correlated.

Another possible avenue of study is to apply the recent progress in
understanding the associated energy in events with jets, which grew out of
earlier work in the study of event shapes in $e^{+}e^{-}$ collisions. \ This
work, by a variety of 
theorists\cite{hep-ph/0501270,hep-ph/0509078,Banfi:2004nk,Appleby:2003sj}, 
will not be reviewed here,
except to suggest its useful further application to jet issues at
the Tevatron and the LHC.

A quite different approach that deserves further study is the possibility that hard
scattering events can be usefully studied without the need to identify a
discrete set of jets, as we have assumed here from the outset. \ The general
idea is that the same information now carried by the jets could instead be encoded
in a distribution describing the energy flow, event-by-event, removing the
need to identify specific jets in each event. \ Recall that this is where the
problems arise in the discussion above. \ The general idea for this approach
was outlined at Snowmass 2001\cite{hep-ph/0202207} and is touched on in some of the
previously mentioned 
references\cite{hep-ph/0501270,hep-ph/0509078,Banfi:2004nk,Appleby:2003sj}.

We close this discussion of the future with a suggestion, which arose during
this Workshop, concerning how the
information currently lost in the dark towers can be preserved and subsequently
used in analyses.  In the study leading to Fig.~\ref{d_vs_z_color} the concepts
of 1st and 2nd pass jets was introduced.  Here we outline a similar 2nd pass jet
algorithm in more detail. In the first step we apply the ``standard'' Midpoint Cone
Algorithm as recommended in the Run II Workshop\cite{Blazey:2000qt} (keeping in mind that 
there are differences in the current implementations).
This step includes the full algorithm, including
the iterative procedures to find stable cone solutions around
the seeds, and later around the midpoints.
It also includes the full split/merge procedure.  The resulting identified jets
are labeled 1st pass jets.  As we have discussed, while this step produces 
well-defined jets, it sometimes
leaves substantial amounts of energy in the event unclustered.
To characterize this unclustered energy we identify it as jets in a second pass
using exactly the same algorithm, but applied to the final 
state after all particles/towers assigned to one of the 1st pass jets
are removed. These new jets are labeled 2nd pass jets.

\begin{figure}[!h]
\centering
\includegraphics[width=\textwidth]{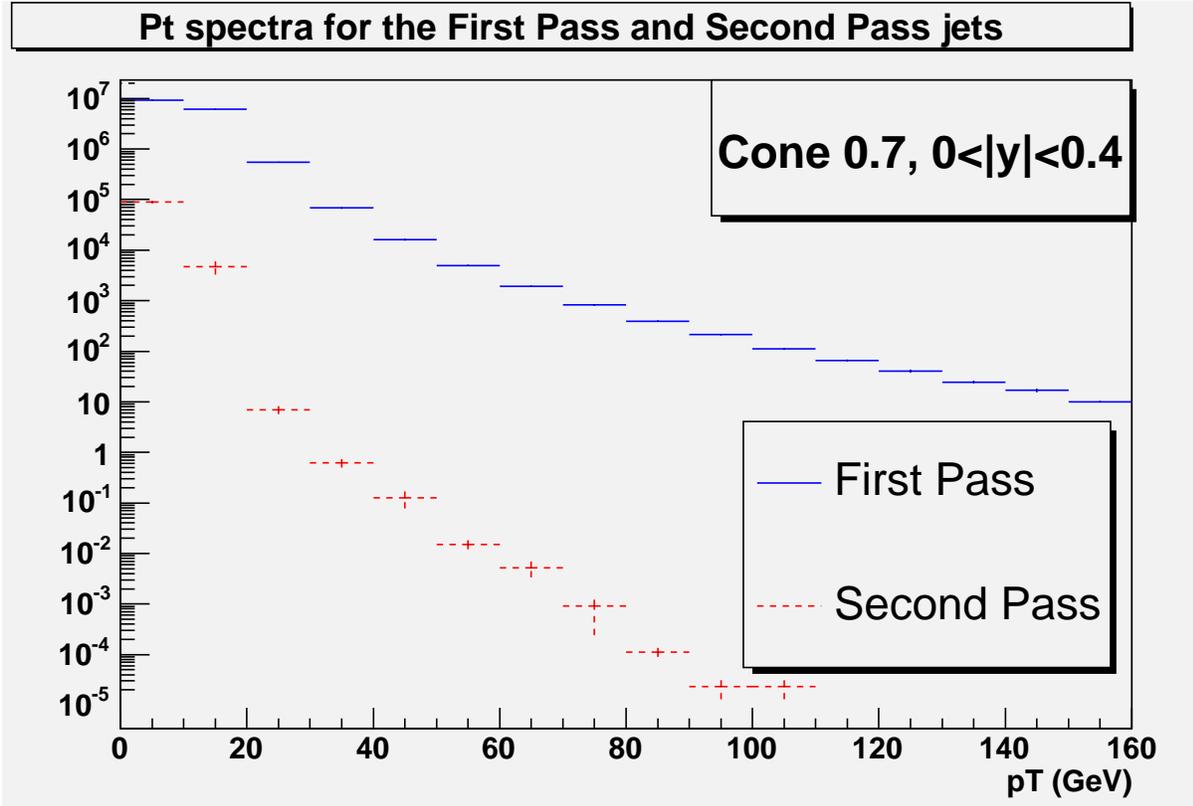}
\caption{The $p_T$ spectra for 1st and 2nd pass jets reconstructed 
in \PY QCD events using the MidPoint cone algorithm with R=0.7.
\label{secondpass}}
\end{figure}

Clearly these 2nd pass jets do not stand on the same footing
as the 1st pass jets.  They do not correspond to stable
cone solutions when considering the full final state.
Therefore there are various possibilities for making use of 
the 2nd pass jets.  One might simply keep the 2nd pass jets
as separate jets, in addition to the 1st pass jets.  Since the 
second pass jets are typically of much smaller energy than 
the 1st pass jets, this
will have little numerical impact on the jet cross section.  
This point is illustrated in Fig.~\ref{secondpass}, where the 
individual $p_T$ spectra of 1st and 2nd pass jets found in the 
fashion described above are displayed.
A more interesting question is whether a 2nd pass jet 
can/should be merged with the nearest 1st pass jet (the 2nd pass 
jets are almost always
produced within $R_{cone} < d < 2R_{cone}$ to a 1st pass jet).  
For example, we could merge according to the NLO structure in 
Fig. \ref{pertthy}, and in this way
define a dark tower correction.  Based on the recent studies of 
the dark tower issue, we expect the correction of the inclusive 
single jet cross section to be of order a 
few percent (\textit{i.e.}, a substantially larger effect than 
simply including the 2nd pass jets
in the single jet sample).  Alternatively,
we might study how the 2nd pass jets could be used in the 
reconstruction of interesting
physics objects, such as highly boosted $Z$'s, $W$'s and top quarks, 
that decay into a 1st pass jets plus a 2nd pass jet.  

We recommend that studies of all of these ideas be carried out so that
quantitative conclusions can be reached as to the best way to make use 
of the information carried by the 2nd pass jets.  Such studies may
benefit from employing the MC@NLO tool mentioned above.

\thissection{Summary}

We have come a long way in our understanding of jet algorithms and their
limitations.  We still have a way to go if we want to guarantee a precision
much better than 10\%. Overall, we expect both cone and $k_{T}$ algorithms to
be useful at the LHC. \ Their differing strengths and weaknesses 
will provide useful cross checks. \ Our conclusions and recommendations
include the following. \ 

\begin{itemize}
\item Seeds introduce undesirable IR sensitivity when used in theoretical
calculations. \ The cone algorithm without seeds is IR-safe. \ Experimental
results for the cone algorithm should be corrected for any use of seeds and
compared to theoretical results without seeds.
\item More study is required to understand the quantitative impact and
possible optimization of the choices for the parameters in the split/merge
step of the cone algorithm, $f_{merge}$ and $p_{T,\min}$. \ As in Run I, CDF
and \dzero \ are currently using different parameter choices and the implied
differences in jet rates are not well documented.  The studies aimed at
finding optimal choices for the split/merge parameters should include 
participation by the LHC collaborations to ensure relevance to the LHC
environment and an enhanced level of commonality in those experiments.
\item The unclustered energy in the dark towers sometimes found when using a
cone algorithm requires further study. \ This issue is now understood to arise
from the smearing effects of showering and hadronization. \ The Search Cone
Algorithm currently employed by CDF is not a satisfactory solution to this
problem and its use should be discontinued.  An alternative approach using 2nd
pass jets is outlined in this report and deserves further study.
\item Most of the challenges found in using cone algorithms are now understood
to arise from the kinematic regime of two nearby (but not collinear)
short-distance partons, especially as this configuration is smeared by
subsequent showering and hadronization. \ It is precisely these effects that
lead to the observation that the \textit{ad hoc} phenomenological parameter
$R_{sep}$ requires a value less than the default value of $2$. \ With the
imminent appearance of MC@NLO code for jets a substantially improved analysis
of this situation will be possible.
\item With the limitations of cone algorithms now fairly well understood and
mostly correctable to the few percent level, it is extremely important that we
achieve a comparable level of maturity in our understanding of the $k_{T}$
algorithm. \ By its nature, the $k_{T}$ algorithm will not suffer from the
same issues as cone algorithms. \ The most pressing question, unanswered by
the use of the $k_{T}$ algorithm at $e^{+}e^{-}$ and $ep$ colliders, is how
the $k_{T}$ algorithm responds to the noisier environment of high energy, high
luminosity $pp$ collisions. \ By definition the $k_{T}$ algorithm clusters
\textit{all }energy into jets and the central issue is whether the
contributions of the underlying event and pile-up will lead to troublesome
fluctuations in the properties of $k_{T}$ algorithm jets primarily associated
with the true short-distance scattering process. \ Thorough studies of the
$k_{T}$ algorithm during Run II are essential to our preparations for the LHC.  These studies should enlist participation from the LHC collaborations to ensure
relevance to the even noisier environment expected at LHC energies and luminosities.
\item Experimental results need to be reported at the hadron level or higher;
corrections between the parton level and hadron level need to be clearly
specified, including uncertainties.
\end{itemize}

\section{Parton Distribution Functions}

\subsection{{Heavy Flavor Parton Distributions and Collider Physics}}

\textbf{Contributed by:  Tung}

\subsection*{Motivation}

Heavy flavor parton distributions represent an important unchartered
territory in the landscape of global QCD analysis of the parton structure of
the nucleon. \ On one hand, since they make relatively small contributions to
the conventional Standard Model processes that contribute to global QCD
analyses, there exist almost no hard experimental constraints on these
distributions. \ On the other hand, their influence on physics analyses of
the next generation of Collider Physics is expected to be increasingly
significant--- directly for Top and Higgs studies, and hence indirectly for
New Physics searches. \cite{Baines:2006uw,Karshon:2006yf}

Conventional global QCD analyses include heavy flavor partons, i.e. charm,
bottom (and, optionally, top), under the key \emph{assumption} that these
partons are \textquotedblleft radiatively generated\textquotedblright\ by
QCD evolution---mainly gluon splitting. The rationale for this assumption
is twofold: heavy quarks should be decoupled at low energy scales where
non-perturbative light parton distributions are normally parametrized; and
if the mass of the quark is much larger then $\Lambda _{QCD}$, then heavy
quark effects should be calculable perturbatively. Thus, in the parton
parameter space, \emph{no degrees of freedom are associated with heavy
flavors} in all conventional analyses. \ While this assumption certainly is
\emph{reasonable} for the top quark, it is obviously questionable for the
charm quark since its mass is comparable to that of the nucleon, the
existence of which is definitely non-perturbative. The bottom quark case
lies in-between.

There are a number of models for heavy flavor parton distributions,
particularly the charm distribution, in the literature. Most anticipate
distinctive non-perturbative components that are significant mainly in the
large-$x$ region. \ However, throughout the history of global QCD analysis
of parton distributions, nature has repeatedly surprised us about the flavor
dependence of the sea-quarks inside the nucleon. In spite of more than 20
years of continuing efforts, large uncertainties remain even for the strange
quark distribution (in addition to the gluon).

It is thus important to follow a model-independent approach in exploring the
heavy flavor frontier, keeping an open mind on the range of
possibilities---not just for the charm, but also for the bottom, which plays
a particularly significant role in Top/Higgs physics and beyond.

\subsection*{Opportunities}

\begin{figure}[!h]
\subfigure[$s(x,Q^2)$]{\includegraphics[width=.5\textwidth]{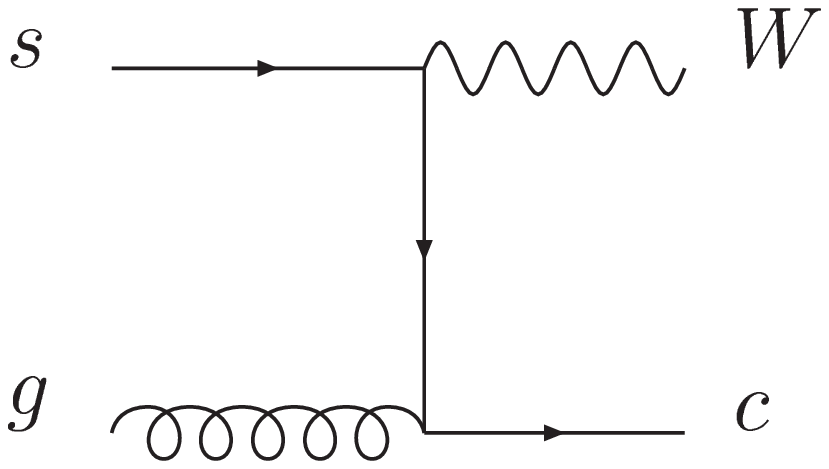}}
\subfigure[$c(x,Q^2)$]{\includegraphics[width=.5\textwidth]{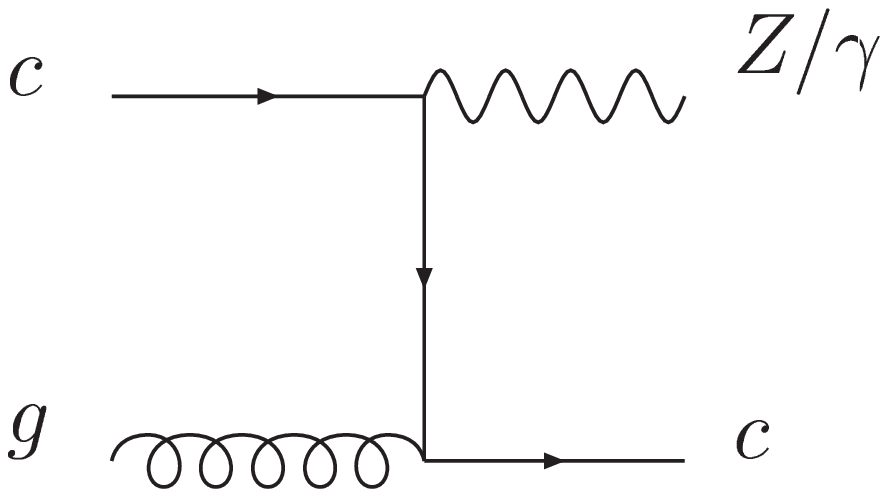}}
\subfigure[$b(x,Q^2)$]{\includegraphics[width=.5\textwidth]{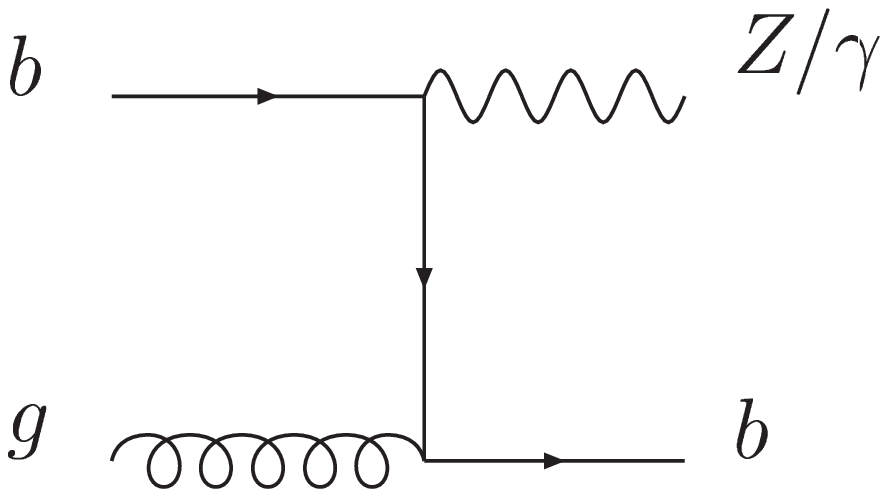}}
\subfigure[$c(x,Q^2)$]{\includegraphics[width=.5\textwidth]{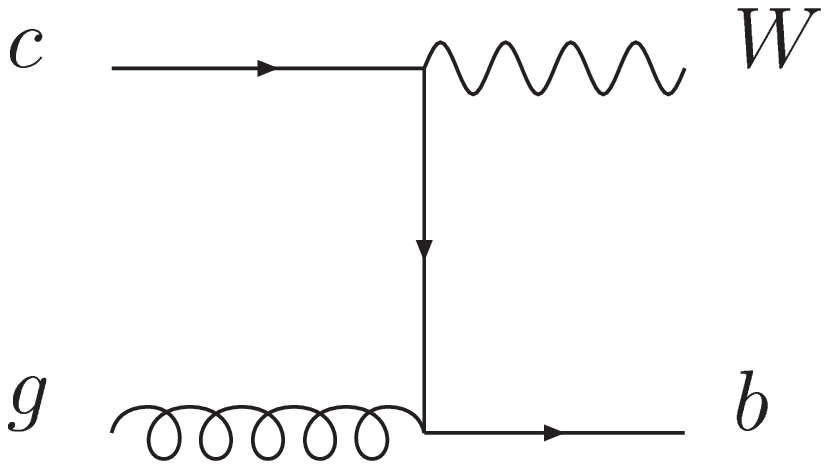}}

\caption{\label{fwk:heavyFlavor}
Processes which can be used to probe the heavy flavor content of the proton.
}
\end{figure}

Available data on deep inelastic scattering and production of Drell-Yan
pairs, jets, and W/Z's---the conventional sources of parton distribution
determination---are not sensitive to the relatively small charm/bottom
constituents of the nucleon. \ Heavy flavor production at HERA offer some
limited constraints. To gain quantitative information, one needs to look at
new channels opened up in the hadron colliders themselves. \ In particular,
it has been known (and repeatedly emphasized, e.g.~\cite{Tung:2004md}) for
some time, final states of $\gamma $/W/Z plus a tagged heavy-quark jet are
directly sensitive to individual $s/c/b$ parton distributions, depending on
which channel is measured. Cf.~\ref{fwk:heavyFlavor}.

At Run II of the Tevatron, and at LHC, these are challenging measurements. \
But they are unique, fundamental processes that contain information about
the heavy flavors not available elsewhere. \ Therefore, these measurements
should command high priority in the overall physics program at both
colliders.

On the theory side, the treatment of heavy quarks in pQCD had followed two
distinct, seemingly contradictory, paths, resulting in considerable
confusion in the field. On one hand, order-by-order calculations of heavy
quark production cross-section were mainly carried out in the so-called
\emph{fixed-flavor-number scheme} (FFNS), based on the premise that the
relevant quark mass is the largest scale in the process. Whereas this
assumption considerably simplifies the calculation, it is clearly an
inappropriate approximation in the high energy regime where the typical
energy scale is larger than the quark mass (for both charm and bottom). \ On
the other hand, most practical parton model calculations (global analyses,
event generators, {\it etc.}) are carried out in the \emph{%
variable-flavor-number scheme} (VFNS), in which charm and bottom are put on
the same footing as the light quarks (i.e.~zero mass) above a scale
comparable to their respective mass. Although this is a reasonable
description over most high energy regime, it becomes untenable at scales
comparable to the mass (where much of the input experimental data for global
PDF analysis lie). \ This dichotomy is naturally resolved in a generalized
pQCD framework, most precisely formulated by Collins
\cite{Collins:1986mp,Aivazis:1993pi,Collins:1998rz}, based on an elegant
composite renormalization scheme (CWZ \cite{Collins:1978wz}, dating back to
the 70's). The extensive recent literature on heavy quark production,
sometime described as \textquotedblleft fixed-order plus
resummation\textquotedblright\ \cite{Baines:2006uw,Karshon:2006yf}, are all
specific implementations of the general principles of this formalism.

Although the theoretical issues have thus been clarified already for quite
some time\cite{Aivazis:1993pi}, and some aspects of the new insight have been
adopted in many recent calculations in a variety of guises
\cite{Karshon:2006yf}, a comprehensive global analysis based on the general
theory incorporating the heavy quark degrees of freedom has not been carried
out. \ However, the importance of the heavy quark sector for LHC physics is
beginning to inspire more focused study on this frontier. \cite{Tung1Dis06}

\subsection*{Strategy and First Results}

The scale (commonly designated as ``$Q$'') dependence of the parton
distributions are governed by the QCD evolution equation; the dynamical
degrees of freedom to be probed reside in the momentum fraction ($x$)
dependence, usually parametrized at some relatively low $Q$, where ample
data exist to experimentally constraint them. \ Since QCD evolution couples
all quark flavors to the gluon and to each other, the determination of the
heavy flavor content of the nucleon must be done within the context of a
comprehensive global analysis. Any viable strategy, thus has to involve the
simultaneous improvement of the currents limits on uncertainties of the
light partons, in particular the strange quark and the gluon.

In order to provide a quantitative basis for studying the potential for
measuring the heavy flavor PDFs in new experiments, such as described above,
one can start by establishing the current limits on these in a dedicated
global QCD analysis without the usual restrictive assumptions on heavy
flavor degrees of freedom, using all available data. \

A necessary step in this direction is the establishment of new analysis
programs that incorporate the generalized QCD framework with non-zero quark
mass effects mentioned in the previous section. This is well underway for the
most important input process to global analysis---deep inelastic scattering.
\ Both the MRST and the CTEQ projects have done this. (Comparable effort for
the other processes, D-Y, jets, {\it etc.}~don't yet exist; but they are less
important because the corresponding experimental errors are larger, and the
scales are higher.) \ The existing implementations by these two groups are
not the same. \ Whereas both are consistent with the general formalism in
principle, MRST \cite{Thorne1Dis06} emphasizes higher order effects, while
CTEQ \cite{Tung1Dis06} emphasizes uniformity and simplicity.

\subsubsection*{First Results}
\begin{figure}[!h]
\includegraphics[width=.95\textwidth]{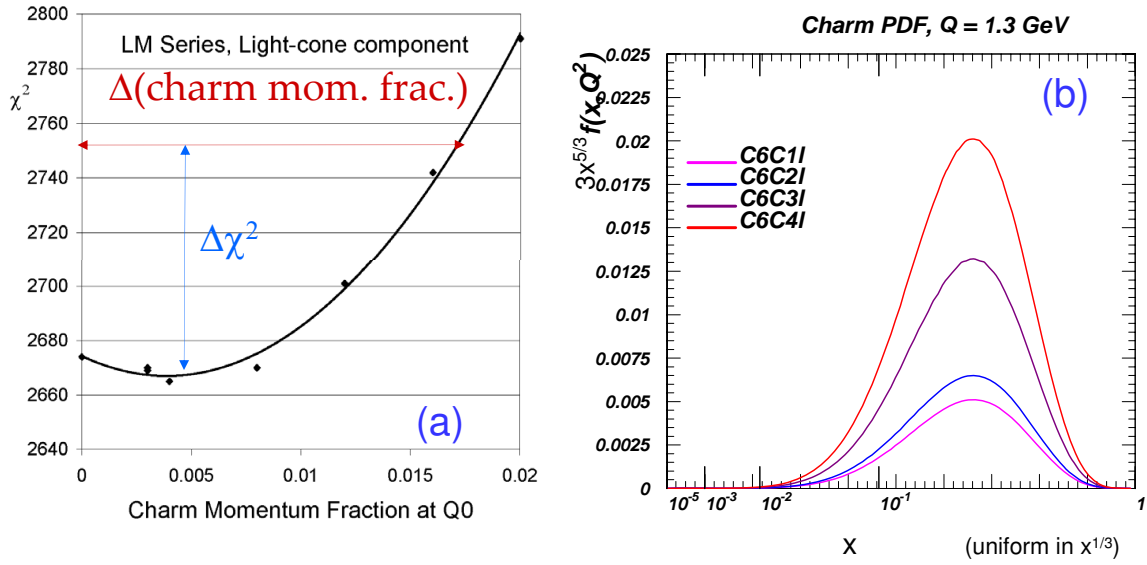}
\caption{(a) Overall $\chi^2$ of global fit vs.~input charm momentum fraction
at $Q_0=m_C=1.3 \mathrm{GeV}$. (b) Shape of charm distribution for the series
of input functions, with increasing amount of charm fraction, used to
generate the curve on the left plot.} \label{fig:charm1}
\end{figure}

\begin{figure}[!h]
\includegraphics[width=.95\textwidth]{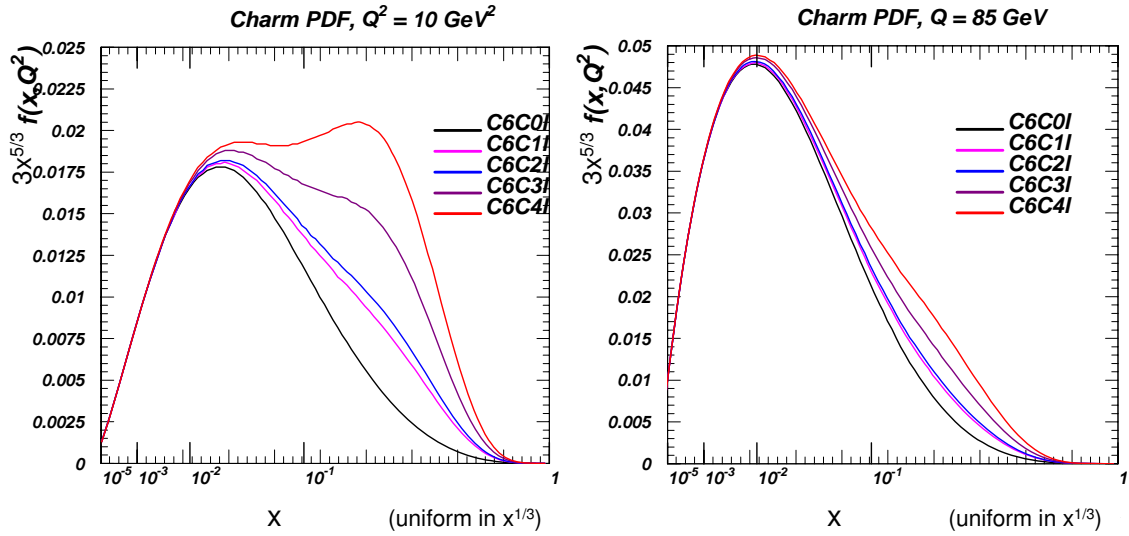}
\caption{Shape of charm distributions for the same series as those of the
previous plot, but at higher energy scale of (a) $Q^2=10 \mathrm{GeV}^2$; and
(b) $Q=80$ GeV---the W/Z mass range.} \label{fig:charm2}
\end{figure}

Figs.\,\ref{fig:charm1} and \ref{fig:charm2} show first results on the charm
degree of freedom in the parton structure of the nucleon obtained by the CTEQ
group. Two scenarios for the input charm distribution at $Q=m_{c}$ are
explored: (i) it has the same shape as the strange distribution
(\textquotedblleft sea-like\textquotedblright ); and (ii) it has a shape
suggested by many models of \textquotedblleft intrinsic
charm\textquotedblright\ based on lightcone wave-function arguments
\cite{Brodsky:1980pb,Brodsky:1981se}. \ Similar conclusions are obtained for
both scenarios, since existing experimental constraints are still relatively
loose. \ We reproduce here only the results of the intrinsic
charm scenario. %
Fig.\,\ref{fig:charm1}a shows the overall $\chi ^{2}$ of the global fit as a
function of the size of the input charm degree of freedom of the nucleon at
$Q=m_{c}$, as measured by the momentum fraction carried by the $c$-quark. We
see that, whereas the lowest $\chi ^{2}$ corresponds to a non-zero charm
fraction, the minimum is a very shallow one. \ By the commonly used tolerance
of $\Delta \chi ^{2}\sim 50-100$ for an acceptable global fit, this
analysis sets an upper limit on the fraction of intrinsic charm at the level
of $1.5-1.8\cdot 10^{-3}$. It is quite interesting that \emph{current global QCD
analysis can, indeed, place a reasonable upper limit on the charm content of
the nucleon}.

Fig.\,\ref{fig:charm1}b shows the shape of the charm distribution for the
series of input functions with increasing amount of charm inside the nucleon
in the intrinsic charm scenario. The horizontal axis scale is
$x^{1/3}$---intermediate between linear and logarithmic---in order to display
both large and small x behaviors clearly. The vertical axis scale is
$3x^{5/3}f(x,Q_0)$, so that the area under the curves is proportional to the
momentum fraction carried by the distribution.

Fig.\,\ref{fig:charm2} shows the shape of charm distributions for the same
series as those of the previous plot, but at higher energy scales. %
At $Q^2=10 \mathrm{GeV}^2$, we see clearly that $c(x,Q)$ has a two-component
form: a radiatively generated component peaking at small $x$; and the evolved
intrinsic component at higher $x$. At $Q=80 \mathrm{GeV}$---around the W/Z mass range,
the radiatively generated component is dominant, but the intrinsic component
can still be seen. The latter can have physically observable effects on
processes that are sensitive to charm in future collider studies, but it
would take dedicated efforts to uncover them.

\subsubsection*{Prospects}

The above results represent only the beginning of the exploration of the
heavy quark sector of the nucleon structure. They can then help set important
benchmarks for new measurements. \ On one hand, one can map out the range of
uncertainties of the predicted cross sections for the proposed measurements.
These are expected to be quite wide, given the paucity of existing
experimental constraints. \ On the other hand, by the same global analysis
tools, one can assume some measurement goals in terms of hoped for accuracy,
and determine how much improvement on our knowledge of the heavy flavor
parton distributions can result from such measurements if the goals can be
achieved. \ Such studies would provide valuable input to the planning of the
real measurements and the physics analysis of the results.

This effort requires close cooperation between theorists and experimentalists.
From the experimental side, it is important to assess the difficulties and
the opportunities. The following article \cite{Abulencia:2006ys} summarizes some of
the CDF measurements involving heavy quark production in the final state,
stating the present status of the analysis, the main sources of systematic
errors and possible improvements with larger statistics.

If the course laid out above is actively pursued at Tevatron Run II with
concerted effort by experimentalists and theorists, enough real progress
might be made to provide valuable input to the execution of Top/Higgs physics
studies at the LHC, as well as further improvements on the measurement of
heavy quark degrees of freedoms at the LHC itself.

\clearpage
\subsection{{Some Extrapolations of Tevatron Measurements and the Impact on 
Heavy Quark PDFs}}
\textbf{Contributed by:  Campanelli}

Most of the measurements at hadron colliders, in particular cross sections, 
are sensitive to parton distribution functions of the colliding protons.
The most uncertain PDF determinations are those referring to heavy quarks,
since very few measurements exist and the present estimation are coming from
NLO evolutions from the gluon PDF's. \par
In the following we summarize some of the CDF measurements involving
heavy quark production in the final state, stating the present status of the
analysis, the main sources of systematic errors and possible improvements
with larger statistics.\par 
When possible, the effect of changing PDF's has also been included, to be 
compared to the present and expected uncertainties. The evaluation of the PDF
effects follows two different approaches, depending on the analysis.
The most used method is to chose a set of PDF's, and vary by 1$\sigma$
errors the eigenvalues used to express them. Another approach is to consider
as an error the maximal variation of the cross section between a given set
of PDF's chosen as standard, and other available sets. Errors obtained using
the first method are more rigorous from the statistical point of view, while
the second is more conservative.\par
In general, sources of systematics common to all analyses are the error
on luminosity, on the energy scale and on b-tagging. Apart from the luminosity
error (that can be reduced using proper normalisation channels), these 
uncertainties are expected to be reduced with the accumulation of new data.
However, this improvement will involve a lot of work to achieve a better 
knowledge of the detector, and will most likely not scale as fast as the
statistical error. Eventually, even if the accumulation of more statistics
is the only way to reduce systematic errors, these are most like to end up 
dominate most of the measurements presented.
\subsubsection*{Heavy quarks and photons}
The present CDF analysis is using data taken with an unbiased photon trigger,
with a 25 GeV cut on the transverse photon energy, and no additional 
requirement on the rest of the event. Offline, the photon is required to
pass strong quality requirements, and an hadronic jet with 20 GeV transverse
energy, tagged as an heavy flavor. The jet tagging is based on finding a 
suitable secondary vertex from the tracks in the jet. It has an efficiency
of about 40\% (jet-\ET dependent) for beauty, few \% for charm and about
0.5\% for light quarks. However, given the relative abundance of these
jets, the tagged sample contains an approximately equal amont of these
three categories; to get the cross sections for production of beauty and charm,
a fit is performed on the invariant mass of the tracks that constitute the 
secondary vertex found.\par
\begin{figure}[!h]
\subfigure[Photon+charm]{\includegraphics[width=.5\textwidth]{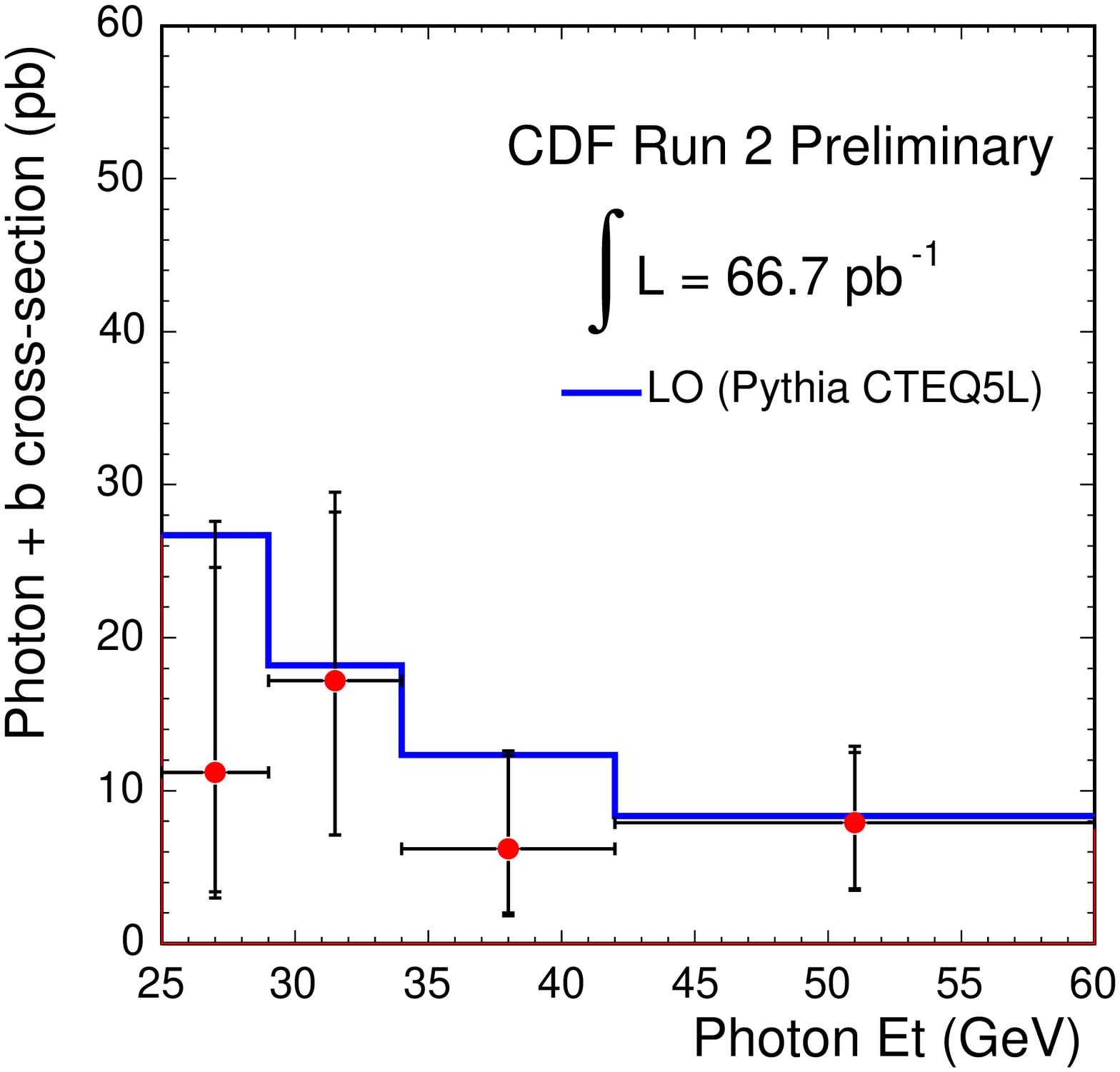}}
\subfigure[Photon+bottom]{\includegraphics[width=.5\textwidth]{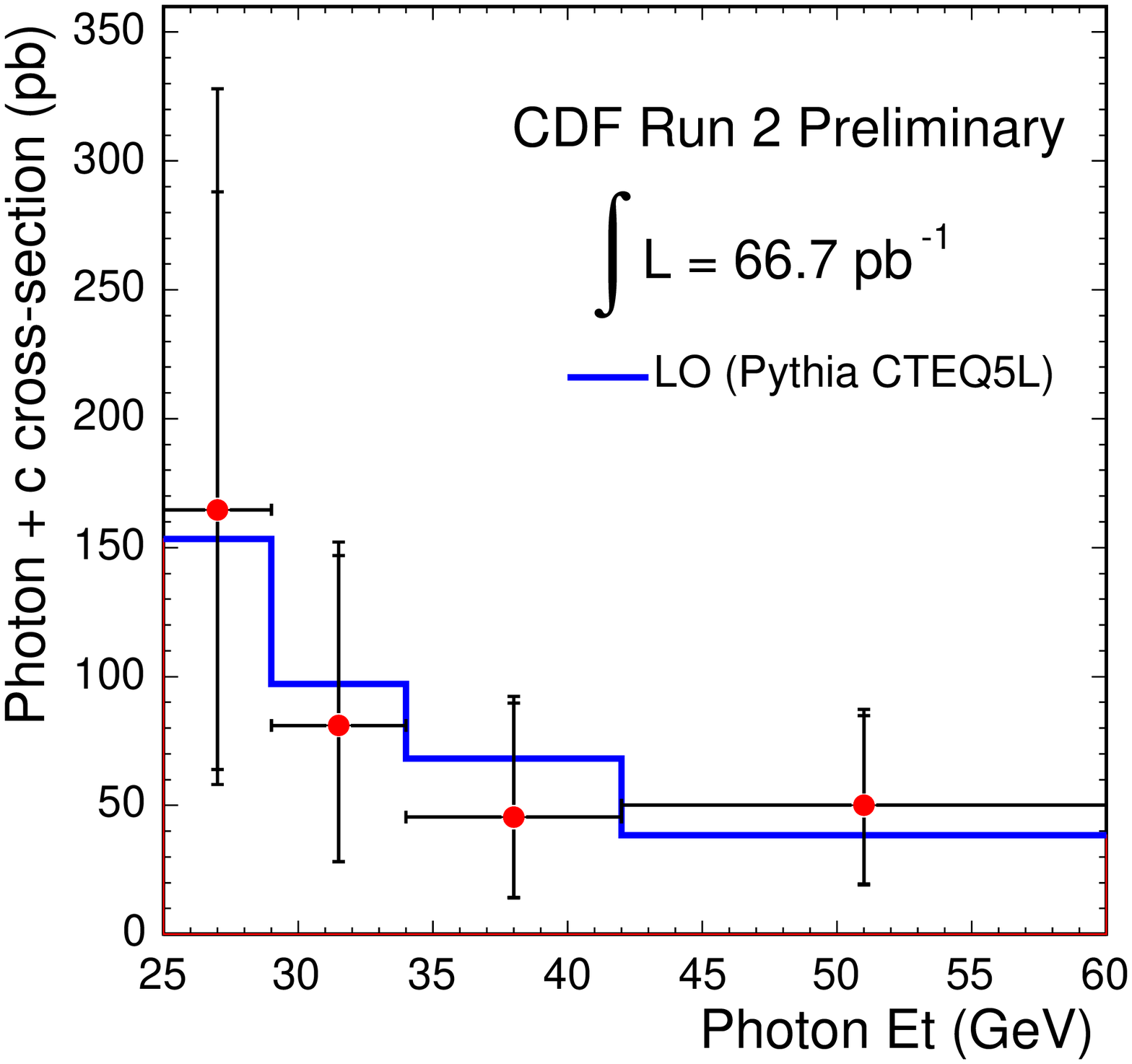}}
\caption{\label{fig:photet}Transverse energy distribution for photons in
events with a tagged heavy quark.  The data is compared to \PY.} 
\end{figure}

The present CDF analysis uses 67~\ipb~of data, and an update to higher
integrated luminosities is in preparation. The photon energy distribution for
selected events is shown in figure \ref{fig:photet}. The different corrections 
for charm and beauty events lead to the cross sections shown on the left and
right plots, respectively.

So far statistical error is larger than the systematics, but analyses on
datasets about ten times as large as those presented here are almost finished.
A table with statistical and systematic errors on the b-photon analysis is 
shown in \ref{tab:errbphot}; the errors for the c-photon case are very similar
in relative importance. 
\begin{table}
\centering
\begin{tabular}{|l|c|c|c|c|}\hline
\ET range (GeV)&25-29&29-34&34-42&42-60\\ \hline
Tag Efficiency&+1.7-1.3&+2.6-2.0&+0.9-0.7&+1.1-0.9\\
Photon Id&$\pm$0.2&$\pm$0.1&$<$0.1&$\pm$0.1\\
Jet correction&0.5&+0.5&+0.1&+0.1\\
Jet energy scale&+3.3-1.4&+2.2-2.1&+0.5-0.3&+0.5-0.4\\
B jet correction&$\pm$0.2&$\pm$0.3&$\pm$0.1&$\pm$0.1\\
CPR fake estimate&+0.1&$<$0.1&$<0.1$&$<0.1$\\
trigger&+2.5-1.7&$<0.1$&$<0.1$&$<0.1$\\
luminosity&$+0.7-0.6$&$+1.1-1.0$&$+0.4-0.3$&$+0.5-0.4$\\ \hline
PDF&$\pm$0.3&$\pm$0.5&$\pm$0.2&$\pm$0.2\\ \hline
Statistical&11.2&17.2&6.2&7.9\\ \hline
Systematics&+16.4-8.2&+12.3-10.1&+6.4-4.4&+5.0-4.1\\ \hline
\end{tabular}
\caption{\label{tab:errbphot}Sources of systematic errors compared to the
statistical one for the b-photon channel with a luminosity of 67~\ipb.}
\end{table}

The dominant systematic errors are the jet energy scale and the tagging 
efficiency; both of them are expected to decrease with luminosity, albeit
not as quick as the statistical error; it is therefore likely that
this measurement will start to be systematics dominated. 
Moreover, while  the statistical errors in the various bins are uncorrelated, 
the effect of a change in PDF's is likely to be a simultaneous shift of 
all bins in the same direction, so the biggest obstacle to PDF's determination
will be global effects like enregy scale, $b$-tagging efficiency and luminosity.
Although they can be certainly be controlled with a precision at least a 
factor of 2-3 better than the present analysis, from the numbers in the
table it is not likely that their precision can be better than the effect
of varying the PDF's within present limits, indicated as the last source
of systematics. This measurement will probably not allow a direct 
determination of the PDF's, however it will provade an extremely valid 
cross-check of the latters, that so far have only been indirectly derived 
from the gluon distribution.
Another experimental approach being pursued by CDF on this measurement is 
the use of a dataset with a lower threshold on the photon at trigger level
(12 GeV), but the requirement for a track with impact parameter measured
on-line. This study will allow adding more high-statistics low-\ET
bins to the measurement, however the question remains if the trigger
efficiencies will be understood at a sufficient level to reach the precision
envisaged to observe effects due to PDF's.

\subsubsection*{Heavy quarks and $Z$ bosons}
\begin{figure}[!h]
\subfigure[Invariant mass of the tracks composing the
secondary vertex]{\includegraphics[width=.5\textwidth]{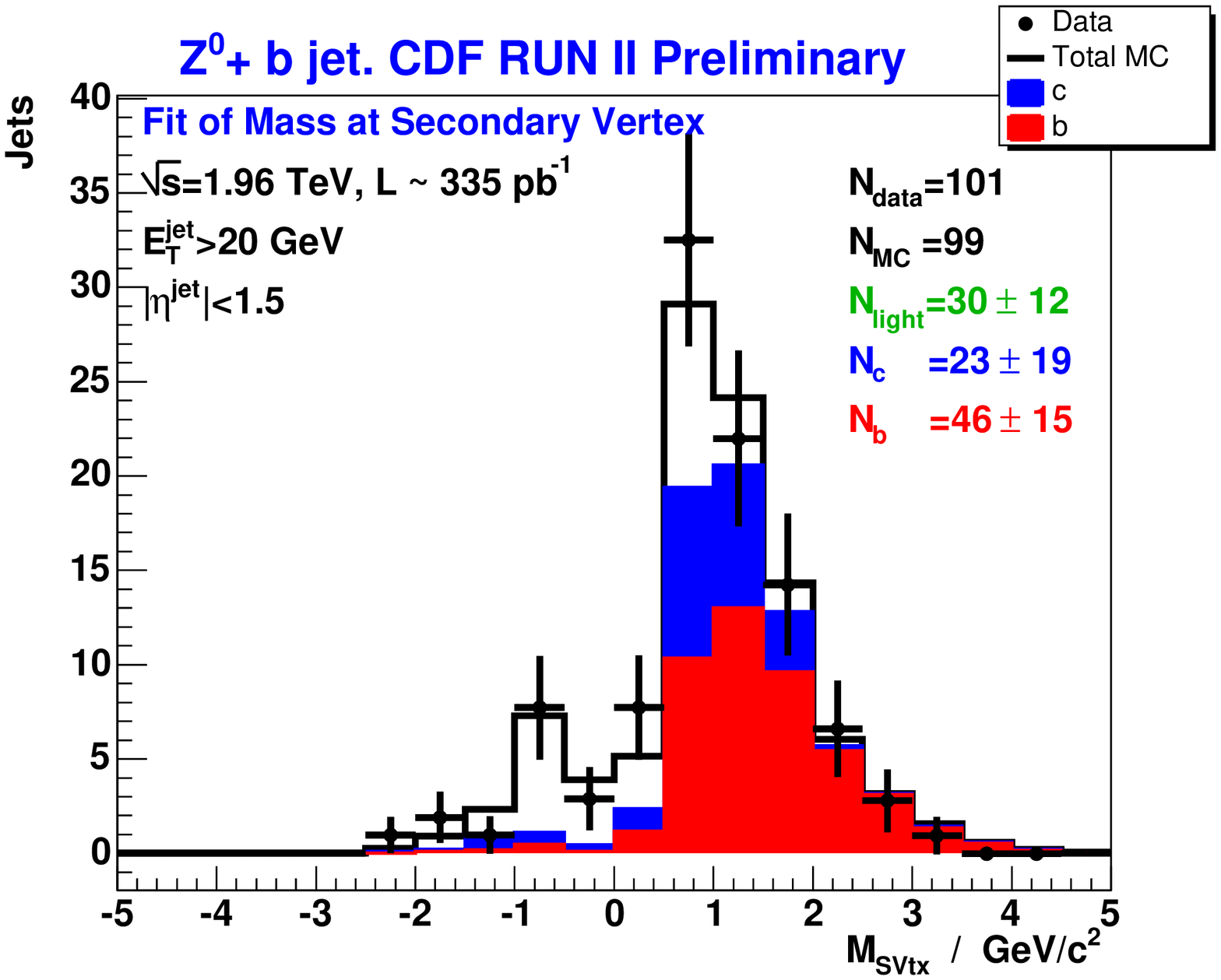}}
\subfigure[$\eta$ distribution of the tagged jets]{\includegraphics[width=.5\textwidth]{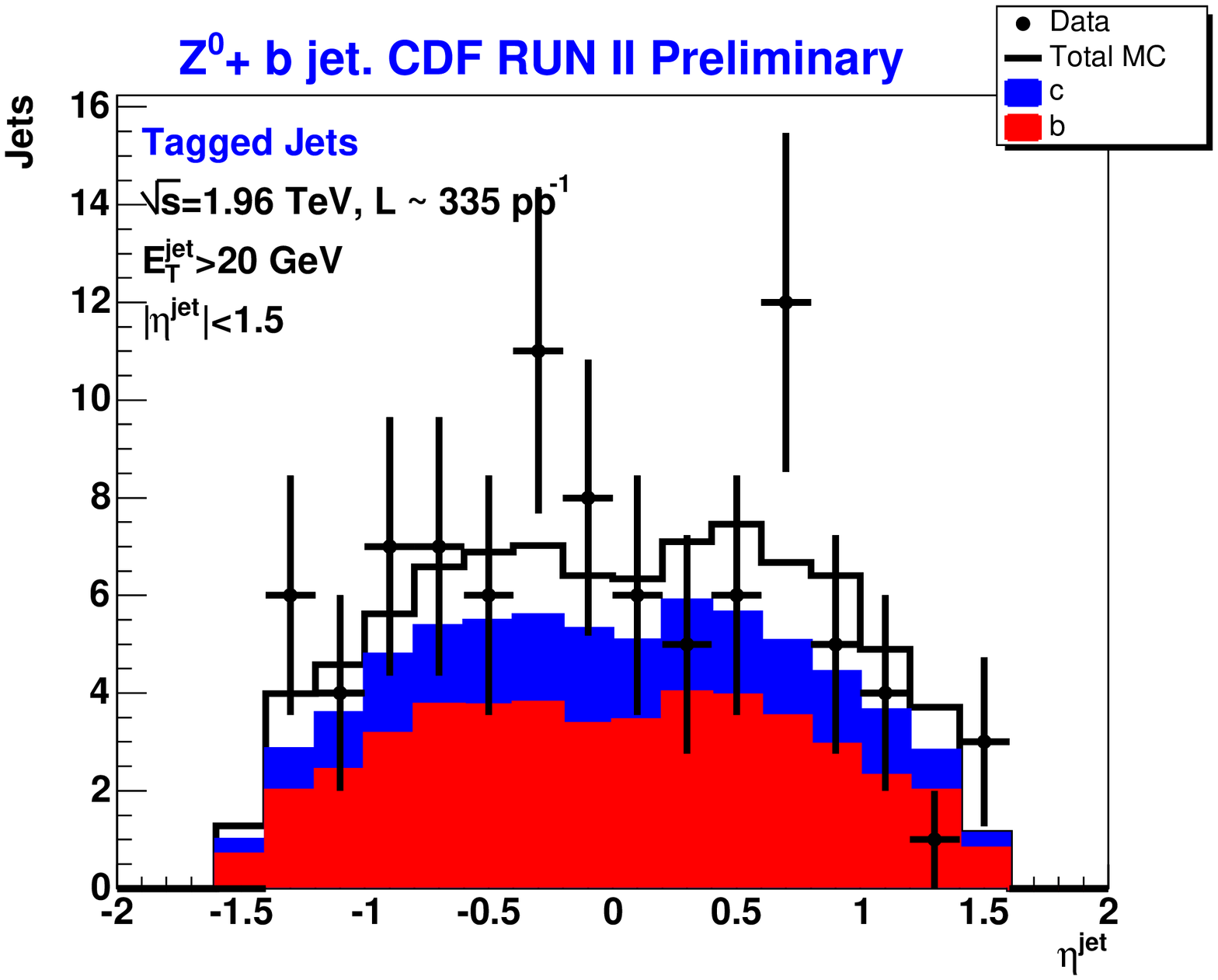}}
\caption{\label{fig:bzed}Distributions for $Zb$ events}
\end{figure}

The production of beauty and charm in association with a $Z$ boson decaying
into electrons and muons is presently measured in CDF for an integrated
luminosity of about 340~\ipb. We require two opposite-charge leptons
to lie inside a $Z$ mass window, and a jet tagged as an heavy quark, with the
same $b$-tagging tagging as the previous analysis. The leptonic $Z$ channel (without
$b$-tagging) is used as a normalization channel, to account from trigger and detector
effects directly from data.
The separate contributions from beauty, charm and light quarks are extracted,
similarly to the previous analysis, from a fit to the vertex mass of the tagged jet 
(figure \ref{fig:bzed}, left). The $\eta$ distribution of the selected quarks is shown in
figure \ref{fig:bzed}, right.
The preliminary measured cross sections and branching fractions 
have presently a statistical
error of about 30\%, and a systematic error about half this value.

We can assume that systematic errors will end up being around 10\%,and that
statistical errors of the same order of magnitude will be obtained with
data already available. A further improvement towards a total (statistics +
systematics) error of the order of 10-15\% could be envisaged for the final
Tevatron dataset.
\subsubsection*{Heavy quarks and $W$ bosons}
\begin{figure}[!h]
\begin{center}
\includegraphics[width=.95\textwidth]{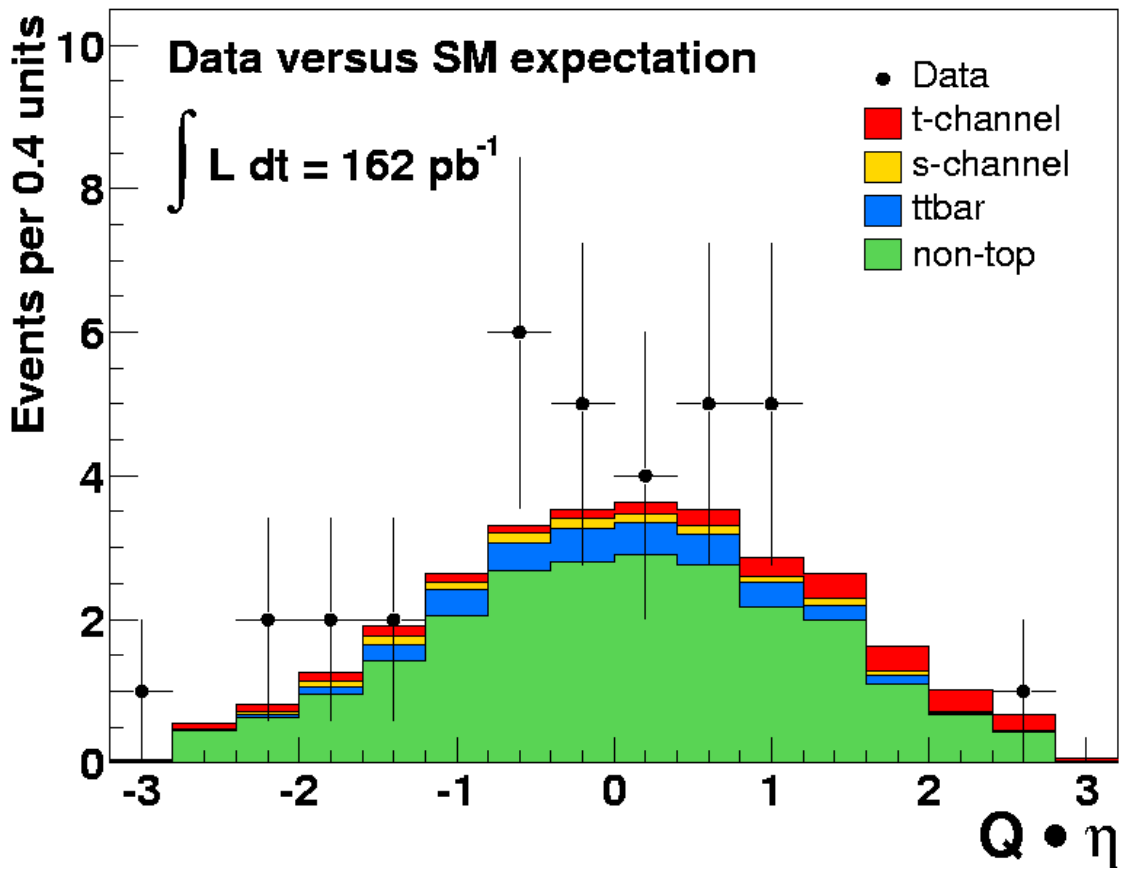}
\end{center}
\caption{\label{fig:eta} $\eta$ of the reconstructed top candidate times its
charge. This variable allows discrimination between $s$- and $t-$ channel production.}
\end{figure}

The signature of a b quark and a $W$ boson is characteristic of single top 
production, a signal long sought after since Run I. The final state searched 
for is an electron or muon plus missing energy, compatible with a W bosons, plus at
least one b-tagged hadronic jet. The latest CDF publication
\cite{Acosta:2004bs} uses a data sample corresponding to an integrated luminosity of
162 pb$^{-1}$, and puts a 95\% C.L. of 17.8 pb for the combined cross section
of s- and t-channel. As expected, from the $\eta$ distribution in figure \ref{fig:eta},
most of the observed data comes from QCD $W + b/c$ production.

Table \ref{tab:syswb} shows the main sources of systematic uncertainty on single
top production. The PDF error is the cross section difference between the 
``standard'' set used in the analysis (CTEQ5L) and the one leading to the 
largest variation (MRST72). Using this conservative method, differences can 
be relevant even with the present limited statistics. Moreover, since the
PDF influence is different for the s- and for the t-channel, the
rapidity distribution, shown above, can yield additional information with
respect to the simple cross section measurement.

\begin{table}
\centering
\begin{tabular}{|l|c|}\hline
Source&Syst. error (\%)\\ \hline
Energy Scale&+0.1-4.3\\
Initial State Radiation&$\pm$1.0\\
Final State Radiation&$\pm$2.6\\
Generator&$\pm$3\\
Top quark mass&-4.4\\
Trigger, lepton ID, Lumi&$\pm$9.8\\ \hline
PDF&$\pm$3.8\\ \hline
\end{tabular}
\caption{\label{tab:syswb}Sources of systematic errors for the single top search
($Wb$ measurement)}
\end{table}

\subsubsection*{Inclusive $b$ cross section}
\begin{figure}[!h]
\begin{center}
\includegraphics[width=\textwidth]{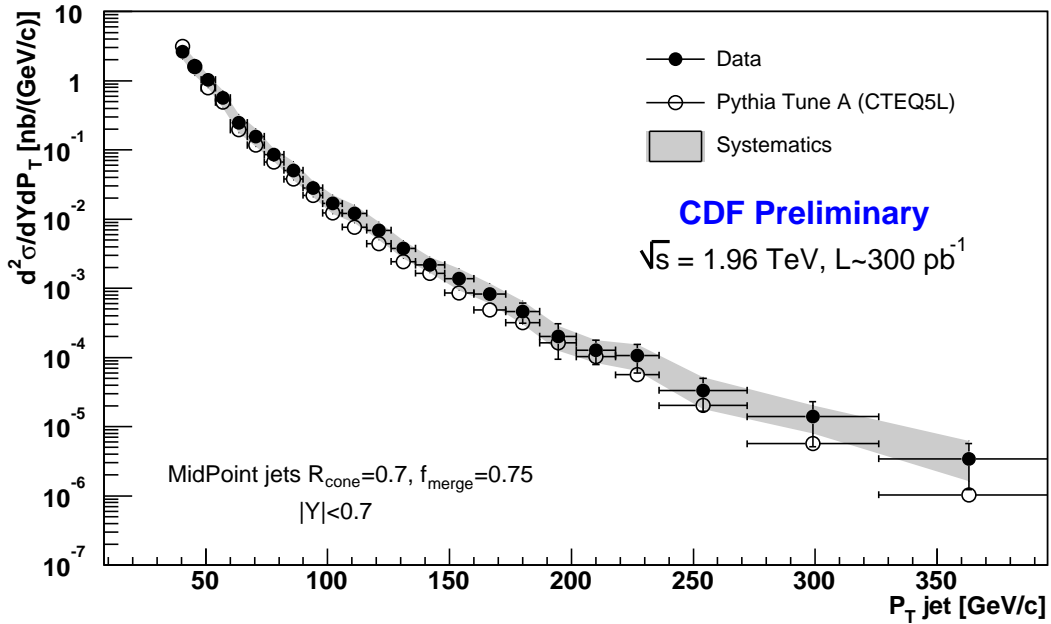}
\end{center}
\caption{\label{fig:binclus} Cross section for inclusive $b$ production for a
luminosity of 300 pb$^{-1}$, compared to \PY Tune A predictions}
\end{figure}

This measurement requires the presence of a tagged hadronic jet in the event,
collected with a series of prescaled triggers with cuts on rising values of
the jet transverse energy. A vertex mass method is used to extract the $b$
fraction, and corrections for the b-tagging efficiency and jet energy scale
are applied. This measurement, performed on an integrated luminosity of
300 pb$^{-1}$ covers a jet \PT range between 38 and 400 GeV,
where the cross section spans over six orders of magnitude. The resulting
cross section is shwn in figure \ref{fig:binclus}.\par
The main systematics for this measurement are summarised in table 
\ref{tab:syswb2}; since systematics are individually computed for each \PT bin,
in the table only an indicative value for the low-\PT and the high-\PT ends
of the spectrum are given.\par

\begin{table}
\centering
\begin{tabular}{|l|c|c|}\hline
Source&Syst. low-\PT(\%)&Syst. high-\PT(\%)\\ \hline
Energy Scale&+10-8&+39-22\\
Energy resolution&$\pm$6&$\pm$6\\
Unfolding&$\pm$5&$\pm$15\\
b fraction&+14-15&+47-50\\
b-tagging eff.&$\pm$7&$\pm$7\\
Luminosity&$\pm$6&$\pm$6\\ \hline
PDF&$\pm$7&$\pm$20\\ \hline
\end{tabular}
\caption{\label{tab:syswb2}Sources of systematic errors for the inclusive $b$ cross section
}
\end{table}
We see that the jet energy scale and the calculation of the b fraction largely dominate
the error, and they increase at high-\PT, where statistics of the control samples is
scarcer. More data can certainly improve these errors, possibly by a factor of 2 in the
low-\PT and intermediate region, and more in the high-energy region. A global fit of
the \PT spectrum, and of the angular distribution will e needed to extract most of the
information about PDF's.
\subsubsection*{Conclusions}
We highlighted some of the $b$ production measurements recently performed in CDF, and
their sensitivity to PDF's measurements. With present measurements we are still far
from observing effects due to uncertainties in PDF's in CDF data. A lot of work will
be needed to reduce the systematic uncertainties, especially those relative to the
jet energy scale and the $b$-tagging efficiency and purity. For doing that, the largest
possible control samples are needed, so these measurements will benefit from as much
data as possible. Even if they will end up being limited by systematics, the only way
to reduce this systematics will be to accumulate more statistics. In any case, even
if it will turn out that none of these single measurements will alone be able to
constraint present errors on PDF models, they will constitute a fundamental direct
cross-check of the validity of these distributions, so far only derived by QCD 
calculations.

\clearpage
\subsection{Issues of QCD
Evolution and Mass Thresholds in Variable Flavor Schemes and
their Impact on Higgs Production in Association with Heavy Quarks}

\textbf{Contributed by: B. Field, Olness, Smith}

{\em We examine some general issues regarding Parton Distribution Functions
(PDF's) involving different numbers of heavy flavors. Specifically we
compare the differences and similarities between $3$, $4$, and $5$
flavor number scheme PDF's for lowest order (LO), next-to-leading order
(NLO), and next-to-next-to-leading order (NNLO) evolutions.  We look
at the implications for these different schemes and orders of
perturbation expansions, and also study the matching conditions at the
threshold. We use the \textsc{cteq}6 data fit as our starting point in
comparing the different evolutions.}

The LHC will span an unprecedented range of energies and open up new
kinematic regions for exploration and discovery. This large energy
range poses a theoretical challenge as we encounter multi-scale processes
involving many distinct energy/momentum scales. Understanding the
data that is to come from the Large Hadron Collider (LHC) in the coming
years will require a detailed and precise understanding of parton
distribution functions through next-to-next-to-leading order (NNLO). 

As we move through these different energy ranges, one can consider
implementing PDF's with varying number of flavors $N_{F}$. If our
experiments were confined to a limited energy range, a single $N_{F}$
flavor scheme would be adequate; however, since the energy scales
are widely varying, it becomes necessary to use a series of different
$N_{F}$ schemes ($N_{F}=\{3,4,5\}$) to accurately describe the entire
energy range. The question of how and where to {}``join'' these
schemes also raises significant issues
\footnote{The recent paper by Thorne\cite{Thorne:2006qt} considers
some of these issues and choices, particularly the issues that arise
at NNLO.}. While it is common to join these schemes at an energy scale
equal to the quark mass (i.e., $\mu=m_{c,b}$) since the PDF's will be
continuous (but with discontinuous derivatives) at LO and NLO with
this matching condition, we see this property is no longer true at
NNLO.

\subsubsection*{Issues of $N_{F}$ Flavor Schemes}

We can illustrate some of these issues by considering the example of
Higgs production via b-quarks in the kinematic range where the b-quark
mass is too heavy to be ignored, but too light to be decoupled from
the hadron dynamics. To work towards a solution which is valid
throughout the full energy spectrum, we start by focusing on the
asymptotic high and low energy regions (where the issues are simpler),
and then try and make these disparate regions match in the (more
difficult) intermediate region.

For example, if the characteristic energy range $\mu$ is small
compared to the b-quark mass $m_{b}$, then the b-quark decouples from
the dynamics and \textit{does not} appear as a partonic constituent of
the hadron; that is, $f_{b}(\mu<m_{b})=0$ and we are working in a
$N_{F}=4$ flavor scheme. In such a scheme, the Higgs is produced in
the ${\mathcal{O}}(\alpha_{s}^{2})$ process $gg\rightarrow H$ with
$b$-pairs in the intermediate state. Calculations in the $N_{F}=4$
flavor scheme have the advantage that they do not need to introduce
the b-quark PDF. If instead we consider energy scales much larger than
the b-quark mass ($\mu\gg m_{b}$), then we work in a $N_{F}=5$ flavor
scheme where the b-quark \textit{does} appear as a partonic
constituent of the hadron {[}$f_{b}(\mu>m_{b})>0${]}. In this regime,
the b-quark mass scale enters as powers of
$\alpha_{s}\ln(\mu^{2}/m_{b}^{2})$ which are resummed via the DGLAP
equations. This scheme has the advantage that they involve lower-order
Feynman graphs, and the $\alpha_{s}\ln(\mu^{2}/m_{b}^{2})$ terms are
resummed. Ideally, there is an intermediate region where both the
4-flavor and 5-flavor schemes are both a good representation of the
physics; in this region we can transition from the low energy 4-flavor
scheme to the high energy 5-flavor scheme thereby obtaining a
description of the physics which is valid throughout the entire energy
range from low to high scales\footnote{We label the 4-flavor and
5-flavor schemes as {}``fixed-flavor-number'' (FFN) schemes since the
number of partons flavors is fixed. The hybrid scheme which combines
these FFS is a {}``variable-flavor-number'' (VFN) scheme since it
transitions from a 4-flavor scheme at low energy to a 5-flavor scheme
at high energy\cite{Aivazis:1993pi,Aivazis:1993kh}.}.

In this report, we will focus on the different PDF's which result from
different orders of evolution (LO, NLO, NNLO) and different numbers of
active flavors ($N_{F}=\{3,4,5\}$).

\subsubsection*{Generation of PDF Sets}

For the purposes of this study, we will start from a given set of PDF's
$f(x,Q_{0})$ at an initial scale $Q_{0}<m_{c}$. We will then evolve
the PDF's from this point and study the effect of the number of active
heavy flavors $N_{f}=\{3,4,5\}$, as well as the order of the
evolution: \{LO, NLO, NNLO\}. No fitting is involved here; the
resulting PDF's are designed to such that they are all related (within
their specific $N_{F}$-scheme and order of evolution) to be related to
the same initial PDF, $f(x,Q_{0})$. In this sense, our comparisons
will be focused on comparing schemes and evolution, rather than
finding accurate fits to data. Were we able to perform an all-orders
calculation, the choice of the number of active heavy flavors
$N_{f}=\{3,4,5\}$would be equivalent; however, since we necessarily
must truncate the perturbation expansion at a finite order, there will
be differences and some choices may converge better than others.

For our initial PDF, $f(x,Q_{0})$, we chose the \textsc{cteq}6
parametrization as given in Appendix A of
Ref.~\cite{Pumplin:2002vw}. Using the evolution program described in
Ref.~\cite{Chuvakin:2001ge}, we created several PDF tables for our
study. Essentially, we explored two-dimensions: 1) the number of
active heavy flavors $N_{f}=\{3,4,5\}$, and 2) the order of the
evolution: \{LO, NLO, NNLO\}; each of these changes effected the
resulting PDF. All the sets were defined to be equivalent at the
initial scale of $Q_{0}=m_{c}=1.3$~GeV. For the $N_{F}=3$ set, the
charm and bottom quarks are never introduced regardless of the energy
scale $\mu$. The $N_{F}=4$ set begins when the charm quark is
introduced at $\mu=m_{c}=1.3$~GeV. The $N_{F}=5$ set begins when the bottom
quark is introduced at $\mu=m_{b}=5$~GeV.

\subsubsection*{Technical Issues:}

Before we proceed to examine the calculations, let's briefly address
two technical issues.

When we evolve the b-quark PDF in the context of the DGLAP evolution
equation $df_{b}\sim P_{b/i}\otimes f_{i}$, we have the option to use
splitting kernels which are either mass-dependent
{[}$P_{b/i}(m_{b}\not=0)${]} or mass-independent
{[}$P_{b/i}(m_{b}=0)${]}. While one might assume that using
$P_{b/i}(m_{b}\not=0)$ yields more accurate results, this is not the
case. The choice of $P_{b/i}(m_{b}\not=0)$ or $P_{b/i}(m_{b}=0)$ is
simply a choice of scheme, and both schemes yield identical results up
to high-order corrections\cite{Olness:1997yn}. For simplicity, it is
common to use the mass-independent scheme since the $P_{b/i}(m_{b}=0)$
coincide with the $\overline{\textrm{MS}}$ kernels.

When the factorization proof of the ACOT scheme was extended to
include massive quarks, it was realized that fermion lines with an
initial or internal {}``cut'' could be taken as
massless\cite{Kramer:2000hn}. This simplification, referred to as the
simplified-ACOT (S-ACOT) scheme, is \textit{not} an approximation; it
is again only a choice of scheme, and both the results of the ACOT and
S-ACOT schemes are identical up to high-order
corrections\cite{Collins:1998rz}. The S-ACOT scheme can lead to
significant technical simplifications by allowing us to ignore the
heavy quark masses in many of the individual Feynman diagrams.  We
show how we exploit this feature in the case of NNLO calculation of
$b\bar{b}\rightarrow H$ in the next section.

\subsubsection*{Consistency Checks}

We first recreated the published \textsc{cteq}6 table to check our
evolution program and found excellent agreement. The evolution program
was also checked against the output described and cataloged in
Ref.~\cite{Giele:2002hx} and was found to be in excellent agreement
(generally five decimal places) for all three orders when run with the
same inputs\footnote{The NNLO results presented here and in
Ref.~\cite{Giele:2002hx} used an approximate form for the three-loop
splitting functions since the exact results were not available when
the original programs were produced\cite{Chuvakin:2001ge}. A
comparison of the NNLO splitting functions finds the approximate quark
distributions underestimate the exact results by at most a few percent
at small x ($x<10^{-3}$), and overestimate the gluon distributions by
about half a percent for $\mu=100$~GeV\cite{Dittmar:2005ed}. This
accuracy is sufficient for our preliminary study; the evolution
program is being updated to include the exact NNLO kernels.}.

\subsubsection*{Matching Conditions}

A common choice for the matching between $N_{F}$ and $N_{F+1}$ schemes
is to perform the transition at $\mu=m$. To be specific, let us
consider the transition between $N_{F}=3$ and $N_{F}=4$ flavors at
$\mu=m_{c}$.  If we focus on the charm ($f_{c}$) and gluon ($f_{g}$)
PDF's, the boundary conditions at NNLO can be written schematically
as\footnote{Here we use the short-hand notation $f^{N_{F}}$ for the
$N_{F}$ flavor PDF.}:
\[
f_{c}^{4}\sim f_{g}^{3}\otimes\left\{
0+\left(\frac{\alpha_{s}}{2\pi}\right)\, P_{g\rightarrow
q}^{(1)}\left(L+a_{g\rightarrow
q}^{1}\right)+\left(\frac{\alpha_{s}}{2\pi}\right)^{2}\,
P_{g\rightarrow q}^{(2)}\left(L^{2}+L+a_{g\rightarrow
q}^{2}\right)+O(\alpha_{s}^{3})\right\} 
\]

\[
f_{g}^{4}\sim f_{g}^{3}\otimes\left\{
1+\left(\frac{\alpha_{s}}{2\pi}\right)\, P_{g\rightarrow
g}^{(1)}\left(L+a_{g\rightarrow
g}^{1}\right)+\left(\frac{\alpha_{s}}{2\pi}\right)^{2}\,
P_{g\rightarrow g}^{(2)}\left(L^{2}+L+a_{g\rightarrow
g}^{2}\right)+O(\alpha_{s}^{3})\right\}
\]
where $L=\ln(\mu^{2}/m_{q}^{2})$ and $m_{q}=m_{c}$. Because the terms
$L=\ln(\mu^{2}/m_{q}^{2})$ vanish when $\mu=m$, the above conditions
are particularly simple at this point.

An explicit calculation shows that $a_{g\rightarrow q}^{1}=0$ and
$a_{g\rightarrow g}^{1}=0$. Consequently, if we perform the matching
at $\mu=m$ where $L=0$, we have the continuity condition
$f_{c}^{4}(x,\mu=m_{c})=f_{c}^{3}(x,\mu=m_{c})=0$ and
$f_{g}^{4}(x,\mu=m_{c})=f_{g}^{3}(x,\mu=m_{c})$. Therefore, the PDF's
will be continuous at LO and NLO.

This is no longer the case at NNLO. Specifically, the ${\cal
O}(\alpha_{s}^{2})$ coefficients $a_{g\rightarrow q}^{2}$ and
$a_{g\rightarrow g}^{2}$ have been calculated in
Ref.~\cite{Buza:1995ie} and found to be non-zero. Therefore we
necessarily will have a discontinuity no matter where we choose the
matching between $N_{3}$ and $N_{4}$ schemes; $\mu=m_{c}$ is no longer
a {}``special'' transition point. This NNLO discontinuity changes the
boundary value of the differential equations that govern the evolution
of the partons densities, thus changing the distributions at all
energy levels; these effects then propagate up to higher scales.

It is interesting to note that there are similar discontinuities in
the fragmentation function appearing at NLO. For example the NLO heavy
quark fragmentation function first calculated by Nason and
Mele\cite{Mele:1990yq}
\[
d_{c\to c}\sim\left\{
\delta(1-x)+\left(\frac{\alpha_{s}}{2\pi}\right)\, P_{c\rightarrow
c}^{(1)}\left(L+a_{c\to c}^{1}\right)+O(\alpha_{s}^{2})\right\} 
\]
found the $a_{c\to c}^{1}$ coefficient was non-zero. Additionally, we
note that $\alpha_{S}(\mu,N_{F})$ is discontinuous across flavor
thresholds at order $\alpha_{S}^{3}$\cite{Eidelman:2004wy}:
\[
\alpha_{S}(m;N_{f})=\alpha_{S}(m;N_{f}-1)-\frac{11}{72\pi^{2}}\,
\alpha_{S}^{3}(m;N_{f}-1)+O\left(\alpha_{S}^{4}(m;N_{f}-1)\right)
\]
Note that the NNLO matching conditions on the running coupling
$\alpha_{s}(N_{F},Q^{2})$ as $Q^{2}$ increases across heavy-flavor
flavor thresholds have been calculated in \cite{Bernreuther:1981sg,
Bernreuther:1983zp} and \cite{Larin:1994va, Chetyrkin:1997sg}.

\subsubsection*{Comparison of 3,4, and 5 Flavor Schemes}

\begin{figure}[h]
\centering
\subfigure[]{\includegraphics[width=0.45\linewidth]{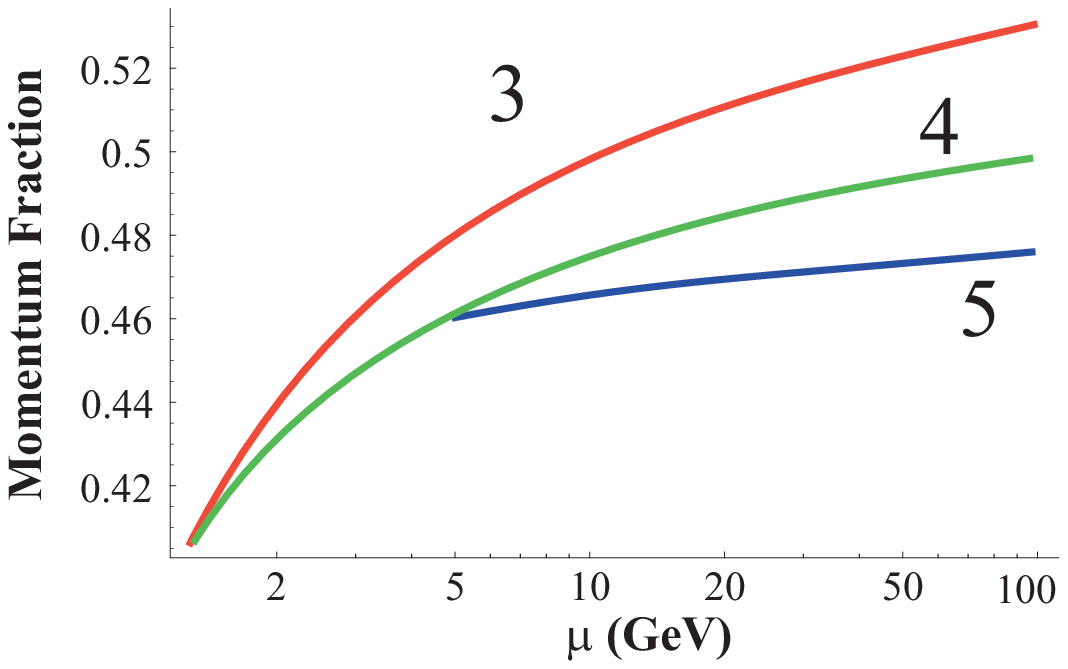}}
\subfigure[]{\includegraphics[width=0.45\linewidth]{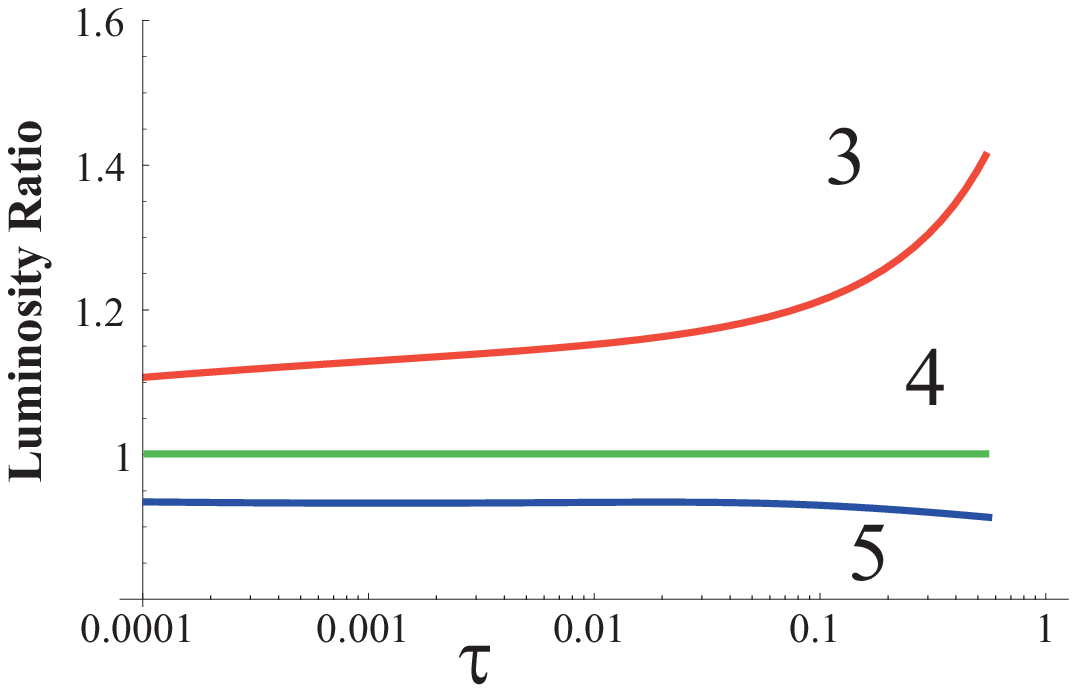}}
\caption{(a) Integrated momentum fraction, $\int_{0}^{1}xf_{g}(x,\mu)\,
dx$ vs. $\mu$ of the gluon for $N_{F}=\{3,4,5\}$ = \{Red, Green,
Blue\}. (b) The ratio of the gluon-gluon luminosity
($d{\mathcal{L}}_{gg}/d\tau$) vs. $\tau$ for $N_{F}=\{3,4,5\}$ =
\{Red, Green, Blue\} as compared with $N_{F}=4$ at $\mu=120$~GeV.}
\label{fig:fredi}
\end{figure}
\begin{figure}
\begin{center}
\includegraphics[width=0.45\linewidth]{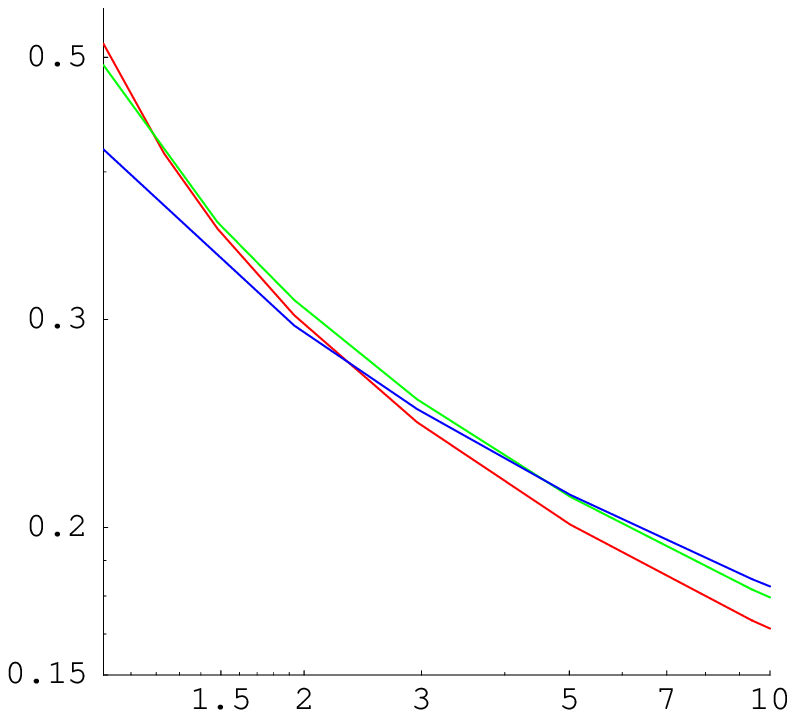}
\includegraphics[width=0.45\linewidth]{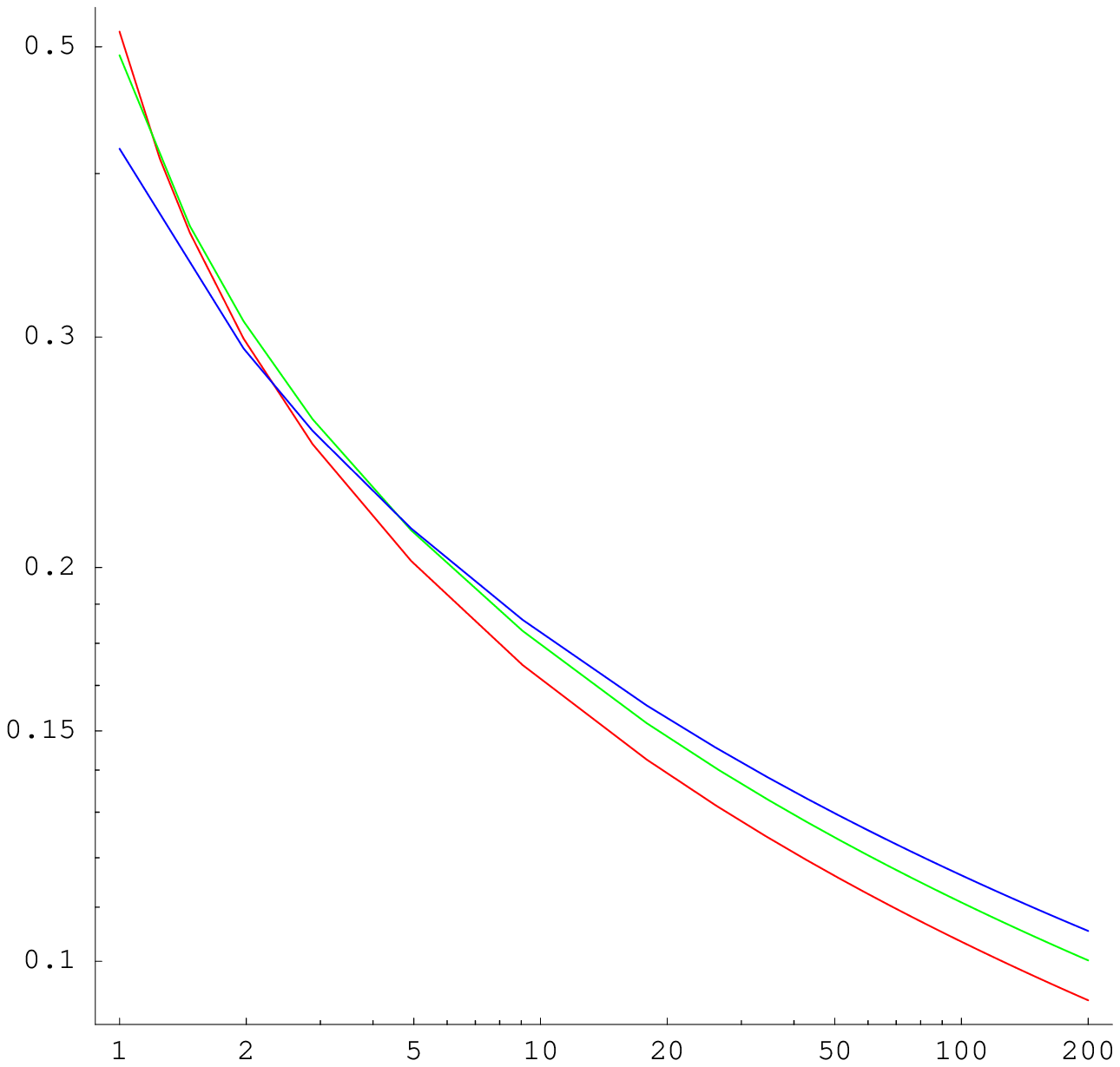}
\end{center}
\caption{$\alpha_{s}$ vs. $\mu$ (in GeV) for $N_{F}=\{3,4,5\}$ (for
large $Q$, reading bottom to top: red, green, blue, respectively)
flavors. Fig. a) illustrates the region where $\mu$ is comparable to
the quark masses to highlight the continuity of $\alpha_{S}$ across the
mass thresholds at $m_c=1.3$~GeV and  $m_b=5$~GeV. 
Fig. b) extrapolates this to larger $\mu$ scale. }
\label{fig:alphas}
\end{figure}

 To illustrate how the active number of {}``heavy'' flavors affects
the {}``light'' partons, in Fig.~\ref{fig:fredi}a) we show the
momentum fraction of the gluon vs. $\mu$. We have started with a
single PDF set at $\mu=1.3$~GeV, and evolved from this scale invoking
the {}``heavy'' flavor thresholds as appropriate for the specified
number of flavors. While all three PDF sets start with the same
initial momentum fraction, once we go above the charm threshold
($m_{c}=1.3$~GeV) the $N_{F}=\{4,5\}$ gluon momentum fractions are
depleted by the onset of a charm quark density. In a similar fashion,
the gluon momentum fraction for $N_{F}=5$ is depleted compared to
$N_{F}=4$ by the onset of a bottom quark density above the bottom
threshold ($m_{b}=5$~GeV).

To gauge the effect of the different number of flavors on the cross
section, we compute the gluon-gluon luminosity which is defined as
$d{\mathcal{L}}_{gg}/d\tau=f_{g}\otimes f_{g}$. We choose a scale of
$\mu=120$~GeV which is characteristic of a light Higgs. In terms of
the luminosity, the cross section is given as
$d\sigma/d\tau\sim[d{\mathcal{L}}_{gg}/d\tau]\,[\hat{\sigma}(\hat{s}=\tau
s)]$ with $\tau=\hat{s}/s=x_{1}x_{2}$.

To highlight the effect of the different $N_{F}$ PDF's, we plot the
ratio of the luminosity as compared to the $N_{F}=4$ case,
\textit{c.f.}, Fig.~\ref{fig:fredi}b). We see that the effects of
Fig.~\ref{fig:fredi}a) are effectively squared (as
expected---$f_{g}\otimes f_{g}$) when examining the thin lines of
Fig.~\ref{fig:fredi}b).

However, this is not the entire story. Since we are interested in
$gg\rightarrow H$ which is an $\alpha_{s}^{2}$ process, we must also
take this factor into account. Therefore we display
$\alpha_{s}^{2}(\mu,N_{F})$ computed at NLO for $N_{f}=\{3,4,5\}$ as a
function of $\mu$ in Fig.~\ref{fig:alphas}. Note at this order,
$\alpha_{s}^{2}(\mu,N_{F})$ is continuous across flavor
boundaries. Fig.~\ref{fig:alphas} explicitly shows that
$\alpha_{s}(m_{c},3)=\alpha_{s}(m_{c},4)$ and
$\alpha_{s}(m_{b},4)=\alpha_{s}(m_{b},5)$.  Comparing
Figs.~\ref{fig:fredi} and~\ref{fig:alphas} we observe that the
combination of the $N_{F}$ and $\alpha_{s}$ effects tend to compensate
each other thereby reducing the difference. While these simple
qualitative calculations give us a general idea how the actual cross
sections might vary, a full analysis of these effects is required to
properly balance all the competing factors. However, there are
additional considerations when choosing the active number of flavors,
as we will highlight in the next section.

\subsubsection*{Resummation}

\begin{figure}
\begin{center}
\includegraphics[width=0.45\linewidth]{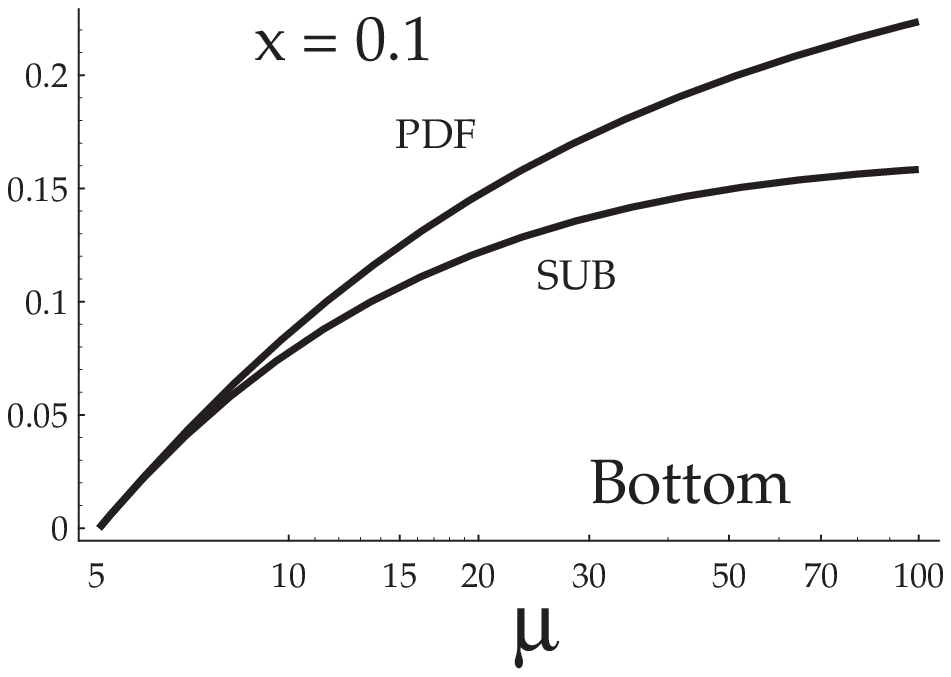}
\includegraphics[width=0.45\linewidth]{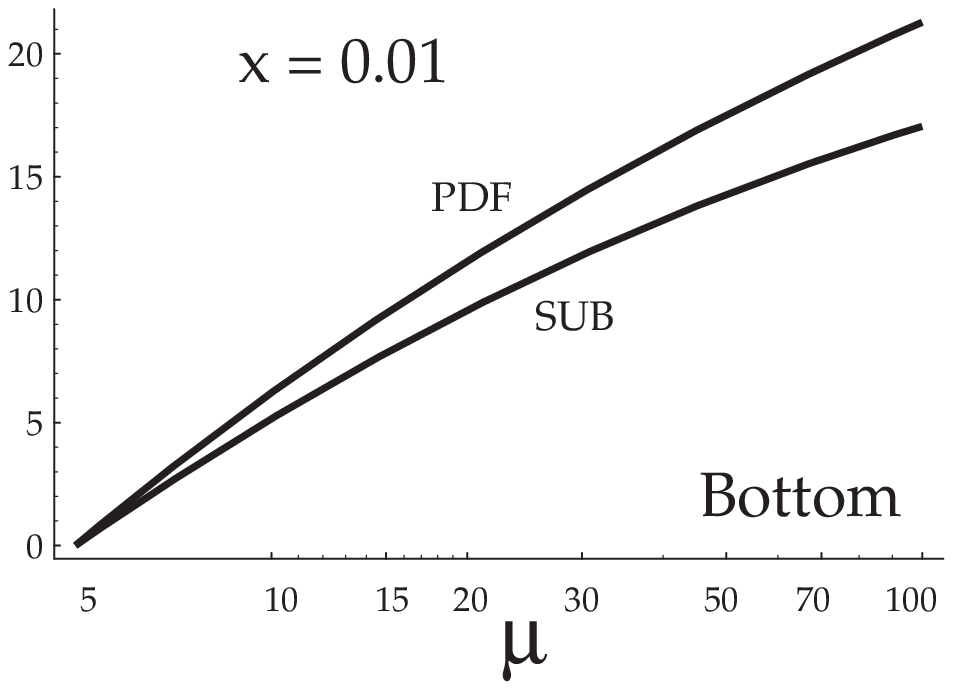}
\end{center}
\caption{Comparison of the evolved PDFs, $f_{b}(x,\mu)$ (labeled PDF),
and perturbative PDFs, $\widetilde{f}_{b}(x,\mu)\sim P_{b/g}\otimes
f_{g}$ (labeled SUB), as a function of the renormalization scale $\mu$
for bottom at a) $x=0.1$ and b) $x=0.01$. Taken from
Ref.~\cite{Olness:1997yc}}
\label{fig:fredii}
\end{figure}

\begin{figure}
\begin{center}
\includegraphics[width=0.45\linewidth]{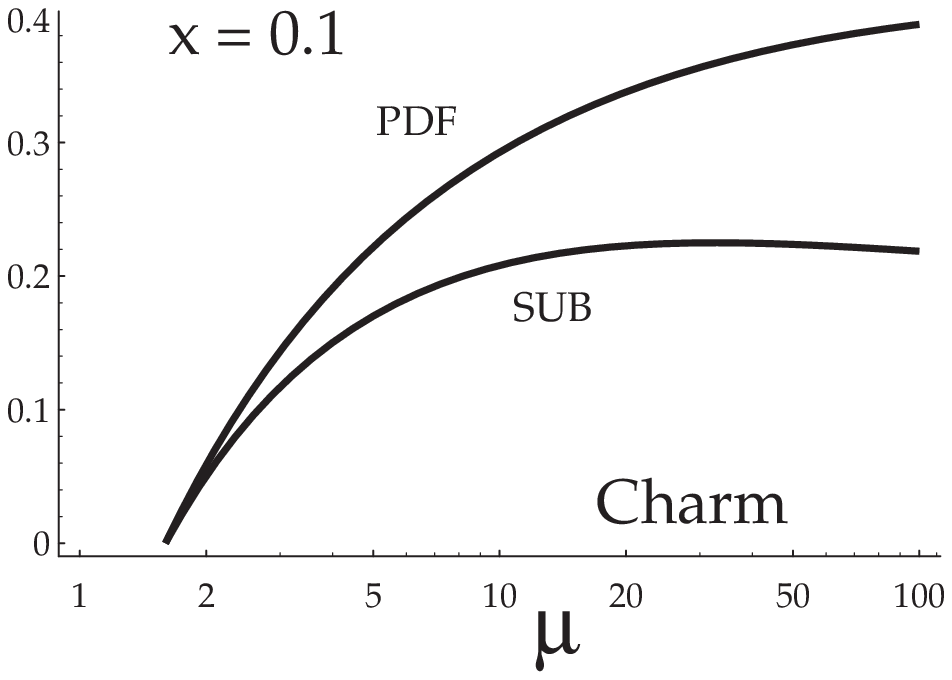}
\includegraphics[width=0.45\linewidth]{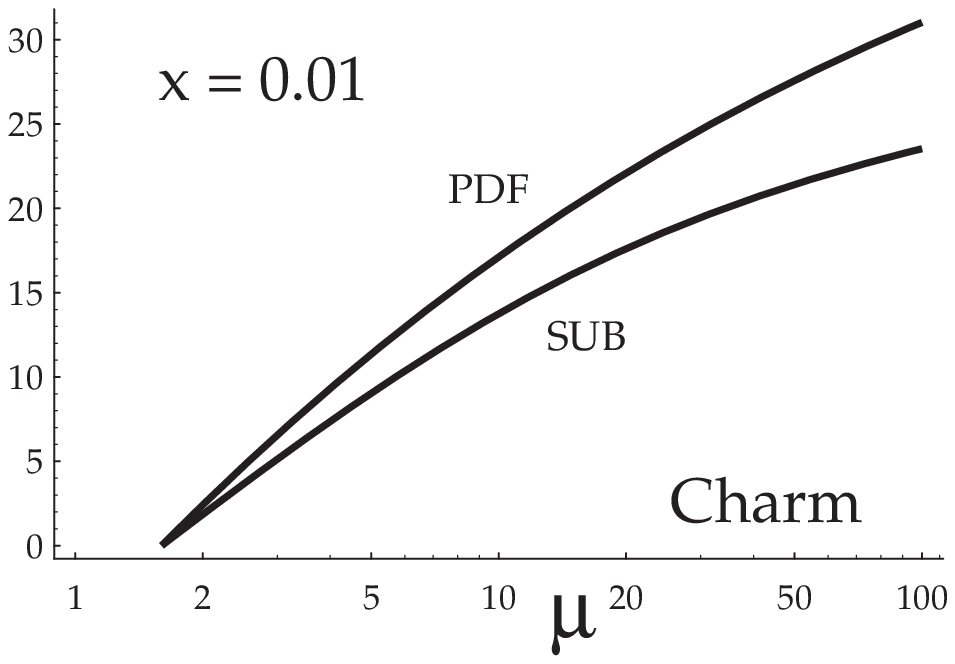}
\end{center}
\caption{Comparison of the evolved PDFs, $f_{c}(x,\mu)$ (labeled PDF),
and perturbative PDFs, $\widetilde{f}_{c}(x,\mu)\sim P_{c/g}\otimes
f_{g}$ (labeled SUB), as a function of the renormalization scale $\mu$
for charm at a) $x=0.1$ and b) $x=0.01$. Taken from
Ref.~\cite{Olness:1997yc} }
\label{fig:fredcharm}
\end{figure}

The fundamental difference between the $gg\rightarrow H$ process and
the $b\bar{b}\rightarrow H$ amounts to whether the radiative
splittings (e.g., $g\rightarrow b\bar{b}$) are computed by the DGLAP
equation as a part of the parton evolution, or whether they are
external to the hadron and computed explicitly. In essence, both
calculations are represented by the same perturbation theory with two
different expansion points; while the full perturbation series will
yield identical answers for both expansion points, there will be
difference in the truncated series.

To understand source of this difference, we examine the contributions
which are resummed into the b-quark PDF by the DGLAP evolution
equation, $df\sim P\otimes f$. Solving this equation perturbatively in
the region of the b-quark threshold, we obtain $\widetilde{f}_{b}\sim
P_{b/g}\otimes f_{g}$.  This term simply represents the first-order
$g\rightarrow b\bar{b}$ splitting which is fully contained in the
${\mathcal{O}}(\alpha_{s}^{2})$ $gg\rightarrow H$ calculation.

In addition to this initial splitting, the DGLAP equation resums an
infinite series of such splittings into the non-perturbative evolved
PDF, $f_{b}$. Both $f_{b}$ and $\widetilde{f}_{b}$ are shown in
Fig.~\ref{fig:fredii} for two choices of $x$.\cite{Olness:1997yc} Near
threshold, we expect $f_{b}$ to be dominated by the single splitting
contribution, and this is verified in the figure. In this region,
$f_{b}$ and $\widetilde{f}_{b}$ are comparable, and we expect the
4-flavor $gg\rightarrow H$ calculation should be reliable in this
region. As we move to larger scales, we see $f_{b}$ and
$\widetilde{f}_{b}$ begin to diverge at a few times $m_{b}$ since
$f_{b}$ includes higher-order splitting such as $\{
P^{2},P^{3},P^{4},...\}$ which are not contained in
$\widetilde{f}_{b}$. In this region, we expect the 5-flavor
$b\bar{b}\rightarrow H$ calculation should be most reliable in this
region since it resums the iterative splittings. For comparison,
$f_{c}$ and $\widetilde{f}_{c}$ are shown in Fig.~\ref{fig:fredcharm}
which have similar behavior.

\subsubsection*{NNLO\label{sec:NNLO}}

The fixed-flavor NLO QCD corrections to charm quark electro-production
were calculated in Ref.~\cite{Laenen:1992zk} in the three-flavor
scheme. The treatment of the heavy quark as a parton density requires
the identification of the large logarithmic terms $\log(Q^{2}/m^{2})$,
which was done in Ref.~\cite{Buza:1995ie} through next-to-next-leading
order (NNLO). Then based on a two-loop analysis of the heavy quark
structure functions from an operator point of view, it was shown in
Refs.~\cite{Buza:1997nv}, \cite{Buza:1996wv} and
\cite{Matiounine:1998ky} how to incorporate these large logarithms
into charm (and bottom) densities. Two different NNLO variable flavor
number schemes were defined in Refs.~\cite{Chuvakin:2000jm} and
\cite{Chuvakin:1999nx}, where it was shown how they could be matched
to the three-flavor scheme at small $Q^{2}$, the four-flavor scheme at
large $Q^{2}$, and the five-flavor scheme at even larger $Q^{2}$.

This NNLO analysis yielded two important results. One was the complete
set of NNLO matching conditions for massless parton evolution between
$N$ and $N+1$ flavor schemes. Unlike the LO and NLO case, the NNLO
matching conditions are discontinuous at these flavor thresholds.
Such matching conditions are necessary for any NNLO calculation at the
LHC, and have already been implemented in parton evolution packages by
\cite{Chuvakin:2001ge}, \cite{Vogt:2004ns}.

\begin{figure}[h]
\begin{center}
\includegraphics[scale=0.3,angle=-90]{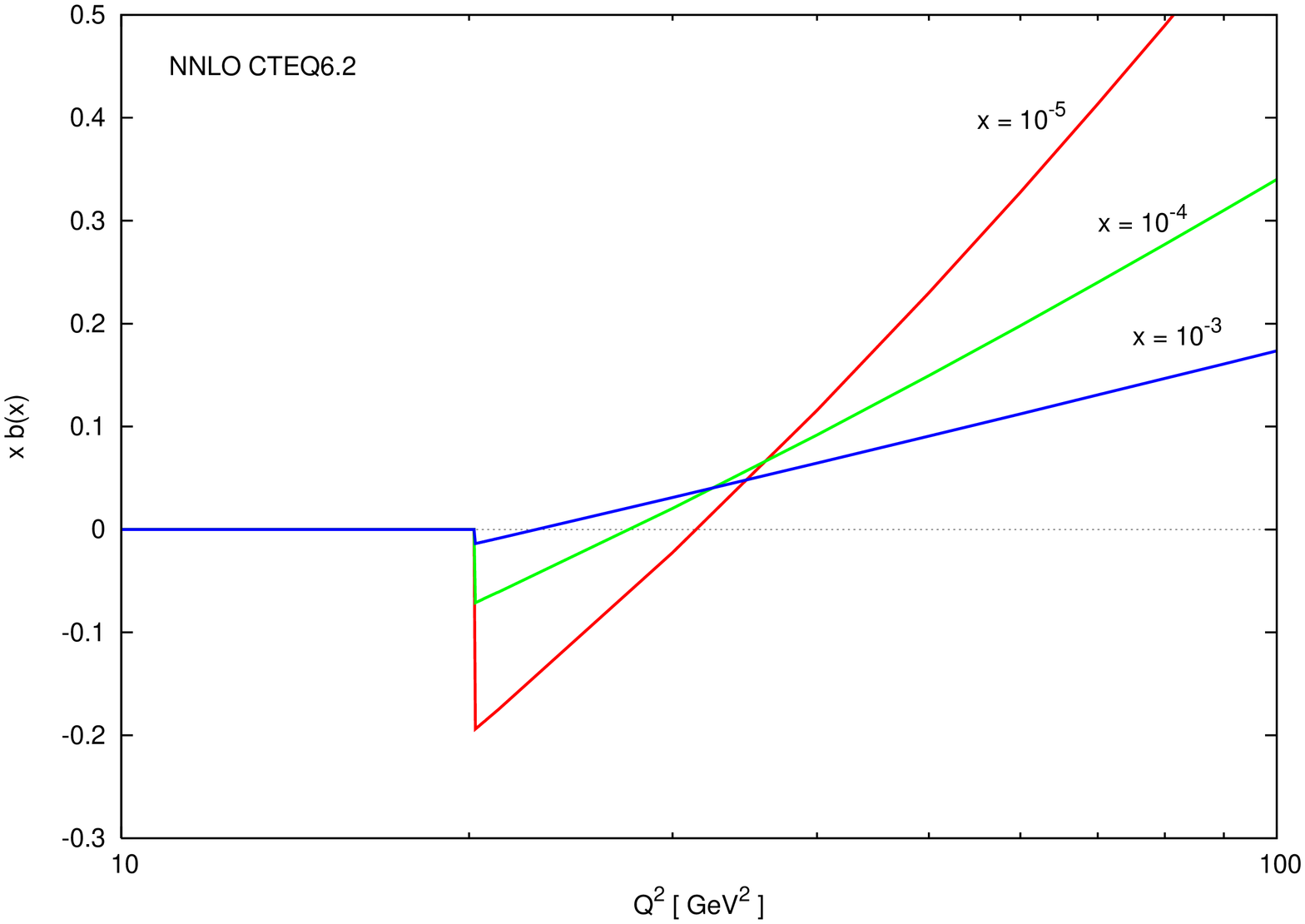}
\includegraphics[scale=0.3,angle=-90]{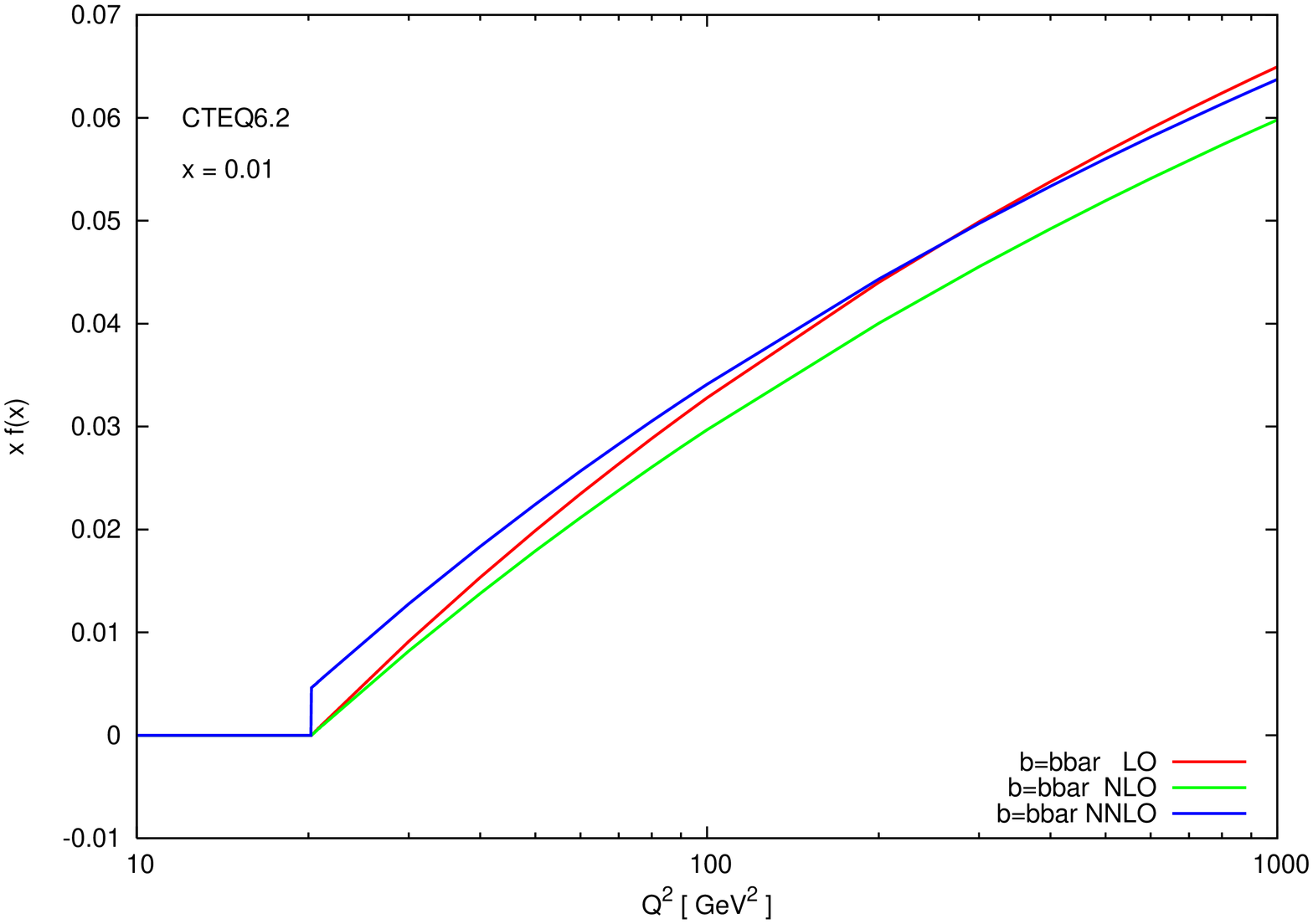}\end{center}
\caption{a) Comparison of the NNLO evolved PDFs, $f_{b}(x,\mu)$
vs. $Q^{2}$ using the NNLO matching conditions at$\mu=m_{b}$ for three
choices of x values: $\{10^{-3},10^{-4},10^{-5}\}$. b) Comparing
$f_{b}(x,\mu)$ vs. $Q^{2}$ for three orders of evolution \{LO, NLO,
NNLO\} at $x=0.01$. In this figure we have set $m_{b}=4.5$~GeV.
\label{fig:bottom}}
\end{figure}

\begin{figure}[h]
\centering
\includegraphics[scale=0.3,angle=-90]{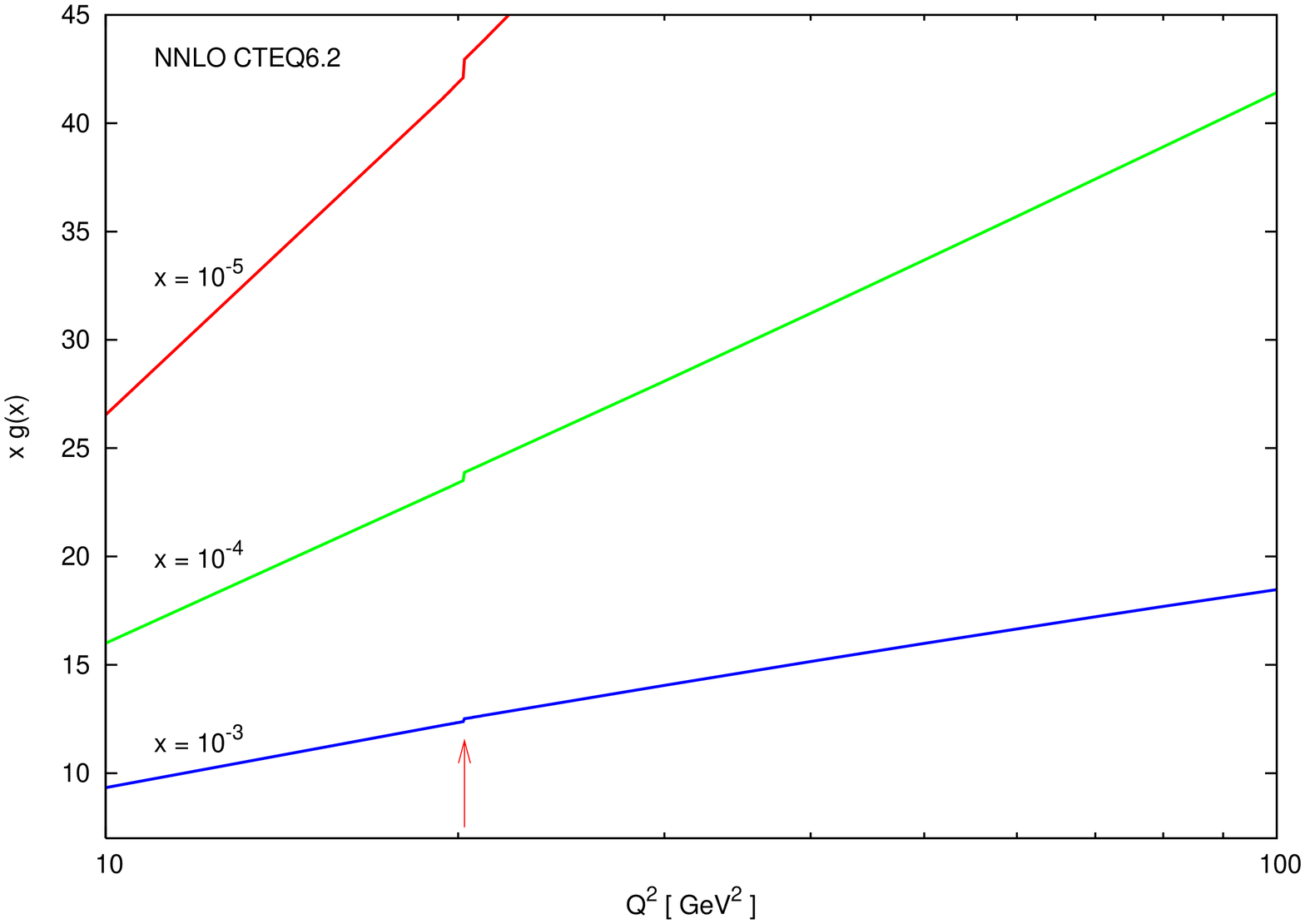}
\includegraphics[scale=0.3,angle=-90]{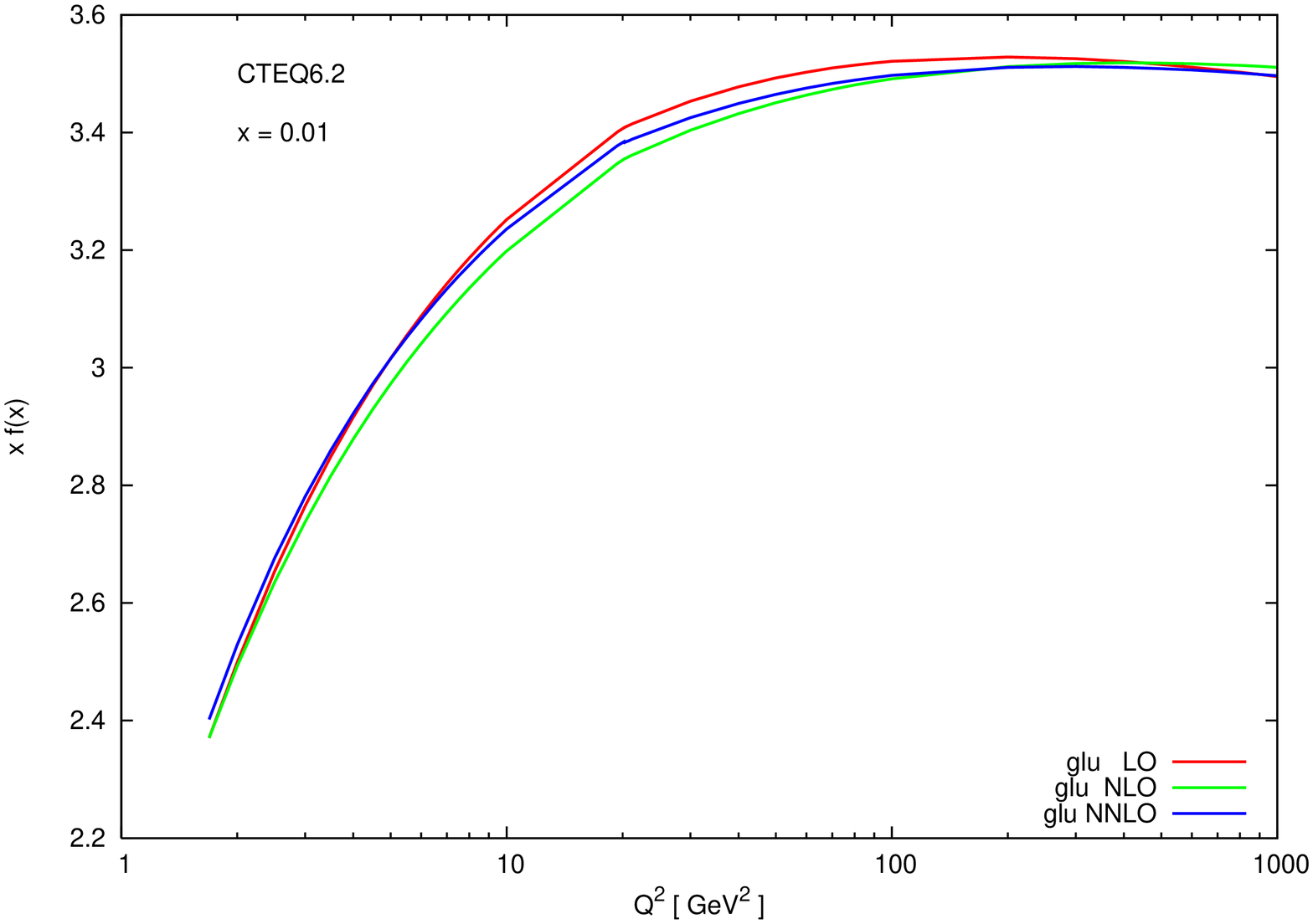}
\caption{a) Comparison of the NNLO evolved PDFs, $f_{g}(x,\mu)$
vs. $Q^{2}$ using the NNLO matching conditions at$\mu=m_{b}$ for three
choices of x values: $\{10^{-3},10^{-4},10^{-5}\}$. b) Comparing
$f_{g}(x,\mu)$ vs. $Q^{2}$ for three orders of evolution \{LO, NLO,
NNLO\} at $x=0.01$. In this figure we have set $m_{b}=4.5$~GeV.
\label{fig:gluon}}
\end{figure}

We illustrate this property in Figs.~\ref{fig:bottom} and
\ref{fig:gluon}.  In Fig.~\ref{fig:bottom}, we see that $f_{b}(x,\mu)$
vanishes for $\mu<m_{b}$; however, due to the non-vanishing NNLO
coefficients, we find $f_{b}(x,\mu)$ is non-zero (and negative) just
above the $m_{b}$ scale. This leads to a ${\cal O}(\alpha_{S}^{2})$
discontinuity in the b-quark PDF when making the transition from the
$N_{F}=4$ to $N_{F}=5$ scheme. Additionally, note that the value of
the discontinuity is $x$-dependent; hence, there is no simple adjustment
that can be made here to restore continuity. We also observe that
there is a discontinuity in the gluon PDF across the $N_{F}=4$ to
$N_{F}=5$ transition. While the PDF's have explicit discontinuities at
${\cal O}(\alpha_{S}^{2})$, the net effect of these NNLO PDF
discontinuities will compensate in any (properly calculated) NNLO
physical observable such that the final result can only have
discontinuities of order ${\cal O}(\alpha_{S}^{3})$.

Finally, we note that the NNLO two-loop calculations above explicitly
showed that the heavy quark structure functions in variable flavor
approaches are not infrared safe. A precise definition of the
heavy-flavor content of the deep inelastic structure function requires
one to either define a heavy quark-jet structure function, or
introduce a fragmentation function to absorb the uncanceled infrared
divergence. In either case, a set of contributions to the inclusive
light parton structure functions must be included at NNLO.

\subsubsection*{Conclusions}

While an exact {}``all-orders'' calculation would be independent of
the number of active flavors, finite order calculations necessarily
will have differences which reflect the higher-order uncalculated
terms. To study these effects, we have generated PDFs for
$N_{F}=\{3,4,5\}$ flavors using \{LO, NLO, NNLO\} evolution to
quantify the magnitude of these different choices. This work
represents an initial step in studying these differences, and
understanding the limitations of each scheme.

\subsubsection*{Acknowledgements}

We thank John Collins and Scott Willenbrock for valuable discussions.
F.I.O acknowledge the hospitality of Fermilab and BNL, where a portion
of this work was performed. This work is supported by the
U.S. Department of Energy under grants DE-FG02-97ER41022 and
DE-FG03-95ER40908, the Lightner-Sams Foundation, and by the National
Science Foundation under grant PHY-0354776.

\subsection{{LHAPDF: PDF Use from the Tevatron to the LHC}} 
\textbf{Contributed by:  Bourilkov, Group, Whalley}

The experimental errors in current and future hadron colliders are expected to decrease to a level that will challenge the uncertainties in theoretical calculations. One important component in the prediction of uncertainties at hadron colliders comes from the Parton Density Functions (PDFs) of the (anti)proton.  

 The highest energy particle colliders in the world currently, and in the near future, collide hadrons.  To make predictions of hadron collisions, the parton cross sections must be folded with the parton density functions:

\begin{equation}
\rm \frac{d\ \sigma}{d\; variable}[pp\rightarrow X] \sim \sum_{ij} \, \left(f_{i/p}(x_1)  f_{j/p}(x_2) +(i \leftrightarrow j)\right)\,  \hat{\sigma}, 
\label{llpdf1}
\end{equation}
with
\begin{trivlist}
\item $\rm \hat{\sigma}$ - cross section for the partonic subprocess $ij\rightarrow X$
\item $\rm x_1$, $\rm x_2$ - parton momentum fractions,
\item $\rm f_{i/p(\bar{p})}(x_i)$ - probability to find a parton $i$ with momentum
fraction $x_i$ in the (anti)proton.
\end{trivlist}

A long standing problem when performing such calculations is to quantify the uncertainty of the results coming from our limited knowledge of the PDFs.  Even if the parton cross section $\rm \hat{\sigma}$ is known very precisely, there may be a sizable error on the hadronic cross section $\sigma$
due to the PDF uncertainty.
The Tevatron can contribute to PDF knowledge in many ways that will benefit the experiments at the LHC.  First, measurements made by the experiments at FNAL will reduce PDF uncertainties by constraining PDF fits.  Perhaps more importantly tools and techniques for propagating PDF uncertainty through to physical observables can be improved and tested at the Tevatron.  

Next-to-leading order (NLO) is the first order at which the normalization of the hard-scattering cross sections has a reasonable uncertainty.  Therefore, this is the first order at which PDF uncertainties are usually applied.  To date, all PDF uncertainties have been calculated in the context of NLO global analysis.  However, useful information can still be obtained from NLO PDF uncertainties with leading order (LO) calculations and parton shower Monte Carlos~\cite{Joey}.  

Techniques and tools for calculating PDF uncertainty in the context of LO parton shower Monte Carlos will be the primary topic of this document.  Examples are provided employing CTEQ6~\cite{Pumplin:2002vw} error sets from LHAPDF and the parton shower Monte Carlo program \PY~\cite{Sjostrand:2003wg}.

\subsubsection*{LHAPDF update}

Historically, the CERN PDFLIB library~\cite{pdflib} has provided a widely used standard FORTRAN interface to PDFs with interpolation grids built into the PDFLIB code itself. However, it was realized that PDFLIB would be increasingly unable to meet the needs of the new generation of PDFs which often involve large numbers of sets ($\approx$20--40) describing the uncertainties on the individual partons from  variations in the fitted parameters.  As a consequence of this, at the Les Houches meeting in 2001~\cite{Giele:2002hx}, the beginnings of a new interface were conceived --- the so-called \underline{L}es \underline{H}ouches \underline{A}ccord \underline{PDF} (LHAPDF).  The LHAGLUE package~\cite{Bourilkov:2003kk} plus a unique PDF numbering scheme enables LHAPDF to be used in the same way as PDFLIB, without requiring {\em any} changes in the \PY or \HW codes.  The evolution of LHAPDF (and LHAGLUE) is well documented~\cite{Whalley:2005nh,Bourilkov:2006cj}.  
  
Recently, LHAPDF has been further improved.  With the release of v4.1 in August of 2005 the installation method has been upgraded to the more conventional {\tt configure; make; make install}.  Version 4.2, released in November of 2005, includes the  new cteq6AB (variable $\alpha(M_Z)$) PDF sets.  It also includes new modifications by the CTEQ group to other cteq code to improve speed.   Some minor bugs were also fixed in this version that affected the a02m$\_$nnlo.LHgrid file (previous one was erroneously the same as LO) and SMRSPI code which was wrongly setting {\tt usea} to zero.    

A v5 version, with the addition of the option to store PDFs from multiple sets
in memory, has been released.  This new functionality speeds up the code by
making it possible to store PDF results from many sets while only generating a
MC sample once without significant loss of speed.

As a technical check, cross sections have been computed, as well as errors where appropriate, for all PDF sets included in LHAPDF. 10,000 events are generated
for each member of a PDF set for both \HW~\cite{Corcella:2002jc} and \PY~\cite{Sjostrand:2003wg},
and at both Tevatron and LHC energies.
As this study serves simply as a technical check of the
interface, no attempt was made to unfold the true PDF error.
The maximum Monte Carlo variance (integration error) in our checks is less than 1 percent.
This has not been subtracted and will result in an overestimate of the true
PDF uncertainty by a factor $\lsim 1.05$ in our analysis.  The results in general show good agreement for most PDFs included in the checks. Overall the consistency is better for Tevatron energies, where we do not have to make large extrapolations to the new energy domain and much broader phase space covered by the LHC.

Two complementary processes are used:
\begin{itemize}
\item Drell--Yan Pairs ($\mu^+\mu^-$):
the Drell--Yan process is chosen here to probe the functionality of the
quark PDFs included in the LHAPDF package.  
\item Higgs Production:
the cross section for $gg \rightarrow H$ probes the gluon PDFs, so this
channel is complementary to the case considered above.
\end{itemize}

\subsubsection*{{PDF uncertainties}}
As stated above, the need to understand and reduce PDF uncertainties in theoretical predictions for collider physics is of paramount importance.  One of the first signs of this necessity was the apparent surplus of high $P_T$ events observed in the inclusive jet cross section in the CDF experiment at FNAL in run I.  Subsequent analysis of the PDF uncertainty in this kinematic region indicated that the deviation was within the range of the PDF dominated theoretical uncertainty on the cross section.  Indeed, when the full jet data from the Tevatron (including the D0 measurement over the full rapidity range) was included in the global PDF analysis, the enhanced high x gluon preferred by CDF jet data from Run I became the central solution.  This was an overwhelming sign that PDF uncertainty needed to be quantified~\cite{Giele:2001mr}. Below, a short review of one approach to quantify these uncertainties called the Hessian matrix method is given, followed by outlines of two methods used to calculate the PDF uncertainty on physical observables. 

 Experimental constraints must be incorporated into the uncertainties of parton distribution functions before these uncertainties can be propagated through to predictions of observables.  The Hessian Method~\cite{Pumplin:2001ct} both constructs a N Eigenvector Basis of PDFs and provides a method from which uncertainties on observables can be calculated.  
 The first step of the Hessian method is to make a fit to data using N free parameters.  The global $\chi ^2$ of this fit is minimized yielding a central or best fit parameter set $S_0$.  Next the global  $\chi ^2$ is increased to form the Hessian error matrix:
\begin{equation}
\Delta \chi ^{2}=\sum_{i=1}^N \sum_{j=1}^N H_{ij}(a_{i}-a_{i}^{0})(a_{j}-a_{j}^{0})
\end{equation}

This matrix can then be diagonalized yielding N (20 for CTEQ6) eigenvectors.  Each eigenvector probes a direction in PDF parameter space that is a combination of the 20 free parameters used in the global fit.  The largest eigenvalues correspond to the best determined directions and the smallest eigenvalues to the worst determined directions in PDF parameter space.  For the CTEQ6 error PDF set, there is a factor of roughly one million between the largest and smallest eigenvectors.  The eigenvectors are numbered from highest eigenvalue to lowest eigenvalue.  Each N eigenvector direction is then varied up and down within tolerance to obtain 2N new parameter sets, $S^{\pm}_i(i=1,..,N)$.  These parameter sets each correspond to a member of the PDF set, $F_{i}^{\pm}=F(x,Q;S_{i}^{\pm})$. The PDF library described above, LHAPDF, provides standard access to these PDF sets.

  Although the variations applied in the eigenvector directions are symmetric by construction, this is not always the case for the result of these variations when propagated through to an observable.  In general the well constrained directions (low eigenvector numbers) tend to have symmetric positive and negative deviations on either side of the central value of the observable (X$_0$).  This can not be counted on in the case of the smaller eigenvalues (larger eigenvector numbers).  The 2N+1 members of the PDF set provide 2N+1 results for any observable of interest.  Two methods for obtaining a set of results are described in detail below.  Once results are obtained they can be used to approximate PDF uncertainty through the use of a 'Master Equation'.  Although many versions of these equations can be found in the literature, the type which considers maximal positive and negative variations of the physical observable separately is preferred~\cite{Nadolsky:2001yg}:

\begin{equation}  
\Delta X^{+}_{max}=\sqrt{\sum_{i=1}^N[max(X^{+}_{i}-X_{0},X^{-}_{i}-X_{0},0)]^{2}}
\end{equation}

\begin{equation}  
\Delta X^{-}_{max}=\sqrt{\sum_{i=1}^N[max(X_{0}-X^{+}_{i},X_{0}-X^{-}_{i},0)]^{2}}
\end{equation}  

Other forms of 'Master' equations with their flaws are summarized:
\begin{itemize}

\item 
$\Delta X_{1}=\frac{1}{2}\sqrt{\sum_{i=1}^N  (X^{+}_{i}-X^{-}_{i})^{2}}$ \\
This is the original CTEQ 'Master Formula'.  It correctly predicts uncertainty on the PDF values since in the PDF basis $X^{+}_{i}$ and  $X^{-}_{i}$ are symmetric by construction.  However, for physical observables this equation will underestimate the uncertainty if $X^{+}_{i}$ and  $X^{-}_{i}$ lie on the same side of $X_{0}$.\\

\item
 $\Delta X_{2}=\frac{1}{2}\sqrt{\sum_{i=1}^{2N}  R_i^{2}}$ ($R_{1}= X^{+}_{1}-X_{0}$,$R_{2}= X^{-}_{1}-X_{0}$, $R_{3}= X^{+}_{2}-X_{0}$ ...)\\
If $X^{+}_{i}$ and  $X^{-}_{i}$ lie on the same side of $X_{0}$ this equation adds contributions from both in quadrature.  NOTE: For symmetric and asymmetric deviations, $\Delta X_{1}$ varies from $0\rightarrow {\sqrt{2}} \Delta X_{2}$\\

\item  positive and negative variations based on eigenvector directions\\
 $\Delta X^{+}=\sqrt{\sum_{i=1}^N(X^{+}_{i}-X_{0})^{2}}$,\ \ $\Delta X^{-}=\sqrt{\sum_{i=1}^N(X^{-}_{i}-X_{0})^{2}}$\\
Since the positive and negative directions defined in the PDF eigenvector space are not always related to positive and negative variations on an observable these equations can not be interpreted as positive and negative errors in the general case.

\end{itemize}

Two main techniques are currently employed to study the effect of PDF uncertainties of physical observables.  Both techniques work with the PDF sets derived from the Hessian method.   

The 'brute force' method simply entails running the MC and obtaining the observable of interest for each PDF in the PDF set.  This method is robust, and theoretically correct.  Unfortunately,  it can require very large CPU time since large statistical samples must be generated in order for the PDF uncertainty to be isolated over statistical variations.  This method generally is unrealistic when detector simulation is desired.   

Because the effect on the uncertainty of the PDF set members is added in quadrature, the uncertainty is often dominated by only a few members of the error set.  In this case, a variation of the 'brute force' method can be applied.  Once the eigenvectors that the observable is most sensitive to are determined, MC samples only need to be generated for the members corresponding to the variation of these eigenvalues.  This method will always slightly underestimate the true uncertainty.

As mentioned above, often it is not possible to generate the desired MC sample many times in order to obtain the uncertainty on the observable due to the PDF.  The 'PDF Weights' method solves this problem~\cite{Joey}.  The idea is that the PDF contribution to Equation~\ref{llpdf1} may be factored out.  That is, for each event generated with the central PDF from the set, a PDF weight 
can be stored for each event.  The PDF weight technique can be summarized as follows... 

\begin{itemize}
\item Only one MC sample is generated but 2N (e.g. 40) PDF weights are obtained for 
\begin{equation} 
W_{n}^{0}=1,W_{n}^{i}=\frac{f(x_{1},Q;S_{i})f(x_{2},Q;S_{i})}{f(x_{1},Q;S_{0})f(x_{2},Q;S_{0})}
\end{equation}

 where $n=1...N_{events},i=1..N_{PDF}$

\item Only one run, so kinematics do not change and there is no residual statistical variation in uncertainty.
\item The observable must be weighted on an event by event basis for each PDF of the set.  One can either store a ntuple of weights to be used 'offline', or fill a set of weighted histograms (one for each PDF in the set).
\end{itemize}

The benefits of the weighting technique are twofold.  First, only one sample of MC must be generated.  Second, since the observable for each PDF member is obtained from the same MC sample there is no residual statistical fluctuation in the estimate of the PDF uncertainty.  One concern involving this method is that re-weighting events does not correctly modify the Sudakov form factors.  However, the difference in this effect due to varying the PDF was shown to be negligible ~\cite{Gieseke:2004tc}.  That is, the initial state parton shower created with the central PDF (CTEQ6.1) also accurately represents the parton shower that would be produced by any other PDF in the error set.
 
The weighting method is only theoretically correct in the limit that all possible initial states are populated.  For this reason, it is important that reasonable statistical samples are generated when using this technique.  Any analysis which is sensitive to the extreme tails of distributions should use this method with caution. 

There are two options for using the PDF weighting technique.  One can either store 2N (e.g. 40 for CTEQ) weights for each event, or store  $X_{1},X_{2},F_{1},F_{2}$, and $Q^{2}$ and calculate the weights 'offline'.  The momentum of the two incoming partons may be obtained from \PY via  {\tt PARI(33)} 
and {\tt PARI(34)}.  
Flavour types of the 2 initial partons are stored in 
$F_{1}={\tt MSTI(15)}$ and $F_{2}={\tt MSTI(16)}$, 
and the numbering scheme is the same as the one used by LHAPDF, Table~\ref{flavour-scheme}, except that the gluon is labeled '21' rather than '0'.  The $Q^2$ of the interaction is stored in  
$Q^{2}={\tt PARI(24)}$.  In theory, this information and access to LHAPDF is all that is needed to use the PDF weights method. This approach has the additional benefit of enabling the 'offline reweighting' with new PDF sets, which have not been used, or even existed, during the MC generation.   We plan to include sample code facilitating the use of PDF weights in future releases of LHAPDF.

\begin{table}[ht]
\begin{center}
\caption{The flavour enumeration scheme used for {\sl f(n)} in LHAPDF}
\label{flavour-scheme}
\begin{tabular}{
|p{1.0cm}
*{13}{|p{0.4cm}}
|}
\hline
parton &
$\bar{t}$ &
$\bar{b}$ &
$\bar{c}$ &
$\bar{d}$ &
$\bar{u}$ &
$\bar{d}$ &
g &
d &
u &
s &
c &
b &
t \\
\hline
{\sl n} & $-6$ & $-5$ & $-4$ & $-3$ & $-2$ & $-1$ & $0$ & $1$ & $2$ & $3$ & $4$ & $5$ & $6$ \\
\hline
\end{tabular}
\end{center}
\end{table}

\subsubsection*{{Example Studies}}

The Drell--Yan process is chosen as an almost ideal test case involving quark PDFs for the different flavours.

\begin{figure*}[htb]
\includegraphics[width=80mm]{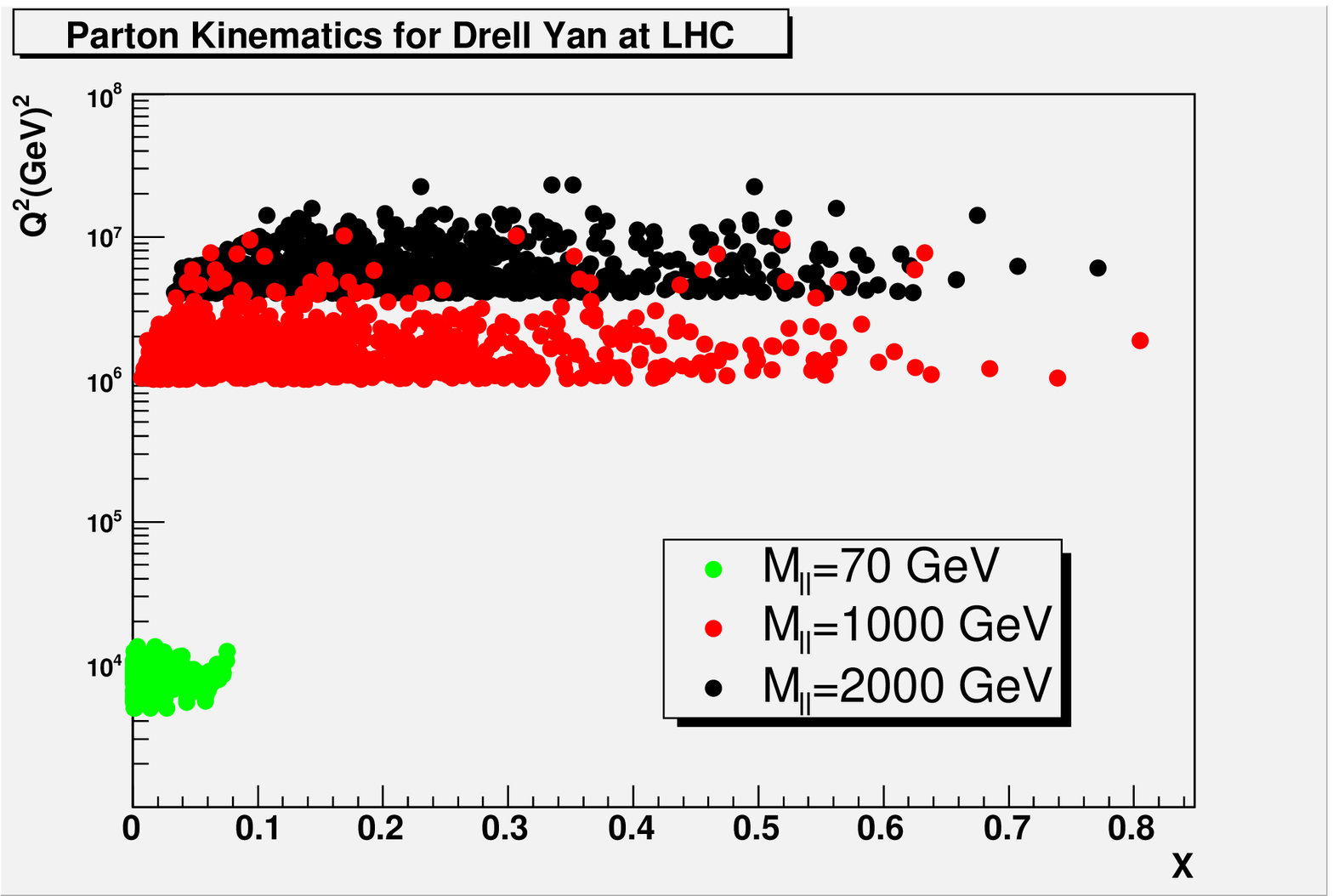}\includegraphics[width=80mm]{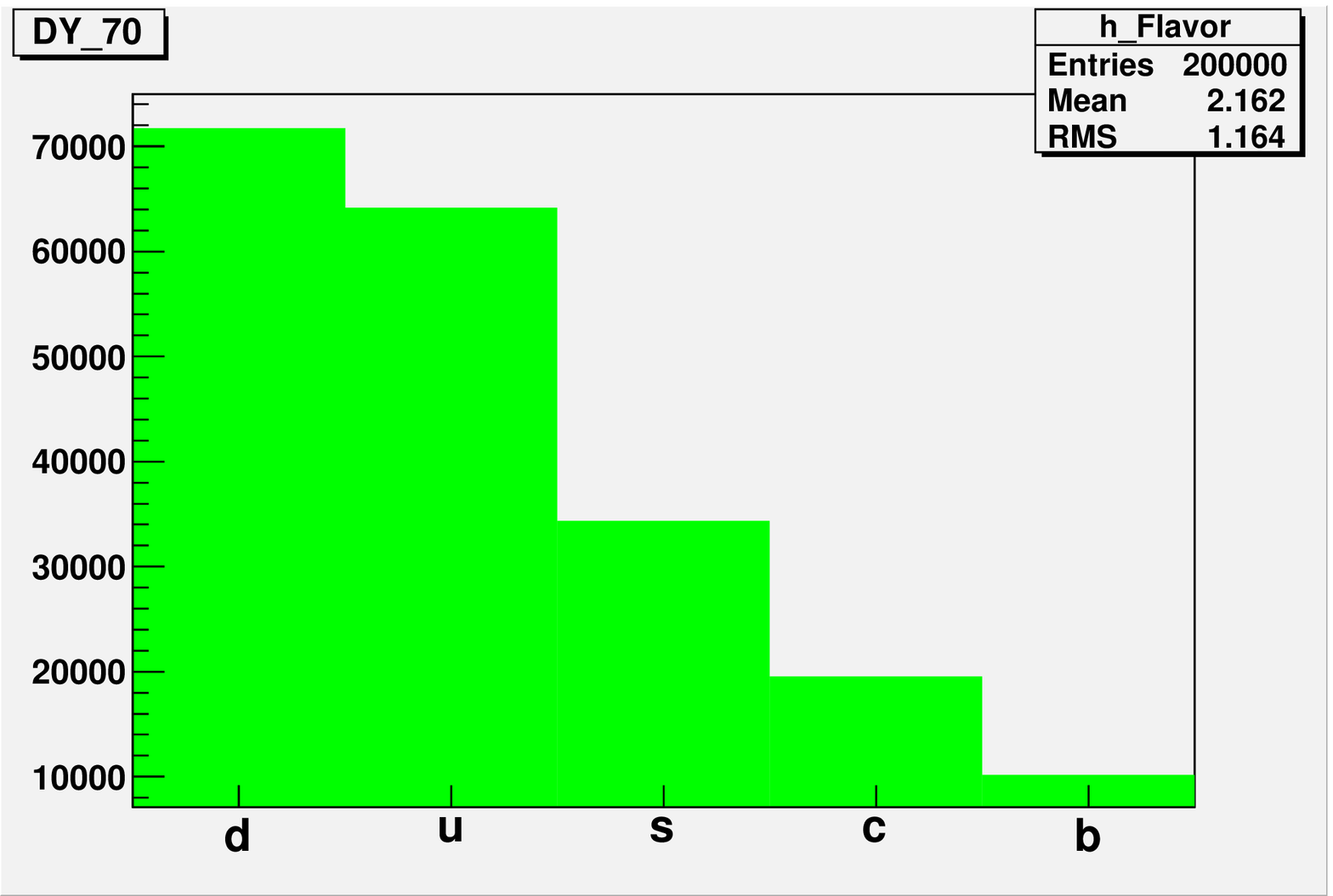} 
\includegraphics[width=80mm]{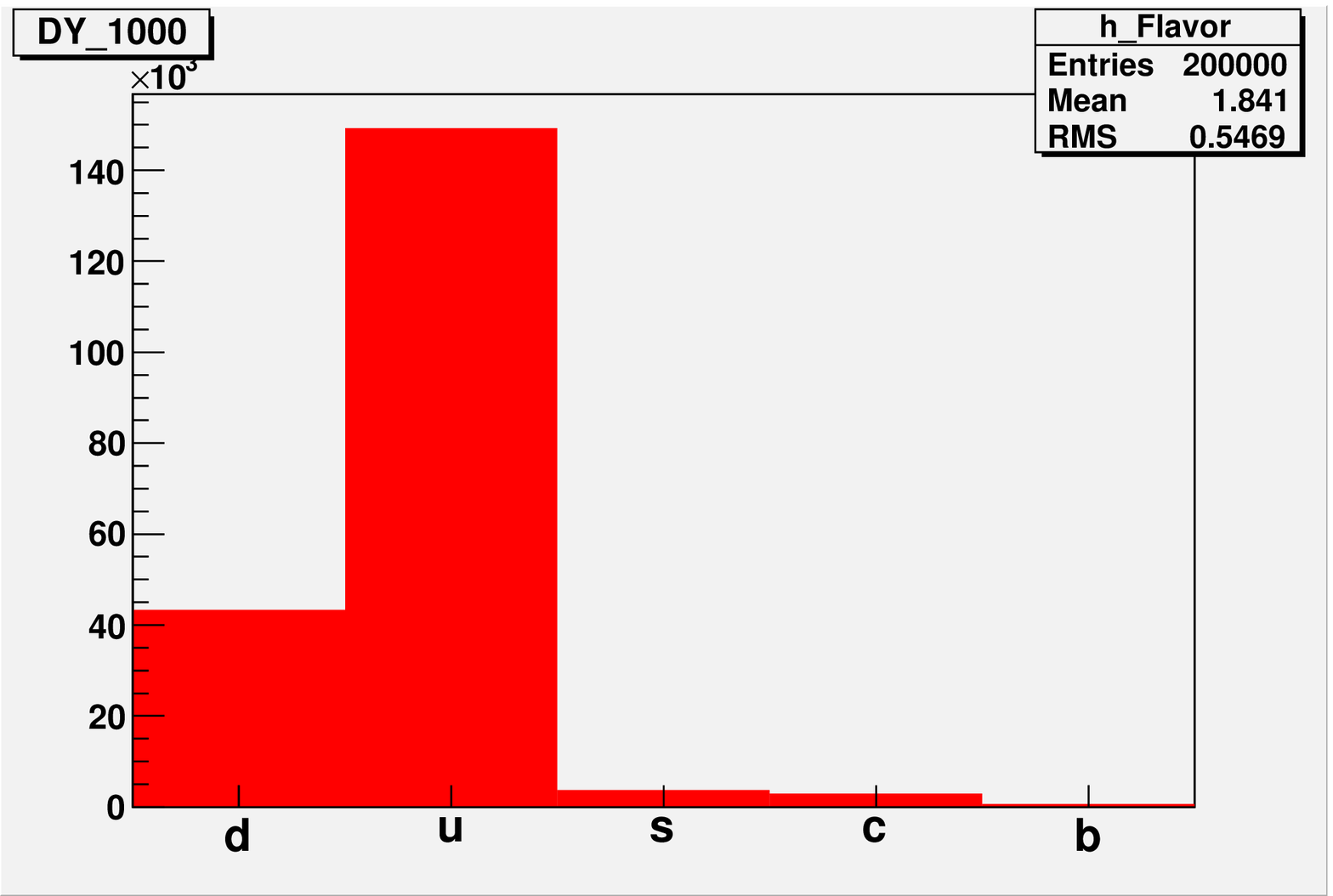}\includegraphics[width=80mm]{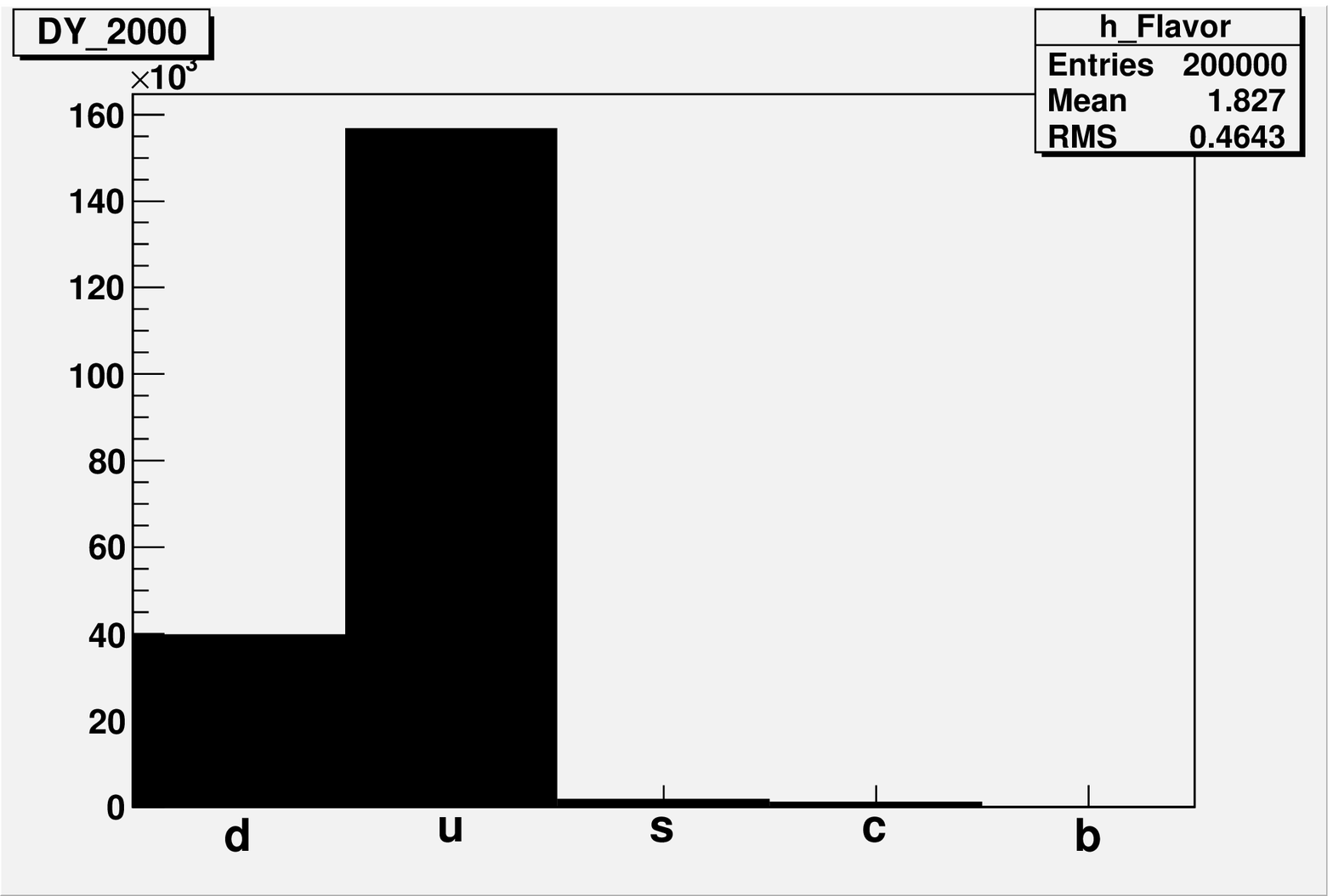} 
\caption{The parton kinematics for Drell--Yan production at the LHC for three Drell--Yan mass choices.  Also the initial parton flavour content for the three cases is shown.}
\label{DY_figs}
\end{figure*}
The initial state parton kinematics and flavour contributions are given in Figure~\ref{DY_figs} for three regions of invariant mass of the final state lepton pair: $70 < M < 120,\ M > 1000,\ M > 2000$~GeV. As we can observe, they cover very wide range in X and Q$^2$. It is interesting to note that the flavour composition around the Z peak contains important contributions from five flavours, while at high mass the u and d quarks (in ratio 4:1) dominate almost completely.

{{Higgs Production in gg$\rightarrow$H at the LHC}}
is chosen as complementary to the first one and contains only contributions from the gluon PDF. A light Higgs mass of 120 GeV is selected.

\begin{figure*}[htb]
\centering
\includegraphics[width=.7\textwidth]{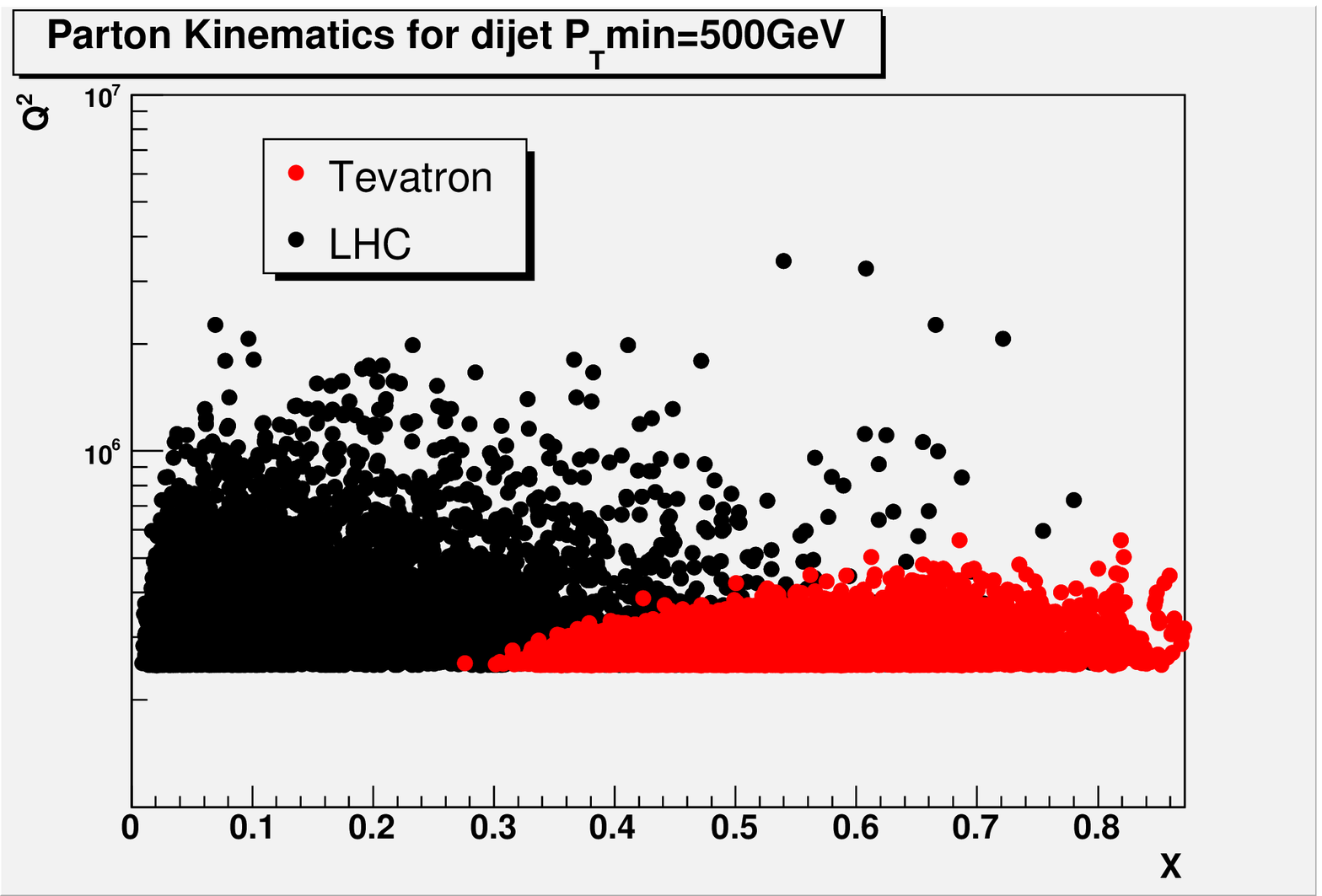} \\
\caption{The partonic jet kinematics for the inclusive jet cross section at the Tevatron and the LHC.}
\label{jet_kinematics}
\end{figure*}

As mentioned above, the inclusive jet cross section was one of the first measurements where the need to quantify PDF uncertainty was evident.  QCD 2-2 processes are studied for $\hat{P_T}\ >\ 500\ GeV$.  The kinematic range probed can be seen in Figure~\ref{jet_kinematics}.

The results for all 3 studies are summarized in Table~\ref{result_table}.
The weighting technique produces the same results as the more elaborate
'brute force' approach for all cases.

\begin{table}[htb]
{
\normalsize
\begin{center}
\caption{Results for the 3 case studies. The central values of the
cross sections in [pb] are shown, followed by the estimates of the
uncertainties for the different master equations and the 'brute force' (B.F.)
and weighting (W) techniques.}
\label{result_table}
\begin{tabular}{|l|c|c|c|c|c||c|}

\hline
Process (method)          & $X_{0}$   &  $\Delta X_{1}$ & $\Delta X_{2}$ & ($+\Delta X^{+}$,$-\Delta X^{-}$) 
& ($+\Delta X_{max}^{+}, -\Delta X_{max}^{-}$) & \\ \hline
DY 70$<$M$<$120 (B.F.)   & 1086      &48   & 42  & $(+55,-63)$   & $(+51,-62)$  & \\ \hline
DY 70$<$M$<$120 (W)     & 1086      &48   & 42  & $(+55,-64)$    &$(+51,-63)$  & \\ \hline \hline
DY M$>$1000 (B.F.)  & 67      & 3.5   & 2.6  & $(+3.5,-3.8)$   &$(+3.4,-3.9)$ & $\cdot 10^{-4}$\\ \hline
DY M$>$1000 (W)     & 67      & 3.5   & 2.6  & $(+3.5,-3.8)$   &$(+3.4,-3.8)$ & $\cdot 10^{-4}$\\ \hline \hline
DY M$>$2000 (B.F.)  & 22      & 1.8   & 1.3  & $(+1.9, -1.9)$  &$(+2.0,-1.7)$ & $\cdot 10^{-5}$ \\ \hline
DY M$>$2000 (W)     & 22      & 1.8   & 1.3  & $(+1.8, -1.9)$  &$(+2.0,-1.7)$ & $\cdot 10^{-5}$ \\ \hline\hline\hline 
$gg\rightarrow H$ (B.F.)  & 17      &.94   & .68  & $(+.82,-1.1)$   &$(+.8,-1.1)$ & \\ \hline
$gg\rightarrow H$ (W)     & 17      &.94   & .68  & $(+.82,-1.1)$    &$(+.8,-1.1)$ & \\ \hline \hline \hline
DJ500 TeV (B.F.)  & 22        & 6.8   & 5.7     &$(4.8,   10)$    &$(11,4.2)$ & $\cdot 10^{-3}$ \\ \hline
DJ500 TeV (W)     & 22        & 6.8   & 5.7     &$(4.8,   10)$    &$(11,4.2)$ & $\cdot 10^{-3}$  \\ \hline\hline
DJ500 LHC (B.F.)  & 880         &63       & 47      &$(56      ,74)$      &$(76        ,53)$ &   \\ \hline
DJ500 LHC (W)     & 880         &63       & 47      &$(57      ,75)$      &$(77        ,53)$ &   \\ \hline
    \end{tabular}
\end{center}
}
\end{table}

\subsubsection*{{Summary}}
In this contribution, new developments of LHAPDF and consistency checks for all PDF sets are described. The approaches to PDF uncertainty analysis are outlined and the modern method of PDF weighting is described in detail and tested in different channels of current interest.  Drell-Yan, gluon fusion to Higgs, and high $P_T$ jet production are studied at the Tevatron and LHC energy scales.  The methods are in agreement in all cases.  Equations for quantifying PDF uncertainty are discussed and the type which relies on maximal positive and negative variations on the observable is considered superior.

\subsubsection*{{Acknowledgments}}

The authors would like to thank Joey Huston for encouraging this study.
DB wishes to thank the United States National Science Foundation
for support from grant NSF 0427110 (UltraLight). CG wishes to thank the US
Department of Energy for support from an Outstanding Junior Investigator award
under grant DE-FG02-97ER41209. MRW wishes to thank the UK PPARC for support
from grant PP/B500590/1.

\clearpage
\subsection{{fastNLO: Fast pQCD Calculations for PDF Fits}} 

\textbf{Contributed by Kluge, Rabbertz, Wobisch}

{\em 
We present a method for very fast repeated computations of higher-order
cross sections in hadron-induced processes
for arbitrary parton density functions.
A full implementation of the method
for computations of jet cross sections
in Deep-Inelastic Scattering and in Hadron-Hadron Collisions
is offered by the ``fastNLO'' project.  
A web-interface for online calculations and user code
can be found at
{\tt http://hepforge.cedar.ac.uk/fastnlo/}.
} \\


The aim of the "fastNLO" project is to make the inclusion of jet data
into global fits of parton density functions (PDFs) feasible. 
Due to the prohibitive computing time required for the jet cross sections 
using standard calculation techniques,
jet data have either been omitted in these fits completely 
or they were included using a simple approximation.
The fastNLO project implements a method that offers exact and 
very fast pQCD calculations
for a large number of jet data sets 
allowing to take full advantage of their direct sensitivity 
to the gluon density in the proton in future PDF fits.
This includes Tevatron jet data beyond
the inclusive jet cross section and also
HERA jet data which have 
been used to determine the proton's gluon 
density~\cite{Adloff:2000tq,Chekanov:2001bw,Chekanov:2002be,Chekanov:2005nn},
but which are ignored in current 
PDF fits~\cite{Alekhin:2005gq,Martin:2004ir,Pumplin:2002vw}.


\subsubsection*{Cross Sections in Perturbative QCD}

Perturbative QCD predictions for observables in 
hadron-induced processes depend on the strong coupling 
constant $\alpha_s$ and on the PDFs of the hadron(s).
Any cross section in hadron-hadron collisions 
can be written as the convolution of 
the strong coupling constant  $\alpha_s$ in order $n$,
the perturbative coefficient $c_{n,i}$ for the partonic
subprocess $i$,
and the corresponding linear combination of PDFs 
from the two hadrons $F_i$
which is a function of the  fractional hadron momenta
$x_{a,b}$ carried by the partons
\begin{equation}
\sigma(\mu_r,\mu_f) = \sum_{n,i}  \, c_{n,i}(x_a, x_b, \mu_r,\mu_f) 
\otimes 
\left[ \alpha_s^n(\mu_r) \cdot F_i(x_a,x_b,\mu_f) \right] \,.
\label{eq:fnmain}
\end{equation}
The PDFs and $\alpha_s$ also depend on the factorization and the 
renormalization scales $\mu_{f,r}$, respectively,
as does the perturbative prediction for the cross section
in finite order $n$.
An iterative PDF fitting procedure
using exact NLO calculations for jet data, 
based on Monte-Carlo integrations of~(\ref{eq:fnmain}), 
is too time-consuming.
Only an approximation of~(\ref{eq:fnmain}) is, therefore,
currently being used in global PDF fits.

\subsubsection*{A Simple Approach}

\begin{figure}[t]
\centerline{
   \psfig{figure=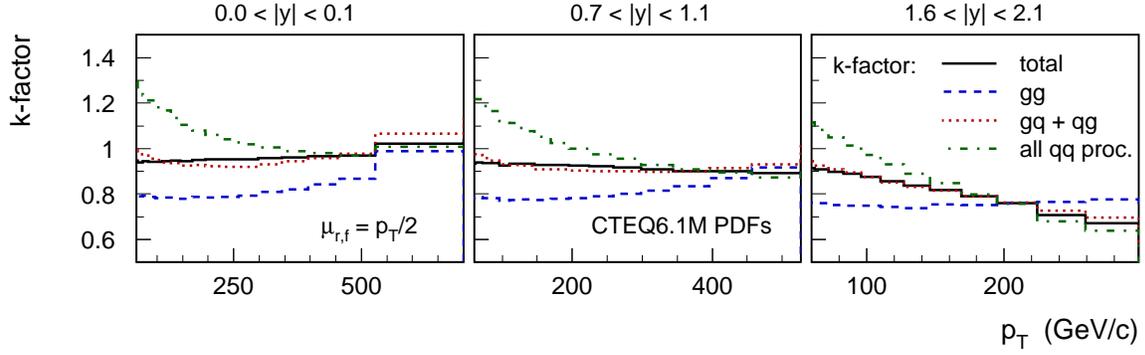,height=4.9cm}
}
   \caption{The $k$-factor for the inclusive $p\bar{p}$ jet cross section 
   at $\sqrt{s}=1.96$\,TeV as a function of $p_T$ at different rapidities $y$
   for the total cross section (solid line) and for different 
   partonic subprocesses:
   gluon-gluon (dashed), gluon-quark (dotted) and the sum of all
   quark and/or anti-quark induced subprocesses (dashed-dotted).
\label{fig:kfactor}}
\end{figure}

The ``$k$-factor approximation''
as used in~\cite{Martin:2004ir,Pumplin:2002vw}
parameterizes higher-order corrections
for each bin of the observable by a factor
$\displaystyle k = \frac{\sigma_{\rm NLO}}{\sigma_{\rm LO}}
= \frac{\sigma_{(2)}+\sigma_{(3)}}{\sigma_{(2)}}$
computed from the contributions 
with $n=2$ ($\sigma_{(2)}$) and $n=3$ ($\sigma_{(3)}$)
for a fixed PDF, averaged over all subprocesses~$i$.
In the iterative fitting procedure
only the LO cross section is computed
and multiplied with $k$ to obtain an estimate of 
the NLO cross section.
This procedure does not take into account that 
different partonic subprocesses can have largely 
different higher-order corrections.
Fig.~\ref{fig:kfactor} shows that the $k$-factors
for quark-only and gluon-only induced subprocesses
can differ by more than $\pm20\%$ from the average.
The $\chi^2$ is therefore minimized under an incorrect assumption
of the true PDF dependence of the cross section.
Further limitations of this approach are:
\begin{itemize}
\item 
   Even the LO Monte-Carlo integration of~(\ref{eq:fnmain})
   is a trade-off between speed  
   and precision. With finite statistical errors,
   however, theory predictions are not ideally smooth
   functions of the fit parameters.
   This contributes to numerical noise in the $\chi^2$ 
   calculations~\cite{Pumplin:2000vx}
   distorting the  $\chi^2$ contour during the  
   PDF error analysis, especially for fit parameters
   with small errors.

\item
   The procedure can only be used for observables for 
   which LO calculations  are fast. 
   Currently, this prevents the global PDF analyses from  
   using Tevatron dijet data and DIS jet data. 
\end{itemize}
In a time when phenomenology is aiming towards 
NNLO precision~\cite{Alekhin:2005gq,Martin:2004ir},
the $k$-factor approximation is clearly not satisfying concerning both
its limitation  in precision and its restrictions concerning data sets.

\subsubsection*{The fastNLO Solution}

\begin{figure}[t]
\centerline{
   \psfig{figure=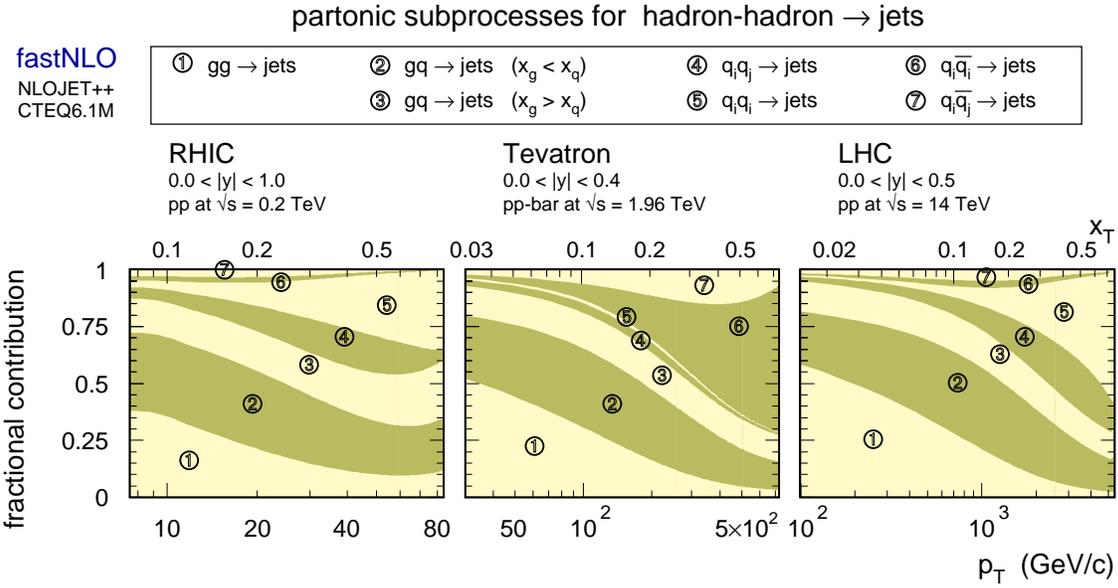,width=15cm}
}
  \caption{Contributions of different partonic subprocesses to 
   the inclusive jet cross section at 
   RHIC (left), the Tevatron (middle) and the LHC (right)
   as a function of $p_T$ and $x_T = 2 p_T/\sqrt{s}$.
   The subprocess $gq \rightarrow {\rm jets}$ has been
   separated into the contributions (2) and (3) where either the 
   quark- or the gluon momentum fraction is larger.
  \label{fig:fnsubprocpp}}
\end{figure}

\begin{figure}[t]
\centerline{
   \psfig{figure=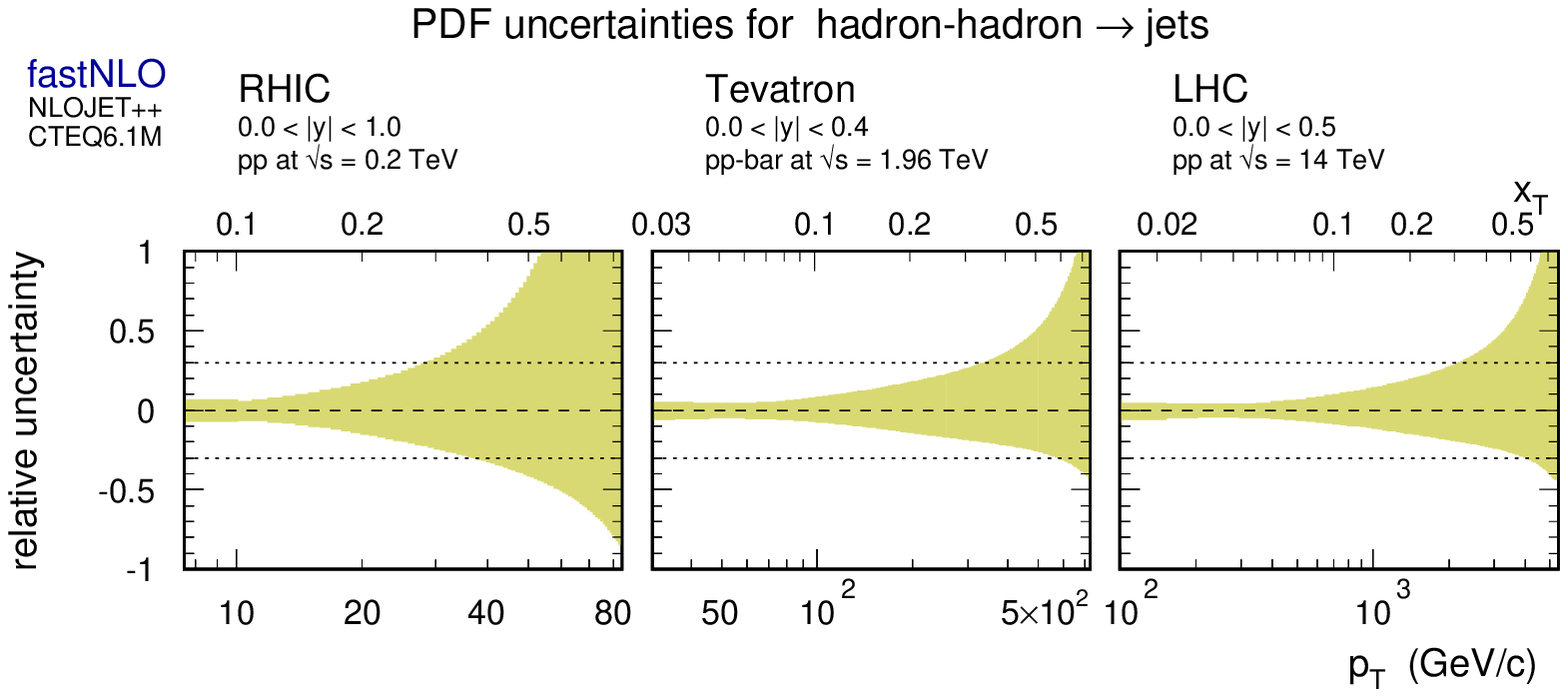,width=15cm}
}
  \caption{Comparison of PDF uncertainties for 
   the inclusive jet cross section at 
   RHIC (left), the Tevatron (middle) and the LHC (right).
   The uncertainty band is obtained for the CTEQ6.1M 
   parton density functions and the results are shown
   as a function of $p_T$ and $x_T = 2 p_T/\sqrt{s}$.
  \label{fig:fnpdfuncpp}}
\end{figure}

A better solution is implemented in the fastNLO project.
The basic idea is to transform the convolution 
in~(\ref{eq:fnmain}) into the factorized expression~(\ref{eq:fnfinal}).
Many proposals for this have been made in the past, originally
related to solving the DGLAP parton evolution equations~\cite{Pascaud:1994vx}
and later to computing of jet cross 
sections~\cite{Lobo:1996,Graudenz:1995sk,Kosower:1997vj,Wobisch:2000dk,Carli:2005ji}.
The fastNLO method is an extension of the 
concepts developed for DIS jet production~\cite{Lobo:1996,Wobisch:2000dk}
which have been applied at HERA
to determine the gluon density in the proton from DIS jet data~\cite{Adloff:2000tq}.
Starting from~(\ref{eq:fnmain}) for  the following discussion the 
renormalization scale is set equal to the factorization scale 
($\mu_{r,f}=\mu$).
The extension to $\mu_r \ne \mu_f$ is, however, trivial.
The $x$ dependence of the PDFs and the 
scale dependence of $\alpha_s^n$ and the PDFs can be approximated 
using an interpolation between sets of fixed values $x^{(k)}$ 
and $\mu^{(m)}$ 
($k=1, \cdots, k_{\rm max}\,;\,  m =1, \cdots, m_{\rm max}$)
\begin{eqnarray}
  &  \alpha^n_s(\mu)&  \cdot \; F_i(x_a,x_b,\mu) \; \simeq 
      \hskip28mm 
{[{\scriptstyle \mbox{``='' is true for 
$k_{\rm max}, l_{\rm max}, m_{\rm max}\rightarrow \infty $} }]}
\nonumber \\
& & 
\sum_{k,l,m}  \alpha^n_s(\mu^{(m)}) \cdot F_i(x_a^{(k)}, x_b^{(l)}, \mu^{(m)}) 
\, \cdot \,  e^{(k)}(x_a) \cdot  e^{(l)}(x_b) \cdot b^{(m)}(\mu)   
\end{eqnarray}
where $e^{(k,l)}(x)$ and $b^{(m)}(\mu)$ are interpolation functions
for the $x$ and the $\mu$ dependence, respectively.
All information of the perturbatively calculable piece
(including phase space restrictions, jet definition, etc.\
but excluding $\alpha_s$ and the PDFs)
is fully contained in the quantity
\begin{equation}
\tilde{\sigma}_{n,i,k,l,m}(\mu) = 
 c_{n,i}(x_a, x_b, \mu) \otimes 
\left[ e^{(k)}(x_a) \cdot e^{(l)}(x_b)  \cdot b^{(m)}(\mu) \right] \, .
\label{eq:sigmatilde}
\end{equation}
In the final prediction for the cross section
the convolution in~(\ref{eq:fnmain}) is then reduced
to a simple product
\begin{equation}
\sigma(\mu) \, \simeq \sum_{n,i,k,l,m} 
\tilde{\sigma}_{n,i,k,l,m}(\mu)  \, \cdot \,
 \alpha^n_s(\mu^{(m)}) \cdot
 F_i(x_a^{(k)}, x_b^{(l)}, \mu^{(m)}) \, .
\label{eq:fnfinal}
\end{equation}
The time-consuming step involving the calculation of the universal
(PDF and $\alpha_s$ independent) $\tilde\sigma$
is therefore factorized and needs to be done only once.
Any further calculation of the pQCD prediction
for arbitrary PDFs and $\alpha_s$ values can later
be done very fast by computing the simple sum of products
in~(\ref{eq:fnfinal}).
While the extension of the method from one 
initial-state hadron~\cite{Wobisch:2000dk}
to two hadrons was conceptually trivial, the case of two hadrons
requires additional efforts to improve the efficiency
and precision of the interpolation.
Both, the efficiency and the precision, are directly related to the 
choices of the points 
$x^{(k,l)}$, $\mu^{(m)}$ and the 
interpolation functions $e(x)$, $b(\mu)$.
The implementation in 
fastNLO achieves a precision of better than $0.1\%$ 
for $k_{\rm max},l_{\rm max} =10$ and $m_{\rm max}\le 4$.
Computation times for cross sections in fastNLO are roughly 
40-200\,$\mu$s per order $\alpha_s$ (depending on 
$m_{\rm max}$).
Further details are given in Ref~\cite{fastnlo}.

The $\tilde{\sigma}$ in~(\ref{eq:sigmatilde}) are computed using 
{\tt NLOJET++}~\cite{Nagy:2003tz,Nagy:2001fj}.
A unique feature in fastNLO is the inclusion of the $O(\alpha_s^4)$
threshold correction terms to the 
inclusive jet cross section~\cite{Kidonakis:2000gi},
a first step towards a full NNLO calculation.

\subsubsection*{Results}

\begin{figure}[!h]
\centerline{
   \includegraphics[width=\textwidth]{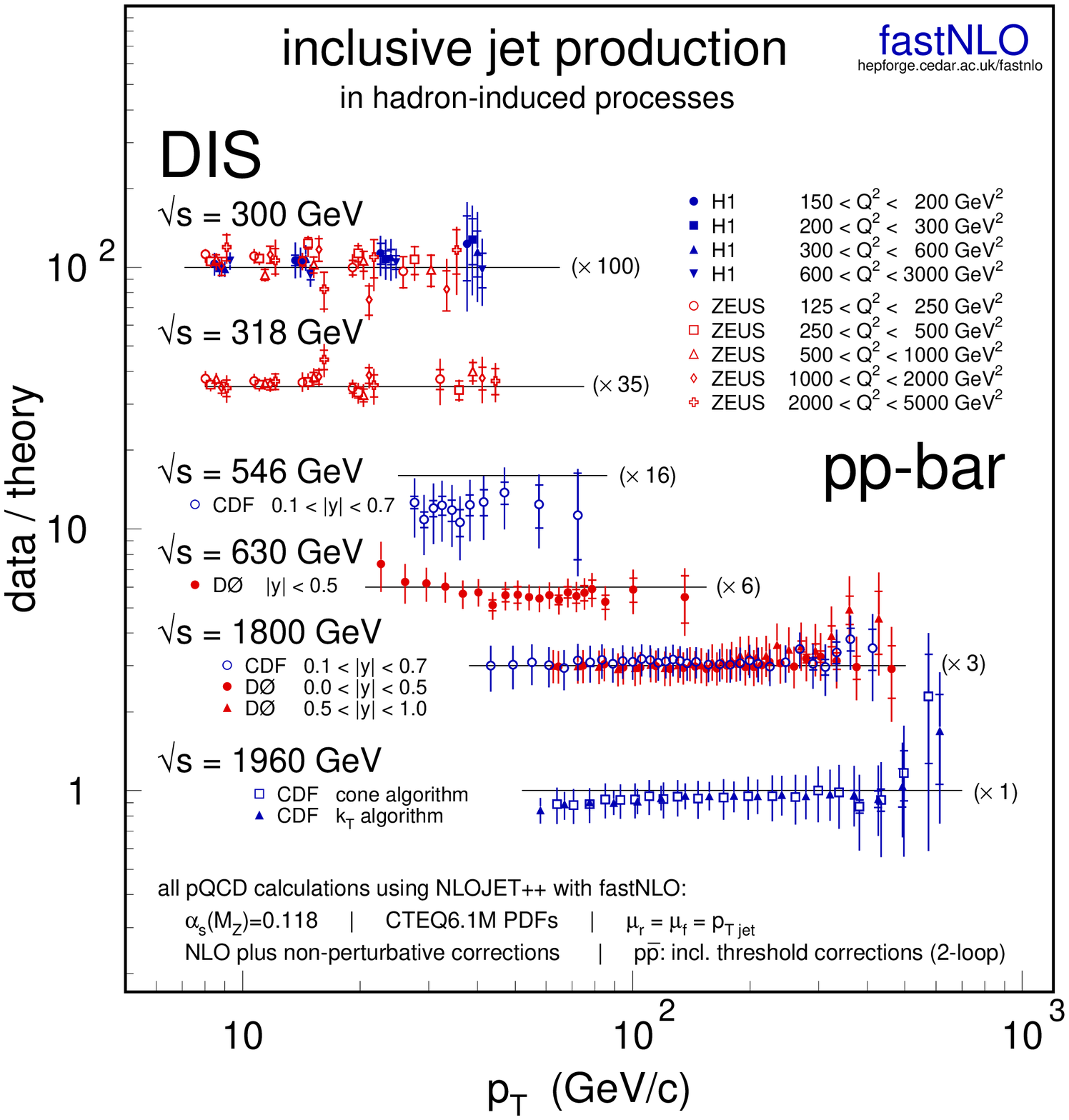}
}
  \caption{An overview of data over theory ratios for 
  inclusive jet cross sections, measured 
  in different processes at different center-of-mass energies.
  The data are compared to calculations obtained by fastNLO
  in NLO precision (for DIS data) and including 
  ${\cal O}(\alpha_s^4)$ threshold  corrections (for $p\bar{p}$ data).
  The inner error bars represent the statistical errors and the
  outer error bars correspond to the quadratic sum of all 
  experimental uncertainties.
  In all cases the perturbative predictions have been 
  corrected for non-perturbative effects.
  \label{fig:fnresults}}
\end{figure}

\begin{figure}[!h]
\centerline{
   \includegraphics[width=\textwidth]{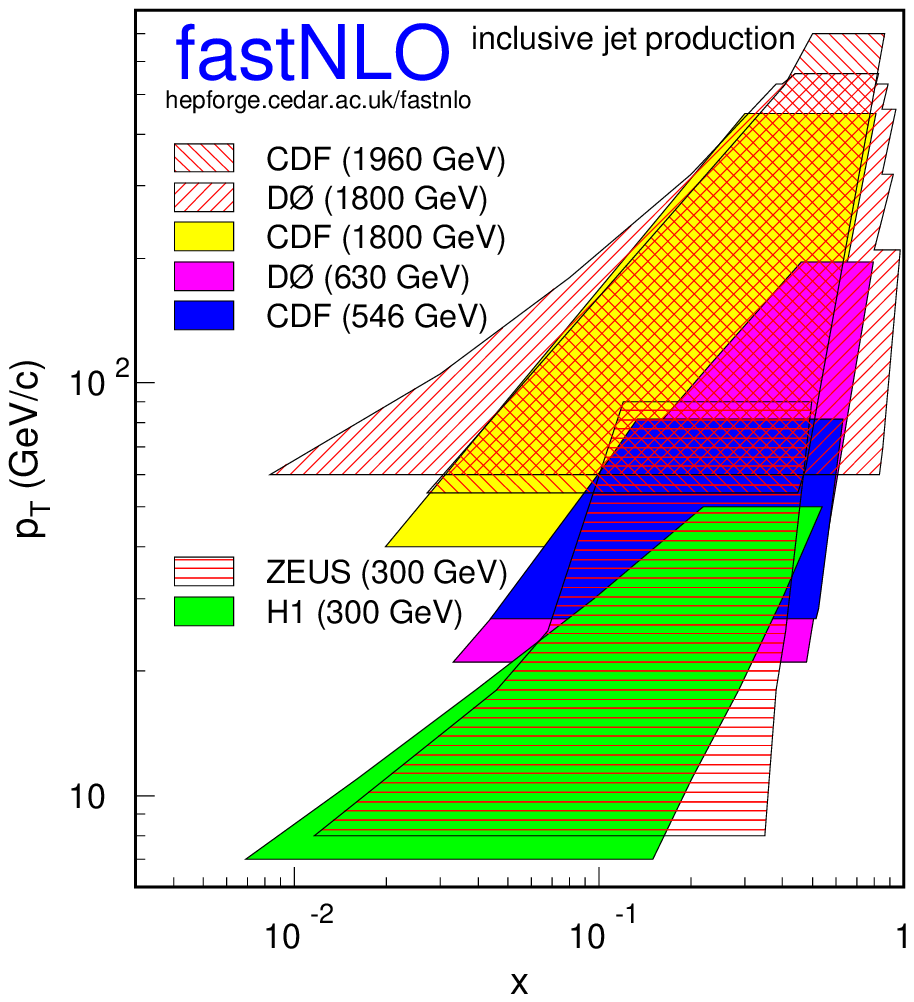}
}
  \caption{The phase space in $x$ and $p_T$
      covered by the data sets shown in the previous figure.
  \label{fig:fnresults2}}
\end{figure}

Calculations by fastNLO
are available at {\tt http://hepforge.cedar.ac.uk/fastnlo}
for a large set of (published, ongoing, or planned) 
jet cross section measurements at 
HERA, RHIC, the Tevatron, and the LHC
(either online or as computer code for inclusion in PDF fits).
Some fastNLO results for the inclusive jet cross section 
in different reactions are shown in this section.
The contributions from different partonic subprocesses
to the central inclusive jet cross section
are compared in Fig.~\ref{fig:fnsubprocpp} for different
colliders: 
For $pp$ collisions at RHIC and the LHC, 
and for $p\bar{p}$ scattering at Tevatron Run II energies.
It is seen that the quark-induced subprocesses are dominated
by the valence quarks:
In proton-proton collisions (RHIC, LHC)
the quark-quark subprocesses (4,5) give much larger 
contributions than the quark-antiquark subprocesses (6,7)
while exactly the opposite is true for proton-antiproton collisions
at the Tevatron.
The contribution from gluon-induced subprocesses is 
significant at all colliders over the whole $p_T$ ranges.
It is interesting to note that at fixed $x_T = 2 p_T/\sqrt{s}$
the gluon contributions are largest at RHIC.
Here, the jet cross section at 
$x_T = 0.5$ still receives $55\%$
contributions from gluon-induced subprocesses,
as compared to only $35\%$ at the Tevatron or $38\%$ at the LHC.
As shown in Fig.~\ref{fig:fnpdfuncpp}, this results in much larger
PDF uncertainties for the high $x_T$ inclusive jet cross section 
at RHIC, as compared to the Tevatron and the LHC for which
PDF uncertainties are roughly 
of the same size (at the same $x_T$).
This indicates that the PDF sensitivity at the same $x_T$
is about the same at the Tevatron and at the LHC, 
while it is much higher at RHIC.

An overview over published measurements of the inclusive
jet cross section in different reactions and at different
center-of-mass energies is given in Fig.~\ref{fig:fnresults}.
The results are shown as ratios of data over theory.
The theory calculations include the best available 
perturbative predictions (NLO for DIS data and NLO + 
${\cal O}(\alpha_s^4)$ threshold corrections for $p\bar{p}$ data)
which have been corrected for non-perturbative effects.
Over the whole phase space of $8 < p_T < 700$\,GeV
jet data in DIS and $p\bar{p}$ collisions are well-described
by the theory predictions using 
CTEQ6.1M PDFs~\cite{Pumplin:2002vw}.
The phase space in $x$ and $p_T$ covered 
by these measurements is shown in Fig.~\ref{fig:fnresults2},
demonstrating what can be gained by using fastNLO 
to include these data sets in future PDF fits.
A first study using fastNLO on the future potential
of LHC jet data has been published in Ref.~\cite{cmsptdrv2}.

\clearpage
\section{Event Generator Tuning}

\subsection{{Dijet Azimuthal Decorrelations and Monte Carlo Tuning}}

\textbf{Contributed by Begel, Wobisch, Zielinski}

{\em Using a recent D\O\ measurement of correlations in the 
dijet azimuthal angle in $p\bar{p}$ collisions,
we investigate the description of data
by  Monte Carlo event generators. 
We analyze the impact of various phenomenological parameters
employed in the generators and demonstrate that the data can
unambiguously constrain the description of Initial State Radiation
(ISR) in \PY. 
Finally, we use the next-to-leading order (NLO) pQCD extrapolation 
to evaluate the description of QCD radiation effects by 
the Monte Carlo tools at the LHC energy.
} \\

\newcommand{\Dphi}{\Delta \phi\,{}_{\rm dijet}}
\newcommand{\ptmax}{p_T^{\rm max}}

The proper description of multi-parton radiation
is crucial for a wide range of precision measurements as
well as for searches for new physical phenomena at the LHC.
Thus, it is essential that the Monte Carlo tools employed in data analyses
accurately describe the observed aspects of such radiation.
While the Monte Carlo generators have been tuned using selected Tevatron
and lower energy data, it is interesting to inquire to what extent 
such tuning will be valid also at the LHC.
In particular, the \PY Monte Carlo~\cite{Sjostrand:2006za} 
has been tuned to describe the underlying event
in hadron collisions at around 
$\sqrt{s} = 2$ TeV~\cite{Affolder:2001xt,Field:2005sa}.
This tuning involved adjusting parameters 
for (``soft'') physics processes 
at small transverse momentum, $p_T$, (hadronization, underlying event)
as well as parameters for high $p_T$ (``hard'') physics processes,
like parton showering.
A significant correlation between the parameters
for the soft  and the hard contributions was noticed,
and the resulting tunes represented
different parameter sets which all gave a good 
description of the underlying-event data~\cite{Field,Field:2005qt}.

In this paper we study a recent measurement from the 
D\O\ collaboration, which allows to isolate the effects of 
hard physics processes.
The D\O\ collaboration has measured distributions of the 
azimuthal difference $\Dphi$ between the two jets with largest
$p_T$ in an event~\cite{Abazov:2004hm}.
This observable provides an effective probe of radiation effects; 
consequently, the D\O\ measurement adds independent information 
to that included in the previous tunes, 
and constrains the effects from high $p_T$
initial-state radiation (ISR) unambiguously.

In the following, we compare the $\Dphi$ data to 
predictions from Monte Carlo event generators,
and investigate the sensitivity of their phenomenological parameters. 
Finally, we use next-to-leading order (NLO) perturbative QCD (pQCD) 
predictions to 
demonstrate that the tuning is successfully transferred 
to LHC energies.

\subsubsection*{The Physics of $\Dphi$ Decorrelations}
Dijet azimuthal decorrelations
in hadron-hadron collisions are sensitive to different
physics processes in different regions of the azimuthal angle
between the two leading jets $\Dphi$.
At Born-level dijet events have two jets with equal $p_T$
with respect to the beam axis and their azimuthal angles
$\phi_{\rm jet}$ are correlated such that
$\Dphi=|\phi_{\rm jet\,1} - \phi_{\rm jet\,2}| = \pi$.
Any deviation from $\Dphi = \pi$ 
is caused by additional radiation in the event which is not
clustered into the two highest $p_T$ jets.
Radiation with small $p_T$ will change $\Dphi$ by a smaller amount,
while hard radiation (with high $p_T$) can reduce  $\Dphi$
significantly, as illustrated in Fig.~\ref{fig:sketch}.
The $\Dphi$ distribution therefore allows to study the continuous 
transition from soft (non-perturbative) to hard (perturbative) 
QCD processes, based on a single observable.
The QCD predictions for the different contributions are 
determined as follows:

\begin{figure}[t]
  \centerline{
   \psfig{figure=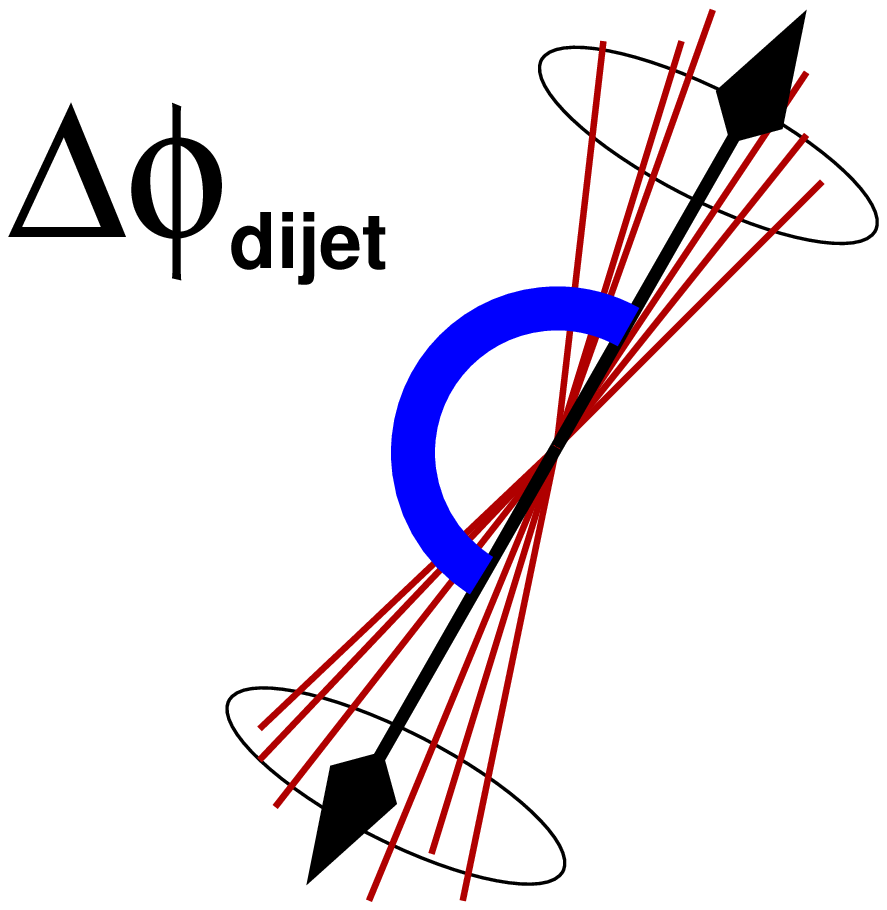,height=4cm}
   \psfig{figure=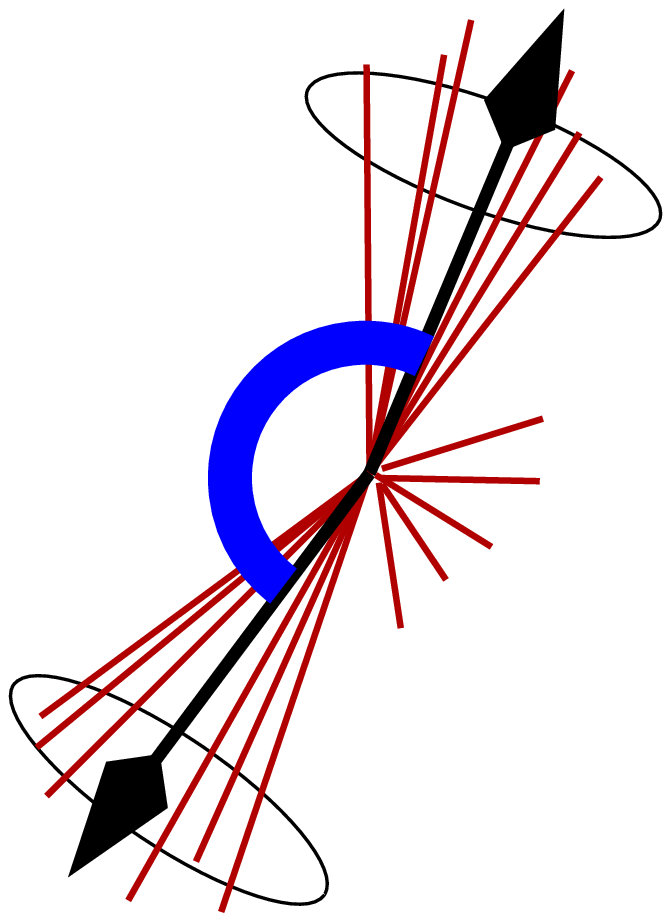,height=4cm}
   \psfig{figure=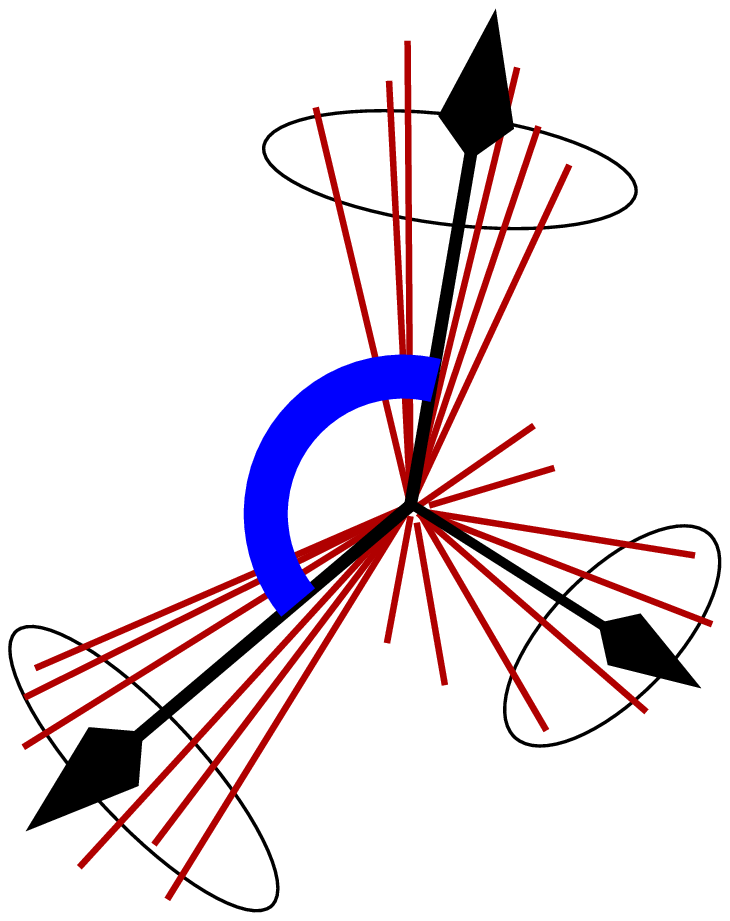,height=4cm}\hskip3mm
   \psfig{figure=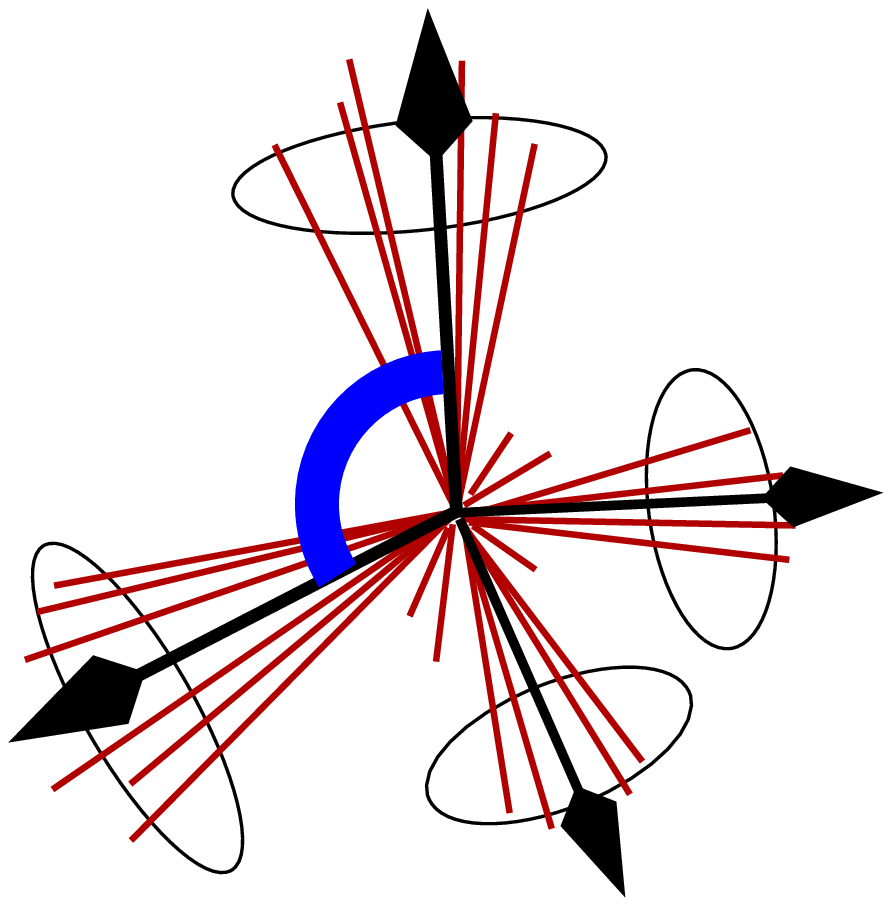,height=4cm}
  }
  \caption{\label{fig:sketch}
   A sketch of the angle $\Dphi$ in dijet events with an increasing  
   amount of additional radiation outside the dijet system.}
\end{figure}

\begin{itemize}
\item  hard perturbative processes \\
Hard emissions which produce additional jets 
have been computed in pQCD in fixed order of the strong coupling constant
$\alpha_s$ up to next-to-leading order (${\cal O}(\alpha_s^4)$)
for the differential $\Dphi$ distribution~\cite{Nagy:2001fj,Nagy:2003tz}.

\item  soft perturbative processes \\
Fixed-order calculations fail in phase space regions which
are dominated by soft multi-parton emissions.
In these regions contributions from logarithmic terms are enhanced 
and need to be resummed to all orders of $\alpha_s$.
Methods for the automated resummation of certain classes of
observables in hadron-hadron collisions have recently become 
available~\cite{Banfi:2004yd,Banfi:2004nk}.
The $\Dphi$ distribution is, however, not a ``global'' 
observable~\footnote{An observable is called ``global''
when it is sensitive to all particles in the event.
The $\Dphi$ distribution is, however, not sensitive
to the particles inside the jets.}
(as defined in~\cite{Banfi:2004nk}).
Therefore these automated resummation methods can not be applied. \\
An alternative description of multi-parton emissions
is given in parton cascade models (parton shower or dipole cascade).
These are implemented in Monte Carlo event generators like
\PY or \HW~\cite{Corcella:2002jc},
where they are matched to the
Born-level matrix elements.

\item  non-perturbative processes \\
Processes like hadronization and activity related to the
beam remnants (``underlying event'') can not be computed
from first principles.
Phenomenological models for these processes, matched to the
parton cascade models, are used in the Monte Carlo event generators.  

\end{itemize}

\subsubsection*{Distributions of $\Dphi$ in Data and Monte Carlo}

The experimental observable has been defined as the differential 
dijet cross section in $\Dphi$, normalized by the dijet cross section 
integrated over $\Dphi$ in the same phase space:
$(1/\sigma_{\rm dijet})\, 
 (d\sigma_{\rm dijet}/ d\Delta\phi)$~\cite{Abazov:2004hm}.
In this ratio theoretical and experimental uncertainties are reduced.
Jets have been defined using an iterative seed-based cone algorithm 
(including mid-points)~\cite{run2cone} with radius $R_{\rm cone}=0.7$ at
parton, particle, and experimental levels.
Four analysis regions have been defined based on the jet with largest
$p_T$ in an event ($\ptmax$).
The second leading-$p_T$ jet in each event is required to have
$p_T>40$~GeV and both jets have central rapidities with  $|y| < 0.5$.

\begin{figure}[t]
  \centerline{
   \psfig{figure=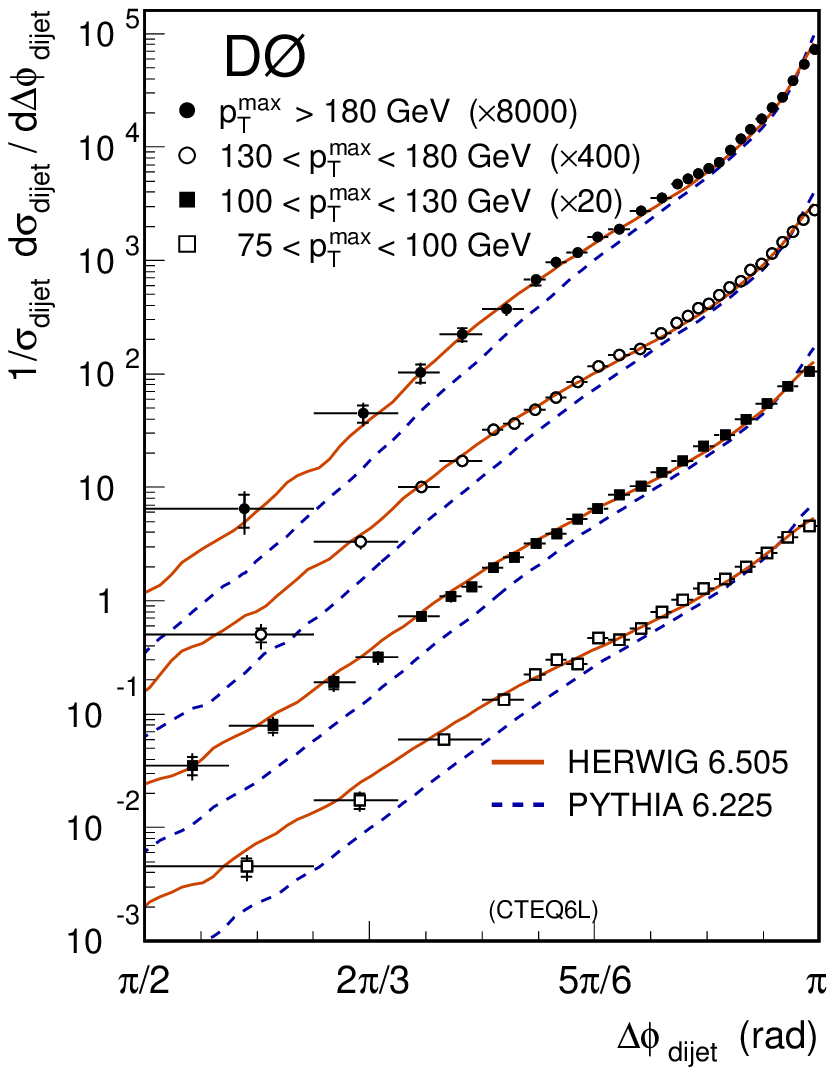,width=7.7cm}
   \psfig{figure=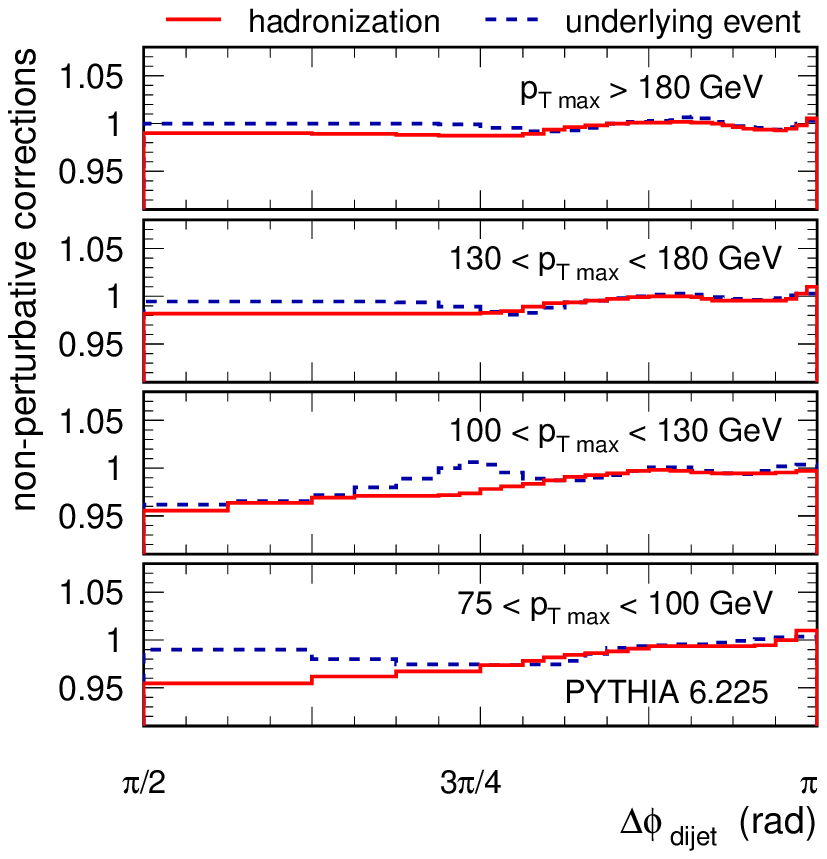,width=7.7cm}
  }
  \caption{\label{fig:fig0}
Left: The $\Dphi$ distributions in different $\ptmax$ ranges.
Data and predictions with $\ptmax > 100$ GeV are scaled by
successive factors of 20 for purposes of presentation.
Results from default versions of \HW and 
\PY are overlaid on the data.
Right:
Model predictions of non-perturbative corrections for the 
$\Dphi$ distribution in four $\ptmax$ regions.
Hadronization corrections (solid line) 
and effects from underlying event (dashed line)
have been determined using \PY.}
\end{figure}

The data are compared to predictions from the \PY and 
\HW generators in Fig.~\ref{fig:fig0} (left).
The observed spectra are strongly peaked at $\Dphi\approx\pi$
and the peaks are narrower at larger values of $\ptmax$.
The predictions of the Monte Carlo event generators
have been obtained using the respective default settings, unless
stated otherwise.
The generators are using the {\sc cteq6l} parton density functions (PDF's).
The $\Lambda_{\rm QCD}$ values in the generators are adjusted such that 
the resulting value of $\alpha_s(M_Z)=0.118$ is consistent with the world 
average and with the value that was used in 
the {\sc CTEQ}6 PDF fit~\cite{Pumplin:2002vw}.
Consistent with the procedure in the PDF fit we are using the
2-loop solution for the renormalization group equation.
This is the default in \HW, but needs to be set in 
\PY using the switch {\tt MSTP(2)=2}.
Below, these settings will be referred to as the ``default'' settings.

The default \HW (version 6.505) gives a good
description of the data over the whole $\Dphi$ range in all $\ptmax$ regions.
It is slightly below the data around $\Dphi \approx 7\pi/8$
and slightly narrower peaked at $\pi$.
The default version of \PY (version 6.225) does not describe the data.
The spectrum is much steeper over the whole $\Dphi$ range, independent
of $\ptmax$.
These deviations will be investigated in the following.

\subsubsection*{Non-Perturbative Contributions}

Before we investigate the contributions from perturbative
QCD processes we study the sensitivity of the $\Dphi$ distribution
to non-perturbative contributions, stemming from the
hadronization process or the underlying event.

Fig.~\ref{fig:fig0} (right, dashed line) shows the underlying event correction,
defined as the ratio of the default \PY
(including underlying event) and  \PY with the
underlying event switched off by {\tt MSTP(81)=0}.
It is apparent that the effects from underlying event are below
four percent.

The hadronization corrections are defined as the ratio
of the observable, on the level of partons
(from the parton shower) and on the level of stable 
particles.
Fig.~\ref{fig:fig0} (right, solid line) shows that these corrections
are below 2-5\% over the whole range.

We conclude that the $\Dphi$ distribution is not sensitive to 
non-perturbative effects and these can not explain the deviations 
between \PY and the data.
Hence we do not attempt to tune the \PY parameters
for the hadronization or the underlying event models.
We also can neglect the non-perturbative effects 
when comparing to the purely perturbative NLO QCD predictions.

\begin{figure}[t]
  \centerline{
   \psfig{figure=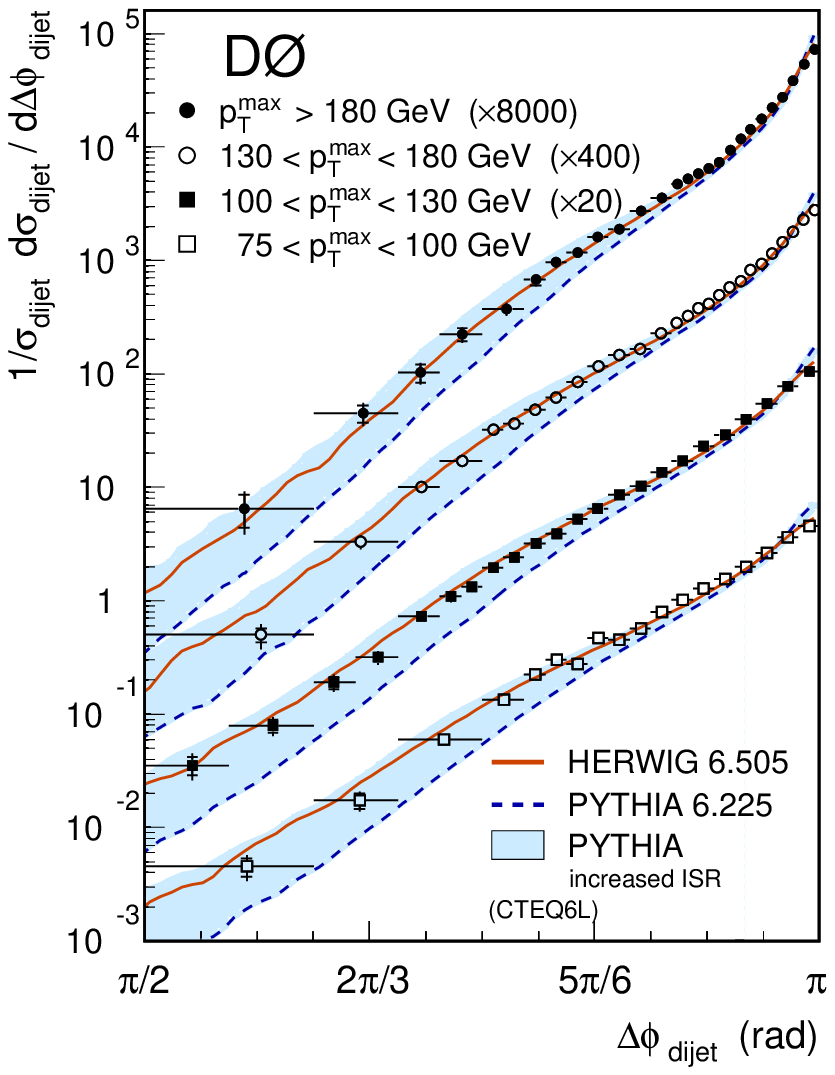,width=7.7cm}
   \psfig{figure=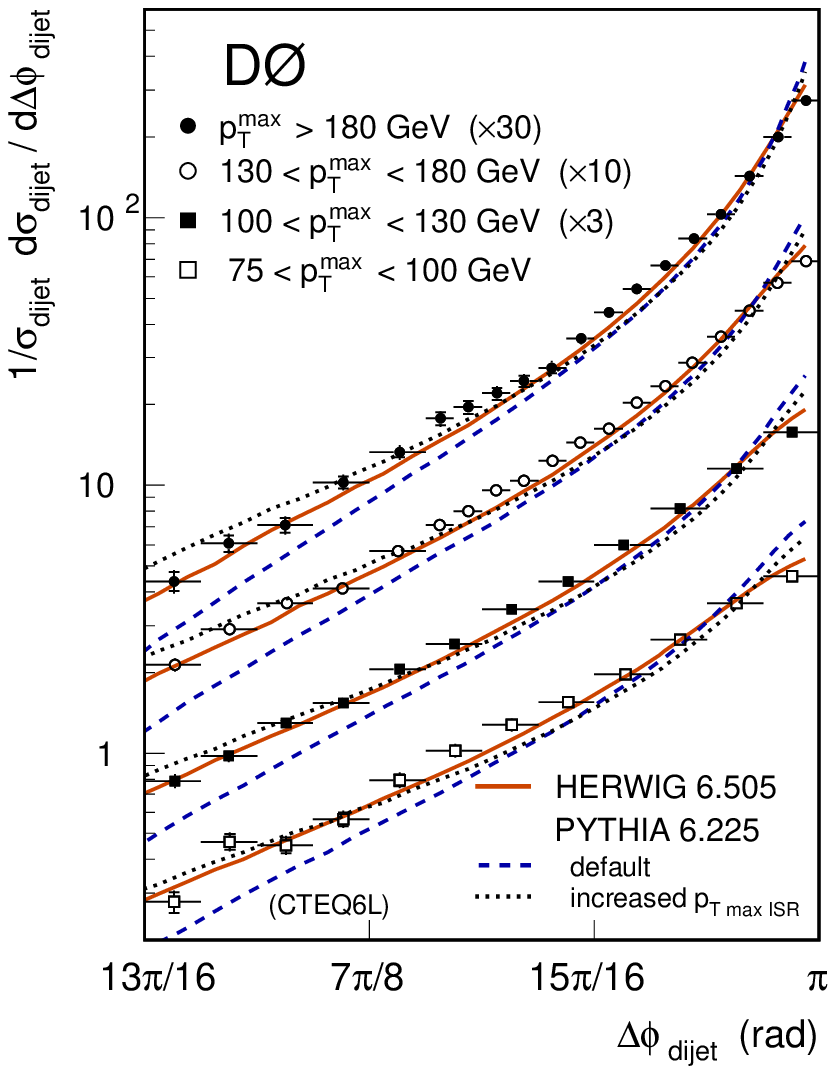,width=7.7cm}
  }
  \caption{\label{fig:mcdef}
Predictions from \HW and \PY are compared to 
the measured $\Dphi$ distributions over the whole range
of $\Dphi$ (left) and in the peak region $\Dphi > 13\pi/16$  (right). 
\PY predictions are shown for a range of settings
of {\tt PARP(67)} between 1.0 and 4.0.}
\end{figure}

\subsubsection*{Parameter Tuning}

To investigate the possibilities of tuning  \PY
we first focus on the impact of the ISR parton shower.
\PY contains various parameters by which the ISR shower
can be adjusted.
The maximum allowed $p_T$, produced by the ISR shower is limited by the
upper cut-off on the parton virtuality. 
This cut-off is controlled by the product of the parameter
{\tt PARP(67)} and the hard scattering scale squared (which is 
equal to $p_T^2$ for massless partons).
Increasing this cut-off by varying {\tt PARP(67)} from its default of 1.0  
to 4.0 leads to significant changes of the \PY prediction for $\Dphi$.
Fig.~\ref{fig:mcdef} shows comparisons of  \PY and \HW 
to data over the whole $\Dphi$ range (left)
and in greater detail in the region $\Dphi > 13\pi/16$ (right).
The increased value of {\tt PARP(67)} in \PY
increases the tail of the distribution strongly, especially at lowest $\Dphi$.
At large $\Dphi$, however, this parameter has not enough effect
to bring \PY close to the data.
The best description at low $\Dphi$ is obtained for {\tt PARP(67) = 2.5}
(referred to as ``TeV-tuned'' in the following) as shown in
Fig.~\ref{fig:dphi-sherpa}.

\begin{figure}[t]
  \centerline{
   \psfig{figure=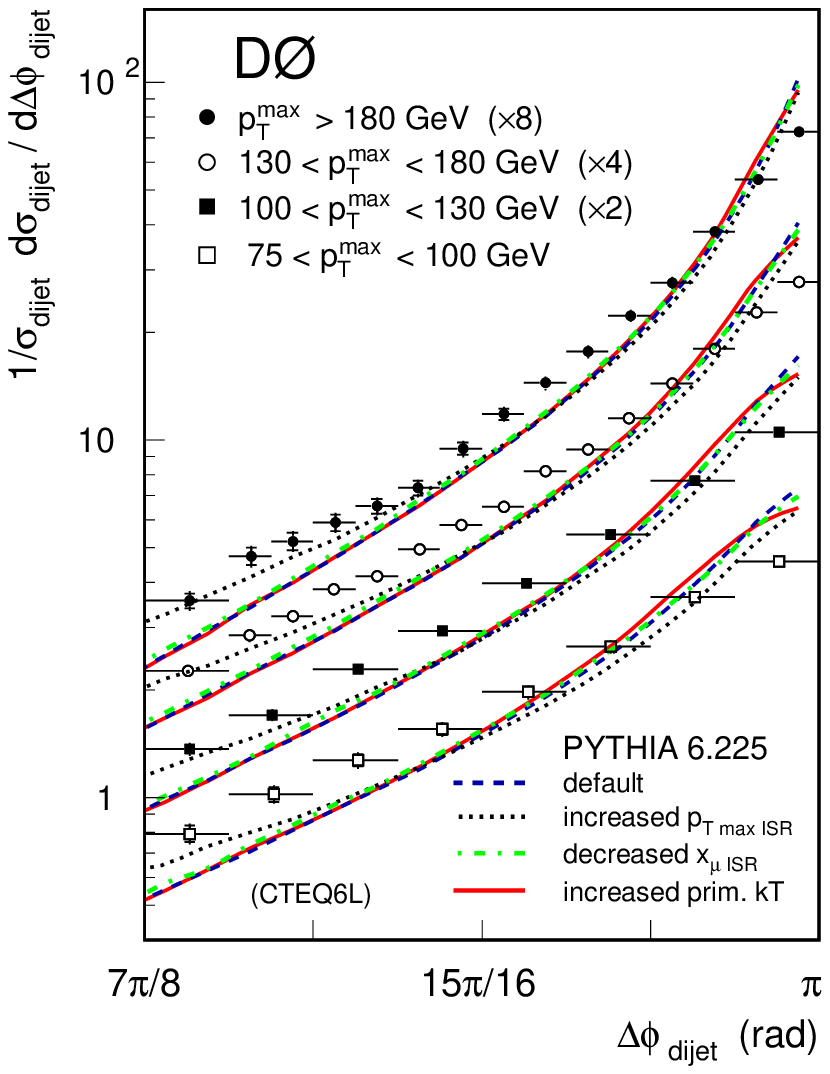,width=7.7cm}
   \psfig{figure=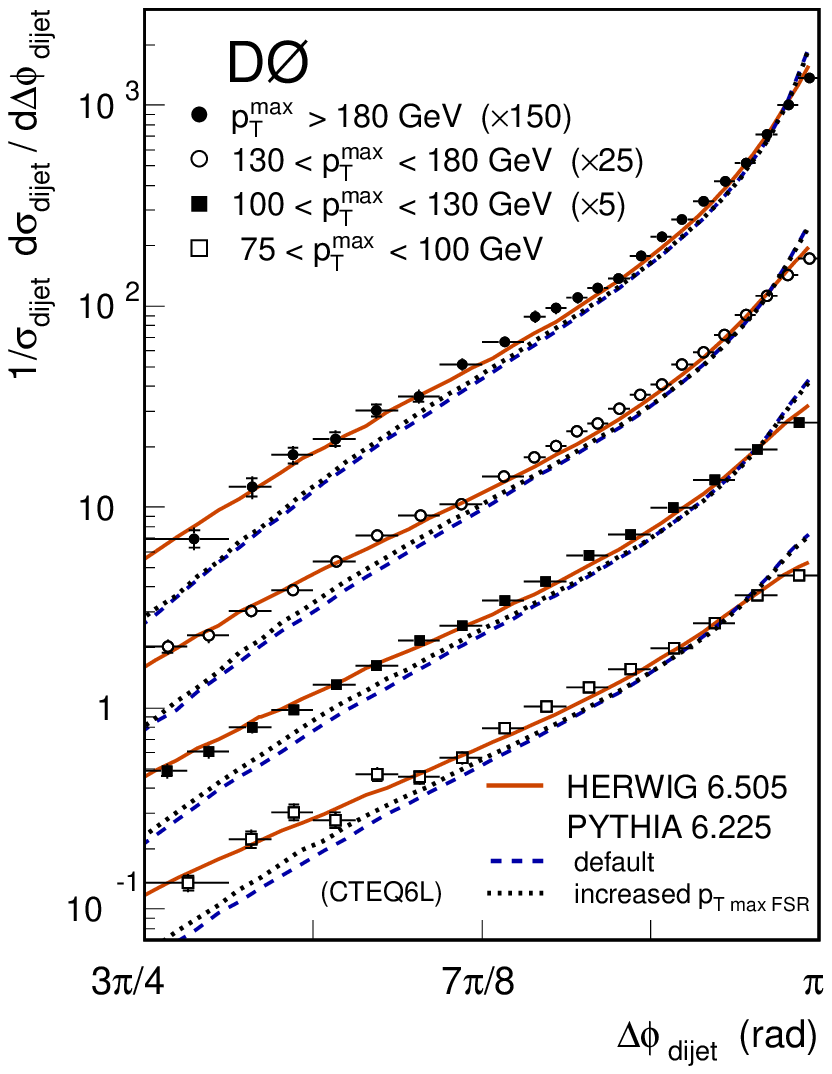,width=7.7cm}
  }
  \caption{\label{fig:mcpyisrfsr}
Left: Predictions from \PY are compared to 
the measured $\Dphi$ distributions for various ISR 
parameter variations.
The comparison is shown in the region of $\Dphi > 7\pi/8$. 
Right: Predictions from \HW and \PY are compared to 
the data at $\Dphi > 3\pi/4$.
In addition to the default \PY version
a prediction with an increased upper limit on the $p_T$ in the
final state parton shower is also shown.}
\end{figure}

In addition, we have tested the impact of other ISR-related parameters 
in \PY.
These are the scale factor ($x_\mu$) for the renormalization scale 
for $\alpha_s$ in the ISR shower, {\tt PARP(64)}, 
and the primordial $k_T$ of partons
in the proton: the central value of the gaussian distribution, 
given by {\tt PARP(91)}, and the upper cut-off, given by {\tt PARP(93)}.
We have lowered the factor for the renormalization scale
from its default of one
to {\tt PARP(64)=0.5} which increases the value of $\alpha_s$.
We have alternatively increased the primordial $k_T$ from 1\,GeV
to 4\,GeV, {\tt PARP(91)=4.0}, and the upper cut-off
of the gaussian distribution from 5\,GeV to 8\,GeV,
{\tt PARP(93)=8.0}.
None of these parameter variations has an appreciable effect 
on the region at low $\Dphi$.
The effects at large $\Dphi$ are shown in Fig.~\ref{fig:mcpyisrfsr} (left).
It is clearly visible that they are very small.
While the scale factor has almost no influence,
there is some small change for the increased primordial $k_T$
which manifests itself only very close to the peak region
and only at lower values of $\ptmax$.

We have also studied the sensitivity of parameters for 
the final-state radiation (FSR) parton shower.
The maximum $p_T$ of partons from FSR is controlled by the 
parameter {\tt PARP(71)} in the same way that {\tt PARP(67)}
controls the maximum $p_T$ from ISR.
We have increased {\tt PARP(71)} from its default value of $4.0$ to $8.0$.
The result is shown in Fig.~\ref{fig:mcpyisrfsr} (right) and compared
to default \PY  and to \HW.
It is seen that the increased $p_T$ in the FSR shower leads only to 
small changes in the range $3\pi/4 < \Dphi < 7\pi/8$, decreasing 
towards higher $\ptmax$.

In conclusion, {\tt PARP(67)} is the only parameter 
we have found in \PY
that has a significant impact on  the $\Dphi$
distribution. While it is not sufficient for a perfect tuning of
\PY to data, this observation can be used for 
an unambiguous determination of the optimal value of this parameter.

\subsubsection*{Matched Monte Carlo Predictions}

\begin{figure}[t]
  \centerline{
   \psfig{figure=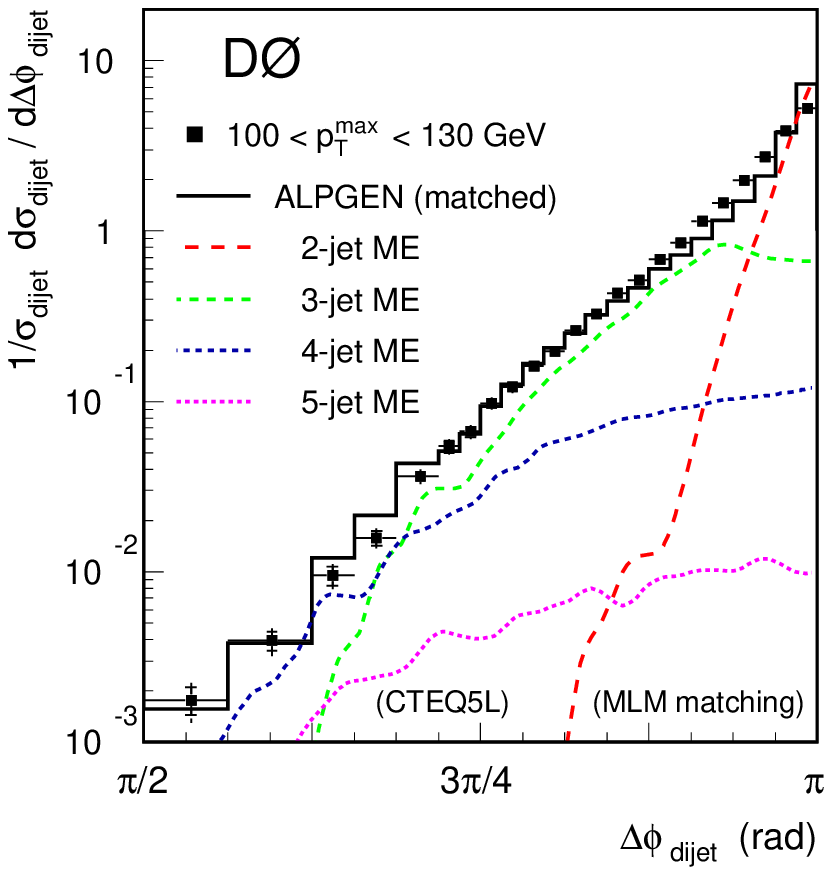,width=7.7cm}
   \psfig{figure=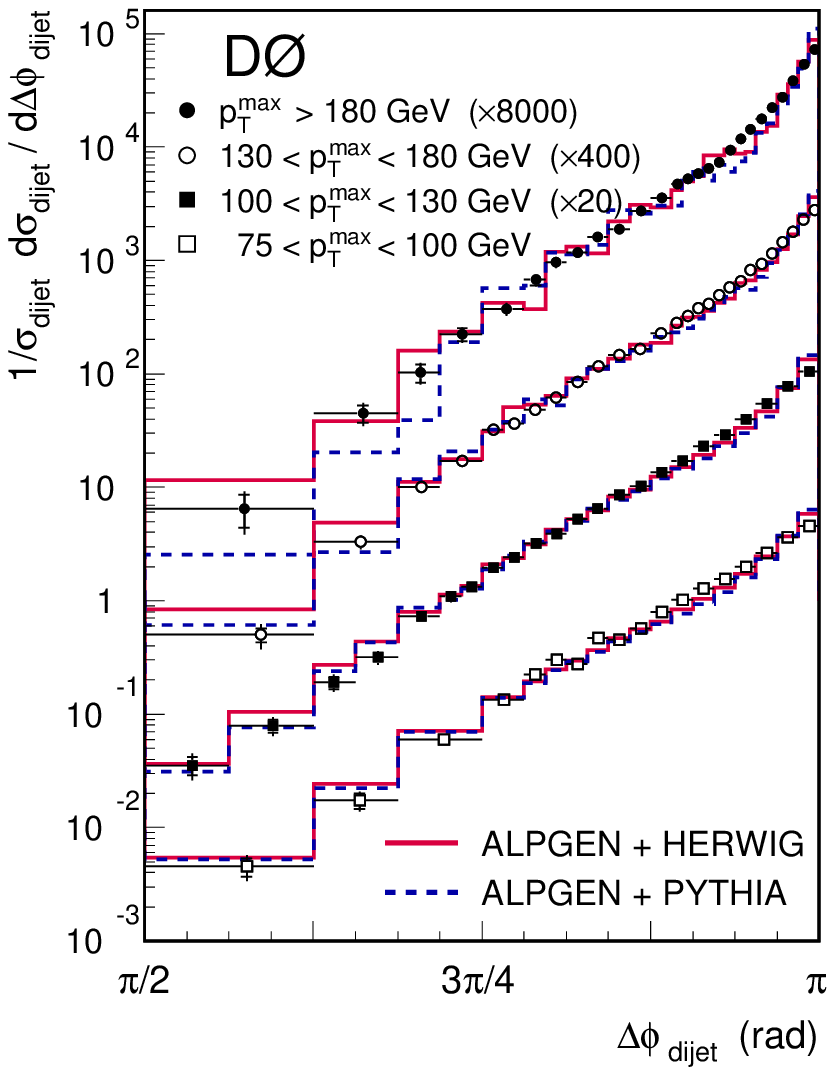,width=7.7cm}
  }
  \caption{\label{fig:dphi-alpgen}
Left: Contributions from different multiplicity bins in ALPGEN
compared to the data in one $\ptmax$ bin.  Right: The $\Dphi$
distributions in four regions of $\ptmax$ overlayed with results
from ALPGEN \& \PY and ALPGEN \& \HW.}
\end{figure}

\begin{figure}[t]
  \centerline{
   \psfig{figure=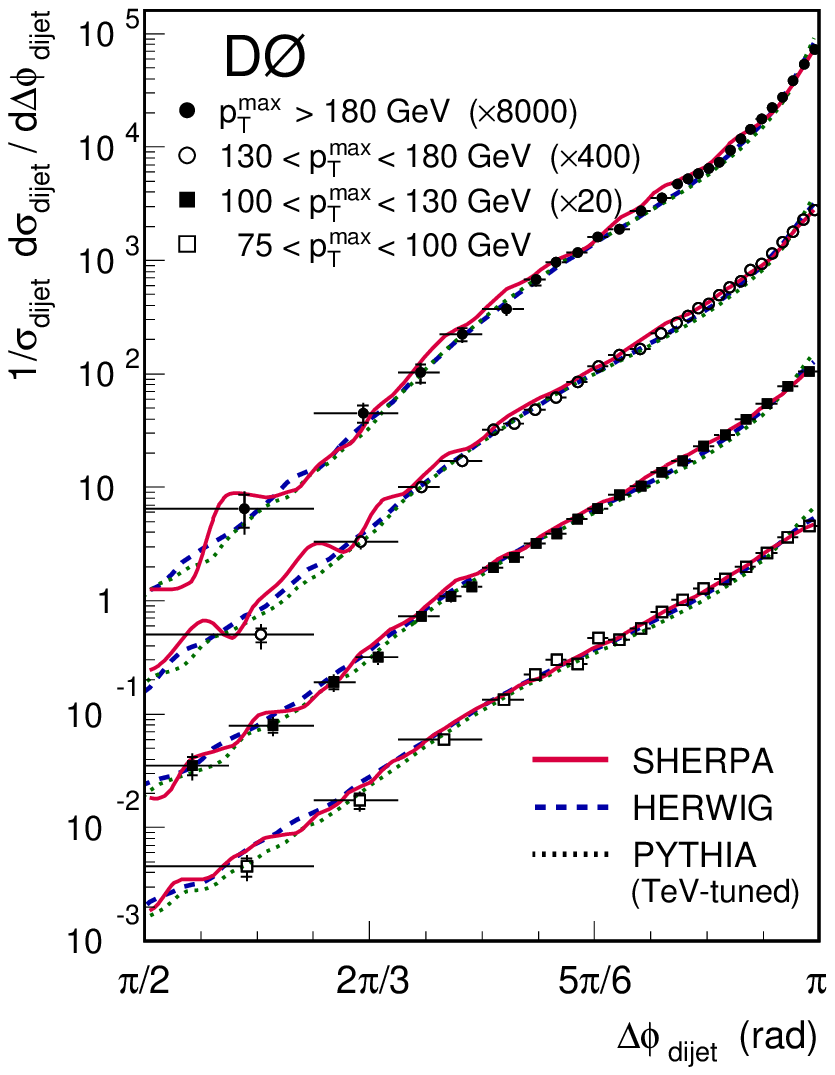,width=7.7cm}
 }
  \caption{\label{fig:dphi-sherpa}
   The $\Dphi$ distributions in four regions of $\ptmax$
   overlayed with results from SHERPA, \HW and the TeV-tuned \PY.}
\end{figure}

Fixed-order matrix-element event generators are used extensively in
studies of top and Higgs production.  Multi-jet configurations are
produced by incorporating high-order tree-level pQCD diagrams with
phenomenological parton-shower models such as those from 
\PY or \HW.  Verification of
their performance using high-statistics QCD processes is of clear
interest for applications that require accurate descriptions of
processes with several jets.  Some of these calculations have
prescriptions to avoid double-counting contributions with equivalent
phase-space configurations~\cite{Mrenna:2003if,Hoche:2006ph}.
$\Dphi$ distributions offer an interesting avenue for testing the
smoothness of matching between matrix-element and parton-shower
contributions as the average jet multiplicity varies across the
$\Dphi$ range.

ALPGEN~\cite{Mangano:2002ea} uses the MLM matching prescription
which rejects events that have reconstructed parton-shower jets that
do not overlap with generated partons, thus excluding those events
where the jets arose from the parton-shower mechanism.  (The highest
multiplicity bin includes these extra jets.)  Samples with different
jet multiplicities are then combined together according to the MLM
scheme into an inclusive sample that can be compared to data.
Fig.~\ref{fig:dphi-alpgen}~(left) shows an example of this scheme for
multi-jet production.  Samples for $2\rightarrow2$, $2\rightarrow3$,
$\ldots$, $2\rightarrow6$ jet production were combined using the MLM
scheme.  Individually, none of the contributions compares favorably
with the data.  However, the combined ALPGEN calculation agrees
reasonably well with the data.  This result does not depend on the
details of the parton-shower model (Fig.~\ref{fig:dphi-alpgen}~right).

SHERPA~\cite{Gleisberg:2003xi}, another tree-level pQCD event
generator, uses the CKKW~\cite{Schalicke:2005nv} matching scheme to
produce multi-jet events.  Here, the parton-shower progression is
pruned so that only allowable configurations are produced.  
SHERPA uses its own parton-shower model; it does not use either 
\PY or \HW.  Fig.~\ref{fig:dphi-sherpa} shows the results
from SHERPA for multi-jet production compared to the D\O\ data
and to the results from \HW and the TeV-tuned \PY.
The results from SHERPA provide a good description of the data.

\subsubsection*{Perturbative QCD Predictions}

\begin{figure}[t]
  \centerline{
   \psfig{figure=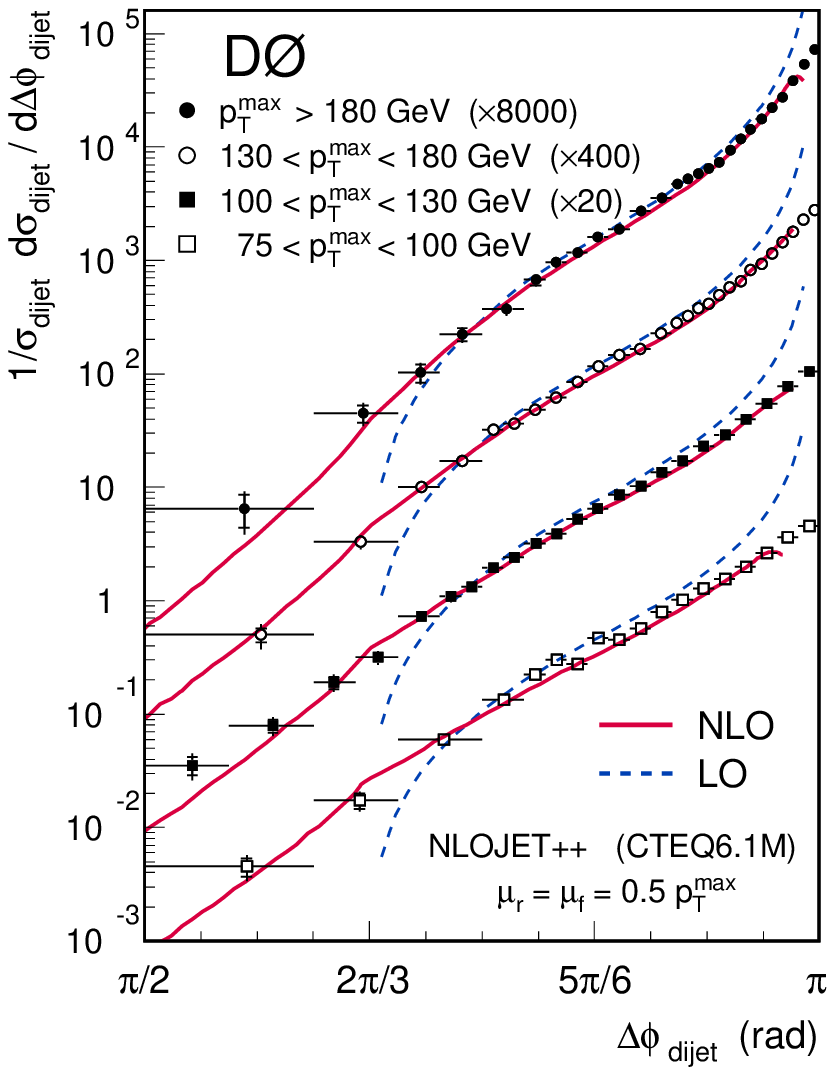,width=7.7cm}
   \psfig{figure=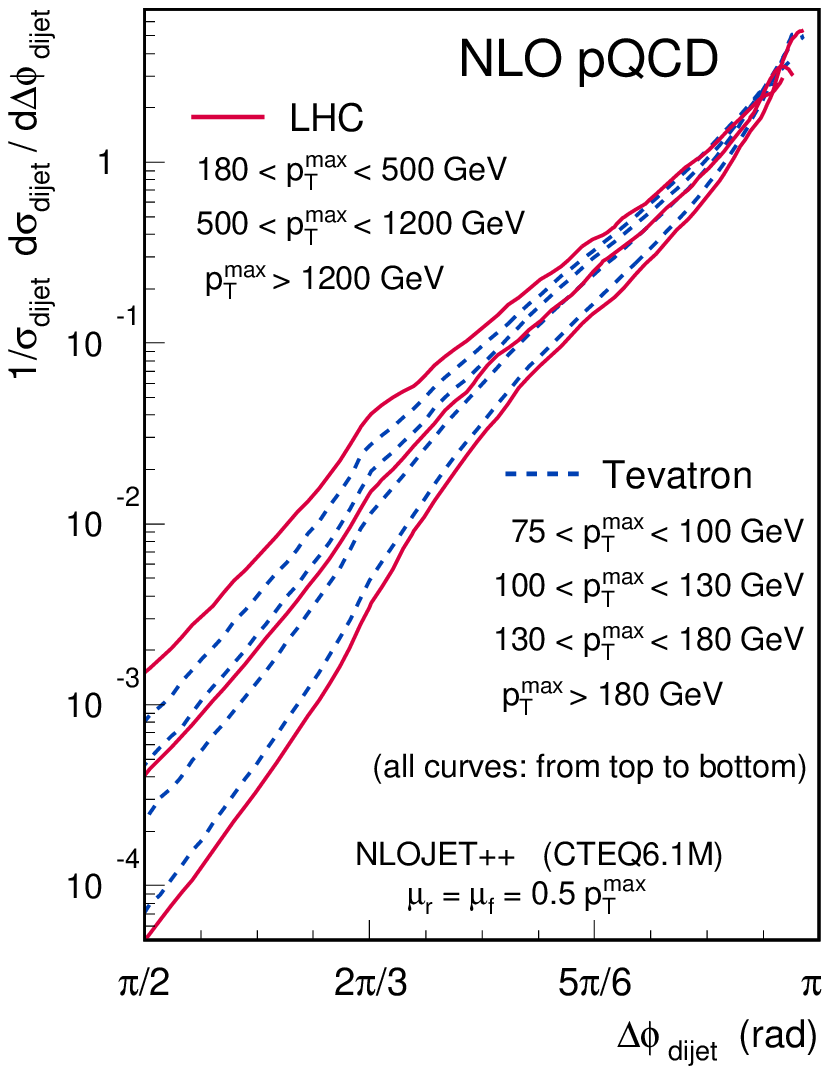,width=7.7cm}
  }
  \caption{\label{fig:NLO}
Left: $\Dphi$ data and pQCD calculations are compared in different 
$\ptmax$ ranges at Tevatron.
The solid (dashed) lines show the NLO (LO) pQCD predictions.
Right:
Comparisons between NLO calculations for $\Dphi$ at the Tevatron and LHC,
for selected $p_T$ ranges.  }
\end{figure}

The $\Dphi$ distributions can be directly employed to test the purely
perturbative QCD predictions since non-perturbative corrections 
can be safely neglected.
Fig.~\ref{fig:NLO} (left) shows the comparison of pQCD
calculations obtained using the parton-level event generator 
{\sc nlojet++}~\cite{Nagy:2001fj,Nagy:2003tz} and the {\sc cteq6.1m} PDF's~\cite{Pumplin:2002vw}
and data.
The integrated dijet cross section and the differential
dijet cross section in $\Dphi$ are computed separately in their
respective LO and NLO.
In all cases the renormalization and factorization scales
are set to $\ptmax / 2$.

The leading-order calculation clearly has a limited applicability.
Due to the limited phase space for three-parton final states
it does not cover the region $\Dphi<2/3\pi$, and towards
$\Dphi = \pi$ it becomes divergent.
NLO pQCD provides a good description of the data over most of the
range of $\Dphi$. 
Only for $\Dphi \approx \pi$ the NLO prediction is insufficient, 
and a resummed calculation is required. 
It would be of great interest to test the resummation techniques 
against the $\Dphi$ data when a resummed result becomes available.

\begin{figure}[t]
  \centerline{
   \psfig{figure=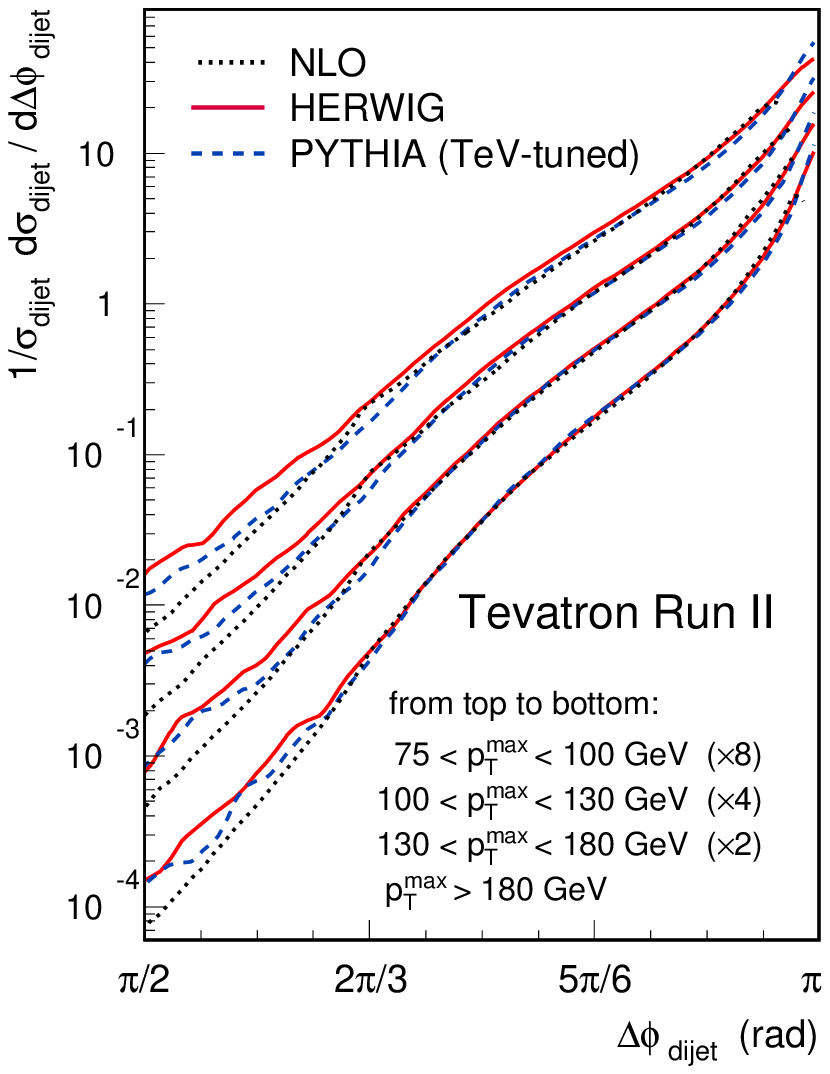,width=7.7cm}
   \psfig{figure=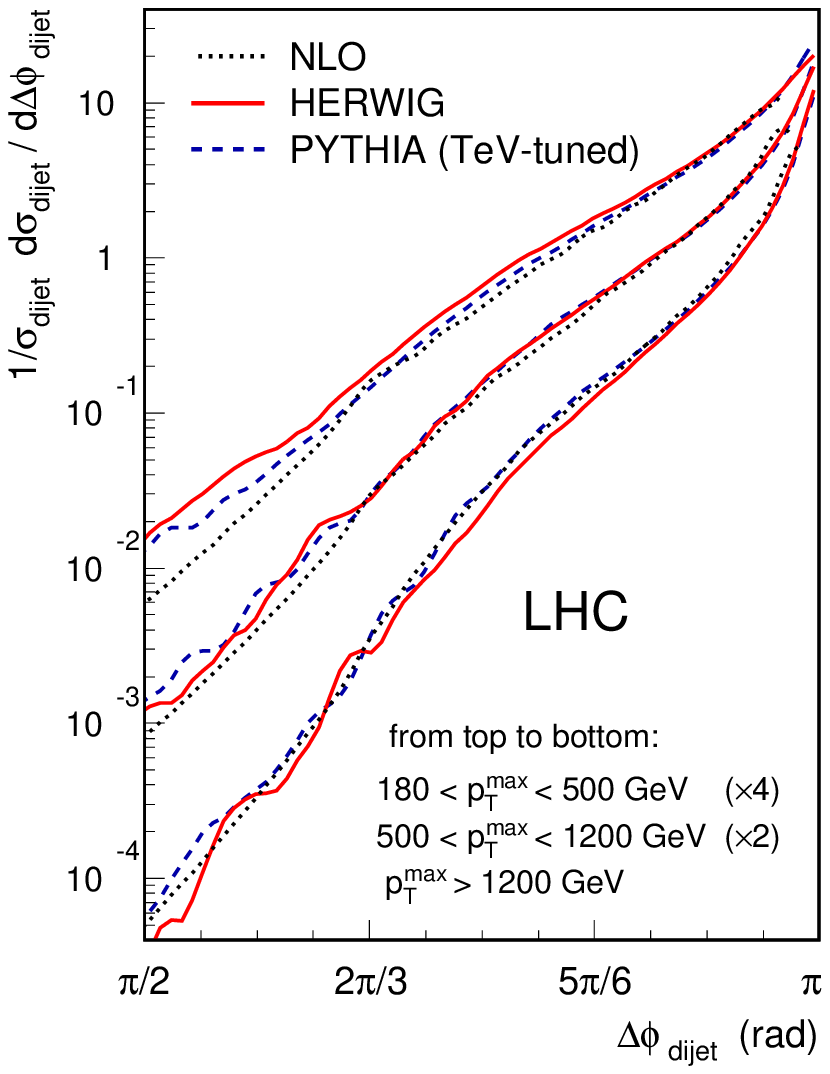,width=7.7cm}
  }
  \caption{\label{fig:TeVLHC} 
Left: Comparisons of  NLO predictions vs. TeV-tuned \PY 
and \HW at the Tevatron.
Right: Analogous comparisons for the LHC.}
\end{figure}

\subsubsection*{Predictions for the LHC}

Having validated the veracity of the NLO calculation
for $p\bar{p}$ Tevatron data at $\sqrt{s}=1.96$\,TeV,
we expect it to provide a reliable extrapolation of
predictions to the LHC energy of $\sqrt{s}=14\,$TeV
for $pp$ collisions. 
To obtain predictions for the LHC, we selected  
$\ptmax$ thresholds of 180, 500 and 1200 GeV.
The second leading-$p_T$ jet in each event is required to have
$p_T>80$~GeV and both jets have central rapidities with
$|y| < 0.5$. 
For these choices, the $\Dphi$ distributions span a similar 
range of values as observed at the Tevatron
(see Fig.~\ref{fig:NLO}, right).

As summarized in Fig.~\ref{fig:TeVLHC} (left), the $\Dphi$
distributions predicted by NLO pQCD, TeV-tuned  \PY
and (default) \HW agree well at Tevatron energy.
This agreement is preserved when extrapolated to the LHC
energy, as demonstrated in Fig.~\ref{fig:TeVLHC} (right).

\subsubsection*{Summary and Outlook}

We conclude that the recent D\O\ measurement of dijet azimuthal 
decorrelations unambiguously constrains the ISR parton 
shower parameters in \PY.
While the default parameters produce insufficient levels
of ISR with high $p_T$, a popular tune 
(tune A which uses {\tt PARP(67)=4.0}~\cite{Field,Field:2005qt}) 
predicts too much ISR.
Our findings provide additional information for 
\PY tuning efforts which so far have been based primarily on 
soft physics in the underlying event.
The re-tuned \PY (with {\tt PARP(67)=2.5})
gives a good description of the
D\O\ data for $\Dphi$ and it also agrees well with NLO pQCD predictions
for this observable.
Extrapolated to LHC energies, the agreement of the re-tuned \PY
with NLO is preserved. 
It is encouraging that Monte Carlo tuning to Tevatron data 
works well also at LHC energies, judging from the comparison to NLO pQCD. 

We believe that it will be worthwhile to investigate the $\Dphi$
distributions at the LHC. 
They can provide an early test of pQCD and Monte Carlo descriptions 
of multi-jet processes.
This is crucial for the understanding of backgrounds affecting discovery 
searches. 
The required dijet data will be accumulated rapidly and with 
virtually no background. 
The reduced sensitivity of the $\Dphi$ measurement to the 
jet energy calibration, normalization and pileup effects promises 
to provide insights into the QCD radiation issues at LHC before 
other multi-jet processes can be measured with sufficient precision.

Thus, the predictions from the TeV-tuned Monte  Carlos and NLO pQCD
for $\Dphi$ distributions can and should be verified quickly with
the first LHC physics-quality data.
Similarly, the expectations from the new Monte Carlo systems,
like ALPGEN and SHERPA, 
currently under development to be among the primary Monte Carlo tools 
at the LHC, can be verified with early data. 
In particular, $\Dphi$ distributions
offer an interesting ground for testing the smoothness
of matching between
Matrix Element and Parton Shower contributions 
as the jet multiplicity varies 
across the $\Dphi$ range. 
These issues have only begun 
to be investigated using the Tevatron data~\cite{match}.

\clearpage
\subsection{Tevatron Run 2 Monte-Carlo Tunes}

\textbf{Contributed by: R. Field}

{\em Several Tevatron Run 2 \PY 6.2 tunes (with multiple parton interactions) are presented and compared 
with \HW (without multiple parton interactions) and with the ATLAS \PY tune (with multiple 
parton interactions).  Predictions are made for the \UE\ in high \pt\ jet production and in 
Drell-Yan lepton-pair production at the Tevatron and the LHC.} \\

\begin{figure}[h]
\begin{center}
\includegraphics[width=\textwidth]{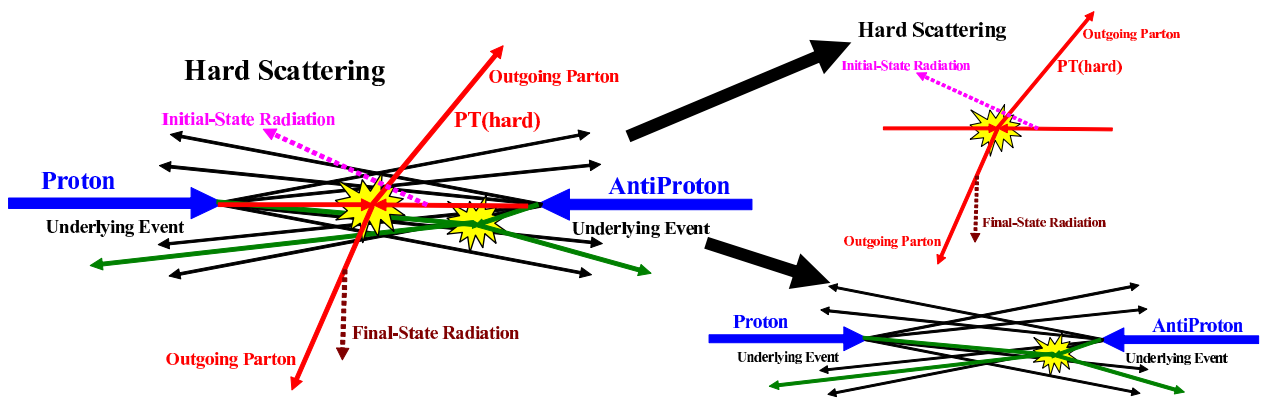}
\caption{\footnotesize 
Illustration of the way QCD Monte-Carlo models simulate a proton-antiproton collision 
in which a ``hard" 2-to-2 parton scattering with transverse momentum, \pthard, has occurred.  
The ``hard scattering" component of the event consists of particles that result from the 
hadronization of the two outgoing partons (\ie the initial two ``jets") plus the particles 
that arise from initial and final state radiation (\ie multijets).  The \UE\ consists of 
particles that arise from the \BBR\ and from multiple parton interactions.
}
\label{RDF_TEV4LHC_fig1}
\end{center}
\end{figure}

In order to find ``new" physics at a hadron-hadron collider it is essential to have Monte-Carlo 
models that simulate accurately the ``ordinary" QCD hard-scattering events.  To do this one must 
not only have a good model of the hard scattering part of the process, but also of the beam-beam 
remnants and the multiple parton interactions. The \UE\ is an unavoidable background 
to most collider observables and a good understanding of it will lead to more precise measurements 
at the Tevatron and the LHC. Fig.~\ref{RDF_TEV4LHC_fig1} illustrates the 
way QCD Monte-Carlo models simulate a 
proton-antiproton collision in which a ``hard" $2$-to-$2$ parton scattering with transverse momentum, \pthard, 
has occurred \cite{Sjostrand:2006za,Corcella:2002jc}.  
The ``hard scattering" component of the event consists of particles that result 
from the hadronization of the two outgoing partons (\ie the initial two ``jets") plus the particles 
that arise from initial and final state radiation (\ie multijets).  The \UE\ consists of 
particles that arise from the \BBR\ and from multiple parton interactions (MPI).  Of course, 
in a given event it is not possible to uniquely determine the origin of the outgoing particles and whatever 
observable one chooses to study inevitably receives contributions from both the hard component and the 
underlying event.  In studying observables that are sensitive to the underlying event one learns not only 
about the beam-beam remnants and multiple parton interactions, but also about hadronization and 
initial and final state radiation.

\begin{figure}[htbp]
\begin{center}
\includegraphics[width=.85\textwidth]{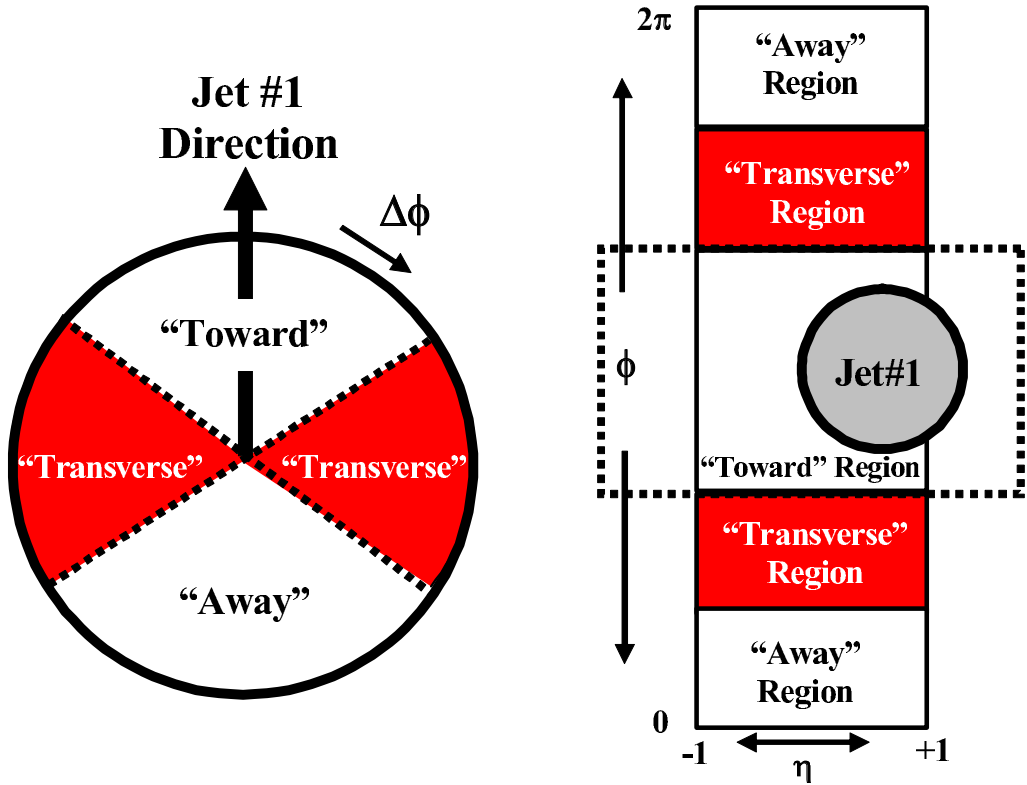}
\caption{\footnotesize
Illustration of correlations in azimuthal angle \delphi\ relative to the direction of the leading 
jet (MidPoint, $R = 0.7$, $f_{merge} = 0.75$) in the event, \Jone.  The angle 
$\Delta\phi=\phi-\phi_{\rm jet\#1}$ is the 
relative azimuthal angle between charged particles and the direction of \Jone. The \TR\ 
region is defined by  $60^\circ<|\Delta\phi|< 120^\circ$ and \etacut.  We examine charged particles in 
the range \etacut\ with \ptlcut\ or \pthcut,  but allow 
the leading jet to be in the region $|\eta(jet\#1)|<2$.
}
\label{RDF_TEV4LHC_fig2}
\end{center}
\end{figure}

In Run2, we are working to understand and model the \UE\ at the Tevatron.  We use the 
topological structure of hadron-hadron collisions to study the underlying event \cite{Affolder:2001xt,Acosta:2004wq,Field:2005yw}.  The 
direction of the leading calorimeter jet is used  to isolate regions of \etaphi\ space that are sensitive 
to the underlying event. As illustrated in Fig.~\ref{RDF_TEV4LHC_fig2}, the direction of the leading jet, \Jone, is 
used to define correlations in the azimuthal angle, \delphi.  The angle $\Delta\phi=\phi-\phi_{\rm jet\#1}$ 
is the relative azimuthal angle between a charged particle and the direction of \Jone.  The \TR\ region is 
almost perpendicular to the plane of the hard $2$-to-$2$ scattering and is therefore very sensitive to the 
underlying event.  Furthermore, we consider two classes of events.  We refer to events in which there 
are no restrictions placed on the second and third highest \pt\ jets (\Jtwo and \Jthree) as \LJ\ events.  
Events with at least two jets with $P_T>15\gev/c$ where the leading two jets are nearly \BB\ (\delphicut) 
with $P_T(jet\#2)/P_T(jet\#1)>0.8$ and $P_T(jet\#3)<15\gevc$ are referred to as \BB\ events.  ``Back-to-back" 
events are a subset of the \LJ\ events.  The idea here is to suppress hard initial and final-state 
radiation thus increasing the sensitivity of the \TR\ region to the  \BBR\ and the multiple 
parton scattering component of the underlying event. 

\begin{figure}[htbp]
\begin{center}
\includegraphics[height=.8\textheight]{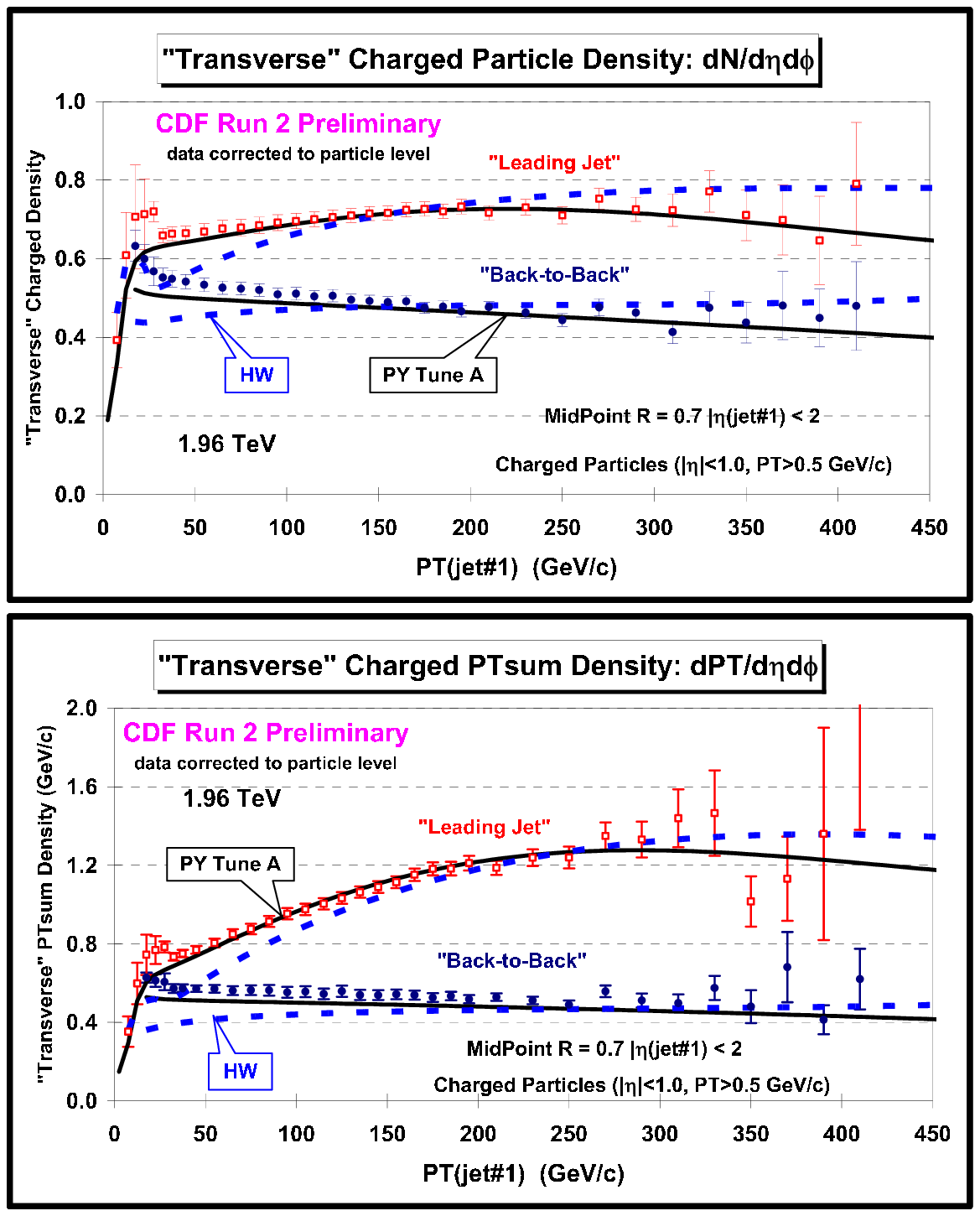}
\caption{\footnotesize
CDF Run 2 data at $1.96\tev$ on the density of charged particles, \nden\ ({\it top}), 
and the charged PTsum density, \ptden\ ({\it bottom}), with \ptlcut\ and \etacut\ in the \TR\ 
region for \LJ\ and \BB\ events as a function of the leading jet \pt\ compared with 
\PY Tune A and \HW.  The data are corrected to the particle level (with errors that 
include both the statistical error and the systematic uncertainty) and compared with the 
theory at the particle level (\ie generator level).
}
\label{RDF_TEV4LHC_fig3}
\end{center}
\end{figure}

Fig.~\ref{RDF_TEV4LHC_fig3} compares the data on the density of charged particles and the charged PTsum density in the \TR\ 
region for \LJ\ and \BB\ events with \PY Tune A (with multiple parton interactions) 
and \HW (without multiple parton interactions).   As expected, the \LJ\ and \BB\ events 
behave quite differently.  For the \LJ\ case the densities rise with increasing $P_T(jet\#1)$, while for 
the \BB\ case they fall slightly with increasing $P_T(jet\#1)$.  The rise in the \LJ\ case 
is, of course, due to hard initial and final-state radiation, which has been suppressed in the \BB\ 
events.  The \BB\ events allow for a more close look at the \BBR\ and multiple parton 
scattering component of the underlying event and \PY Tune A does a better job describing the data than \HW.  
\PY Tune A was determined by fitting the CDF Run 1 \UE\ data \cite{Field:2005qt}.

\begin{figure}[htbp]
\begin{center}
\includegraphics[width=\textwidth]{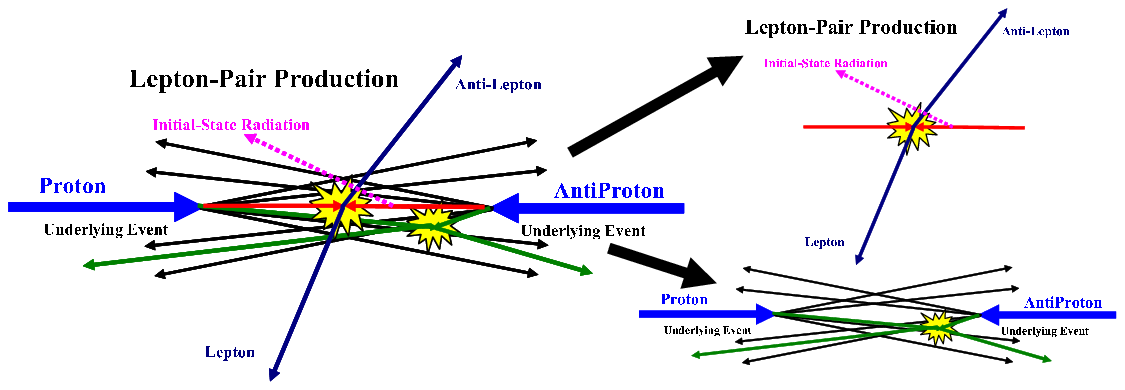}
\caption{\footnotesize
Illustration of the way QCD Monte-Carlo models simulate Drell-Yan lepton-pair production.  
The ``hard scattering" component of the event consists of the two outgoing leptons plus particles that 
result from initial-state radiation.  The ``underlying event'' consists of particles that arise from 
the \BBR\ and from multiple parton interactions.
}
\label{RDF_TEV4LHC_fig4}
\end{center}
\end{figure}

\begin{figure}[!h]
\begin{center}
\includegraphics[width=.6\textwidth]{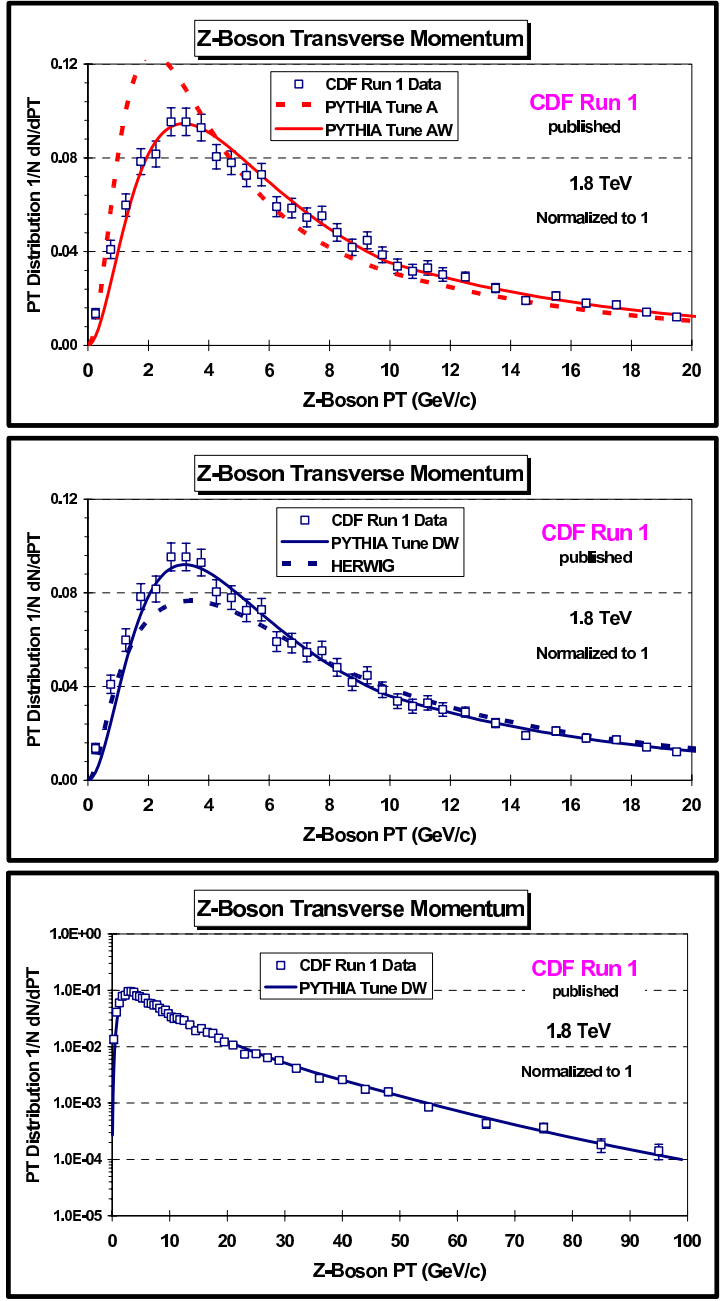}
\caption{\footnotesize
CDF Run 1 data on the Z-boson \pt\ distribution compared with \PY Tune A, Tune AW, Tune DW, and \HW.
}
\label{RDF_TEV4LHC_fig5}
\end{center}
\end{figure}

\begin{table}[!h]
\small
\centering
\begin{tabular}{||c|c|c|c|c|c|c|c||} \hline \hline
 {\bf Parameter}  & {\bf  A}  & {\bf  AW}  & {\bf  DW}  & {\bf  DWT } & {\bf  BW} & {\bf ATLAS} & {\bf  QW} \\ \hline\hline
  CTEQ   & 5L & 5L & 5L & 5L & 5L & 5L & 6.1 \\
  MSTP(81)      & $1$ & $1$ & $1$ & $1$ & $1$ & $1$ & $1$ \\
  MSTP(82)      & $4$ & $4$ & $4$ & $4$ & $4$ & $4$ & $4$ \\
  PARP(82)      & $2.0$ & $2.0$ & $1.9$ & $1.9409$ & $1.8$ & $1.8$ & $1.1$ \\
  PARP(83)      & $0.5$ & $0.5$ & $0.5$ & $0.5$ & $0.5$ & $0.5$ & $0.5$ \\
  PARP(84)      & $0.4$ & $0.4$ & $0.4$ & $0.4$ & $0.4$ & $0.5$ & $0.4$ \\
  PARP(85)      & $0.9$ & $0.9$ & $1.0$ & $1.0$ & $1.0$ & $0.33$ & $1.0$  \\
  PARP(86)      & $0.95$ & $0.95$ & $1.0$ & $1.0$ & $1.0$ & $0.66$ & $1.0$ \\
  PARP(89)      & $1800$ & $1800$ & $1800$ & $1960$ & $1800$ & $1000$ & $1800$ \\
  PARP(90)      & $0.25$ & $0.25$ & $0.25$ & $0.16$ & $0.25$ & $0.16$ & $0.25$ \\
  PARP(62)      & $1.0$ & $1.25$ & $1.25$ & $1.25$ & $1.25$ & $1.0$ & $1.25$ \\
  PARP(64)      & $1.0$ & $0.2$ & $0.2$ & $0.2$ & $0.2$ & $1.0$ & $0.2$ \\
  PARP(67)      & $4.0$ & $4.0$ & $2.5$ & $2.5$ & $1.0$ & $1.0$ & $2.5$ \\
  MSTP(91)      & $1$ & $1$ & $1$ & $1$ & $1$ & $1$ & $1$ \\
  PARP(91)      & $1.0$ & $2.1$ & $2.1$ & $2.1$ & $2.1$ & $1.0$ & $2.1$ \\
  PARP(93)      & $5.0$ & $15.0$ & $15.0$ & $15.0$ & $15.0$ & $5.0$ & $15.0$ \\ \hline\hline
\end{tabular}
\caption{\footnotesize 
Parameters for several \PY 6.2 tunes.  Tune A is a CDF Run 1 \UE\ tune.  Tune AW, DW, DWT, 
and BW are CDF Run 2 tunes which fit the existing Run 2 \UE\ data and fit the Run 1 $Z$-boson \pt\ distribution. 
Tune QW is vary similar to Tune DW except that it uses the next-to-leading order structure function CTEQ6.1. 
The ATLAS Tune is the default tune currently used by ATLAS at the LHC.
}
\label{table1}
\end{table}

\begin{table}[htbp]
\caption{\footnotesize 
Shows the computed value of the multiple parton scattering cross section for 
the various \PY $6.2$ tunes.
}
\label{table2}
\centering
\begin{tabular}{||c|c|c||} \hline \hline
 {\bf Tune}  & {\bf $\sigma(MPI)$ at $1.96\tev$} & {\bf $\sigma(MPI)$ at $14\tev$}\\ \hline\hline
  A,AW  & $309.7$ mb & $484.0$ mb \\
  DW            & $351.7$ mb & $549.2$ mb \\
  DWT           & $351.7$ mb & $829.1$ mb \\
  BW            & $401.7$ mb & $624.8$ mb \\
  QW            & $296.5$ mb & $568.7$ mb \\
  ATLAS & $324.5$ mb & $768.0$ mb \\ \hline\hline
\end{tabular}
\end{table}

\begin{figure}[h]
\centering
\subfigure[\PY~Tunes A, AW, BW, and DW.]{\includegraphics[width=.45\textwidth]{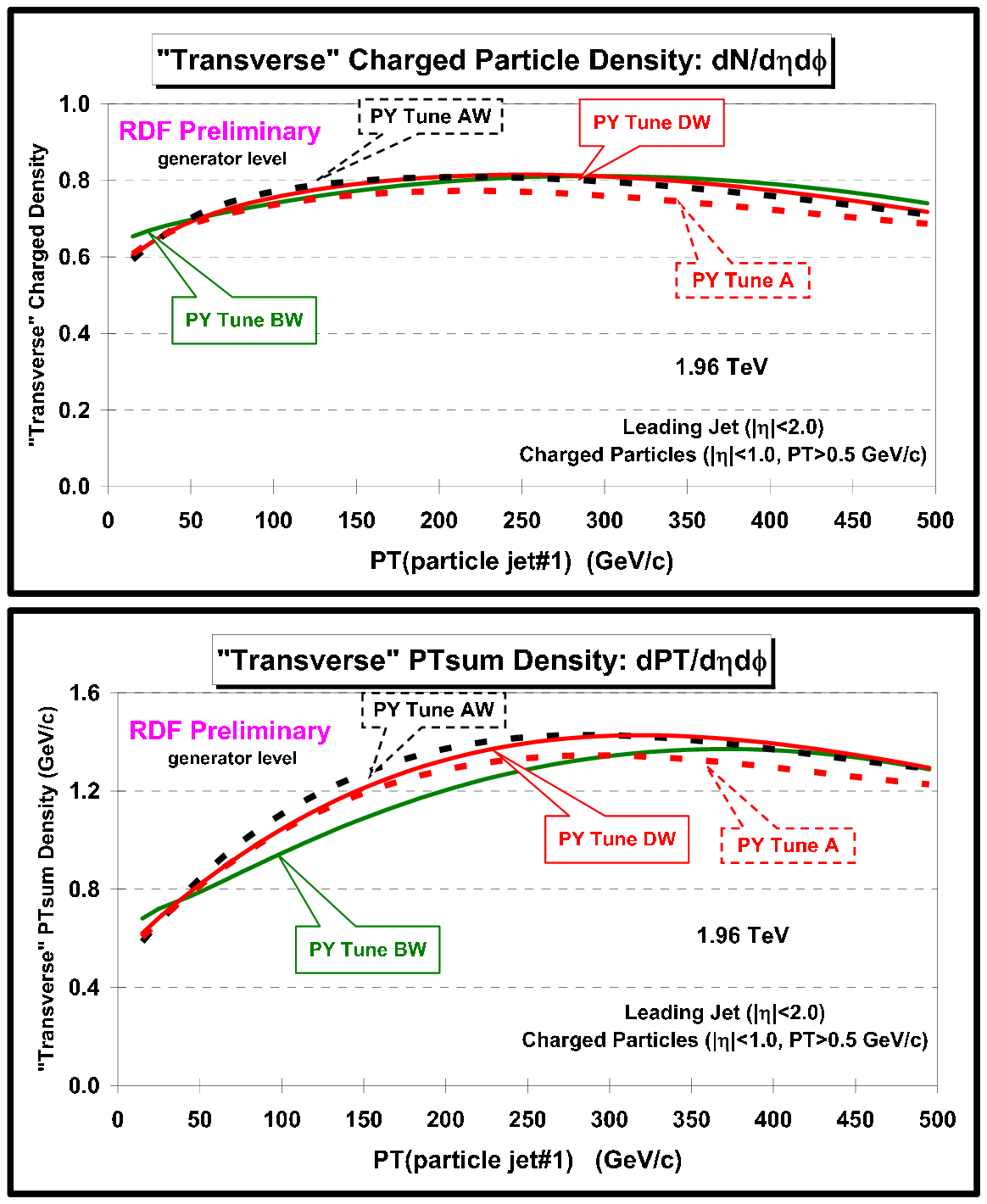}}
\subfigure[\PY~Tune DW (DWT), \HW, and the ATLAS Tune.]{
\includegraphics[width=.45\textwidth]{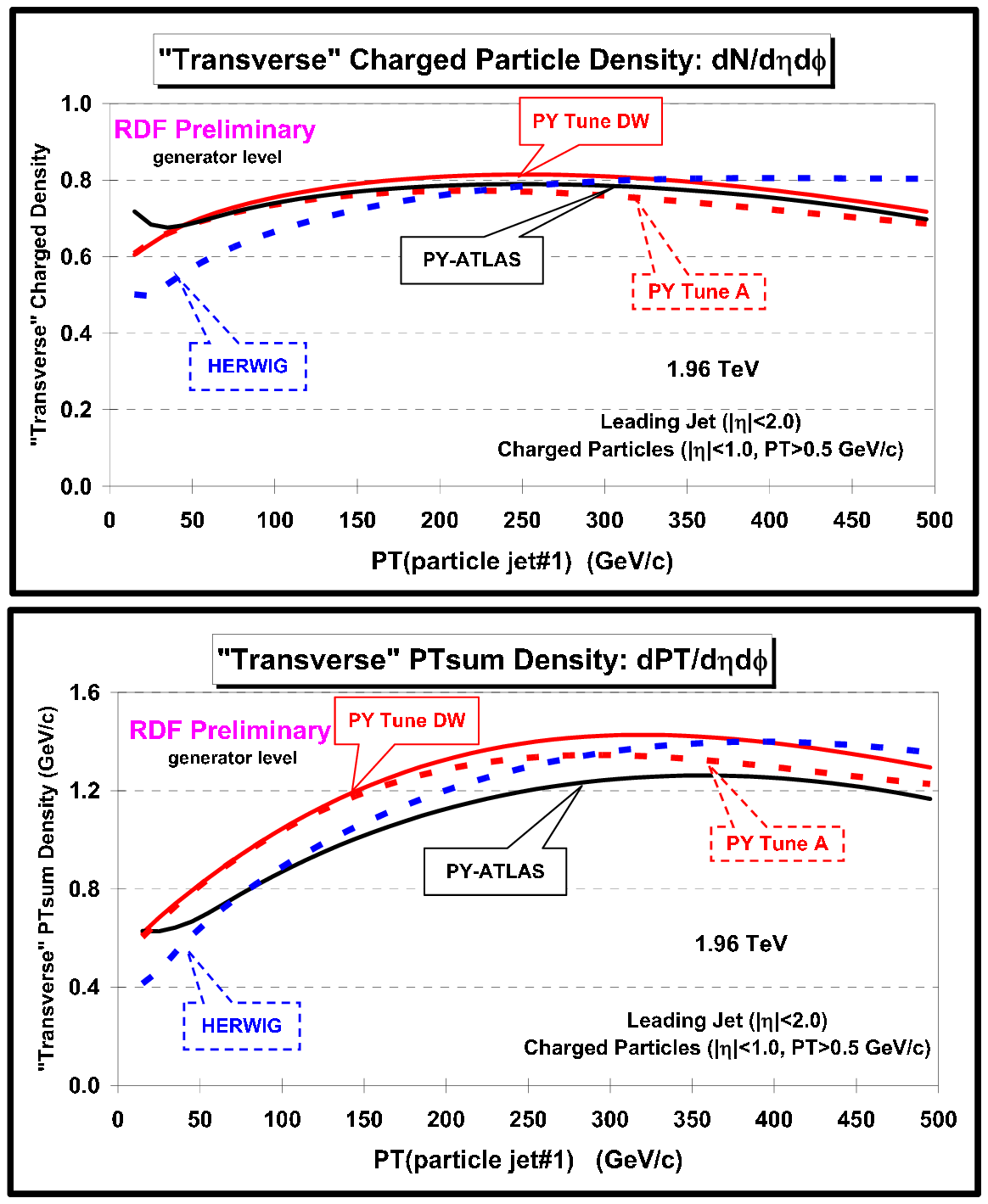}}
\caption{\footnotesize
Predictions at $1.96\tev$
of the density of charged 
particles, \nden\ ({\it top}), and the charged PTsum density, \ptden\ ({\it bottom}), with \ptlcut\ 
and \etacut\ in the \TR\ region for \LJ\ events as a function of the leading jet \pt.
\label{RDF_TEV4LHC_fig6}
}
\end{figure}

\begin{figure}[h]
\centering
\subfigure[]{\includegraphics[width=.45\textwidth]{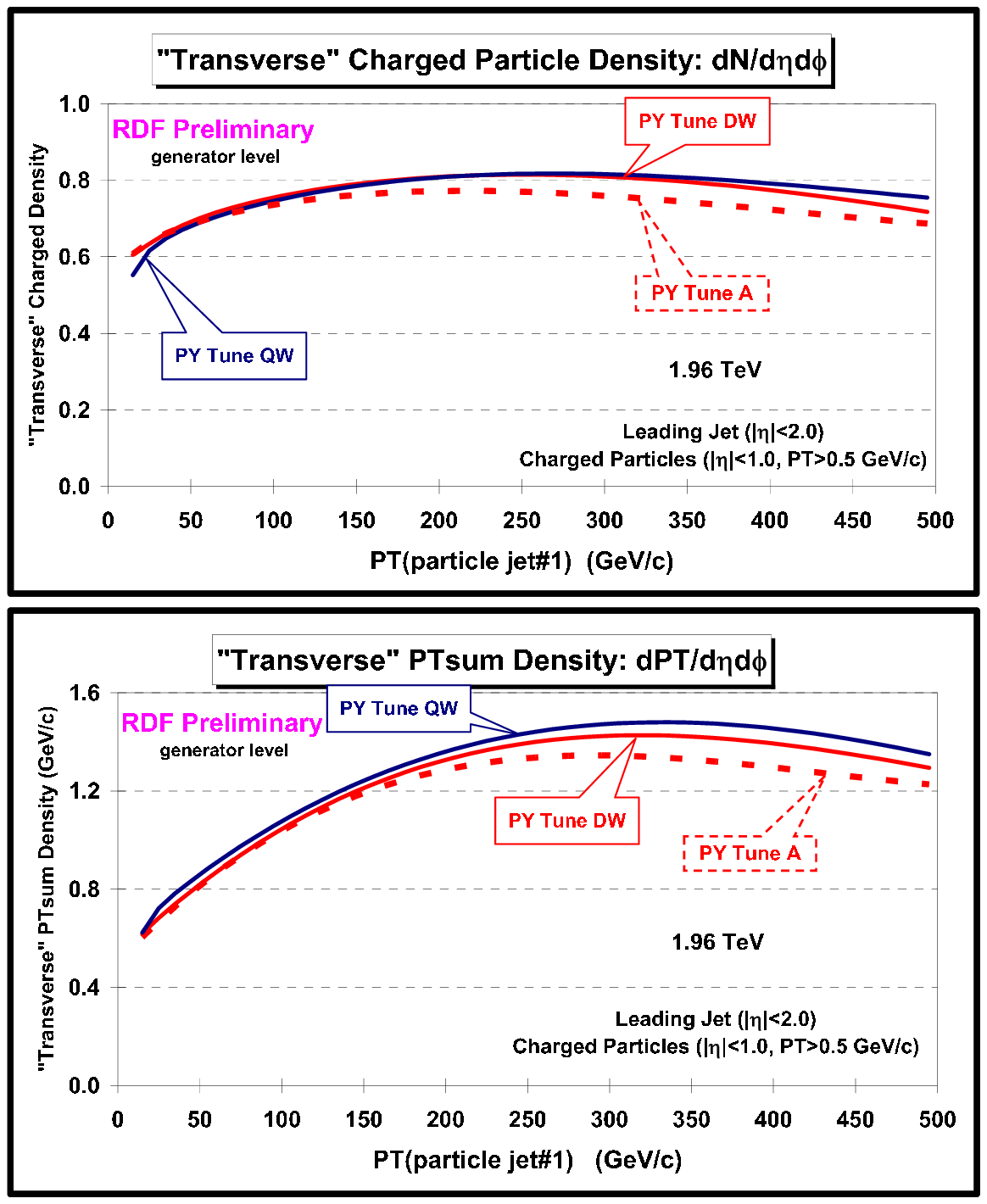}}
\subfigure[]{\includegraphics[width=.45\textwidth]{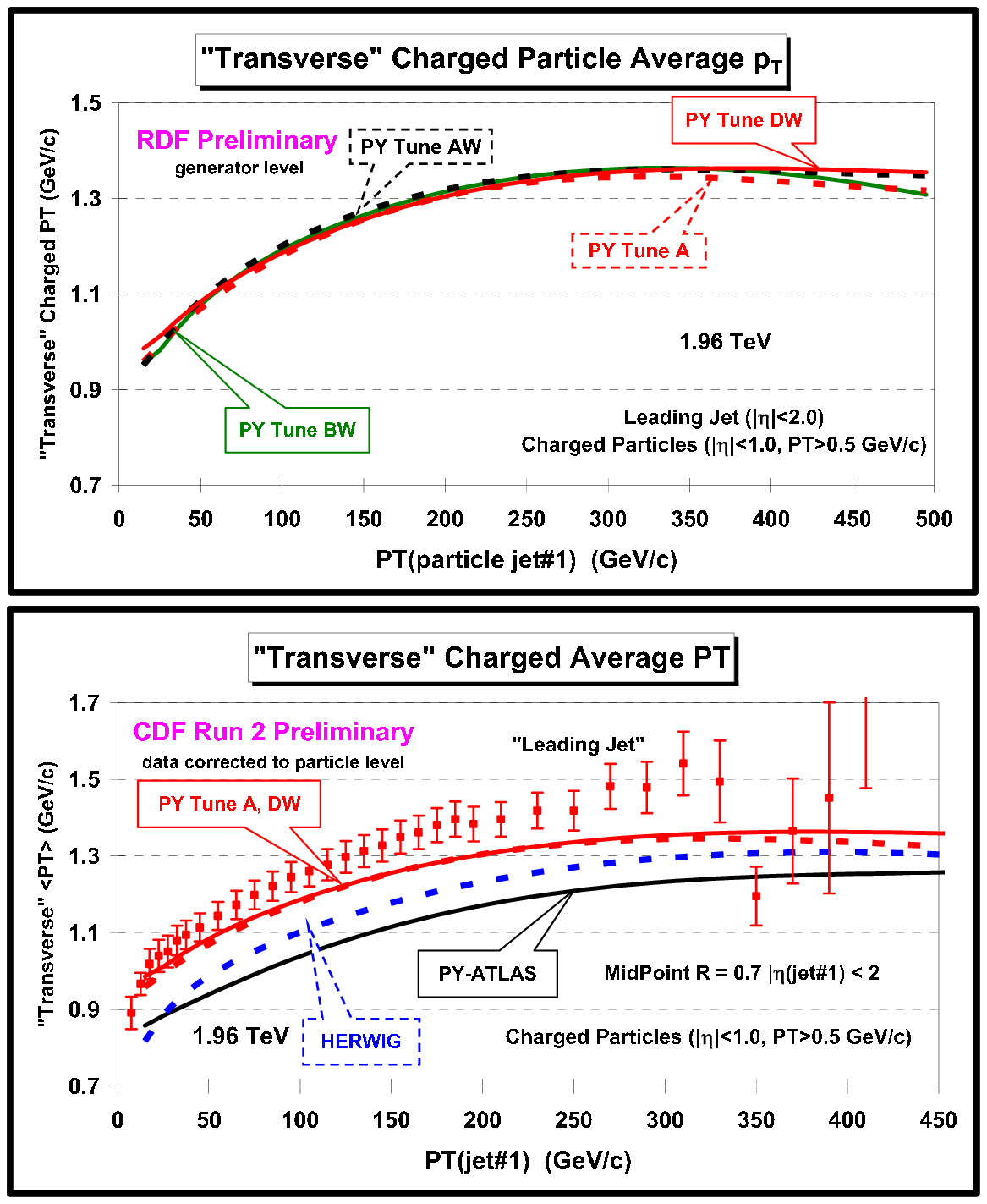}}
\caption{\footnotesize (a) Similar to Fig.~\ref{RDF_TEV4LHC_fig6} for \PY~Tune A, DW, and QW.
(b) {\it (top)} Predictions of \PY Tune A, Tune AW, Tune BW, and Tune DW for average \pt\ of charged 
particles with  \ptlcut\ and \etacut\ in the \TR\ region for \LJ\ events at $1.96\tev$ as a 
function of the leading jet \pt\.   ({\it bottom}) CDF Run 2 data at $1.96\tev$ on the 
average \pt\ of charged particles with \ptlcut\ and \etacut\ in the \TR\ region for \LJ\ events 
as a function of the leading jet \pt\ compared with \PY Tune A, Tune DW, \HW, and the ATLAS \PY Tune.
\label{RDF_TEV4LHC_fig8}
}
\end{figure}

\begin{figure}[htbp]
\centering
\subfigure[]{\includegraphics[width=.45\textwidth]{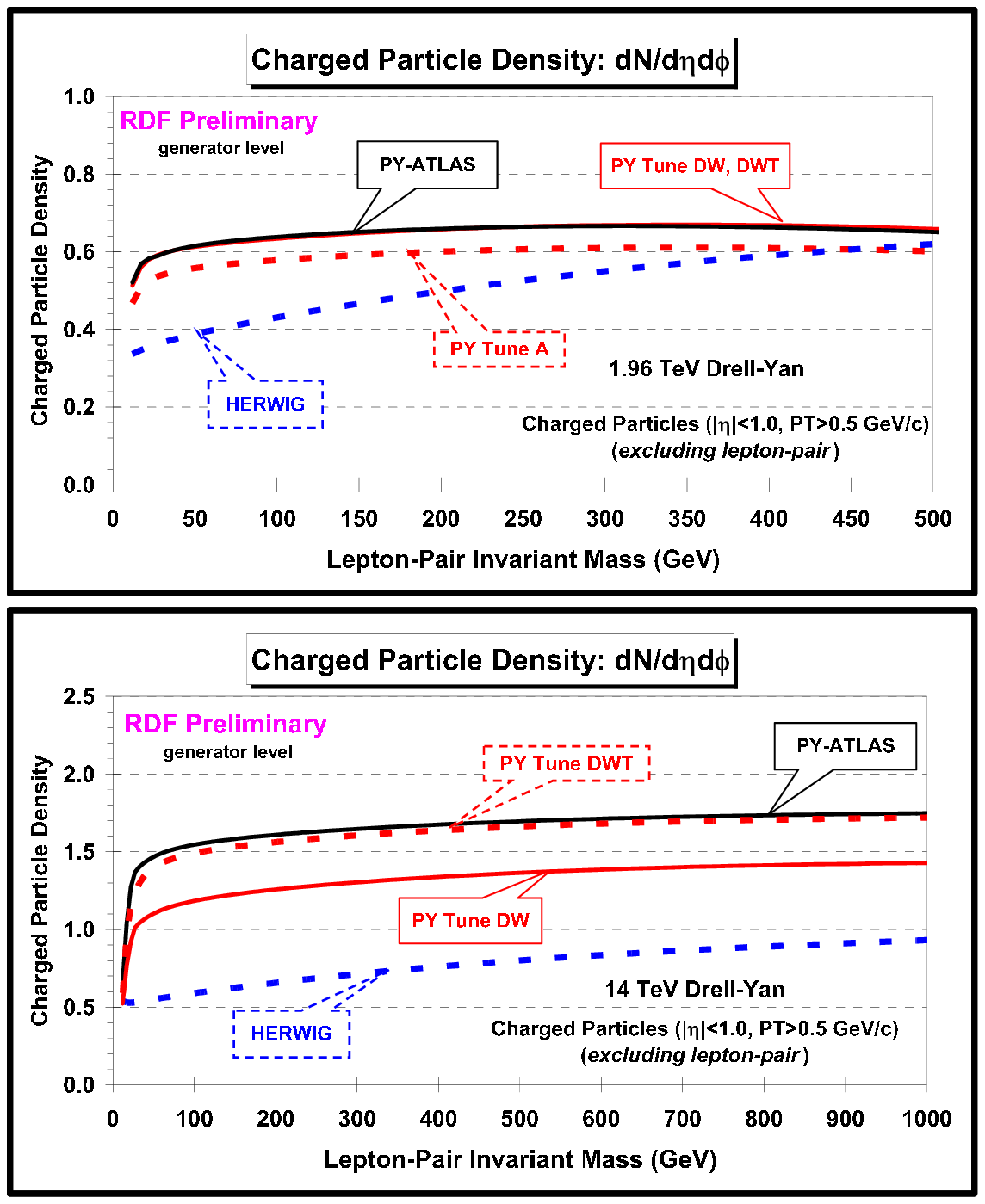}}
\subfigure[]{\includegraphics[width=.45\textwidth]{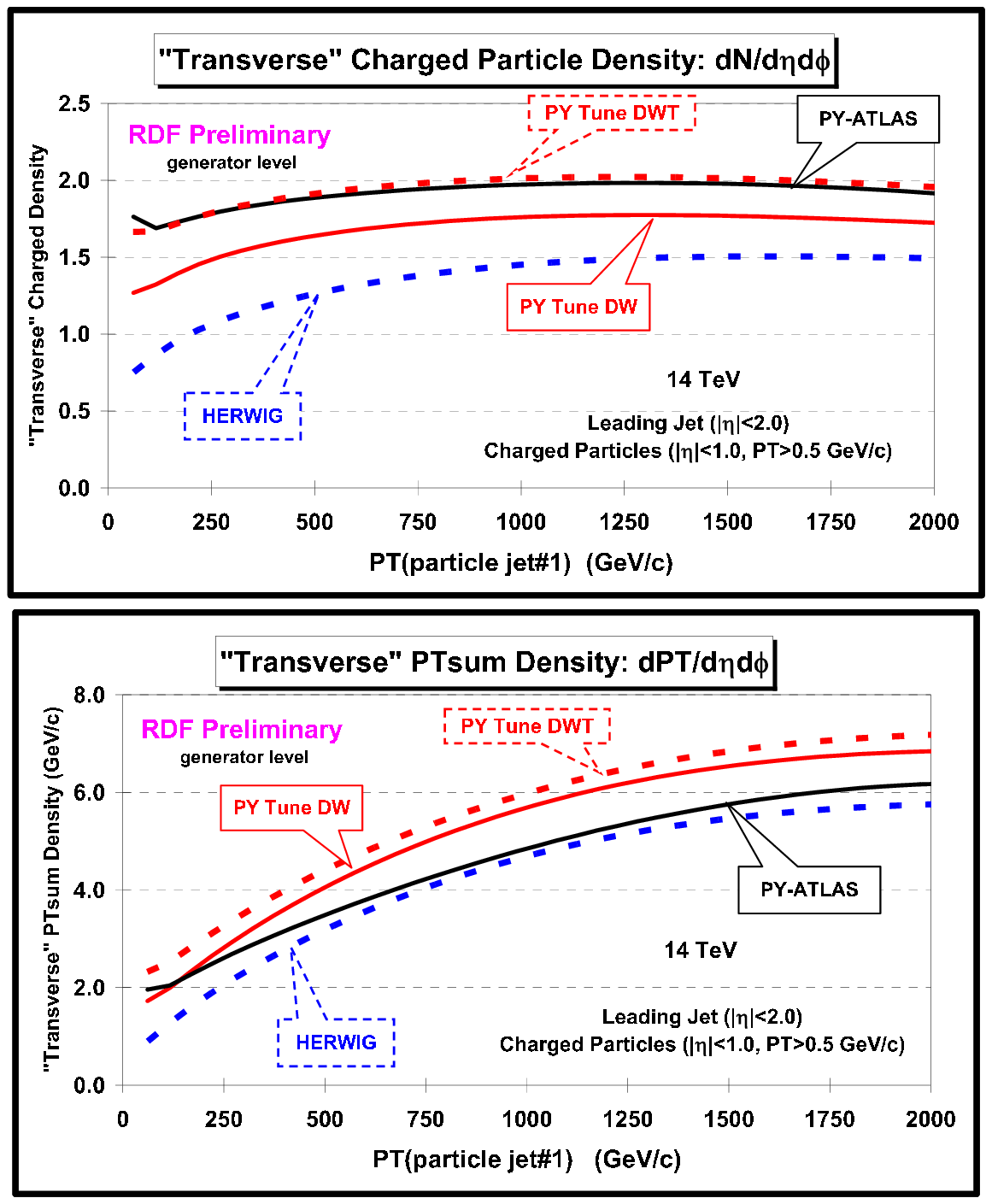}}
\caption{\footnotesize
(a) Predictions of \PY Tune A, Tune DW, Tune DWT, \HW, and the ATLAS \PY Tune for the density of 
charged particles, \nden\, with \ptlcut\ and \etacut\ in Drell-Yan lepton-pair production (excluding the 
lepton-pair) at $1.96\tev$ ({\it top}) and 14 TeV ({\it bottom}) as a function of the invariant mass of the 
lepton pair.  Tune DW and Tune DWT are identical at $1.96\tev$. 
(b) Predictions at 14 TeV of \PY Tune DW, Tune DWT, \HW, and the ATLAS Tune for the density of charged 
particles, \nden\ ({\it top}), and the charged PTsum density, \ptden\ ({\it bottom}), with \ptlcut\ 
and \etacut\ in the \TR\ region for \LJ\ events as a function of the leading jet \pt.
}
\label{RDF_TEV4LHC_fig10}
\end{figure}

\begin{figure}[htbp]
\begin{center}
\includegraphics[width=.8\textwidth]{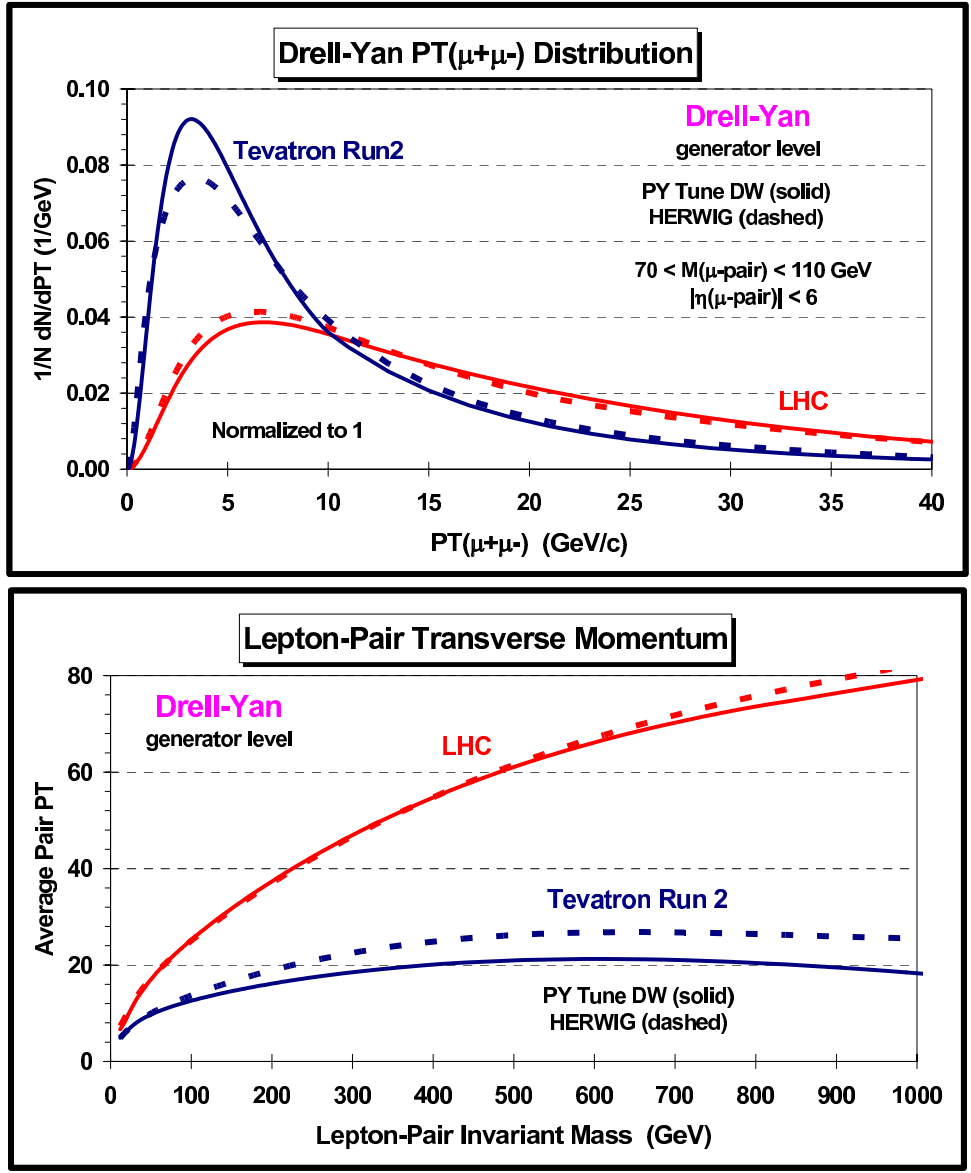}
\caption{\footnotesize
Predictions at $1.96\tev$ (Tevatron Run 2) and 14 TeV (LHC) of \PY Tune DW and \HW 
for ({\it top}) the lepton-pair \pt\ distribution at the Z-boson mass and ({\it bottom}) the 
average lepton-pair \pt\ versus the lepton pair invariant mass.
}
\label{RDF_TEV4LHC_fig12}
\end{center}
\end{figure}

\begin{figure}[htbp]
\begin{center}
\includegraphics[height=.85\textheight]{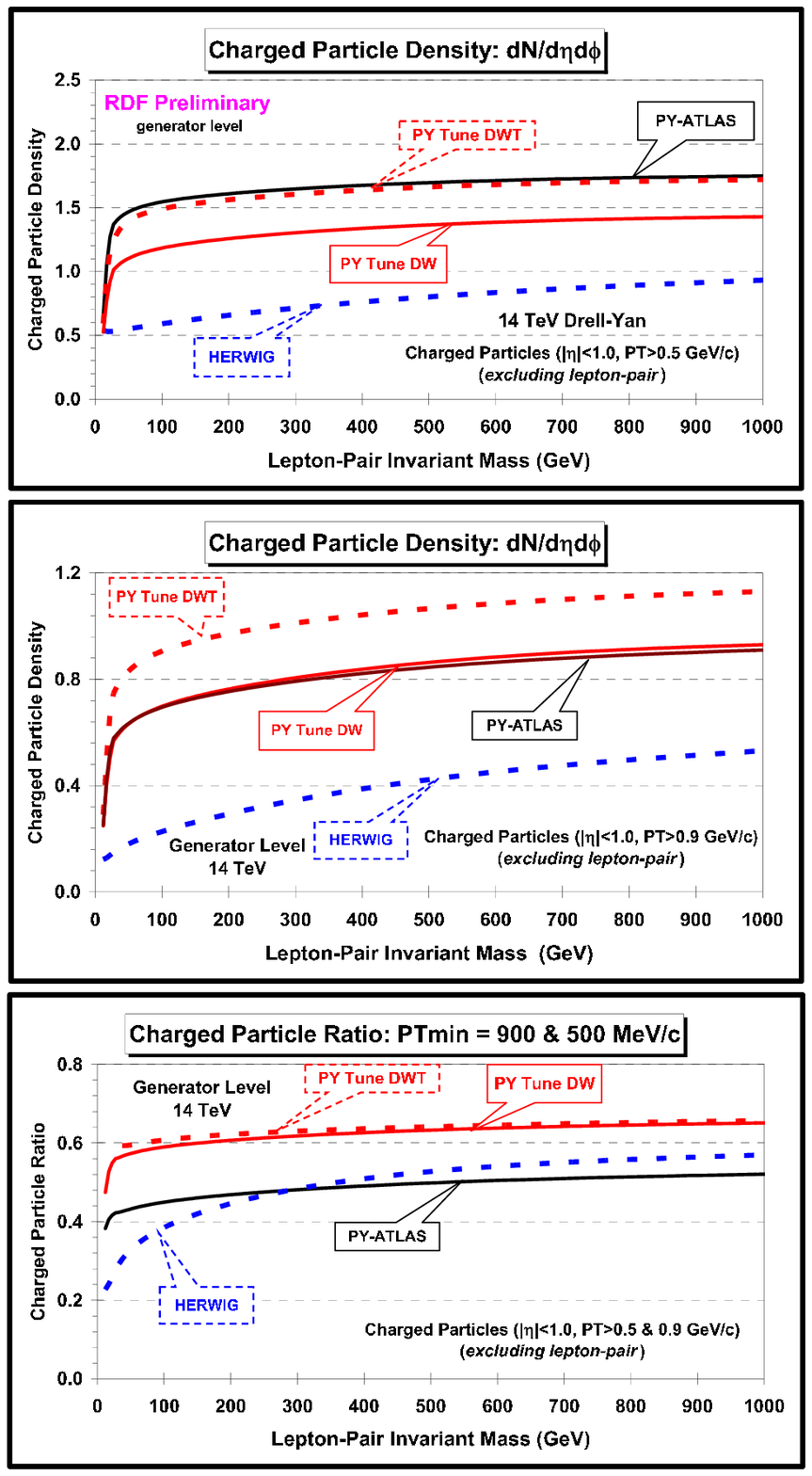}
\caption{\footnotesize
Predictions at 14 TeV of \PY Tune DW, Tune DWT, \HW, and the ATLAS Tune for the density of charged 
particles, \nden\, with \etacut\ and \ptlcut\ ({\it top}) and  \pthcut\ ({\it middle}) for Drell-Yan lepton-pair 
production (excluding the lepton-pair) as a function of the lepton-pair invariant mass. ({\it bottom}) The 
ratio of the charged particle density with \pthcut\ and \ptlcut.
}
\label{RDF_TEV4LHC_fig13}
\end{center}
\end{figure}

As illustrated in Fig.~\ref{RDF_TEV4LHC_fig4}, 
Drell-Yan lepton-pair production provides an excellent place to study the underlying event.  
Here one studies the outgoing charged particles (excluding the lepton pair) as a function of the lepton-pair invariant 
mass.  After removing the lepton-pair everything else results from the beam-beam remnants, multiple parton 
interactions, and initial-state radiation.  Unlike high \pt\ jet production (Fig.~1) for lepton-pair production 
there is no final-state gluon radiation.

Fig.~\ref{RDF_TEV4LHC_fig5} 
shows that \PY Tune A does not fit the CDF Run 1 $Z$-boson \pt\ distribution 
\cite{Phys.Rev.Lett.67.2937}.  \PY Tune A was 
determined by fitting the Run 1 \UE\ data and, at that time, we did not consider the $Z$-boson data. 
\PY Tune AW fits the Z-boson \pt\ distribution as well as the \UE\ 
at the Tevatron \footnote{The values of {\tt PARP(62)}, {\tt PARP(64)}, and 
{\tt PARP(91)} were determined by CDF Electroweak Group.  The {\it W} in Tune AW, BW, DW, DWT, QW stands for {\it Willis}.  
I combined the Willis tune with Tune A, etc.}
\PY Tune DW is very similar to Tune AW except {\tt PARP(67) = 2.5}, 
which is the preferred value determined by D\O~ in 
fitting their dijet \delphi\ distribution \cite{Phys.Rev.Lett.94.221801}.  \HW does a fairly good job fitting the $Z$-boson \pt\ distribution 
without additional tuning, but does not fit the CDF \UE\ data.  

Table~\ref{table1} shows the parameters for several \PY $6.2$ tunes.  Tune BW is a tune with ${\tt PARP(67)} = 1.0$.  Tune DW and 
Tune DWT are identical at $1.96\tev$, but Tune DW and DWT extrapolate differently to the LHC.  Tune DWT uses the 
ATLAS energy dependence, ${\tt PARP(90}) = 0.16$, while Tune DW uses the 
Tune A value of ${\tt PARP(90)} = 0.25$.  The ATLAS Tune 
is the default tune currently used by ATLAS at the LHC.   All the tunes except Tune QW use CTEQ5L.  Tune QW uses CTEQ6.1 
which is a next-to-leading order structure function.   However, Tune QW uses leading order QCD coupling, $\alpha_s$, 
with $\Lambda=0.192\gev$.  Note that Tune QW has a much smaller 
value of ${\tt PARP(82)}$ (\ie the MPI cut-off).  This is 
due to the change in the low x gluon distribution in going from CTEQ5L to CTEQ6.1. Table~\ref{table2} shows the computed 
value of the multiple parton scattering cross section for the various tunes.  The multiple parton scattering 
cross section (divided by the total inelastic cross section) determines the average number of multiple parton 
collisions per event. 

As can be seen in Figs.~\ref{RDF_TEV4LHC_fig6} and \ref{RDF_TEV4LHC_fig8}(a),
\PY Tune A, AW, DW, DWT, and QW have been adjusted to give similar results 
for the charged particle density and the PTsum density in the \TR\ region with \ptlcut\ 
and \etacut\ for \LJ\ events at $1.96\tev$.  \PY Tune A fits the CDF Run 2 \UE\ data 
for \LJ\ events and Tune AW, BW, DW, and QW roughly agree with Tune A. 
Fig.~\ref{RDF_TEV4LHC_fig8}(b) shows that \PY Tune A, 
Tune DW, and the ATLAS \PY Tune predict about the same density of charged particles in the \TR\ region 
with \ptlcut\ for \LJ\ events at the Tevatron.  However, the ATLAS Tune has a much softer \pt\ 
distribution of charged particles resulting in a much smaller average \pt\ per particles.  
Fig.~\ref{RDF_TEV4LHC_fig8}(b) shows that the 
softer \pt\ distribution of the ATLAS Tune does not agree with the CDF data.  

The predictions of \PY Tune A, Tune DW, Tune DWT, \HW, and the ATLAS \PY Tune for the density of 
charged particles with \ptlcut\ and \etacut\ for Drell-Yan lepton-pair production at $1.96\tev$ and $14\tev$ 
are shown in Fig.~\ref{RDF_TEV4LHC_fig10}(a).  The ATLAS Tune and Tune DW predict about the same charged particle density with \ptlcut\ 
at the Tevatron, and the ATLAS Tune and Tune DWT predict about the same charged particle density with \ptlcut\ 
at the LHC.  However, the ATLAS Tune has a much softer \pt\ distribution of particles, both at the Tevatron and the 
LHC.  We are working to compare the CDF Run 2 data on Drell-Yan production with the QCD Monte-Carlo models and 
hope to have results soon.

Fig.~\ref{RDF_TEV4LHC_fig10}(b) shows the predictions of \PY Tune DW, Tune DWT, \HW, and the ATLAS Tune for the density of 
charged particles and the PTsum density in the \TR\ region for \LJ\ production at the LHC.  
The \PY Tunes (with multiple parton interactions) predict a large increase in the charged particle density 
in going from the Tevatron (Fig.~\ref{RDF_TEV4LHC_fig6}) to the 
LHC (Fig.~\ref{RDF_TEV4LHC_fig10}(b)).  \HW (without multiple parton interactions) does 
not increase as much.   \PY Tune DWT and the ATLAS Tune both predict about the same charged particle density 
with \ptlcut, however, the ATLAS Tune predicts a smaller PTsum density than Tune DWT (\ie the ATLAS Tune 
produces a softer \pt\ distribution).

The increased amount of initial-state radiation at the LHC results in a broader lepton-pair \pt\ distribution 
compared to the Tevatron.  As can be seen in Fig.~\ref{RDF_TEV4LHC_fig12}, even at the $Z$-boson mass the lepton-pair \pt\ distribution 
is predicted to be much broader at the LHC.   This is indirectly related to the underlying event. More 
initial-state radiation results in a more active underlying event.  

Fig.~\ref{RDF_TEV4LHC_fig13} shows the predictions at $14\tev$ of \PY Tune DW, Tune DWT, \HW, and the ATLAS Tune for the density 
of charged particles with \etacut\ and \ptlcut\ and  \pthcut\ for Drell-Yan lepton-pair 
production (excluding the lepton-pair) as a function of the lepton-pair invariant mass.  The ratio of the 
two \pt\ thresholds clearly shows that the ATLAS tune is has a much softer \pt\ distribution than the CDF tunes. 
We do not know what to expect at the LHC.  For now I prefer \PY Tune DW or Tune DWT over the ATLAS Tune 
because these tunes fit the CDF Run 2 data much better than the ATLAS Tune.

In my opinion the best \PY $6.2$ tune at present is Tune DW or DWT.  These tunes are identical at the $1.96\tev$ 
and they do a good job fitting the CDF Run 2 \UE\ data.  I expect they will do a good job in 
describing the underlying event in Drell-Yan lepton-pair production at the Tevatron (but we will have to wait 
for the data).  More work will have to be done in studying the ``universality" of these tunes.  For example, we 
do not know if Tune DW will correctly describe the underlying event in top quark production.  Tune QW (or the 
corresponding Tune QWT) is very similar to Tune DW (or Tune DWT) except that it uses the next-to-leading order 
structure function CTEQ6.1. Many Monte-Carlo based  analyses use the $40$ error PDF's associated with CTEQ6.1 and 
it is useful to have a tune using the central fit (\ie CTEQ6.1).

\clearpage
\subsection{Underlying Event Tunes for the LHC}

\textbf{Contributed by:  Moraes}

Hard interactions at hadron-hadron colliders consist of a hard collision
of two incoming partons along with softer interactions from the
remaining partons in the colliding hadrons (``the underlying event
energy''). Minimum bias events are the type of events which would be
observed with a very inclusive experimental trigger, and consist
primarily of the softer interactions resulting from the collision of two
hadrons. What is meant by a minimum bias event is somewhat murky, and
the exact definition will depend on the trigger of each experiment. The
description of the underlying event energy and of minimum bias events
requires a non-perturbative phenomenological model. There are currently
a number of models available, primarily inside parton shower Monte Carlo
programs, to predict both of these processes. We discuss several of the
popular models below. An understanding of this soft physics is
interesting in its own right but is also essential for precision
measurements of hard interactions where the soft physics effects need to
be subtracted. This is true at  the Tevatron and will be even more  so
at the LHC where  the high luminosity running will bring a large number
of additional minimum bias interactions per crossing.

Perhaps the simplest model for the underlying event is the uncorrelated
soft scattering model present in \HW. Basically, the model is a
parametrization of the minimum bias data taken by the UA5 experiment at
the CERN $p\overline{p}$
Collider. The model tends to predict underlying event distributions
softer than measured at the Tevatron and has a questionable
extrapolation to higher center-of-mass energies. A newer model for the
underlying event in \HW is termed \JM  and describes the underlying
event in terms of multiple parton interactions at a scale lower than the
hard scale and with the number of such parton scatterings depending on
the impact parameter overlap of the two colliding hadrons.

\JM 4.1 linked to \HW 6.507 has been tuned to describe the underlying 
event as measured by CDF \cite{Affolder:2001xt,Acosta:2004wq} and the resulting 
set of parameters, labeled UE, is shown in Table
\ref{tab:JIMMY-tunings}. The tuned settings were obtained for
CTEQ6L. The default parameters are also included in table
\ref{tab:JIMMY-tunings} for comparison. 
\begin{table}[!h]
\begin{center}
\begin{tabular}{|c|c|c|}
\hline
\footnotesize \textbf{Default} & \footnotesize \textbf{\JM 4.1 - UE} & 
\footnotesize \textbf{Comments}\\  
\hline \hline
\scriptsize JMUEO=1 & \scriptsize JMUEO=1 & \scriptsize multiparton
  interaction \\
  & & \scriptsize model\\
\hline
\scriptsize PTMIN=10.0 & \scriptsize PTMIN=10.0 & \scriptsize minimum
  p$_{T}$ in \\
  & & \scriptsize hadronic jet production \\
\hline
\scriptsize PTJIM=3.0 & \scriptsize PTJIM=$2.8 \times \left( \frac{\sqrt{s}}{1.8
  ~\text{TeV}} \right)^{0.274} $ & \scriptsize minimum p$_{T}$ of
  secondary \\
  & & \scriptsize scatters when JMUEO=1 or 2 \\
\hline
\scriptsize JMRAD(73)=0.71 & \scriptsize JMRAD(73)=1.8 & \scriptsize
  inverse proton \\
  & & \scriptsize radius squared \\
\hline
\scriptsize PRSOF=1.0 & \scriptsize PRSOF=0.0 & \scriptsize
  probability of a soft \\
  & & \scriptsize underlying event \\
\hline
\end{tabular}
\caption{\JM 4.1 default and \textit{UE} parameters for the underlying
  event.}  
\label{tab:JIMMY-tunings}
\end{center}
\end{table}         
JMRAD(75) should also be changed to the same value used for 
JMRAD(73) when antiprotons are used in the simulation (e.g. Tevatron events).

Notice that an energy dependent term has been introduced in PTJIM for
the UE tuning. This leads to a value of PTJIM=2.1 for collisions at
$\sqrt{\text{s}}$ = 630 GeV and PTJIM=4.9 for the LHC centre-of-mass
energy in pp collisions.

The \PY model for the underlying event also utilizes a multiple
parton interaction framework with the total rate for parton-parton
interactions assumed to be given by perturbative QCD. A cutoff,
$p_{Tmin}$, is introduced to regularize the divergence as the transverse
momentum of the scattering goes to zero. The rate for multiple parton
interactions depend strongly on the value of the gluon distribution at
low  $x$.  The cutoff, $p_{Tmin}$,  is the main free parameter of the
model and basically corresponds to an inverse color screening distance.
A tuning of the \PY underlying event parameters (Tune A) was
discussed earlier and basically succeeds in describing all
of the global event properties in events at the Tevatron. With the new
version of \PY (version 6.3), a new model for the underlying event is
available, similar  in spirit to the old multiple parton interaction
model but with more attention being a more sophisticated treatment of
color, flavor and momentum correlations in the remnants.
Table \ref{tab:PYTHIA-tunings} shows the relevant \PY 6.3 parameters 
tuned to the underlying event data \cite{Affolder:2001xt,Acosta:2004wq}. For 
the purpose of comparison, the corresponding default values in
\PY 6.323 \cite{Sjostrand:2003wg} are also shown.

\begin{table}[!h]
\begin{center}
\begin{tabular}{|c|c|c|c|}
\hline
\footnotesize \textbf{Parameter} & \footnotesize \textbf{Default} & \footnotesize \textbf{UE} & 
\footnotesize \textbf{Comment} \\
\hline\hline
\scriptsize MSTP(51) & \scriptsize 7 (5L) & \scriptsize 10042 (6L) & \scriptsize CTEQ PDF \\ 
\scriptsize MSTP(52) &     & \scriptsize 2 &  \\ \hline
\scriptsize MSTP(68) &  \scriptsize 3 & \scriptsize 1 & \scriptsize max. virtuality scale \\
 & & & \scriptsize and ME matching for ISR  \\ \hline
\scriptsize MSTP(70)& \scriptsize  1 & \scriptsize 2 & \scriptsize regul. scheme for ISR \\ \hline
\scriptsize MSTP(82)& \scriptsize 3 &  \scriptsize 4 & \scriptsize complex scenario and
 double\\
&  & & \scriptsize Gaussian matter distribution \\ \hline
\scriptsize PARP(82)& \scriptsize 2.0 & 
\scriptsize 2.6 & \scriptsize
p$_{t_{\text{min}}}$ \scriptsize parameter \\ \hline
\scriptsize PARP(84)&\scriptsize 0.4 & 
\scriptsize 0.3 & \scriptsize hadronic core
 radius \\ 
 & & & \scriptsize (only for MSTP(82)=4) \\ \hline
\scriptsize PARP(89) & \scriptsize 1.8  &
\scriptsize 1.8 & \scriptsize energy scale
 (TeV) used to \\
& & & \scriptsize calculate p$_{t_{\text{min}}}$ \\ \hline
\scriptsize PARP(90) & \scriptsize 0.25 & 
\scriptsize 0.24 & \scriptsize power of the
 p$_{t_{\text{min}}}$  \\
 &  & & \scriptsize energy dependence  \\[0.1cm]
\hline
\end{tabular}
\caption{\PY 6.323 default \cite{Sjostrand:2003wg} and UE parameters.}
\label{tab:PYTHIA-tunings}
\end{center}
\end{table}         

\subsubsection*{Predictions vs. underlying event data}

Based on CDF measurements \cite{Affolder:2001xt}, the UE is defined as the
angular region in $\phi$ which is transverse to the leading charged
particle jet.  

\begin{figure}
\subfigure[]{\includegraphics[width=.5\textwidth]{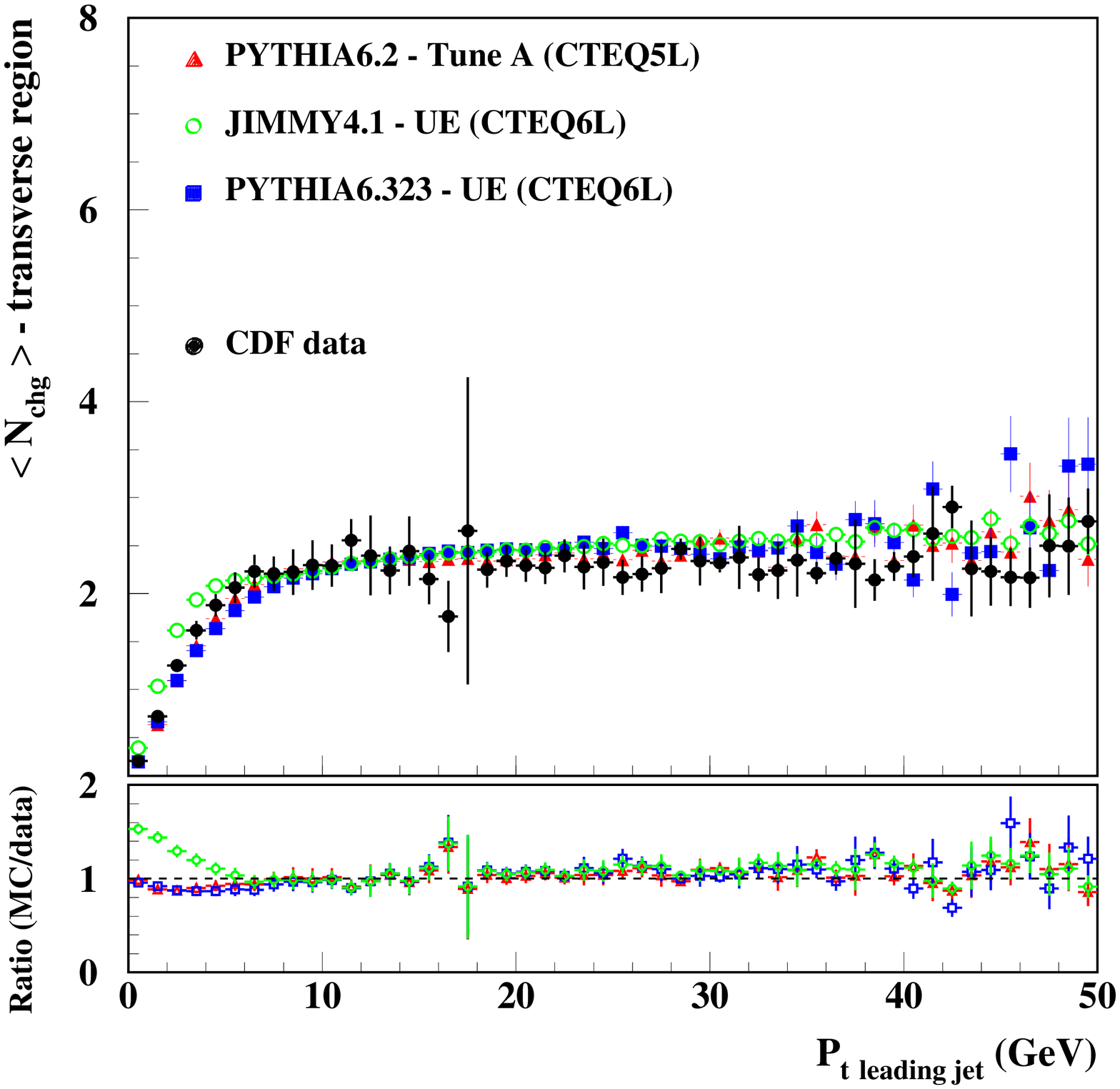}} 
\subfigure[]{\includegraphics[width=.5\textwidth]{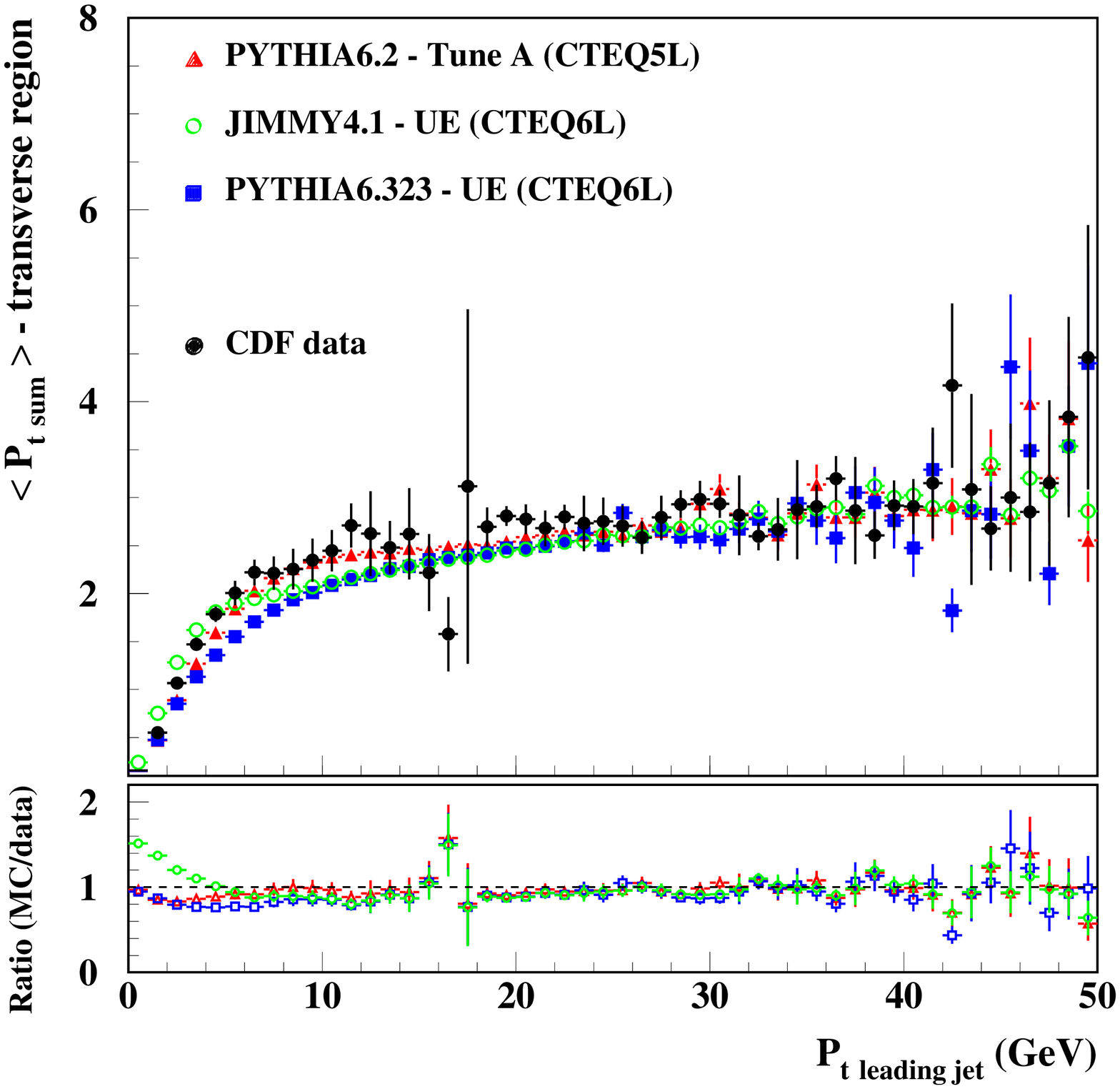}} 
\caption{\PY 6.2 - Tune A, \PY 6.323 - UE and \JM 4.1 - UE predictions 
for the underlying event compared to CDF data: (a) Average charged 
particles multiplicity and (b) average p$_{t}$ sum in the underlying event.} 
\label{fig:ue-tev-rick}
\end{figure}

Figure \ref{fig:ue-tev-rick} shows \JM 4.1 - UE (table
\ref{tab:JIMMY-tunings}) and \PY 6.323 - UE
(table \ref{tab:PYTHIA-tunings}) predictions for the underlying 
event compared to CDF data \cite{Affolder:2001xt} for the average 
charged particle ($p_{t} >0.5~$GeV and $\left| \eta \right| <1$) 
multiplicity (a) and the average p$_{t}$ sum in the underlying event (b). 
Distributions generated with \PY 6.2 - Tune A are also included in the 
plots for comparison. There is a 
reasonably good agreement between the proposed tunings and the data. 
The distribution shapes are slightly different in the region of 
P$_{t_{\text{ljet}}} \lesssim 15$ GeV. \PY 6.323 - UE underestimates 
the data while \JM 4.1 - UE overestimates it. 

\begin{figure}[h]
\subfigure[]{\includegraphics[width=.5\textwidth]{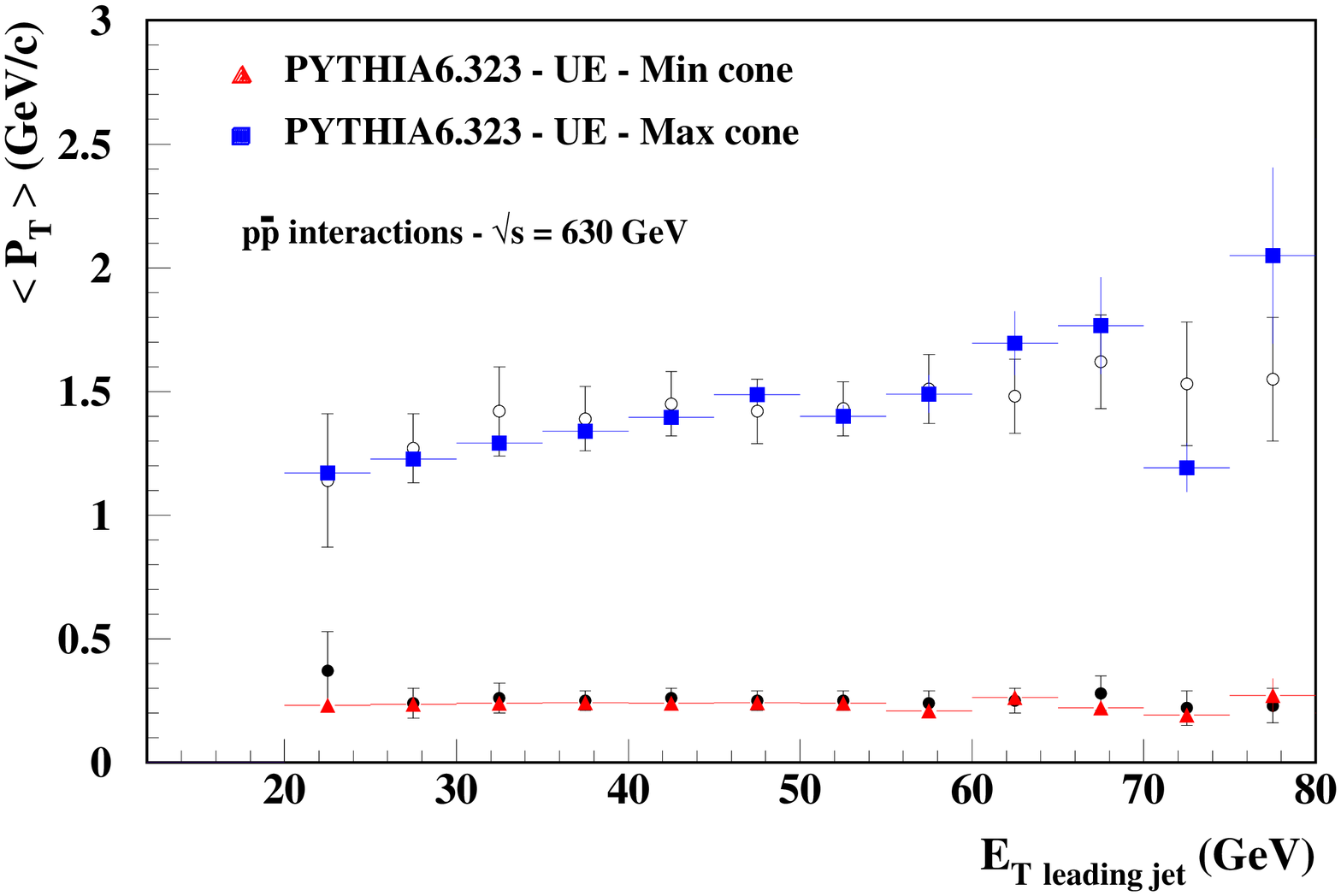}} 
\subfigure[]{\includegraphics[width=.5\textwidth]{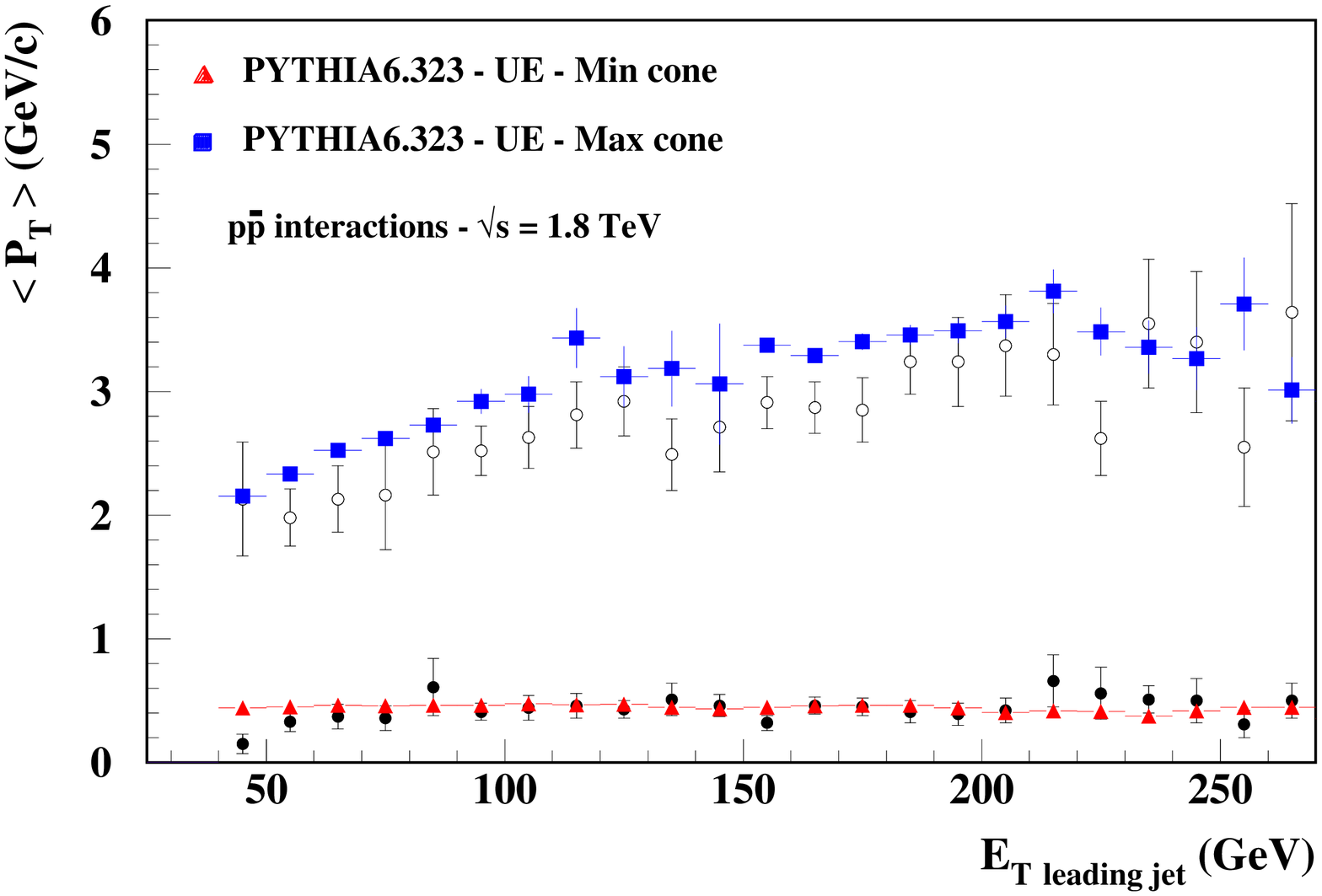}} 
\caption{\PY 6.323 - UE predictions for the underlying event
  compared to the  $<p_{t}>$ in MAX and MIN
  cones for (a) p$\overline{\text{p}}$ collisions at $\sqrt{\text{s}}$
  = 630 GeV  and (b) 1.8 TeV.} 
\label{fig:maxmin-pythia}
\end{figure}

\begin{figure}[h]
\subfigure[]{\includegraphics[width=.5\textwidth]{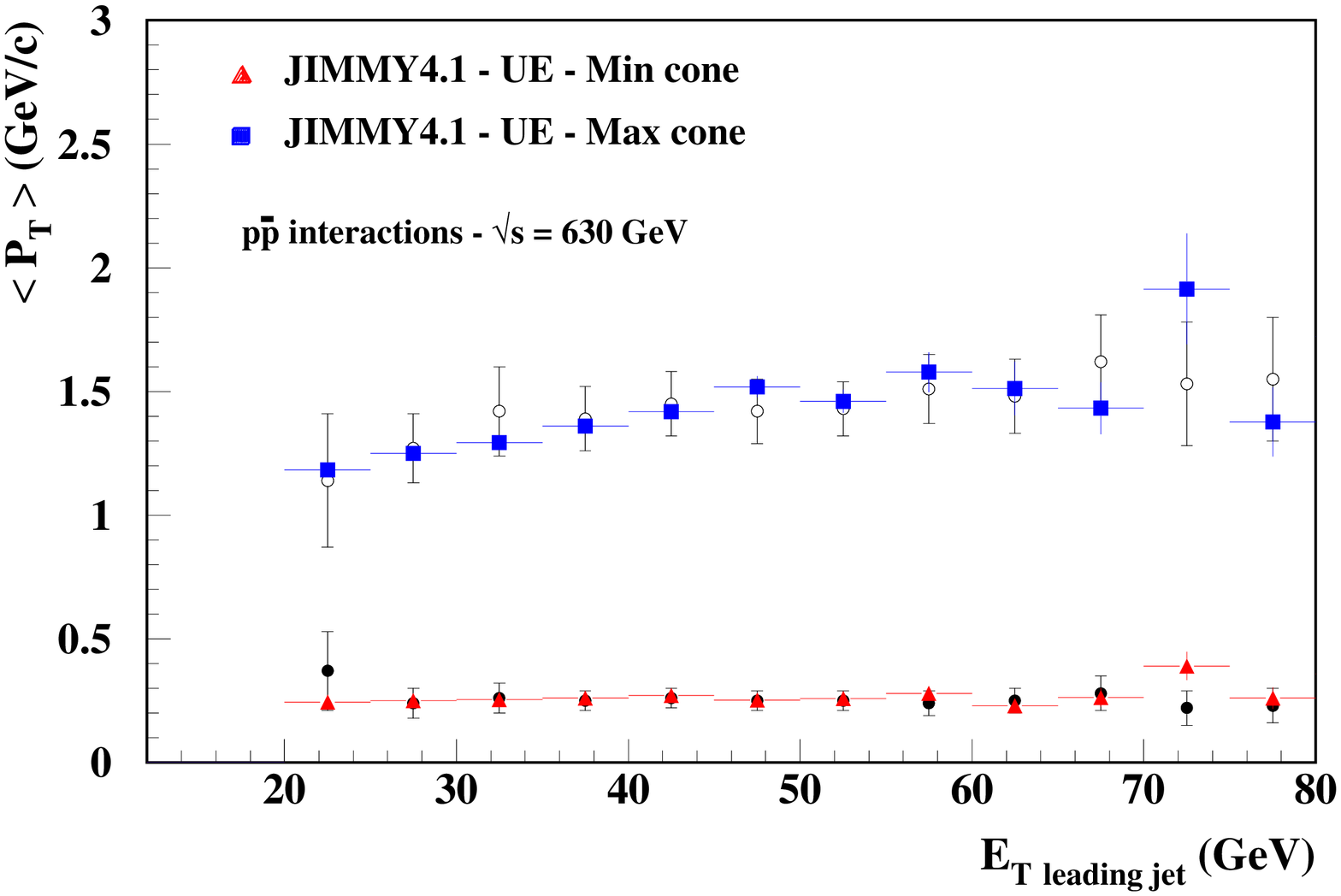}} 
\subfigure[]{\includegraphics[width=.5\textwidth]{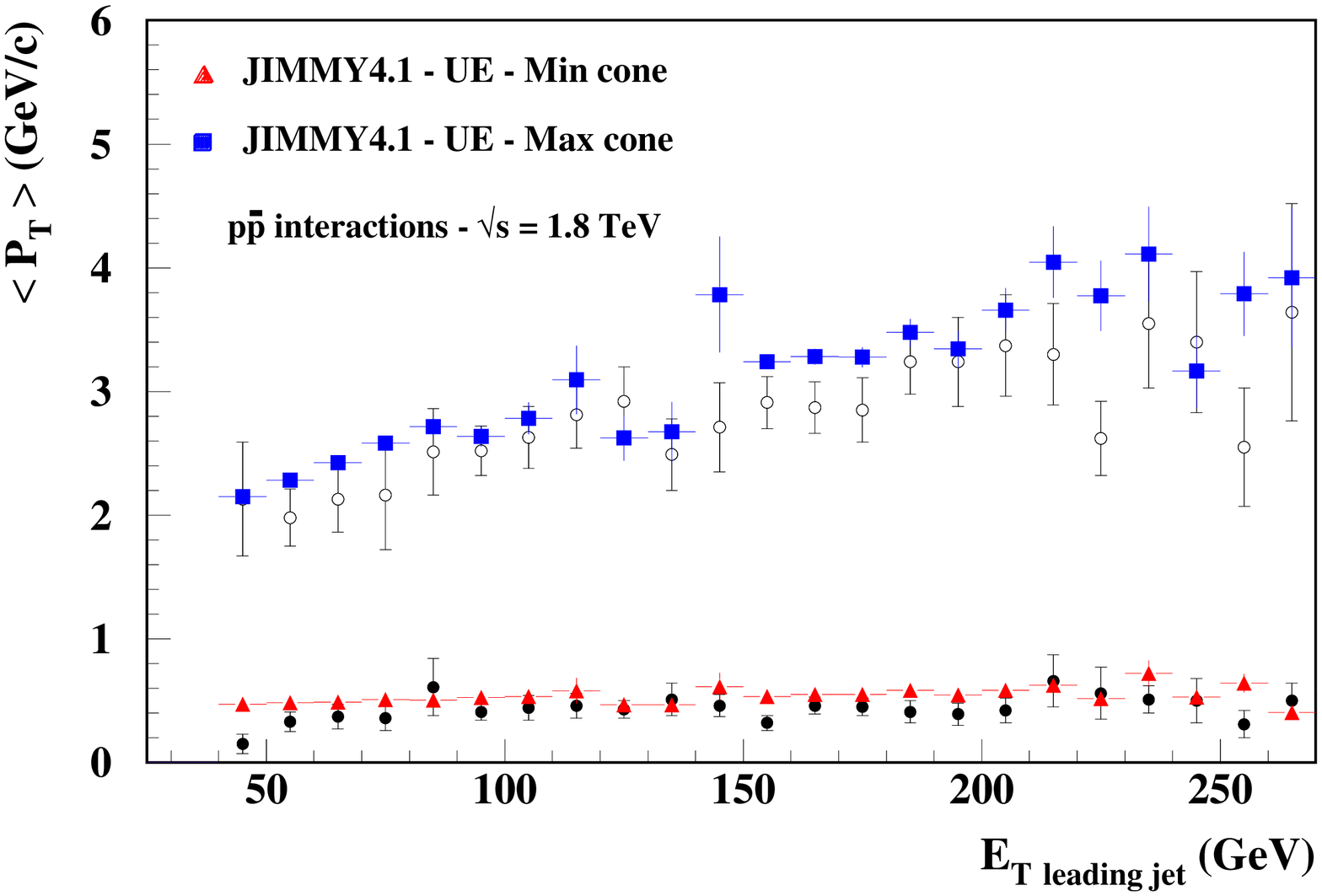}} 
\caption{\JM 4.1 - UE predictions for the underlying event
  compared to the $<p_{t}>$ in MAX and MIN
  cones for (a) p$\overline{\text{p}}$ collisions at $\sqrt{\text{s}}$
  = 630 GeV  and (b) 1.8 TeV.} 
\label{fig:maxmin-jmy}
\end{figure}

Another measurement of the underlying event was made by defining two 
cones in $\eta - \phi$ space, at the same pseudorapidity $\eta$ as the leading
E$_{T}$ jet (calorimeter jet) and $\pm 90^{\circ}$ in the azimuthal
direction, $\phi$ \cite{Acosta:2004wq}. The total charged track momentum
inside each of the two cones is then measured and the higher of the
two values defines the ``MAX'' cone, with the remaining cone being
labeled ``MIN'' cone. 
Figure \ref{fig:maxmin-pythia} shows \PY 6.323 - UE 
predictions for the UE compared to CDF data \cite{Acosta:2004wq} for the
$<p_{t}>$ of charged particles in the MAX and MIN cones for
p$\overline{\text{p}}$ collisions at (a) $\sqrt{\text{s}}$ = 630 GeV
and (b) 1.8 TeV. \JM 4.1 - UE predictions are compared to the data
in fig. \ref{fig:maxmin-jmy}. 
Both tunings describe the data with good agreement, however, this only
became possible by tuning the parameters of the minimum p$_{t}$ cut-off to 
include the correct energy dependence in both generators (PARP(82), (89) and 
(90) for \PY 6.3 and PTJIM for \JM 4.1). 

\begin{figure}[h!]
\begin{center}
\includegraphics[width=.8\textwidth]{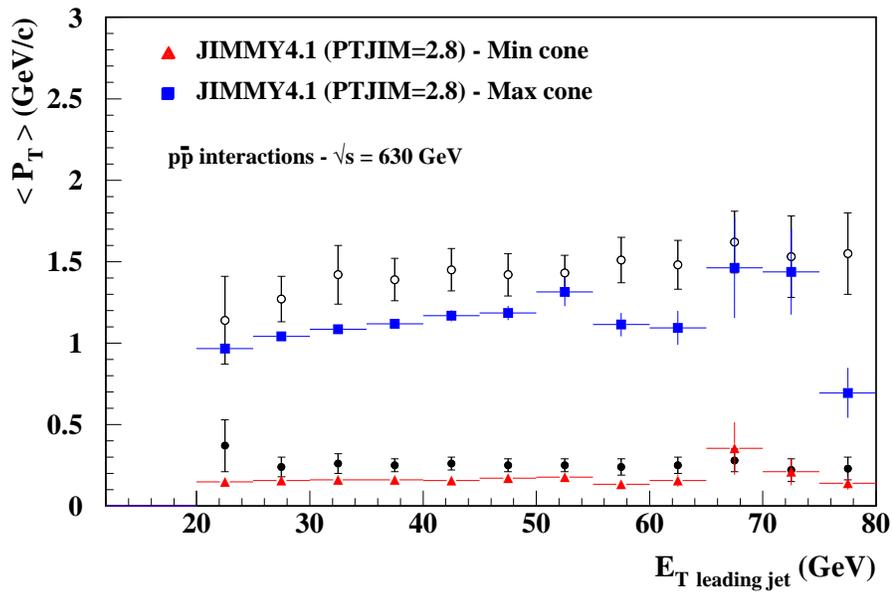}
\caption{\JM 4.1 - PTJIM=2.8 (fixed for comparisons at
  $\sqrt{\text{s}}$ = 1.8 TeV), JMRAD(73,75)=1.8 - predictions for the
  UE compared to the $<p_{t}>$ in MAX and MIN cones for
  p$\overline{\text{p}}$ collisions at $\sqrt{\text{s}}$ = 630 GeV.}  
\label{fig:maxmin-ptjim28}
\end{center}
\end{figure}

Tuning the JIMMY parameter PTJIM to include an energy dependent factor
made it possible to describe the MAX-MIN $<p_{t}>$ distributions at
different energies. Just to
illustrate what would be the result of not adding the energy dependent
factor in PTJIM, in fig. \ref{fig:maxmin-ptjim28}, JIMMY4.1 with PTJIM
fixed by comparisons to the $\sqrt{\text{s}}$ = 1.8 TeV distributions
to PTJIM=2.8, is compared to the $\sqrt{\text{s}}$ = 630 GeV MAX-MIN
data. The predictions underestimate the data, indicating that PTJIM
has to be reduced in order to describe the data. 

\begin{figure}[h!]
\subfigure[]{\includegraphics[width=.5\textwidth]{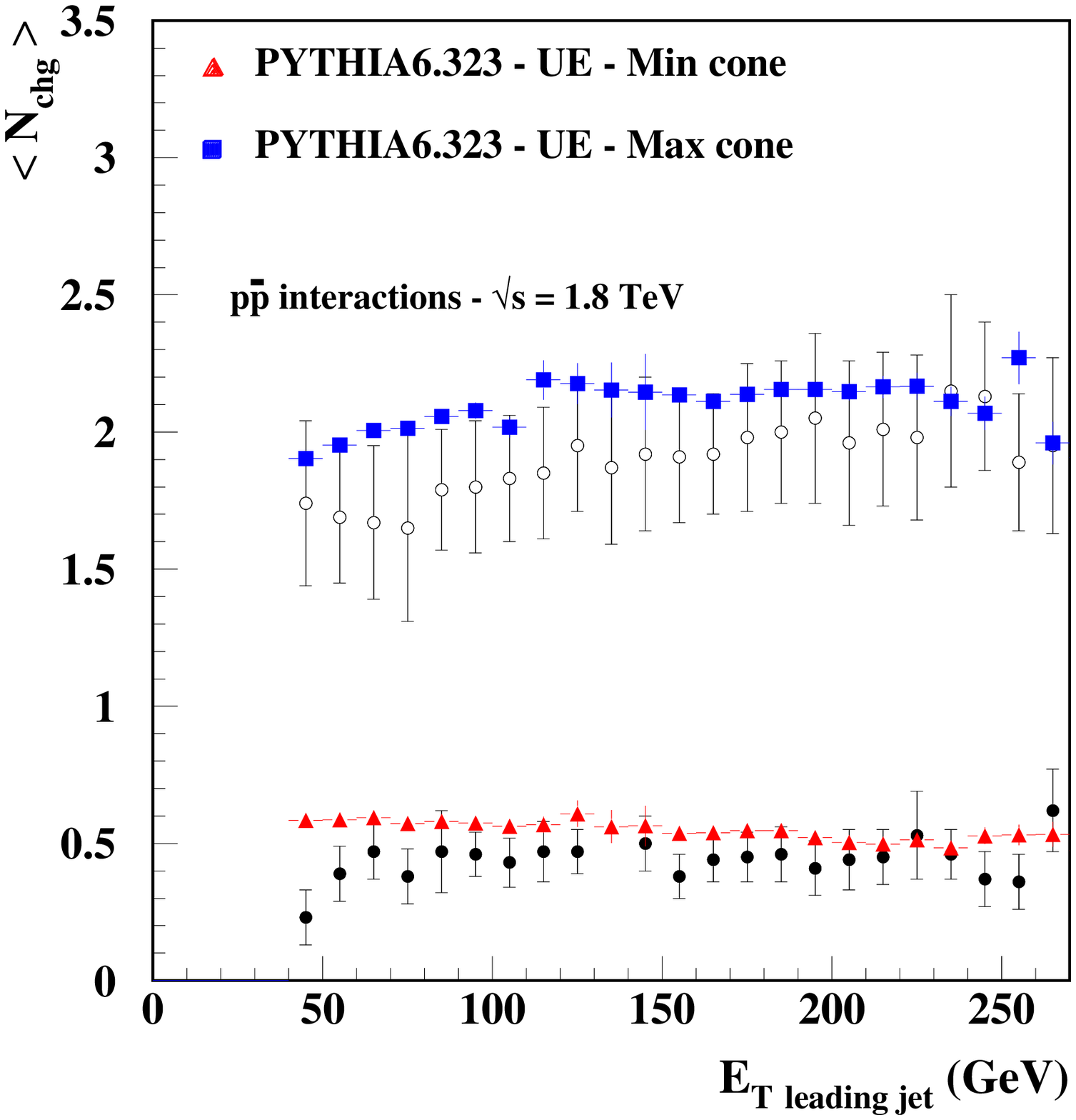}} 
\subfigure[]{\includegraphics[width=.5\textwidth]{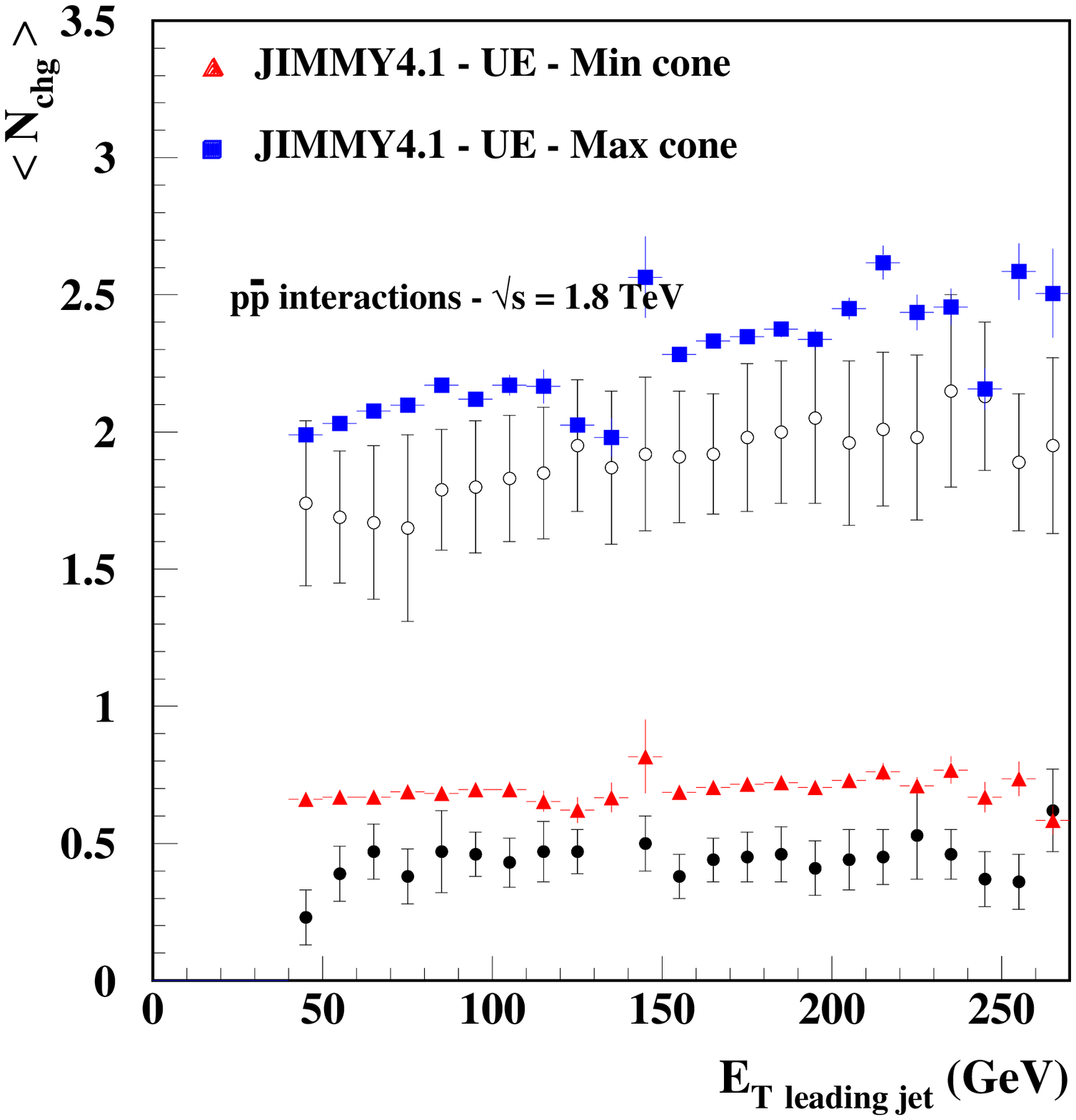}} 
\caption{\PY 6.323 - UE (a) and \JM 4.1 - UE (b) predictions for the 
underlying event compared to the $<N_{chg}>$ in MAX and MIN cones for 
p$\overline{\text{p}}$ collisions at $\sqrt{\text{s}}$ = 1.8 TeV.} 
\label{fig:maxmin-nchg}
\end{figure}

The agreement between predictions and data seen in figs. \ref{fig:maxmin-pythia}
and \ref{fig:maxmin-jmy} for the $<p_{t}>$ in MAX and MIN cones is not
reproduced in the comparisons of $<N_{chg}>$ distributions for
p$\overline{\text{p}}$ collisions at $\sqrt{\text{s}}$ = 1.8 TeV shown
in fig. \ref{fig:maxmin-nchg}. There is no data available for the
$<N_{chg}>$ distributions for p$\overline{\text{p}}$ collisions at
$\sqrt{\text{s}}$ = 630 GeV. Both \PY 6.323 - UE and \JM 4.1 -
UE overestimate the data. This indicates that neither model is
describing the ratio $<p_{t}>$/$<N_{chg}>$ as seen in the
data. This certainly needs to be improved in future tunings.

\subsubsection*{LHC predictions}

\begin{figure}[h!]
\begin{center}
\includegraphics[width=.85\textwidth]{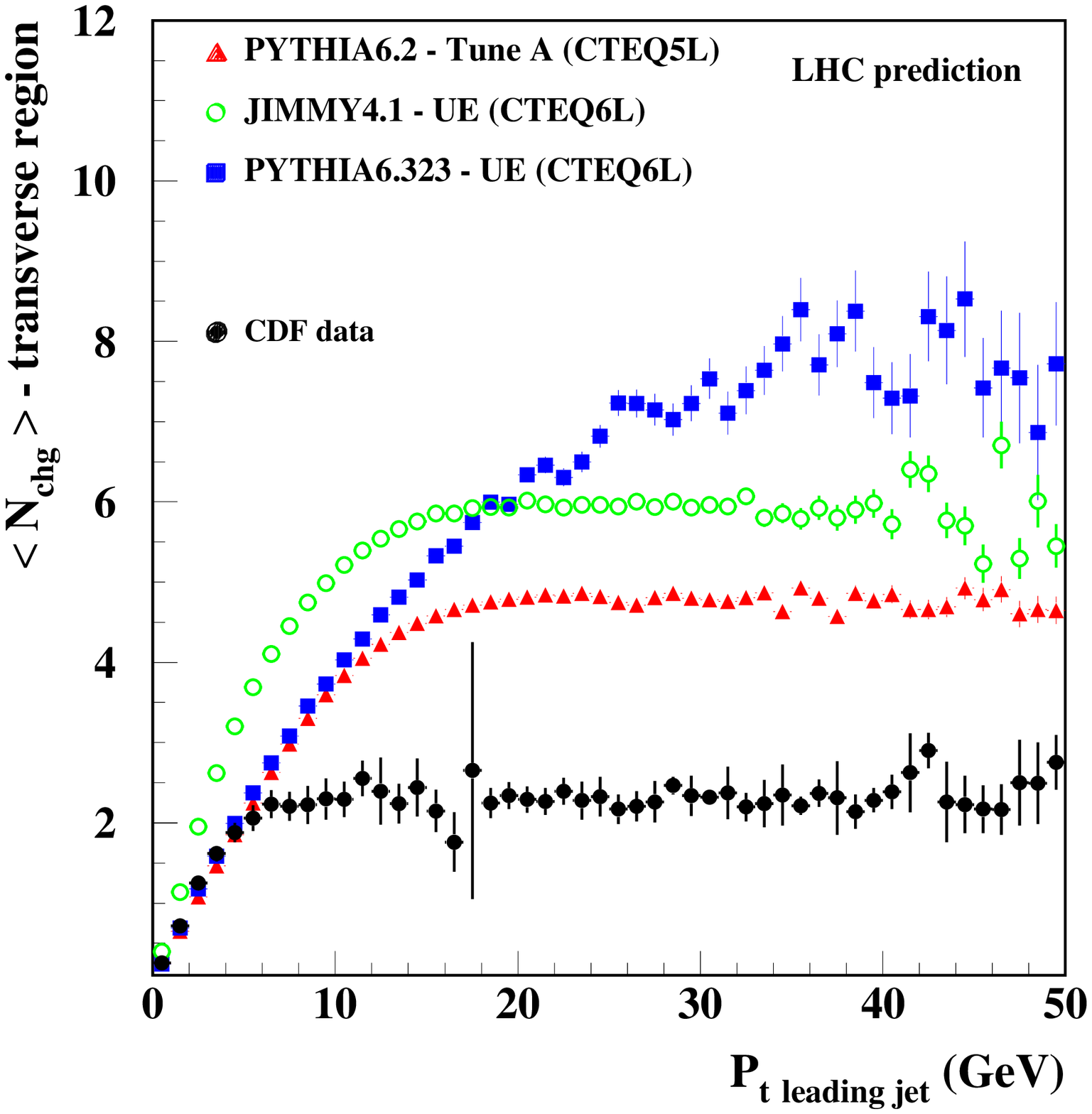}
\caption{\PY 6.2 - Tune A, \JM 4.1 - UE and \PY 6.323 - UE
predictions for the average charged multiplicity in the underlying 
event for LHC pp collisions.}    
\label{fig:ue-lhc}
\end{center}
\end{figure}

Some predictions for the underlying event energy at the LHC are shown in
Fig. \ref{fig:ue-lhc}. It shows \PY 6.323 - UE
(table \ref{tab:PYTHIA-tunings}), \JM 4.1 - UE (table
\ref{tab:JIMMY-tunings}) and \PY 6.2 - Tune A predictions for the
average multiplicity in the underlying event for LHC pp collisions. 
The CDF data (p$\overline{\text{p}}$ collisions at $\sqrt{\text{s}}$ 
= 1.8 TeV.) for the average multiplicity in the UE is also included in
fig. \ref{fig:ue-lhc}. 

A close inspection of predictions for the underlying event given in
fig. \ref{fig:ue-lhc} shows that the average charged particle
multiplicity in the underlying event for leading jets with 
P$_{t_{\text{ljet}}} > 20$ GeV reaches a plateau at $\sim 4.7$ charged 
particles according to \PY 6.2 - Tune A, $\sim 6$ for \JM 4.1 - UE and 
$\sim 7.5$ according to \PY 6.323 - UE. Expressed as particle densities per
unit $\eta - \phi$, where the underlying event phase-space is given by 
$\Delta \eta \Delta \phi = 4 \pi / 3$ \cite{Affolder:2001xt,Field:2005qt}, these
multiplicities correspond to 1.12, 1.43 and 1.79 charged particles per
unit $\eta - \phi$ (p$_{t} >0.5~$GeV), as predicted by \PY 6.2 - Tune A, 
\JM 4.1 - UE, and \PY 6.323 - UE, respectively. 
The shape of the distributions also shows significant differences
between the model predictions. The shape of the multiplicity
distribution generated by \PY 6.323 - UE is considerably different
from the other two models in the region of P$_{t_{\text{ljet}}}
\lesssim 25$ GeV. 

It is clear that (1) all predictions lead to a substantially
larger underlying event energy at the LHC than at the Tevatron and (2)
there are large differences among the predictions from the various
models. Investigations are continuing trying to reduce the energy 
extrapolation uncertainty of these models. This measurement will be one
of the first to be performed at the LHC 
and will be used for subsequent Monte Carlo tunings for the LHC.

\clearpage

\section{Diffractive Physics}
\subsection{Large Multigap Diffraction at LHC}
\newcommand{\pom}{I\!\! P}
\newcommand{\reg}{I\!\! R}

\textbf{Contributed by:  Goulianos}

The large rapidity interval available at the Large Hadron Collider 
offers an arena in which the QCD aspects of diffraction may be explored 
in an environment free of gap survival 
complications using events with multiple rapidity gaps.

\subsubsection*{Soft Diffraction}
Diffractive processes are characterized 
by large rapidity gaps, defined as regions of 
(pseudo)rapidity~\footnote{We use {\em pseudorapidity}, $\eta=-\ln\tan\frac{\theta}{2}$, and
{\em rapidity}, $y=\frac{1}{2}\frac{E+p_L}{E-p_L}$, interchangeably.}
in which there is no particle production. 
Diffractive gaps are presumed to be produced by the exchange of  
a color singlet quark/gluon object 
with vacuum quantum numbers referred to as 
the {\em Pomeron}~\cite{hep-ph/0407035,Goulianos:2005xj} (the present paper contains 
excerpts from these two references). 

Traditionally diffraction had been treated in Regge theory using 
an amplitude based on a simple Pomeron pole and factorization.
This approach was successful at $\sqrt s$ 
energies below $\sim 50$ GeV~\cite{Goulianos:1982vk}, but as collision
energies increased to reach $\sqrt s=$1800 GeV at the Fermilab Tevatron 
the SD cross section was found to be suppressed by a factor of 
$\sim {\cal{O}}(10)$ relative to the Regge-based prediction~\cite{Abe:1993wu}. 
This blatant breakdown of factorization was traced back to 
the energy dependence of the Regge theory  
$\sigma_{sd}^{tot}(s)$,
\begin{equation}
d\sigma_{sd}(s,M^2)/dM^2\sim s^{2\epsilon}/(M^2)^{1+\epsilon},
\label{eq:reggeM2}
\end{equation} 
which is faster than that of $\sigma^{tot}(s)\sim s^\epsilon$, 
so that at high $\sqrt s$ unitarity would have to be 
violated if factorization held. 

In contrast to the Regge theory prediction of Eq.~(\ref{eq:reggeM2}), the
measured SD $M^2$-distribution shows no explicit
$s$-dependence ($M^2$-scaling) over a region of $s$ spanning six orders of 
magnitude~\cite{Phys.Rev.D59.114017}. Thus, factorization appears to {\em yield} to 
$M^2$-scaling. This is a property built into the
{\em Renormalization Model} of hadronic diffraction, in which the 
Regge theory Pomeron flux is renormalized to unity~\cite{Phys.Lett.B358.379}.
\begin{figure}[h]
\vspace{-3em}
\unitlength 0.95in
\thicklines
\begin{center}
\begin{picture}(6,1)(1,0)
\put(1,0){\line(2,0){6}}
\multiput(2,0)(2,0){3}{\oval(0.8,0.5)[t]}
\put(1.9,-0.25){$\eta_1'$}
\put(2.9,-0.25){$\eta_2$}
\put(3.9,-0.25){$\eta_2'$}
\put(4.9,-0.25){$\eta_3$}
\put(5.9,-0.25){$\eta_3'$}
\put(2.8,0.5){$\Delta \eta_2$}
\put(4.8,0.5){$\Delta \eta_3$}
\put(1.25,-0.5){$t_1$}
\put(2.9,-0.5){$t_2$}
\put(4.9,-0.5){$t_3$}
\put(6.7,-0.5){$t_4$}
\put(1.15,0.5){$\Delta \eta_1$}
\put(1.85,0.5){$\Delta \eta'_1$}
\put(3.85,0.5){$\Delta \eta'_2$}
\put(5.85,0.5){$\Delta \eta'_3$}
\put(6.6,0.5){$\Delta \eta_4$}
\end{picture}
\end{center}
\vspace{1em}
\caption{Average multiplicity $dN/d\eta$ vs $\eta$ for a process with four rapidity gaps $\Delta \eta_{i=1-4}$.} 
\label{fig:soft}
\end{figure}
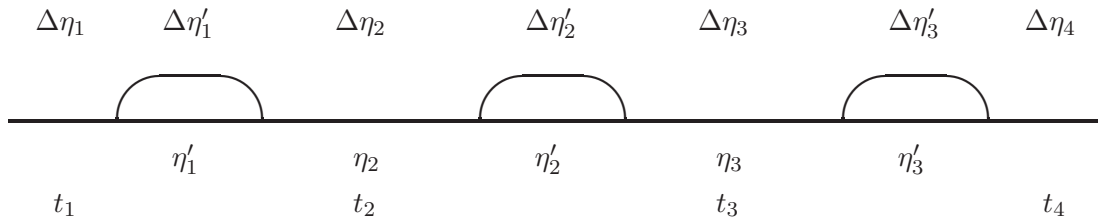
In a QCD inspired approach, the renormalization model was extended
to central and multigap diffractive processes~\cite{hep-ph/0203141}, an example of  
which is the four-gap process shown schematically 
in Fig~\ref{fig:soft}.
In this approach cross sections depend on the 
number of wee partons~\cite{Levin:1998pk} and therefore the $pp$ total cross section 
is given by 
\begin{equation}
\sigma_{pp}^{tot}=\sigma_0\cdot e^{\epsilon\Delta\eta'},
\label{totDeta}
\end{equation}
where $\Delta \eta'$ is the rapidity region in which there is particle production. Since, from the optical theorem, $\sigma^{tot}\sim {\rm Im\,f^{el}}(t=0)$, 
the full parton model amplitude may be written as 
\begin{equation}
{\rm Im\,f^{el}}(t,\Delta\eta)\sim e^{({\epsilon}+\alpha't)\Delta \eta},
\label{eq:fPM}
\end{equation}
\noindent where $\alpha't$ is 
a simple parameterization of the $t$-dependence of the amplitude. 
On the basis of this amplitude, the
cross section of the four-gap process of Fig.~\ref{fig:soft} takes the form 

\begin{equation}\
\frac{d^{10}\sigma^D}{\Pi_{i=1}^{10}dV_i}=
N^{-1}_{gap}\; 
\underbrace{
F^2_p(t_1)F^2_p(t_4)
\Pi_{i=1}^4\left\{e^{[\epsilon+\alpha't_i]\Delta\eta_i}\right\}^2
}_{\hbox{gap probability}}\;
\times\kappa^4\left[\sigma_0\,e^{\epsilon\sum_{i=1}^3\Delta\eta'_i}\right],
\label{eq:diffPM}
\end{equation}
\noindent where the term in square brackets is the $pp$  
total cross section at the reduced $s$-value, defined 
through $\ln (s'/s_0)=\sum_i\Delta \eta_i'$, 
$\kappa$ (one for each gap) is the QCD color factor for gap formation, 
the gap probability is the amplitude squared for elastic scattering 
between two diffractive clusters or between a diffractive cluster and a 
surviving proton with form factor $F^2_p(t)$,
and $N_{gap}$ is the (re)normalization factor defined as 
the gap probability integrated 
over all 10 independent variables $t_i$, $\eta_i$, $\eta'_i$, and 
$\Delta\eta\equiv \sum_{i=1}^4\Delta\eta_i$. 

The renormalization 
factor $N_{gap}$ is a function of $s$ only.
The color factors are $c_g=(N_c^2-1)^{-1}$ 
and $c_q=1/N_c$ for gluon and quark color-singlet exchange, respectively.  
Since the reduced energy cross section is properly normalized, the 
gap probability is (re) normalized to unity. The quark to gluon 
fraction, and thereby the Pomeron intercept parameter $\epsilon$ 
may be obtained from the inclusive parton distribution 
functions (PDFs)~\cite{hep-ph/0407035}. Thus, normalized differential multigap cross 
sections at $t=0$ may be fully derived from inclusive PDFs and QCD color 
factors without any free parameters.

The exponential dependence of the cross section on $\Delta\eta_i$ leads to 
a renormalization factor $\sim s^{2\epsilon}$ independent of the number of 
gaps in the process. This remarkable property of the renormalization model, 
which was confirmed in two-gap to one-gap cross section ratios measured by 
the CDF Collaboration (see~\cite{hep-ph/0407035}), suggests that multigap
diffraction can be used as a tool for exploring the QCD aspects of 
diffraction in an environment free of rapidity gap suppression effects.
The LHC with its large rapidity coverage provides the ideal arena 
for such studies.  

\subsubsection*{Hard Diffraction}
Hard diffraction processes are those in which there is a hard 
partonic scattering in addition to the diffractive rapidity gap. 
SD/ND ratios for $W$, dijet, $b$-quark, and $J/\psi$ production 
at $\sqrt s=$1800 GeV measured by the CDF Collaboration are 
approximately equal ($\sim 1\%$), indicating that the rapidity gap 
formation probability is largely {\em flavor independent}.
However, the SD structure function measured from dijet production 
is suppressed by $\sim{\cal{O}}(10)$ relative to expectations based
on diffractive PDFs measured from 
diffractive DIS at HERA.
 
A modified version of our QCD approach to soft diffraction 
can be used to describe hard diffractive processes and has been  
applied to diffractive DIS at HERA, 
$\gamma^*+p\rightarrow p+Jet+X$, and diffractive dijet production at the 
Tevatron, $\bar p+p\rightarrow\bar p+\hbox{dijet}+X$ in~\cite{Goulianos:2005xj}. 
The hard process generally involves several color ``emissions''
from the surviving proton, the sum of which comprises 
a color singlet exchange with vacuum quantum numbers. 
Two of these emissions are of special interest, 
one at $x=x_{Bj}$ from the proton's hard PDF at scale $Q^2$, 
which causes the hard scattering, and another at $x=\xi$ 
(fractional momentum loss of the diffracted nucleon) from the soft 
PDF at $Q^2\approx 1$ GeV$^2$, which neutralizes the exchanged color 
and forms the rapidity gap. Neglecting the $t$-dependence, 
the diffractive structure function could then be 
expressed as the product of the inclusive 
hard structure function and the soft parton 
density at $x=\xi$,

\begin{equation}F^D(\xi,x,Q^2)=
\frac{A_{\rm norm}}{\xi^{1+\epsilon}}\cdot c_{g,q}\cdot F(x,Q^2)
\Rightarrow
\frac{A_{\rm norm}}{\xi^{1+\epsilon+\lambda(Q^2)}}
\cdot c_{g,q}\cdot \frac{C(Q^2)}{\beta^{\lambda(Q^2)}},
\label{eq:F2D3}
\end{equation}
where $c_{g,q}$ are QCD color factors, 
$\lambda$ is the parameter of a power law fit to the hard structure 
function in the region $x<0.1$,
$A_{\rm norm}$ is a normalization factor, and $\beta\equiv x/\xi$.

At high $Q^2$ at HERA, where factorization is expected to 
hold~\cite{Phys.Lett.B358.379,Collins:2001ga}, $A_{\rm norm}$ is the nominal normalization 
factor of the soft PDF. This factor is constant, leading to two 
important predictions, which are confirmed by the data:

i) The Pomeron intercept in diffractive DIS (DDIS) 
is $Q^2$-dependent and equals 
the average value of the soft and hard intercepts:
\begin{equation}
\alpha^{DIS}_{\pom}=1+\lambda(Q^2),\;\;\;\;
\alpha^{DDIS}_{\pom}=1+\frac{1}{2}\left[\epsilon+\lambda(Q^2)\right]\nonumber
\label{eq:heraintercept}
\end{equation}

ii) The ratio of DDIS to DIS structure functions at fixed $\xi$ 
is independent of $x$ and $Q^2$:
\hspace*{-5em}\begin{equation}
R\left[\frac{F^D(\xi,x,Q^2)}{F^{ND}(x,Q^2)}\right]_{\rm HERA}=
\frac{A_{\rm norm}\cdot c_q}{\xi^{1+\epsilon}}=
\frac{\rm const}{\xi^{1+\epsilon}}
\label{eq:Rhera}
\end{equation}     

At the Tevatron, where high soft parton densities lead to saturation,
$A_{\rm norm}$ must be renormalized to
\begin{equation}
A^{\rm Tevatron}_{\rm renorm}=1/\int^{\xi=0.1}_{\xi_{min}}
\frac{d\xi}{\xi^{1+\epsilon+\lambda}}
\propto
\left(\frac{1}{\beta\cdot s}\right)^{\epsilon+\lambda},
\label{eq:tevrenorm}
\end{equation}
where $\xi_{min}=x_{min}/\beta$ and $x_{min}\propto 1/s$.
Thus, the diffractive structure function acquires a term
$\sim (1/\beta)^{\epsilon+\lambda}$, and the ratio of the 
diffractive to inclusive 
structure functions a term $\sim (1/x)^{\epsilon+\lambda}$.
This prediction is confirmed by CDF data,
where the $x$-dependence of the diffractive to inclusive ratio 
was measured to be $\sim 1/x^{0.45}$ (see~\cite{hep-ph/0407035}). 

A comparison~\footnote{Performed by the author and K. Hatakeyama (see~\cite{hep-ph/0407035}) using CDF published data and preliminary H1 diffractive parton densities~\cite{heraglue}.}
between the diffractive structure function measured on the proton side 
in events with a leading antiproton to 
expectations from diffractive DIS at HERA showed 
approximate agreement, indicating that factorization is largely 
restored for events that already have a rapidity gap.
Thus, as already mentioned for soft diffraction, 
events triggered on a leading proton at LHC provide an environment 
in which the QCD aspects of diffraction may be explored without 
complications arising from rapidity gap survival.    
\subsubsection*{Proposed program of multigap diffraction at LHC}
The rapidity span at LHC running at $\sqrt s=14$ TeV is $\Delta\eta=19$ as 
compared to $\Delta\eta=15$ at the Tevatron. This suggests the following 
program for studies of non-suppressed diffraction:
\begin{itemize}
\item Trigger on two forward rapidity gaps  of $\Delta\eta_F\geq 2$ 
(one on each side of the 
interaction point), or equivalently on 
forward protons of fractional longitudinal momentum loss 
$\xi=\Delta p_L/p_L\leq 0.1$, and explore the central rapidity region of 
$|\Delta\eta|\leq 7.5$, which has the same width 
as the entire rapidity region of the Tevatron.
In such an environment, the ratio of the rate of 
dijet events with a gap between jets to that without a gap,  
$gap$+[jet-gap-jet]+$gap$ to $gap$+[jet-jet]+$gap$, should rise 
from its value of $\sim 1$\% at the Tevatron to $\sim 5$\%. 
\item Trigger on one forward gap of $\Delta\eta_F\geq 2$ or on a proton of $\xi<0.1$,
in which case the rapidity gap available for non-suppressed diffractive 
studies rises to 17 units.
\end{itemize}

\clearpage
\subsection{Hard diffraction at the LHC and the Tevatron using double pomeron exchange}

\newcommand{\be}{\begin{equation}}
\newcommand{\ee}{\end{equation}}
\newcommand{\bL}{\begin{Large}}
\newcommand{\eL}{\end{Large}}
\newcommand{\bc}{\begin{center}}
\newcommand{\ec}{\end{center}}
\newcommand{\bfig}{\begin{figure}}
\newcommand{\efig}{\end{figure}}
\newcommand{\g}{\gamma}
\newcommand{\om}{\omega}

\newcommand{\la}{\label}
\newcommand{\no}{\nonumber \\}
\newcommand{\lra}{\longrightarrow}
\newcommand{\cor}[1]{\left\langle{#1}\right\rangle}

\renewcommand{\th}{\theta}
\newcommand{\sg}{\sigma}
\newcommand{\eps}{\epsilon}
\newcommand{\dl}{\delta}
\newcommand{\bal}{\bar \alpha}
\newcommand{\tg}{\tan \th}
\newcommand{\tgd}{\tan^2 \th}

\newcommand{\qb}{\bar{q}}

\newcommand{\rr}[4]{#1, {\it #2 \/}{\bf #3} #4}
\newcommand{\pref}{\f{1}{2\pi \alpha'}}
\newcommand{\ttl}{\f{\tau^2 \th^2}{L^2}}
\newcommand{\fl}{\f{L^2}{\chi_{L0}}}
\newcommand{\tcl}{\f{\tau^2 \chi_{L0}^2}{L^2}}
\newcommand{\Ttl}{\f{T^2 \th^2}{L^2}}
\newcommand{\Tcl}{\f{T^2 \chi_{L0}^2}{L^2}}
\newcommand{\qqqq}{\quad\quad\quad} 
\newcommand{\xpr}{x_\perp} 
\newcommand{\rrr}{\mathbb{R}} 
 
\newcommand{\numero}[1]{\noindent{\bf #1.}~} 
\newcommand{\remarki}[1]{\noindent{\bf [#1]}} 
\newcommand{\cc}{\chi(\gamma_c)} 
\newcommand{\ccc}{\chi^{\prime\prime}(\gamma_c)}

\textbf{Contributed by:  Royon}

Hard diffraction at the LHC has brought much interest recently related
to diffractive Higgs and SUSY production~\cite{Royon:2003ng}.
It is thus important
that the different models available can be tested at the Tevatron before the
start of the LHC. In this contribution, we will consider only one model based on
the Bialas-Landshoff approach \cite{Boonekamp:2001vk,Boonekamp:2003wm}, and more details about other models and their
implications can be found in \cite{Royon:2003ng} and the references therein.

\subsubsection*{Theoretical framework}

We distinguish in the following the so called {\it inclusive} and 
{\it exclusive} models for
diffraction. We call exclusive models the models where almost the full energy available
in the center of mass is used to produce the heavy object (dijets, Higgs,
diphoton, $W$ {\it etc.}). In other words, we get in the final state the diffractive
protons (which can be detected in roman pot detectors) and the heavy state
which decays in the main detector. The inclusive diffraction corresponds to
events where only part of the available energy is used to produce the heavy
object diffractively. For this model, we assume the pomeron is made of quarks
and gluons (we take the gluon and quark densities from the HERA measurements 
in shape and the normalisation from Tevatron data), and a quark or a gluon from
the pomeron is used to produce the heavy state. Thus the exclusive model appears
to be the limit where the gluon in the pomeron is a $\delta$ distribution in this
framework or in other words, there is no pomeron remnants in exclusive events.
We will see in the following that this distinction is quite relevant for
experimental applications.

\subsubsection*{Exclusive model}
Let us first introduce the model \cite{Bialas:1991wj} we shall use for describing 
exclusive production. In \cite{Bialas:1991wj}, the diffractive mechanism is based on two-gluon 
exchange between the 
two incoming protons. The soft pomeron is seen as a  pair of gluons 
non-perturbatively coupled  to the proton. One of the gluons is then coupled 
perturbatively to the hard process while the other one plays the r\^ole of a 
soft screening of colour, allowing for diffraction to occur. We will give here the
formulae for either the SUSY Higgs boson, or 
the $\tilde t \bar{\tilde t}$ pairs production and other formulae for standard
model Higgs bosons, $t \bar{t}$, diphoton or dijet production can be found
in \cite{Boonekamp:2001vk,Boonekamp:2003wm}.
The corresponding cross-sections for Higgs bosons and $\tilde t \bar{\tilde t}$ 
production read:

\begin{eqnarray}
 d\sigma_{h}^{exc}(s) &=& C_{h}\left(\frac{s}{M_{h}^{2}}\right)^{2\epsilon} 
\delta\left(\xi_{1}\xi_{2}-\frac{M_{h}^{2}}{s}\right)
\prod_{i=1,2} \left\{ d^{2}v_{i} \frac{d\xi_{i}}{1-\xi_{i}} \right.
\left. \xi_{i}^{2\alpha'v_{i}^{2}} \exp(-2\lambda_{h} v_{i}^{2})\right\} 
\sigma (g g \rightarrow h)
 \nonumber \\
d\sigma_{\tilde t \tilde{\bar{t}}}^{exc}(s) &=& C_{\tilde t \tilde {\bar{t}}} 
\left(\frac{s}{M_{\tilde t \tilde{\bar{t}}^{2}}}\right)^{2\epsilon}
\delta\left( \sum_{i=1,2} (v_{i} + k_{i}) \right)
\prod_{i=1,2} \left\{ d^{2}v_{i} d^{2}k_{i} d\xi_{i} \right.
d\eta_{i}\  \xi_{i}^{2\alpha'v_{i}^{2}}\! 
\left. \exp(-2\lambda_{\tilde t\tilde{\bar{t}}} v_{i}^{2})\right\} {\sigma} 
(gg\to {\tilde t \tilde{\bar{t}}}\ ) \nonumber
\label{exclusif}
\end{eqnarray}
where, in both equations,  the variables $v_{i}$ and $\xi_{i}$ denote respectively 
the transverse
momenta and fractional momentum losses of the outgoing protons. In the second 
equation, 
$k_{i}$ and $\eta_{i}$ are respectively the squark  transverse 
momenta and  rapidities. $\sigma (g g \rightarrow H),  {\sigma} (gg\to {\tilde t 
\tilde{\bar{t}}}\ )$ are the hard  production cross-sections which are given 
later on. The model normalisation constants  $C_{h}, C_{\tilde t \tilde 
{\bar{t}}}$ are fixed from the fit to dijet diffractive production,
while the ratio is fixed theoretically \cite{Boonekamp:2001vk,Boonekamp:2003wm}.

In the model,  the soft pomeron 
trajectory is
taken 
from the standard 
Donnachie-Landshoff   parametrisation \cite{Donnachie:1992ny},
 namely $\alpha(t) = 1 + 
\epsilon + \alpha't$, with
$\epsilon \approx 0.08$ and $\alpha' \approx 0.25 
\mathrm{GeV^{-2}}$. 
$\lambda_{h}, \lambda_{\tilde t \tilde{\bar{t}}}$ are  
kept as in  the original paper \cite{Bialas:1991wj} for the SM Higgs and $q 
\bar q$ pairs.  Note that, in this model, the 
strong (non perturbative) 
coupling constant is fixed to a reference value 
$G^2/4\pi,$ which will be taken 
from the fit to the observed centrally produced diffractive dijets.

In order to select exclusive diffractive states,  it is required to take into 
account the corrections from soft hadronic scattering. Indeed, the soft 
scattering  between incident particles tends to mask the genuine
hard diffractive interactions at 
hadronic colliders. The formulation of this 
correction \cite{Kupco:2004fw} to the
scattering amplitudes consists in considering a gap 
survival  probability.
The correction factor is commonly evaluated to be of order $0.03$ for the QCD 
exclusive diffractive processes at the LHC.

More details about the theoretical model and its phenomenological
applications can be found in Refs. \cite{Boonekamp:2004nu} and \cite{Boonekamp:2001vk,Boonekamp:2003wm}. In the following,
we use the Bialas Landshoff model for exclusive Higgs production recently implemented in
a Monte-Carlo generator \cite{Boonekamp:2004nu}. 

\subsubsection*{Inclusive model}
Let us now discuss the inclusive models. We first notice that both models are
related, since they are both based on the Bialas Landshoff formalism.
The main difference, as we already mentionned, is that the exclusive model is a
limit of the inclusive model where the full energy available is used in the
interaction. The inclusive models implies the knowledge of the gluon and quark
densities in the pomeron. Whereas exclusive events are still to be observed,
inclusive diffraction has been studied already in detail at UA8 and then at HERA
and Tevatron. 

The inclusive mechanism is based on the 
idea that a Pomeron is a composite system, made itself from 
quarks and 
gluons. In our model, we thus apply the concept of Pomeron structure functions 
to 
compute the inclusive diffractive Higgs boson cross-section. The H1 
measurement of the diffractive structure function \cite{Adloff:1997sc} and the 
corresponding quark and gluon 
densities are used for this purpose. This implies the existence of
Pomeron remnants and QCD radiation, as is the case for the proton. This 
assumption comes
from {\it QCD factorisation} of hard processes. 
However, and this is also an important issue, we do not assume {\it Regge 
factorisation} at the proton vertices, {\it i.e.} we do not use the H1 Pomeron 
flux factors in 
the proton or antiproton.

Regge factorisation is known to be violated between HERA and the
Tevatron. Moreover, we want to use the same physical idea as in the 
exclusive model \cite{Bialas:1991wj}, namely that a non perturbative gluon exchange 
describes the 
soft interaction between the incident particles. In 
practice, the 
Regge factorisation breaking appears in three ways in our model:

{\bf i)} We keep as in the original model of Ref \cite{Bialas:1991wj} the soft Pomeron
trajectory with an intercept value of 1.08.

{\bf ii)}  We normalize our
predictions to the CDF Run I measurements, allowing for factorisation breaking 
of the Pomeron flux factors in the normalisation between the HERA and hadron 
colliders \footnote{Indeed, recent results from a QCD fit to the diffractive 
structure 
function in H1 \cite{paplaurent} show that the discrepancy between the gluonic 
content of
the Pomeron at HERA and Tevatron appears mainly in
normalisation.}.

{\bf iii)} The color factor derives from the non-factorizable 
character of the model, since it stems from the gluon exchange between the 
incident hadrons. We will see later the difference between this and the 
factorizable
case.

The formulae for the inclusive production processes considered here follow. We 
have, 
for dijet production\footnote{We call ``dijets'' the produced quark and gluon 
pairs.}, 
considering only the dominant gluon-initiated hard processes:

\begin{eqnarray*}
d\sigma_{JJ}^{incl} = C_{JJ}\left(\frac {x^g_1x^g_2 s }{M_{J
J}^2}\right)^{2\epsilon}\!\! \! \! \delta ^{(2)}\! 
\left( \sum _{i=1,2}
v_i\!+\!k_i\right) \prod _{i=1,2} \!\!\left\{{d\xi_i}  {d\eta_i}
d^2v_i d^2k_i {\xi _i}^{2\alpha' v_i^2}\!
\exp \left(-2 v_i^2\lambda_{JJ}\right)\right\} \times\nonumber \\
\times\left\{{\sigma_{JJ}} G_P(x^g_1,\mu) G_P(x^g_2,\mu) \right\};
\label{dinclujj}
\end{eqnarray*}

\noindent and for Higgs boson production:

\begin{eqnarray*}
d\sigma_H^{incl} = C_{H}\left(\frac {x^g_1x^g_2 s
}{M_{H}^2}\right)^{2\epsilon} \!\!\delta \left(\xi _1 \xi
_2\!-\!\frac{M_{H}^2}{x^g_1x^g_2 s} \right) \!\!\prod _{i=1,2}
\left\{G_P(x^g_i,\mu)\ dx^g_i  d^2v_i\ \frac {d\xi _i}{1\!-\!\xi 
_i}\
{\xi _i}^{2\alpha' v_i^2}\ \exp \left(-2 
v_i^2\lambda_H\right)\right\};
\label{dincluH}
\end{eqnarray*}

\noindent In the above, the $G_P$ (resp. $Q_P$) are the Pomeron gluon (resp. 
quark) 
densities, and $x^{g}_{i}$ (resp. $x^{q}_{i}$) are the Pomeron's momentum 
fractions carried by 
the gluons (resp. quarks) involved in the hard process. We 
use as 
parametrizations of the Pomeron structure functions the fits to the diffractive 
HERA data 
performed in \cite{Royon:2000wg, paplaurent}. Additional formulae concerning for instance
inclusive diffractive production of dileptons or diphotons are given in
\cite{Boonekamp:2001vk,Boonekamp:2003wm}.

Both the inclusive and exclusive productions have been implemented in a
generator called DPEMC, which has been interfaced with a fast simulation of the
D\O\ , CDF, ATLAS and CMS detectors.

\subsubsection*{Experimental context}

In this section, we discuss mainly the parameters which we use to simulate the
detectors at the LHC. The simulation will be valid for both CMS and ATLAS
detectors.
The analysis is based on a fast simulation of the CMS detector at the LHC.
The calorimetric coverage of the CMS experiment ranges up to a pseudorapidity 
of $|\eta|\sim 5$. 
The region devoted
to precision measurements lies within $|\eta|\leq 3$, with a typical 
resolution on jet energy measurement of $\sim\!50\%
/\sqrt{E}$,
where $E$ is in GeV, and a granularity in pseudorapidity and azimuth of 
$\Delta\eta\times\Delta\Phi \sim 0.1\times 0.1$. 

In addition to the central CMS detector, the existence of roman pot detectors
allowing to tag diffractively produced protons,
located on both $p$ sides, is assumed \cite{helsinki}. The $\xi$ acceptance and 
resolution have been derived for each device using a complete simulation
of the LHC beam parameters. The combined $\xi$ acceptance is $\sim 100\%
$ 
for $\xi$ ranging from $0.002$ to $0.1$, where
$\xi$ is 
the proton fractional momentum loss. The acceptance limit of the device 
closest to the interaction point
is $\xi > \xi_{min}=$0.02. 

In exclusive double Pomeron exchange, the mass of the central 
heavy object is given by $M^2 = \xi_1\xi_2 s$, where $\xi_1$ and $\xi_2$ are
the proton fractional momentum losses measured in the roman pot detectors.
At this level, we already see the advantages of the exclusive events. Since,
there is no energy loss due to additional radiation or pomeron remnants, we can
reconstruct the total diffractive mass, which means the mass of the
diffractively produced object (the Higgs, dijets, $t \bar{t}$, 
$t \tilde{\bar{t}}$, events, $W$ pairs...), very precisely using the kinematical
measurements from the roman pot detectors. The mass resolution is thus coming
directly from the $\xi$ resolution which is expected to be of the order of 1\%.
For inclusive events, the mass resolution will not be so good since part of the
energy is lost in radiation, which means that we measure the mass of the heavy
object produced diffractively and the pomeron remnants together very precisely.
To get a good mass resolution using inclusive events requires a good
measurement
of the pomeron remnants and soft radiation and being able to veto on it.

\subsubsection*{Existence of exclusive events}
While inclusive diffraction has already been observed at many colliders,
the question arises whether exclusive events exist or not since they have never been
observed so far. This is definitely an area where the Tevatron experiments can
help to test the models and show evidence for the existence of exclusive events
if any. It is crucial to be able to test the different models before the 
start of the LHC.
The D\O\ and CDF experiments 
at the Tevatron (and the LHC experiments) are ideal places to look for
exclusive events in dijet or $\chi_C$ channels for instance
where exclusive events are expected to occur at high dijet mass
fraction.
So far, no evidence of the existence of exclusive events has been found.
The best way to show evidence of the existence of exclusive events would be the
measurement of the ratio of the diphoton to the dilepton cross sections
as a function of the diphoton/dilepton mass ratio (the diphoton-dilepton mass
ratio being defined as the diphoton-dilepton mass divided by the total
diffractive mass).
The reason is quite simple: it is possible to produce exclusively diphoton but
not dilepton directly since ($g g \rightarrow \gamma \gamma$) 
is possible but not  ($g g \rightarrow l^+ l^-$) directly at leading
order. The ratio of
the diphoton to the dilepton cross section should show a bump or a change of
slope towards high diphoton-dilepton masses if exclusive events exist.
Unfortunately, the production cross section of such events is small and it will
probably not be possible to perform this study before the start of the LHC.

Another easier way to show the existence of such events would be to study the
correlation between the gap size measured in both $p$ and $\bar{p}$ directions
and the value of $\log 1/\xi$ measured using roman pot detectors, which can be
performed in the D\O\ experiment. The gap size between the
pomeron remnant and the protons detected in roman pot detector 
is of the order of 
$log 1/\xi$ for usual diffractive events (the measurement giving a slightly
smaller value to be in the acceptance of the forward detectors) while
exclusive events show a much higher value for the rapidity gap since the gap
occurs between the jets (or the $\chi_C$) and the proton detected in roman
pot detectors (in other words, there is no pomeron remnant)
\footnote{To distinguish between pure exclusive and
quasi-exclusive events (defined as inclusive diffractive events where little
energy is taken away by the pomeron remnants, or in other words, events
where the mass of the heavy object produced diffractively is almost equal
to the total diffractive mass), other observables such as
the ratio of the cross sections of double diffractive
production of diphoton and dilepton, or the $b$-jets to all jets 
are needed \cite{Boonekamp:2001vk,Boonekamp:2003wm}}. Another observable leading
to the same conclusion would be the correlation between $\xi$ computed
using roman pot detectors and using only the central detector.

Another way to access the existence of exclusive events would be via QCD
evolution. If one assumes that the DGLAP evolution equations work for parton
densities in the pomeron, it is natural to compare the predictions of
perturbative QCD with for instance dijet production in double pomeron exchange
as a function of the dijet mass fraction (defined as the ratio of the dijet mass
divided by the total diffractive mass) for different domains in diffractive
mass. It has been shown that the dependence of the exclusive production cross
section as a function of the dijet mass is much larger than the one of the
inclusive processes. In other words, if exclusive events exist, it is expected
that the evolution of the dijet cross section in double pomeron exchanges  as a
function of dijet mass fraction in bins of dijet masses will be incompatible
with standard QCD DGLAP evolution, and will require an additional contribution,
namely the exclusive ones \footnote{Let us note that one should also
distinguish this effect from higher order corrections, and also from higher
twist effects, which needs further studies.}. 
It will be quite interesting to perform such
an analysis at the Tevatron if statistics allows.

\subsubsection*{Results on diffractive Higgs production}
\begin{figure}[htb]
\begin{center}
\subfigure[]{\includegraphics[width=.49\textwidth,trim= 10 50 50 0]{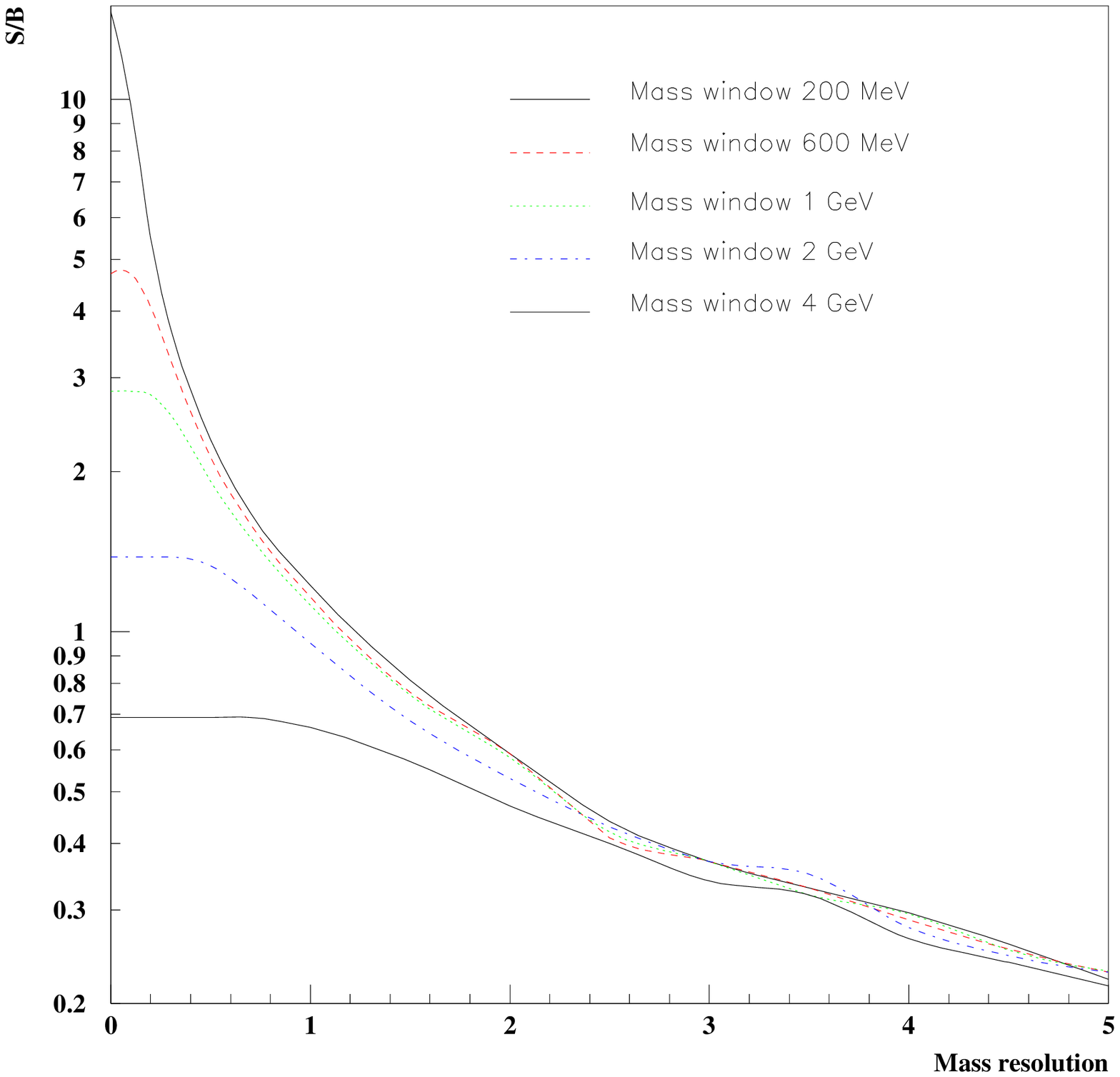}}
\subfigure[]{\includegraphics[width=.49\textwidth,trim= 10 50 50 0]{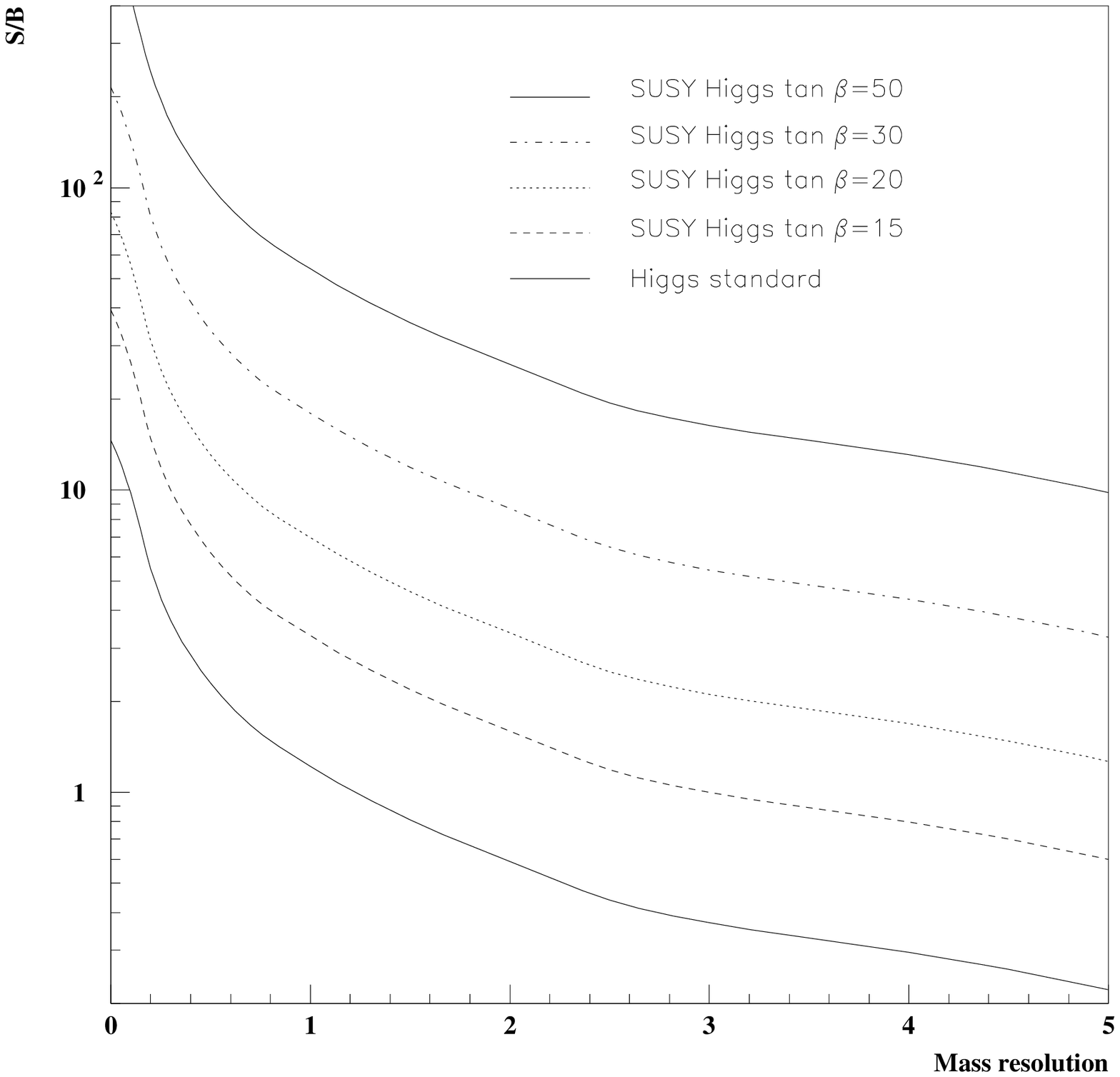}}
\caption{(a) Standard Model Higgs boson signal to background ratio as a function 
  of the resolution on
        the missing mass, in GeV. This figure assumes a Higgs
        boson mass of 120 GeV.
(b) similar to (a) for a SUSY Higgs boson with a mass of 120 GeV.}
\label{royon:fig2}
\end{center}
\end{figure}

Results are given in Fig.~\ref{royon:fig2}(a) for a Higgs mass of 120 GeV, 
in terms of the signal to background 
ratio S/B, as a function of the Higgs boson mass resolution.
Let us notice that the background is mainly due the exclusive $b \bar{b}$
production. However the tail of the inclusive $b \bar{b}$ production can also be
a relevant contribution and this is related to the high $\beta$ gluon
density which is badly known as present. It is thus quite important to constrain
these distributions using Tevatron data as suggested in a next section.

In order to obtain an S/B of 3 (resp. 1, 0.5), a mass resolution of about
0.3 GeV (resp. 1.2, 2.3 GeV) is needed. The forward detector design of 
\cite{helsinki} 
claims a resolution of about 2.-2.5 GeV, which leads to a S/B of about 
0.4-0.6. Improvements in this design
would increase the S/B ratio as indicated on the figure.
As usual, this number is enhanced by a large factor if one considers 
supersymmetric Higgs boson 
production with favorable Higgs or squark field mixing parameters.

The cross sections obtained after applying the survival probability of 3\% at
the LHC as well as the S/B ratios are given in Table \ref{sb} if one assumes a
resolution on the missing mass of about 1 GeV (which is the most optimistic
scenario). The acceptances of the roman pot detectors as well as the simulation
of the CMS detectors have been taken into account in these results. 

Let us also notice that the missing mass method will allow to perform a $W$ 
mass measurement using exclusive (or quasi-exclusive) $WW$ 
events in double Pomeron exchanges, and QED processes
\cite{Boonekamp:2005yi}. The advantage of the
QED processes is that their cross section is perfectly known and that this
measurement only depends on the mass resolution and the roman pot acceptance.
In the same way, it is possible to measure the mass of the top quark in
$t \bar{t}$ events in double Pomeron exchanges \cite{Boonekamp:2005yi} 
as we will see in the following.

\begin{table}
\begin{center}
\begin{tabular}{|c||c|c|c|c|c|} \hline
$M_{Higgs}$& cross & signal & backg. & S/B & $\sigma$  \\
 & section &  & & & \\
\hline\hline
120 & 3.9 & 27.1 & 28.5 & 0.95 & 5.1  \\
130 & 3.1 & 20.6 & 18.8 & 1.10 & 4.8  \\
140 & 2.0 & 12.6 & 11.7 & 1.08 & 3.7  \\ 
\hline
\end{tabular}
\caption{Exclusive Higgs production cross section for different Higgs masses,
number of signal and background events for 100 fb$^{-1}$, ratio, and number of
standard deviations ($\sigma$).}
\label{sb}
\end{center}
\end{table}

The diffractive SUSY Higgs boson production cross section is noticeably enhanced 
at high values of $\tan \beta$ and since we look for Higgs decaying into $b
\bar{b}$, it is possible to benefit directly from the enhancement of the cross
section contrary to the non diffractive case. A signal-over-background up to a
factor 50 can be reached for 100 fb$^{-1}$ for $\tan \beta \sim 50$
\cite{Boonekamp:2005up}. We give in Fig.~\ref{royon:fig2}(b) the
signal-over-background ratio for different values of $\tan \beta$ for a Higgs
boson mass of 120 GeV.

\subsubsection*{Threshold scan method: $W$, top and stop mass measurements}
We propose a new method to measure heavy particle properties via double 
photon and double pomeron exchange (DPE), at the LHC \cite{Boonekamp:2005yi}. In this category of events, the heavy objects 
are produced in pairs, whereas the beam particles
often leave the interaction region intact, and can be measured using very forward detectors.

Pair production of $WW$ bosons and top quarks in QED and  double pomeron exchange are described in detail in this section. 
$WW$ pairs are produced in photon-mediated processes, which are exactly calculable in QED. There is 
basically no uncertainty concerning the possibility of measuring these processes
at the LHC. On the contrary, $t \bar{t}$ events, produced in 
exclusive double pomeron exchange, suffer from theoretical uncertainties since 
exclusive diffractive production is still to be observed at the Tevatron, 
and other models lead to different cross sections, and thus to a different
potential for the top quark mass measurement. However, since the exclusive 
kinematics are simple, the model dependence will be essentially reflected by 
a factor in the effective luminosity for such events.

\subsubsection*{Explanation of the methods}
We study two different methods to reconstruct the mass of heavy objects
double diffractively produced at the LHC. The method is
based on a fit to the turn-on point of the missing mass distribution at 
threshold. 

One proposed method (the ``histogram'' method) corresponds to the comparison of 
the mass distribution in data with some reference distributions following
 a Monte Carlo simulation of the detector with different input masses
corresponding to the data luminosity. As an example, we can produce 
a data sample for 100 fb$^{-1}$ with a top mass of 174 GeV, and a few 
MC samples corresponding to top masses between 150 and 200 GeV by steps of. 
For each Monte Carlo sample, a $\chi^2$ value corresponding to the 
population difference in each bin between data and MC is computed. The mass point 
where
the $\chi^2$ is minimum corresponds to the mass of the produced object in data.
This method has the advantage of being easy but requires a good
simulation of the detector.

The other proposed method (the ``turn-on fit'' method) is less sensitive to the MC 
simulation of the
detectors. As mentioned earlier, the threshold scan is directly sensitive to
the mass of the diffractively produced object (in the $WW$W case for instance, it
is sensitive to twice the $WW$ mass). The idea is thus to fit the turn-on
point of the missing mass distribution which leads directly to the mass 
of the produced object, the $WW$ boson. Due to its robustness,
this method is considered as the ``default" one in the following.

\subsubsection*{Results}

To illustrate the principle of these methods and their achievements,
we  apply them to the 
$WW$ boson and the top quark mass measurements in the
following, and obtain the reaches at the LHC. They can be applied to other 
threshold scans as well.
The precision of the $WW$ mass measurement (0.3 GeV for 300 fb$^{-1}$) is not competitive with other 
methods, but provides a very precise calibration 
of the roman pot detectors. The precision of
the top mass measurement is however competitive, with an expected precision 
better than 1 GeV at high luminosity. The resolution on the top mass is given
in Fig.~\ref{topmassres}(a) as a function of luminosity for different resolutions of the roman
pot detectors.

The other application is to use the so-called ``threshold-scan method"
to measure the stop mass in {\it exclusive} events. The idea is straightforward: 
one
measures the turn-on point in the missing mass distribution at about twice
the stop mass. After taking into account the stop width, we obtain a resolution
on the stop mass of 0.4, 0.7 and 4.3 GeV for a stop mass of 174.3, 210 and 393
GeV for a luminosity (divided by the signal efficiency) of 100 fb$^{-1}$. We
notice that one can expect to reach typical mass resolutions which can be obtained at a linear
collider. The process is thus similar to  those at linear colliders (all final 
states
are detected) without the initial state radiation problem. 

The caveat is  of course that production via diffractive 
{\it exclusive} processes is model dependent, and definitely needs
the Tevatron data to test the models. It will allow to determine more precisely 
the production cross section by testing and measuring at the Tevatron the jet 
and photon production for high masses and high dijet or diphoton mass fraction.

\subsubsection*{How to constrain the high $\beta$ gluon using Tevatron and LHC data?}

In this section, we would like to discuss how we can measure the gluon density
in the pomeron, especially at high $\beta$ since the gluon in this kinematical 
domain shows large uncertainties and this is where the exclusive contributions
should show up if they exist. To take into account, the high-$\beta$
uncertainties of the gluon distribution, we chose to multiply the gluon density
in the pomeron measured at HERA by a factor $(1-\beta)^{\nu}$  where $\nu$
varies between -1.0 and 1.0. If $\nu$ is negative, we enhance the gluon density
at high $\beta$ by definition, especially at low $Q^2$.

The measurement of the dijet mass fraction at the Tevatron for two jets with a
$p_T$ greater than 25 GeV for instance in double pomeron exchange is indeed
sensitive to these variations in the gluon distribution. The dijet mass fraction
is given in Fig 3
and 4 which shows how the Tevatron data can
effectively constrain the gluon density in the pomeron \cite{paplaurent}.
Another possibility to measure precisely the gluon distribution in the pomeron
at high $\beta$ would be at the LHC the measurement of the $t \bar{t}$ cross
section in double pomeron exchange in inclusive events
\cite{papttbarincl}. By requiring the
production of high mass objects, it is possible to assess directly the tails of
the gluon distribution. In Fig.~\ref{ttbarincl}, we give the total mass
reconstructed in roman pot detectors for double tagged events in double pomeron
exchanges and the sensitivity to the gluon in the pomeron.

\begin{figure}[htb]
\begin{center}
\subfigure[]{\includegraphics[width=.49\textwidth,trim= 20 0 20 0]{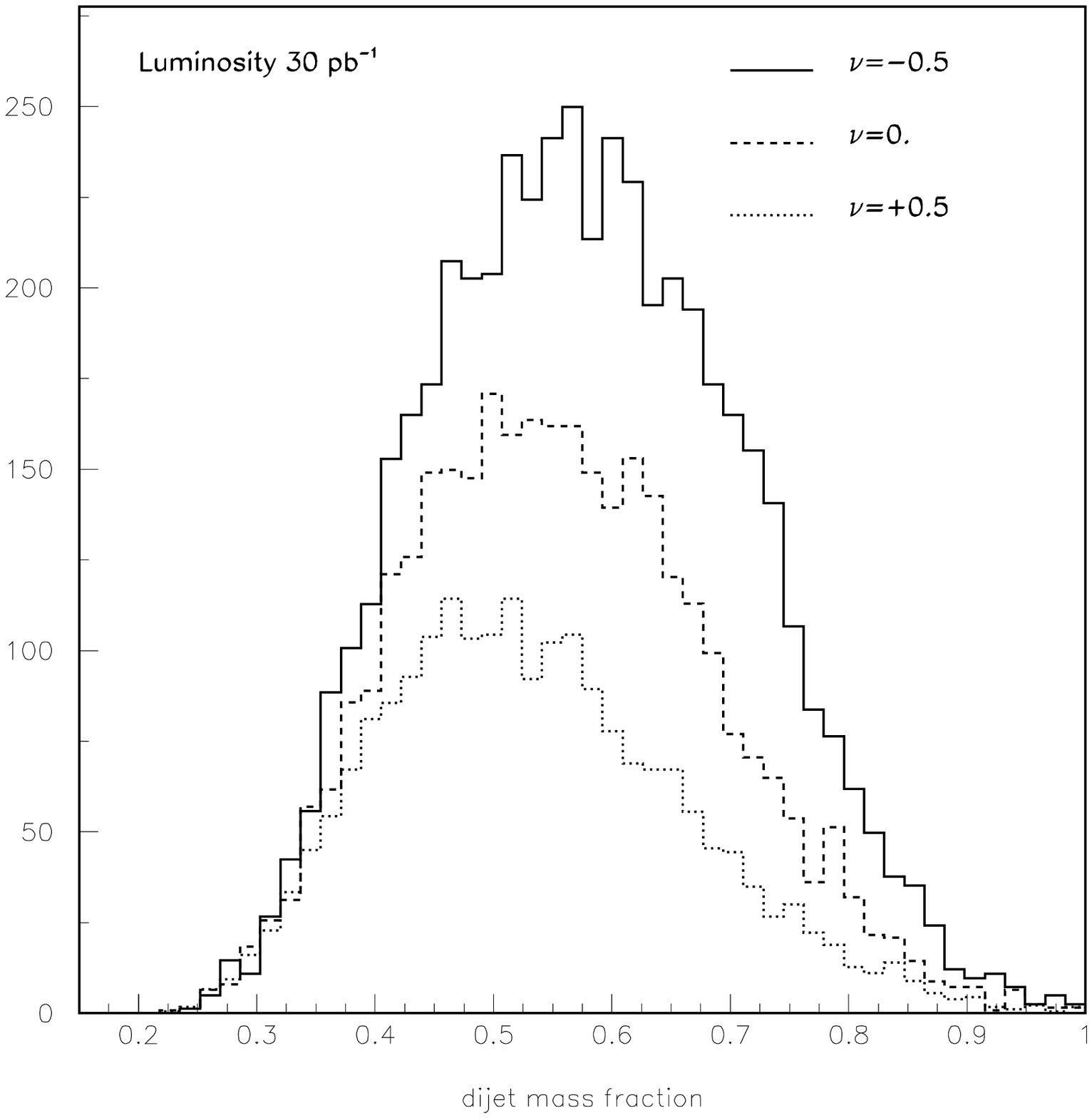}}
\subfigure[]{\includegraphics[width=.49\textwidth,trim= 20 0 20 0]{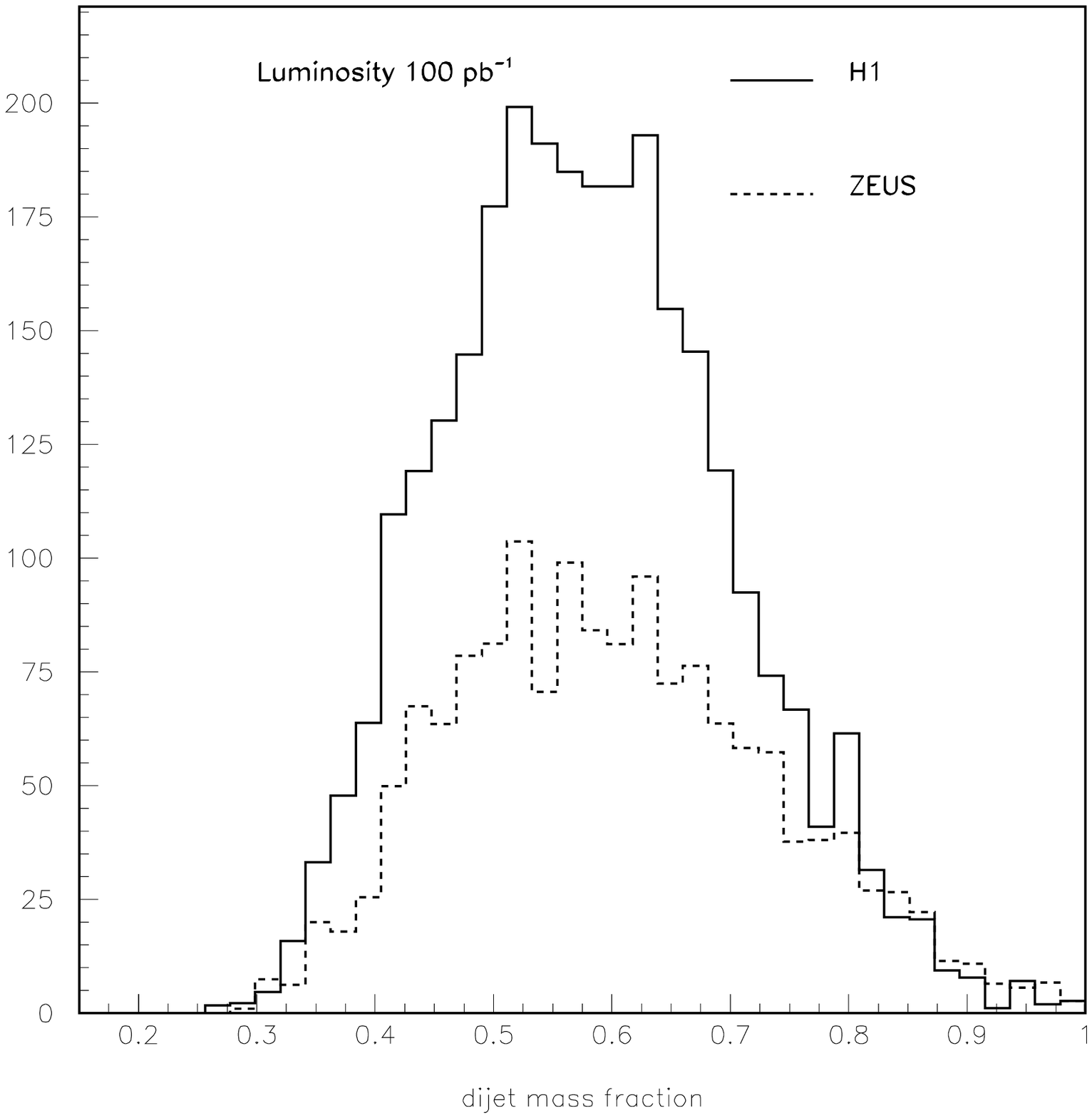}}
\caption{(a) Dijet mass fraction at the Tevatron at generator level when the gluon
density measured in the H1 experiment is used 
\cite{Royon:2000wg, paplaurent} and multiplied by $(1 -
\beta)^{\nu}$;
(b) Dijet mass fraction at the Tevatron at generator level when the gluon
density measured in the H1 or the ZEUS experiment is used \cite{Royon:2000wg, paplaurent}
and one tag
is asked in the roman pot acceptance of the D\O\ or the CDF collaboration in the
$\bar{p}$ direction. We notice the sensitivity of the measurements on the gluon 
density.}
\label{dijetmassfraction}
\end{center}
\end{figure}

\begin{figure}[htb]
\begin{center}
\includegraphics[width=.85\textwidth,clip=true]{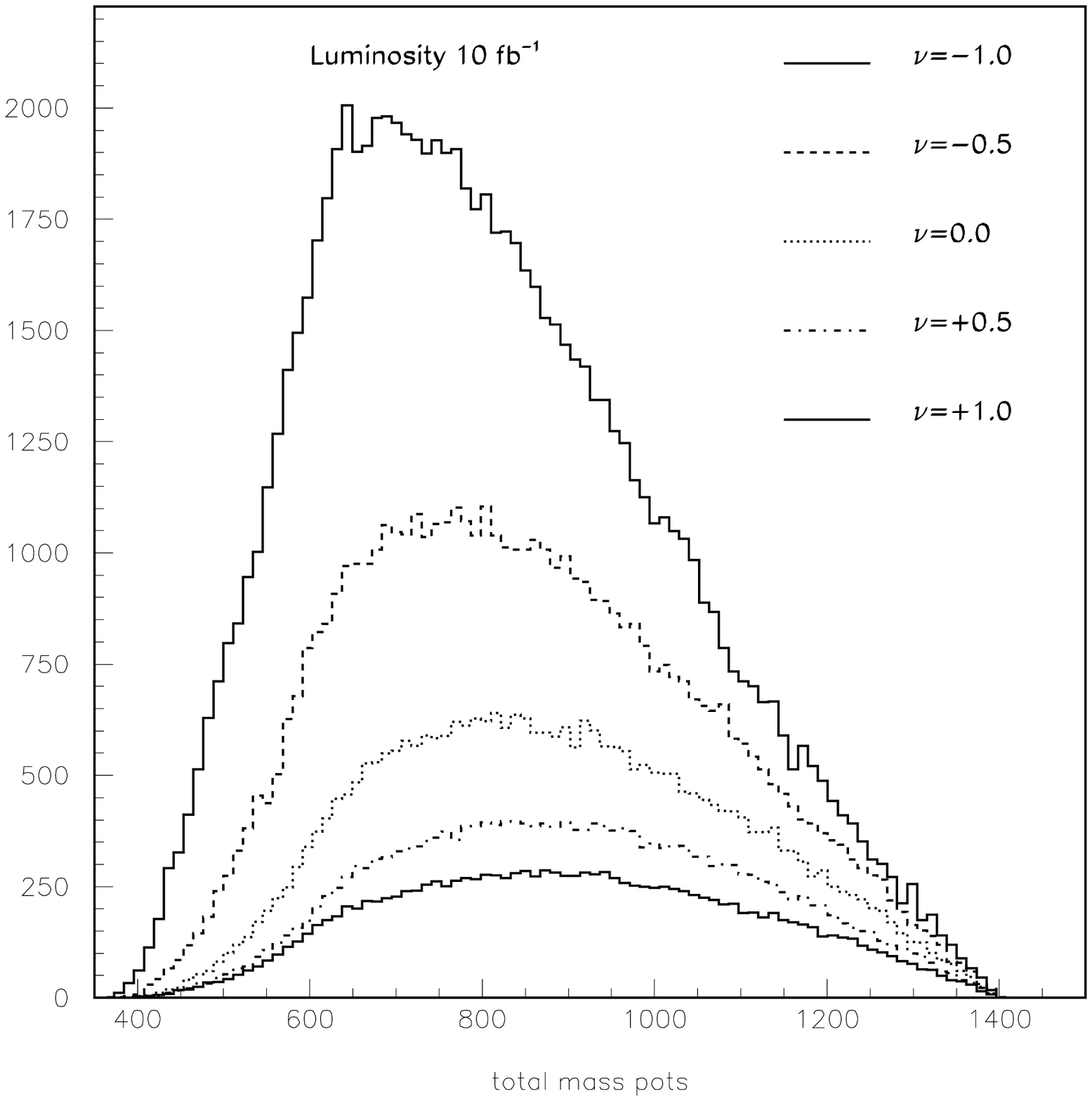}
\caption{Total diffractive mass reconstructed 
for $t \bar{t}$ inclusive events in double pomeron exchanges
using roman detectors at the LHC. We use the gluon density in the pomeron
measured in the H1 experiment \cite{Royon:2000wg, paplaurent} and we multiply it by $(1 -
\beta)^{\nu}$ to show the sensitivity on the gluon density at high $\beta$.}
\label{ttbarincl}
\end{center}
\end{figure}

\subsubsection*{New possible measurement of survival probabilities in the D\O\
experiment}

We propose a new measurement to be performed at the Tevatron,
in the D\O\ experiment \cite{Kupco:2004fw}, which can be
decisive to distinguish between Pomeron-based and soft colour interaction
models of hard diffractive scattering. This measurement allows
to test directly the survival probability parameters as well which is
fundamental to predict correctly the exclusive diffractive Higgs production at
the LHC.
The discriminative potential of our proposal takes its origin  in the 
factorization breaking properties which were already observed at the 
Tevatron. The explanation given to this factorization 
breaking in Pomeron-based models is the occurrence of large corrections 
from the  survival 
probabilities, which is the probability to keep a diffractive event
signed either by tagging the proton in the final state or by requiring the
existence of a rapidity gap in the event. By contrast with Pomeron models, 
the soft color interaction models are by nature, non factorizable. The initial 
hard interaction is the generic standard dijet production, accompanied by 
the full radiative partons. Then, a phenomenological soft color interaction is assumed 
to modify the overall color content, allowing for a probability of color singlet 
exchange and thus diffraction.

The forward detector apparatus 
in the D\O\ experiment at the Tevatron, Fermilab, has the 
potential to discriminate between the predictions of the two approaches in  
hard ``double'' diffractive production, e.g. of  centrally produced dijets, 
by looking 
to the azimuthal 
distributions of the outgoing proton and antiproton with respect to the 
beam direction. This measurement relies on tagging  
both outgoing particles in roman pot detectors
installed by  the D0 experiment. 

The FPD consists of 
eight {\it quadrupole} spectrometers, four being located on the
outgoing proton side, the other four on the antiproton side. On each side,  
the quadrupole spectrometers are
placed both in the inner (Q-IN), and outer (Q-OUT) sides of the accelerator ring,
as well as in the upper (Q-UP) and lower (Q-DOWN) directions.
They provide almost full coverage in azimuthal angle $\Phi$.
The {\it dipole} spectrometer, marked as D-IN, is placed in the inner side of 
the ring, 
in the direction of outgoing antiprotons.

Each spectrometer allows to reconstruct the
trajectories of outgoing protons and antiprotons near the beam pipe
and thus to measure their energies and scattering angles. Spectrometers provide
high precision measurement in $t = -p_{T}^2$ and $\xi = 1 - p^\prime /
p$ variables,
where $p^\prime$ and $p_T$ are the total and transverse momenta of the outgoing 
proton or antiproton, and $p$ is the beam energy.
The dipole detectors show a good acceptance
down to $t=0$  for $\xi > 3. 10^{-2}$ and the quadrupole detectors
are sensitive on outgoing particles down to $|t|= 0.6$ GeV$^2$ for
$\xi < 3. 10^{-2}$, which allows to get a good acceptance for
high mass objects produced diffractively in the D\O\ main detector. 
For our analysis, we use a full simulation of 
the FPD acceptance in $\xi$ and $t$ \cite{fpdaccep}.

We suggest to count the number of events
with tagged $p$ and $\bar{p}$ for different combinations of FPD spectrometers.
For this purpose, we define the following
configurations for dipole-quadrupole tags
(see Fig. 2): same side (corresponding to D-IN on $\bar{p}$ side
and Q-IN on $p$ side and thus to $\Delta \Phi < 45$ degrees), 
opposite side (corresponding to D-IN on $\bar{p}$ side
and Q-OUT on $p$ side, and thus to $\Delta \Phi > 135$ degrees),
and middle side (corresponding to D-IN on $\bar{p}$ side
and Q-UP or Q-DOWN on $p$ side and thus to $45 < \Delta \Phi < 135$ degrees).
We define the same kinds of configurations for quadrupole-quadrupole tags
(for instance, the same side configuration corresponds to sum of the four
possibilities: both protons and antiprotons tagged in Q-UP, Q-DOWN,
Q-IN or Q-OUT).

In Table 2, we give the ratios $middle/(2 \times same)$ and $opposite/same$ 
(note that we divide $middle$ by 2 to get the same domain size in
$\Phi$) for the different models. In order to obtain these predictions, we used
the full acceptance in $t$ and $\xi$ of the FPD detector \cite{fpdaccep}.
Moreover we computed the ratios for two  different tagging configurations
namely for $\bar{p}$ tagged
in dipole detectors, and $p$ in quadrupoles, or for both $p$ and $\bar{p}$ 
tagged in quadrupole detectors.

In Table 2, we notice that the $\Phi$ dependence of the
event rate ratio for the SCI \cite{Edin:1995gi,Cox:2000jt} model 
is weak, whereas for the POMWIG \cite{Edin:1995gi,Cox:2000jt} models the result show important differences 
specially when both $p$ and $\bar{p}$ are tagged in quadrupole
detectors. This measurement can be performed even at low luminosity (1 week of
data) 
if two jets with a transverse momentum greater than 5 GeV are required.

With more luminosity, we also propose to measure directly the $\Delta \Phi$ dependence
between the outgoing protons and antiprotons using the good coverage of the
quadrupole detectors in $\Phi$ which will allow to perform a more
precise test of the models.

\begin{table}
\begin{center}
\begin{tabular}{|c|c||c|c|} \hline
 Config. & model & midd./ & opp./ \\ 
         &       &  same  & same \\
\hline\hline
Quad.  & SCI & 1.3 & 1.1 \\
 $+$ Dipole              & Pom. 1 & 0.36 & 0.18 \\
               & Pom. 2 & 0.47 & 0.20 \\ \hline
Quad.  & SCI & 1.4 & 1.2 \\
$+$ Quad.               & Pom. 1 & 0.14 & 0.31 \\
               & Pom. 2 & 0.20 &  0.049     
\\
\hline
\end{tabular}
\end{center}
\caption{Predictions for a proposed measurement of diffractive 
cross section ratios in different regions of $\Delta \Phi$ at the Tevatron
(see text for the definition of middle, same and opposite).
The first (resp. second) measurement involves the quadrupole and dipole
detectors (resp. quadrupole detectors only) leading to asymmetric (resp.
symmetric) cuts on $t$.
We notice that the SCI models do not predict any significant dependence
on $\Delta \Phi$ whereas the Pomeron-based models show large variations. }
\end{table}

\begin{figure}[htb]
\centering
\includegraphics[width=.85\textwidth,clip=true]{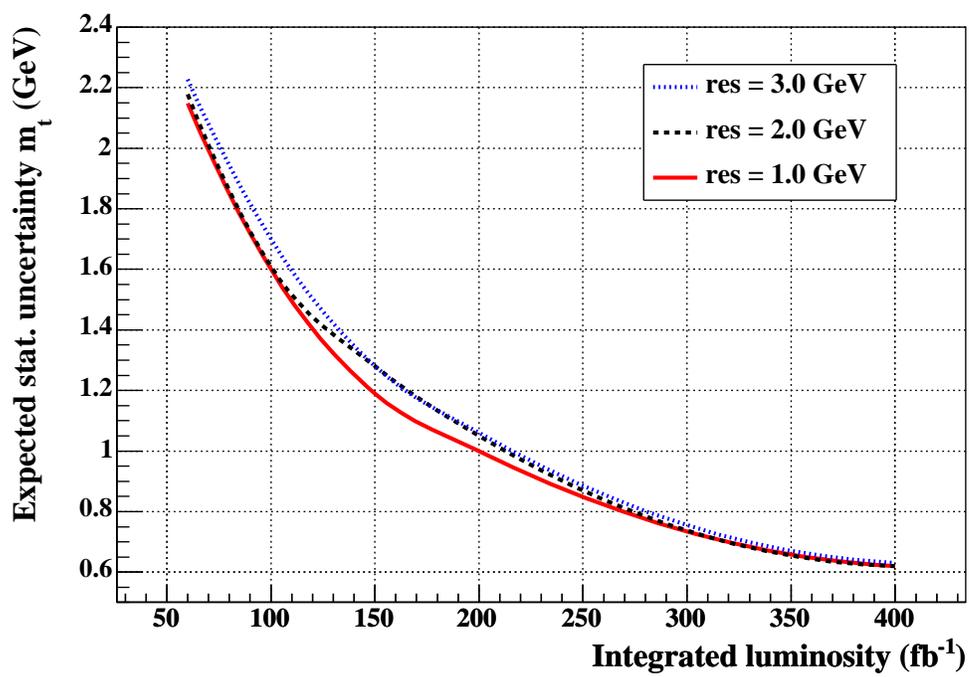}
\caption{Expected statistical precision of the top mass
    as a function of the integrated luminosity for various resolutions
    of the roman pot detectors (full line: resolution of 1 GeV, dashed line: 2
    GeV, dotted line: 3 GeV).}
\label{topmassres}
\end{figure}

\subsubsection*{Acknowledgments}
These results come from a fruitful collaboration with M. Boonekamp, J. Cammin,
A. Kupco,
S. Lavignac, R. Peschanski and L. Schoeffel.

\clearpage
\subsection{{Diffraction and Central Exclusive Production}} 
\textbf{Contributed by:  Albrow, De Roeck}

Diffractive physics covers the class of interactions that 
contain large rapidity gaps (typically $>$ 4 units)
with no hadrons. This implies color singlet exchange, requiring two or more
gluons with a (minor) contribution of $q\bar{q}$. This is a frontier of
QCD, not fully understood but where much progress has been made through
experiments at the Tevatron and HERA. Here we consider two areas in hadron-hadron collisions,
\emph{diffraction} and a special subset \emph{central exclusive production}. The latter has become very topical
as a window on the Higgs sector and BSM physics at the LHC. To this end the FP420 R\&D collaboration~\cite{fp420} aims to add
  high precision forward proton detectors to CMS and/or ATLAS.
  
Central exclusive processes are defined as $pp \rightarrow p \oplus X \oplus p$ where $X$ is a fully measured simple
state such as $\chi_c, \chi_b, \gamma\gamma, Jet+Jet, e^+e^-, \mu^+\mu^-, H, W^+W^-, ZZ$ and $\oplus$ represents a large
($>\approx$ 4) rapidity gap. As there are no additional particle produced, precise measurements of both forward protons
give the mass of the central exclusive state. There are several other important advantages in these processes, as
described below, and the Higgs and di-boson sectors are most interesting. Measurements of all the other listed processes
tend to be directed towards understanding these electroweak processes better, although they are at the same time probes
of QCD in an important region (the perturbative-non-perturbative boundary). The $e^+e^-$ and $\mu^+\mu^-$ states are
purely QED with negligible QCD corrections.

If the reaction $pp \rightarrow p \oplus H \oplus p$ is seen at the LHC, precise proton
  measurements ($\frac{dp}{p} \approx 10^{-4}$) allow one to measure the Higgs mass with
  $\sigma(M_H) \approx$ 2 GeV per event, independent of the decay mode (e.g. $b\bar{b}, W^+W^-, ZZ$), as discussed in the next section.
  The signal: background can be $\approx$ 1:1, 
even considerably larger for MSSM Higgs. The Higgs
  quantum numbers (Is it a scalar? Is CP = ++?) can be determined from 
the azimuthal $pp$ correlations.
  The key question is : ``What is the cross section for $pp \rightarrow p \oplus H \oplus p$"?
  We proposed~\cite{Albrow:2005fw} that $p\bar{p} \rightarrow p \oplus \gamma\gamma \oplus \bar{p}$ has an
  identical QCD structure, might
  be measurable at the Tevatron and, if seen, would confirm that $pp \rightarrow p\oplus H \oplus
  p$ must occur and ``calibrate" the theory. The Durham group (see e.g. Ref~\cite{hep-ph/0311023,hep-ph/0409037,hep-ph/0508274}) calculated
  the cross sections and they have been incorporated into the ExHume~\cite{Monk:2005ji} generator.
  CDF has now observed the $\gamma\gamma$ process\cite{cdfgg}, confirming
  that the exclusive cross section for (SM) $M(H) \approx$ 130 GeV is $\approx$ 3 fb or perhaps a
  factor $\approx$ 2-3 higher, which is very encouraging.

  In the MSSM the Higgs cross
  section can be an order of magnitude higher than in the SM, depending on tan($\beta$).  In addition to $H$ observations,
  exclusive central $W^+W^-$ produced by 2-photon exchange should be seen, $\sigma(pp \rightarrow
  p \oplus W^+W^- \oplus p) \approx 100$ fb, and final state interactions between the $W$'s can be
  studied. Exclusive $\mu^+\mu^-$ and $e^+e^-$ have
  recently been observed in CDF\cite{cdfgg}, the first time $\gamma\gamma \rightarrow X$ processes have been
  seen in hadron-hadron collisions. The $pp \rightarrow p \oplus \mu^+\mu^- \oplus p$ reaction is
  important at the LHC for two reasons: (1) The cross section is very well known (QED) and it can
  be used to calibrate luminosity monitors, perhaps to $\approx 2$\%. The dominant uncertainty would be knowledge of the trigger,
  acceptance and efficiencies. (2) The forward proton
  momenta are very well known and can calibrate those spectrometers; the central mass $M(X)$ 
  scale is calibrated with the precision $\sigma(M_{\mu\mu})$. Once the forward proton spectrometers are well calibrated
  it should be possible to use reactions such as $pp \rightarrow p \oplus Jet+Jet \oplus p$,
  when there is only one interaction per crossing, to calibrate the full CMS/ATLAS calorimeter (i.e. find a global energy
  scale factor).
 Hence it is important to study
  these processes at the Tevatron. For example, how cleanly can one select the exclusive
  $\mu^+\mu^-$ with pile up, by requiring no other tracks on the $\mu^+\mu^-$ vertex, $\Delta\phi(\mu\mu)
  = 180 ^\circ$, and $p_T(\mu^+) = p_T(\mu^-)$? Unfortunately the 
  $p\bar{p} \rightarrow p \oplus \gamma\gamma \oplus \bar{p}$  reaction can only be seen in the
  absence of pile-up, requiring luminosity less than about 5.10$^{31}$ cm$^{-2}$ s$^{-1}$, which
  is becoming rare, so little more can be done.
  
  Other exclusive processes which can be studied at the Tevatron and are related to exclusive $H$
  production are $p\bar{p} \rightarrow p \oplus \chi_{c(b)} \oplus \bar{p}$ and 
  $p\bar{p} \rightarrow p \oplus Jet+Jet \oplus \bar{p}$

\subsubsection*{Exclusive Higgs Production}

\begin{figure}[htb] 
\centering
\subfigure[]{\includegraphics[width=.49\textwidth]{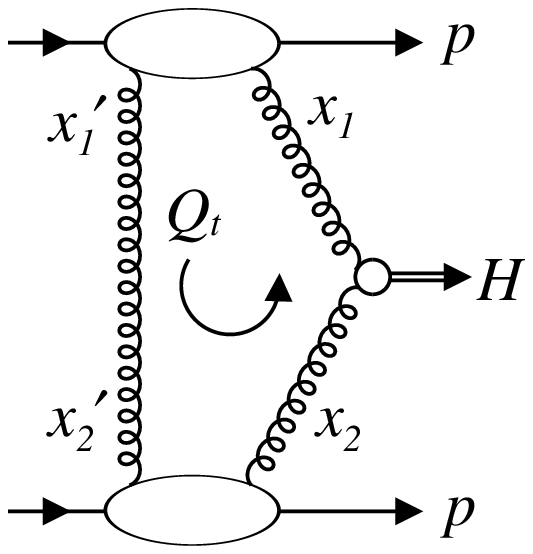}}
\subfigure[]{\includegraphics[width=.49\textwidth]{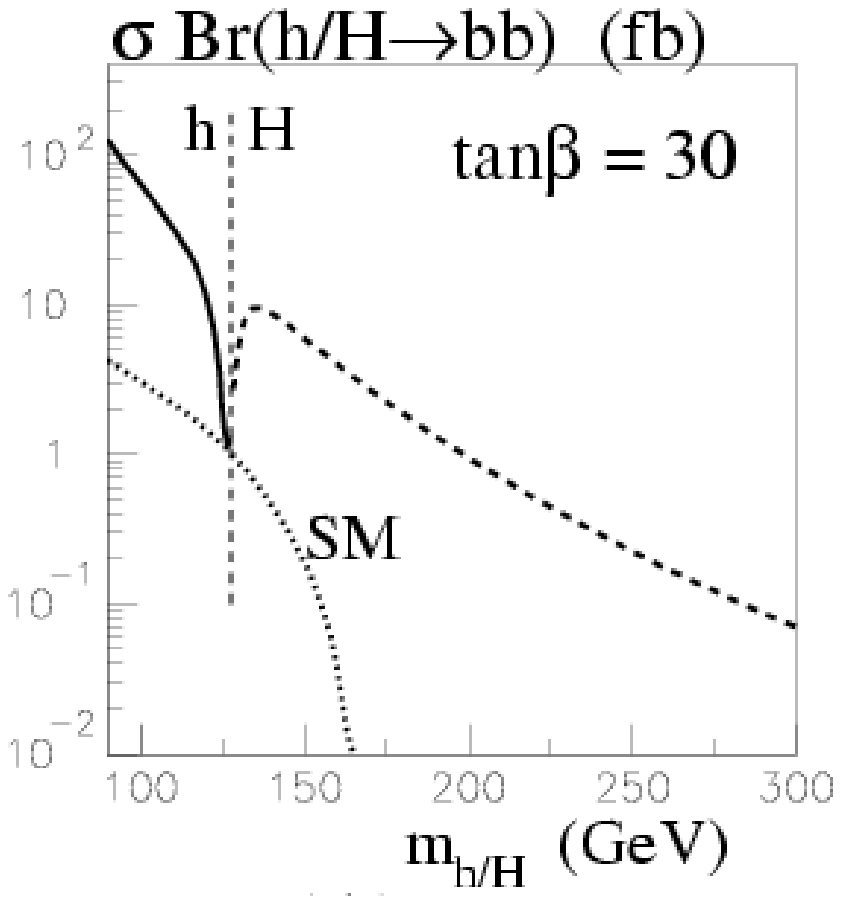}}
\label{higgs_process} 
\caption{ (a) Diagram for the exclusive production of the Higgs 
particle
in $pp$ interactions;
(b) 
The cross section times $b\overline{b}$ branching ratio predicted for
the central exclusive production of MSSM Higgs bosons (for $\tan
\beta$=30)
at the LHC compared with the SM result.}
\end{figure}

A recent development in the study of
rapidity gap phenomena is  the 
search for and measurements of the  Higgs particle in 
central exclusive production events,
shown in Fig.~\ref{higgs_process} (left).
The process was first proposed for the Tevatron 
collider~\cite{Albrow:2000na}, pointing out that when using the missing mass 
calculated with respect to the two outgoing protons, a Higgs mass 
resolution of the order of 250 MeV could be achieved. As it turned out the 
cross section at the CM energy of the Tevatron is too low to be measured.
The mass resolution scales approximately with $\sqrt{s}$, and can be $\approx$ 2 GeV
at the LHC.

A 
calculation of the cross section for DPE exclusive 
Higgs production
$pp \rightarrow p H p$ with $H\rightarrow b\overline{b}$ 
at the LHC gives about 3 fb for a SM Higgs with mass of 
120 GeV~\cite{Khoze:2002nf}. After exeprimental cuts (not optimized) about 10 
signal events will be reconstructed for 30 fb$^{-1}$, with a similar
amount of background events expected in a 2 GeV mass bin.
The cross sections can be up to a factor 10-20 larger for MSSM, see 
Fig.~\ref{higgs_process} (right).
Backgrounds in the $b\overline{b}$ channel are suppressed at LO
due to the $J_Z=0$ spin selection rule. 
Reconstructing the Higgs mass from the missing mass to the protons 
$$
M_H^2= (p_1+p_2-p^{'}_1-p^{'}_2)^2
$$
with $p_1,p_2$ the four momenta of the incoming beam particles and 
$p^{'},1-p^{'}_2$ the ones of the outgoing protons,
will 
allow to 
measure the Higgs mass via the missing mass technique with a 
resolution of $\approx$ 2 GeV, independent of the decay mode.
For the $b\bar{b}$ and $W^+W^-$ channels this is greatly superior to any other technique.

Recently also the decay $H\rightarrow WW$ has been studied.
With experimental acceptances and cuts about 7 signal events are expected with 
3 background events for 30 fb$^{-1}$ and a S.M. Higgs mass of 160 GeV
with detector cuts\cite{Cox:2005if}.
More channels are being studied and the phenomenology is moving forward fast.

The predictability of the cross section has been a long debate 
and (after some selection, when e.g. including the present limits on the 
exclusive di-jet production at the Tevatron) 
factors differences of an order of magnitude have been reported.
During the HERA/LHC workshop the so called ``Durham'' calculations have
been scrutinized and have been confirmed by other groups. Hence within the 
pQCD picture used it seems calculations can be used, and  
sophisticated predictions~\cite{Khoze:2002nf} claim an uncertainty of only 
a factor 2 - 3, mainly due to PDFs, but there is still
some controversy. 
Crucial information will 
have to come from similar measurements to test these calculations with data.
In particular exclusive di-jets, exclusive 2 photon and $\chi_c$ production 
are candidate processes to test the theory and already now at the Tevatron 
such measurements are being performed, as will be discussed 
further in this report.

Besides a possible discovery channel for the MSSM Higgs, and a channel to 
measure the $H\rightarrow b\bar{b}$ decay mode, the central exclusive Higgs 
production also
allow for CP studies via the azimuthal correlations~\cite{Khoze:2004rc}.
The exclusive system is so constrained that it produces predominantly 
spin 0 or 2 states, and these impose different
azimuthal correlations on the protons. Thus it can confirm the scalar nature of a new 
particle discovered and called Higgs at the LHC, a measurement otherwise 
difficult to make at the LHC, especially for a light Higgs.
Hence this measurement will have a strong added value to the LHC
physics program.

\begin{figure}[htb] 
\centering 
\includegraphics[width=.6\textwidth]{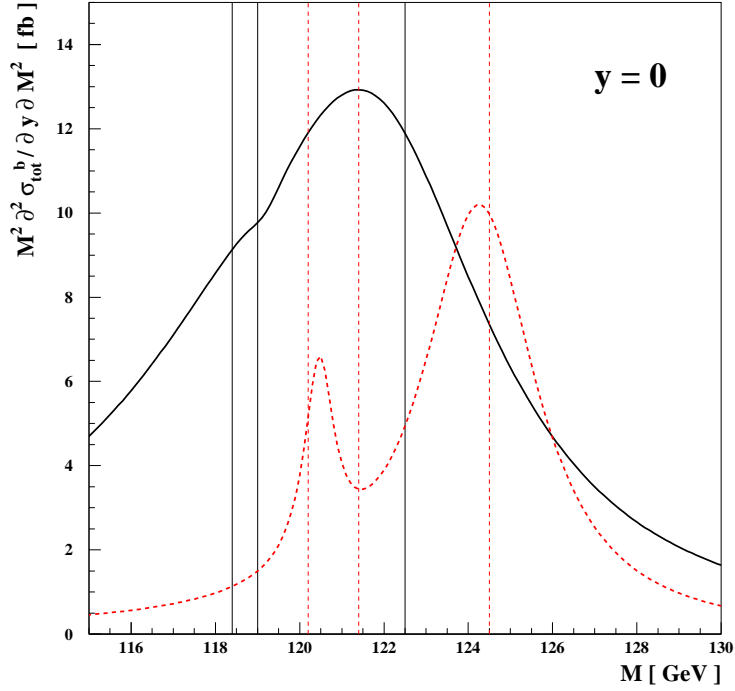}
\label{cphiggs} 
\caption{ The hadronic level cross section when the produced Higgs boson 
decays into b quarks, for two tri-mixing scenarios as detailed in 
\protect \cite{Ellis:2005fp}. The vertical lines indicate the three Higgs-boson pole
mass positions.}
\end{figure}

Recently it was also shown 
in \cite{Ellis:2005fp} that the forward tagging of protons would be important 
for CP violating MSSM Higgs scenarios, where the three 
neutral Higgses are nearly degenerate in mass and can mix.
This will lead to structures as shown in Fig.~\ref{cphiggs}, and will 
need an experimental tool that can scan the Higgs mass region with a 
mass precision of about 1 GeV. The tagged protons can be such a tool, perhaps uniquely.

\subsubsection*{Forward Detectors at the Tevatron}

At the Tevatron both experiments, CDF and D0, are equipped with very forward detectors to measure diffractively
scattered protons and large rapidity particles, specifically as rapidity gap detectors. Very early
(1989) CDF installed  forward proton trackers in roman pots (moveable vacuum chambers that allow
detectors to move very close to the beam during a store) and measured the total cross section
$\sigma_T$, elastic scattering $\frac{d\sigma}{dt}$ and single diffractive excitation
$\frac{d\sigma}{dtdM^2}$.  Elastic scattering $\frac{d\sigma}{dt}$ and the total cross section $\sigma_T$ are basic
 properties of $p\bar{p}/pp$ interactions, which will be measured at the LHC by the TOTEM
 experiment. Unfortunately they are not very well known at the Tevatron, with three inconsistent
 ($>3\sigma$) measurements of $\sigma_T$, and only one measurement of $\frac{d\sigma}{dt}$ into
 the Coulomb region and none into the large $|t|$ region beyond 2 GeV$^2$ (interesting
 from a perturbative point of view). There are no measurements yet at $\sqrt{s}$ = 1960 GeV, although the
 LHC could run down to that $\sqrt{s}$ to make a comparison of $pp$ and $p\bar{p}$.  The first generation CDF roman pots were removed after these measurements, and for a period diffractive
physics was done using large rapidity gaps, without seeing the scattered proton(s). One could
demonstrate diffractive signatures for production of jets, heavy flavors ($b, J/\psi$) and even $W, Z$ as a
distinct class of events with 3-4 units of rapidity devoid of hadrons. (e.g. The distribution of
charged hadron multiplicity or $\Sigma E_T$ in a forward region shows a clear peak at zero). Both
experiments also discovered the phenomenon of ``gaps between jets" (JGJ) in which two high $E_T$ jets
(10-50 GeV) at large opposite rapidity have a gap $\Delta\eta >\approx$ 4 units in between. The
4-momentum transfer$^2$, $t \approx E_T^2$, is huge (of order $10^3$ GeV$^2$). This is presumably not
the regime of a ``pomeron", but is best described as perturbative $q\bar{q},qg,gg$ scattering accompanied by
other gluon exchange(s) to cancel the color exchange and leave a gap. The relationship between these
descriptions is a topical issue.

Shortly before the end of Run 1, in 1995, CDF installed a new set of roman pots on one arm ($\bar{p}$)
specifically to study high mass diffraction in more detail. These were a triplet of pots over about
2m in $z$, 53m from the beam intersection. The detectors were arrays of square section scintillating
fibers to give $x,y$ coordinates with about 100$\mu m$ resolution over 2cm $\times$ 2cm, backed up by
scintillation counters (one in each pot, put in coincidence for a trigger). Level 1 triggers required
a 3-fold ``pot scintillator" coincidence together with central jets or leptons. Also the forward proton detectors were read out for
\emph{all} events, a principle which should be followed in CMS and ATLAS.
These detectors remained in Run 2 (2003 on), supplemented by Beam Shower Counters (BSC1,2,3) which
were scintillation counters tightly around the beam pipe wherever they could be fitted, covering $5.4
\leq |\eta| \leq 7.4$. BSC1 had 3$X_o$ of lead in front to convert photons, the others were in the
shadow of material (beam pipe, flanges) and detected mostly showers. These counters are used as
``rapidity gap detectors", sometimes in a L1 trigger. Another innovation in Run 2 was the ``Miniplug"
calorimeter covering $3.6 \leq |\eta| \leq 5.2$ consisting of a cylindrical tank of liquid scintillator with lead
plates. The wavelength-shifting fiber readout was arranged to have very high $\eta,\phi$ granularity. This is used for
rapidity gap physics and for triggering on very forward jets (especially for JGJ studies).

A lesson from the CDF Run 2 diffractive studies is that it is very important to cover as completely as possible the
forward and very forward $\eta,\phi$ region for charged hadrons and photons, with the ability to trigger on and find
forward rapidity gaps with little background (e.g. from particles that can miss these detectors). This was crucial in
CDF e.g. for showing that the exclusive 2-photon events were really exclusive, and hopefully this will also be done at
the LHC.

In Run 1 D0 studied diffractive physics only with (pseudo-)rapidity gaps, 
but for Run 2 they installed a system of roman
pots (Forward Proton Detectors, FPD) on both arms for $p$ and $\bar{p}$ detection. Three pots detected $\bar{p}$ at
53m after three dipoles, almost identical to the CDF pots, with square scintillating fiber hodoscopes. Others, on both
$p$ and $\bar{p}$ arms, were placed behind low-$\beta$ quadrupoles (not dipoles). These have acceptance for normal
low-$\beta$ running, down to $|t|_{min} \approx$ 0.7 GeV$^2$ (which misses much of the cross section for rare processes)
and the momentum (and $\xi = 1.0 - \frac{p_{out}}{p_{in}}$) resolution is considerably worse than the $\bar{p}$ dipole
spectrometers. However because both $p$ and $\bar{p}$ can be detected, elastic scattering is measurable as
well as a new
measurement of the total cross section (for the first time at $\sqrt{s} = 1960$ GeV). These measurements needed a special high-$\beta$ run to
reach a $|t|_{min} \approx 0.10-0.15$ GeV$^2$. The two-arm FPD also allows studies of double pomeron
exchange with both protons measured. This has never yet been studied at the Tevatron. Low mass ($\leq$ 5 GeV) states
such as $\phi\phi$ and $K^+K^-\pi^+\pi^-$ can provide very interesting 
glueball and hybrid searches, as the soft
pomeron is glue-rich and the forward proton correlations can determine central quantum numbers (spin, CP). The highest
previous energy at which this was done at $\sqrt{s}$ = 63 GeV at the ISR, and showed interesting structures in
exclusive channels (e.g. $pp \rightarrow p \oplus \pi^+\pi^-\oplus p$). This is a potentially rich field, both for
 studying diffractive mechanisms and for spectroscopy ($X$ is rich in glueball and hybrid states). There was almost no background from
non-pomeron exchange. Although for this low mass DPE there would seem to be no real advantage in much higher collision
energies, depending on what D0 is able to achieve there could be an interesting program at the LHC, with high-$\beta$
necessary to get low $\xi,t$ acceptance in the TOTEM (and possible ATLAS?) pots. For cleanliness this needs
single interactions/crossing. In D0 the forward coverage in $\eta,\phi$ is limited by their liquid argon calorimeter
($\eta_{max} \approx$ 4.5), so unfortunately some 3 units of forward rapidity are not covered (except for the roman
pots).

\subsubsection*{Forward Physics Measurements at the Tevatron}

Considering only inelastic diffraction, there have been three phases of measurements at the Tevatron. 

In 1994 CDF
published a study of single diffractive excitation SDE with the scattered $\bar{p}$ measured in pots with drift
chambers and silicon counters. Data were taken at both $\sqrt{s}$ = 546 GeV (to equal the CERN $Sp\bar{p}S$) 
and 1800 GeV. The diffractive peak extends to $\xi$ = 0.05 (as at the ISR) corresponding to an excited mass $M_X \approx$ 
300 GeV. Note that at the LHC the corresponding mass reach is $\approx$ 2000 GeV, well above the $t\bar{t}$ threshold
and perhaps into the realm of new BSM physics. The differential cross section $\frac{d\sigma}{dtdM^2}$ was measured and compared to
parametrizations. The integrated cross sections were
$\sigma_{SD}$(546) = 7.9 $\pm$ 0.3 mb, $\sigma_{SD}$(1800) = 9.5 $\pm$ 0.4 mb, about 10\% of the total cross section. 

In the second phase the emphasis was on rapidity gap physics (the older pots were removed). Both CDF and D0
discovered large rapidity gaps between high $E_T$ jets, and both found high-$E_T$ dijet production in SDE.
CDF presented evidence for diffractive $W$ production and later D0 published both $W$ and $Z$ in SDE with higher
statistics. As a rule-of-thumb, about 1\% of hard processes (jets, $W$, $Z$) are diffractive. CDF also measured diffractive production of $b$-quarks and $J/\psi$. If there is a rapidity gap which
extends into the instrumented acceptance, the momentum loss fraction $\xi$ of the leading baryons can be
calculated from: \[\xi(p,(\bar{p})) = \frac{1}{\sqrt{s}}\sum p_T e^{+(-)\eta}\] where the sum is over all particles.
(This follows from [$E,p_z$] conservation.) From the relative cross sections for diffractive jets, heavy flavors
(predominantly from gluons) and $W$ (predominantly from $q\bar{q}$ annihilation) it was possible, in a model in which
the pomeron has constituents, to conclude that about 60\% of its momentum is carried by gluons. As these
measurements are at moderately high $Q^2$, typically 1000 GeV$^2$, they are not incompatible with a mainly gluonic
pomeron at low $Q^2$. One could also derive a \emph{diffractive structure function} $F^D(x,Q^2,\xi,t)$, which is the standard
structure function $F_2(x,Q^2)$ conditional on a rapidity gap (or diffractive proton). Comparisons with $ep$ data (HERA) showed
it to be lower by an order of magnitude at the Tevatron, interpreted as a much smaller gap survival probability
in $p\bar{p}$ collisions which have additional parton-parton interactions. This is a breakdown of
\emph{factorization}. Double pomeron interactions have two
large rapidity gaps (and two leading protons). CDF found that the probability of a second gap, given one, is
substantially larger than the probability of one gap in inelastic collisions. This is understood: in one-gap
events there is no gap-spoiling additional interaction, so a second gap is not suppressed. About $10^{-3}$ of hard processes
have two large rapidity gaps (Double Pomeron Exchange, DPE).
Single diffractive excitation of low mass and high mass (di-jets, $W$, $Z$, heavy flavors) has been measured, but there
 is a case for a more complete systematic study, e.g. $\frac{d\sigma}{dtdM^2}$ conditional on such massive final states,
 at different $\sqrt{s}$ values. From the $s$-dependence at fixed ($t,M^2$) one could derive a ``hard pomeron" trajectory
 to extrapolate to the LHC. Monte Carlo event generators which have $p\bar{p}$ interactions and
 include diffraction, such as \HW~\cite{Corcella:2002jc} and \PY~\cite{Sjostrand:2006za} could then be tested and
 tuned, to improve predictions for the LHC. A new series of studies with measurement of forward $\bar{p}$ in roman pots
 is described in another note\cite{dino}.
 
The third phase of inelastic diffraction, in Run 2, has again been rapidity gap physics but with an emphasis on
\emph{exclusive processes} in which the central state is simple and completely measured. This is described in the
next section.

\subsubsection*{Central Exclusive Measurements at the Tevatron}

Central exclusive production studies at the Tevatron could have a powerful impact on the LHC program. Most
interesting and very important LHC processes are exclusive Higgs boson and vector boson pair ($W^+W^-,ZZ$)
production, $pp \rightarrow p \oplus H \oplus p$, $pp \rightarrow p \oplus 
[W^+W^- or ZZ] \oplus p$ with other exotic 
BSM possibilities. No hadrons are produced. Measurements of the forward protons allow very good mass measurements
($\sigma(M) \approx$ 2 GeV per event) for the central state, a good signal: background ($\approx$ 1:1) for a SM Higgs (higher
in MSSM scenarios), and determination of the central quantum numbers. $W$-pairs and $Z$-pairs can of course
come from Higgs decay, $W$-pairs (but not $Z$-pairs) can come from two-photon collisions, and both $W^+W^-$  and
$ZZ$ could be produced with an unexpectedly high rate in some BSM models (e.g. the white pomeron~\cite{hep-ph/0510034,Phys.Rev.D72.036007}). 
The two-photon $pp \rightarrow p
W^+W^- p$ cross section by two-photon exchange is about 100 fb, and $W^+W^-$ final state interactions can be studied beyond the LEP-2 range. In the
absence of a Higgs this could be particularly interesting. The 4-momentum constraints in exclusive processes allow
reconstruction of all $W^+W^-$ final states except perhaps 4-jets where the background may be too high.

It was mentioned before that there is still some level of uncertainty and 
perhaps even controversy on 
the predictions of the cross sections for central exclusive Higgs 
production. It is therefore 
very important to be able to use the Tevatron experiments to reduce the 
uncertainty, i.e.
to \emph{calibrate} the predictions. The exclusive Higgs diagram has $gg \rightarrow H$ through a top loop, with
an additional gluon exchange to cancel the color and allow the protons to remain unexcited. Very similar diagrams
with a $c(b)$-loop can produce an exclusive $\chi_{c(b)}$, probably best detectable through radiative decay:
\[p\bar{p}\rightarrow p \oplus \chi_{c(b)} \oplus \bar{p} \rightarrow p \oplus J/\psi(\Upsilon) \gamma \oplus
\bar{p}\] These have a large enough cross section to be detectable at the Tevatron. CDF has preliminary evidence for
exclusive $\chi_c$ production (and probably also exclusive $J/\psi$ photoproduction), with much more data currently
being analyzed. The $\chi_b$ is more difficult, partly because an efficient trigger was not installed early
and the cross section is much smaller, and
now the luminosity is typically too high to give clean single single interactions. These processes are probably not
detectable in the presence of pile-up, at least not without measuring the 
forward protons. (The existing pots do not
have good acceptance for these low mass states.) Of course the $\chi_Q$ are hadrons, unlike the H, so one may worry that
these reactions do not have \emph{identical} QCD amplitudes.

Exclusive di-jets $p\bar{p} \rightarrow p \oplus JJ \oplus \bar{p}$ provide another way of testing the theoretical
calculations. CDF has triggered on events with a diffractive $\bar{p}$ and two central jets and then selected events
with a rapidity gap on the opposite ($p$) side (DPE candidates). They then study the distribution of
$R_{JJ} = \frac{M_{JJ}}{M_X}$, where $M_X$ is the total central mass, which would be near 1.0 if all the central hadrons were in just two jets. There is no
\emph{peak} near 1.0, but the monotonically falling distribution may have a \emph{shoulder}, being a smeared-out
indication of exclusivity. If one (somewhat arbitrarily) takes the cross section for events with $R_{JJ} > 0.8$ it
compares reasonably with the theoretical expectations. However the idea of ``exclusive dijets" is not well defined.
A high $E_T$ jet is dependent on a choice of algorithm, e.g. with a cone (or $k_T$) jet algorithm hadrons at an angle
(or $p_T(rel)$) exceeding a cut are considered outside the jets and spoil the exclusivity. By these criteria 
most LEP $Z\rightarrow q\bar{q}$ events would not be classed as exclusive.
Exclusive $q\bar{q}$ di-jets should be particularly suppressed (by a $J_z = 0$ spin selection rule)
when $Q^2 \gg m_q^2$. (So this process could provide a clean sample of gluon jets.) CDF attempts to exploit this
by studying the  $R_{JJ} = \frac{M_{JJ}}{M_X}$ distribution for $b$-tagged jets. A preliminary analysis shows a drop in the fraction of $b$-jets as $X \rightarrow$ 1, intriguing but needing more data.
A suppression of $b\bar{b}$ dijets as $R_{JJ} \rightarrow$ 1 is just what is needed to reduce di-jet background in
exclusive $H$ production (for $M_H < \approx 130$ GeV).

\begin{figure}[htdb]        
\label{dijets}
\subfigure[]{\includegraphics[width=.5\textwidth]{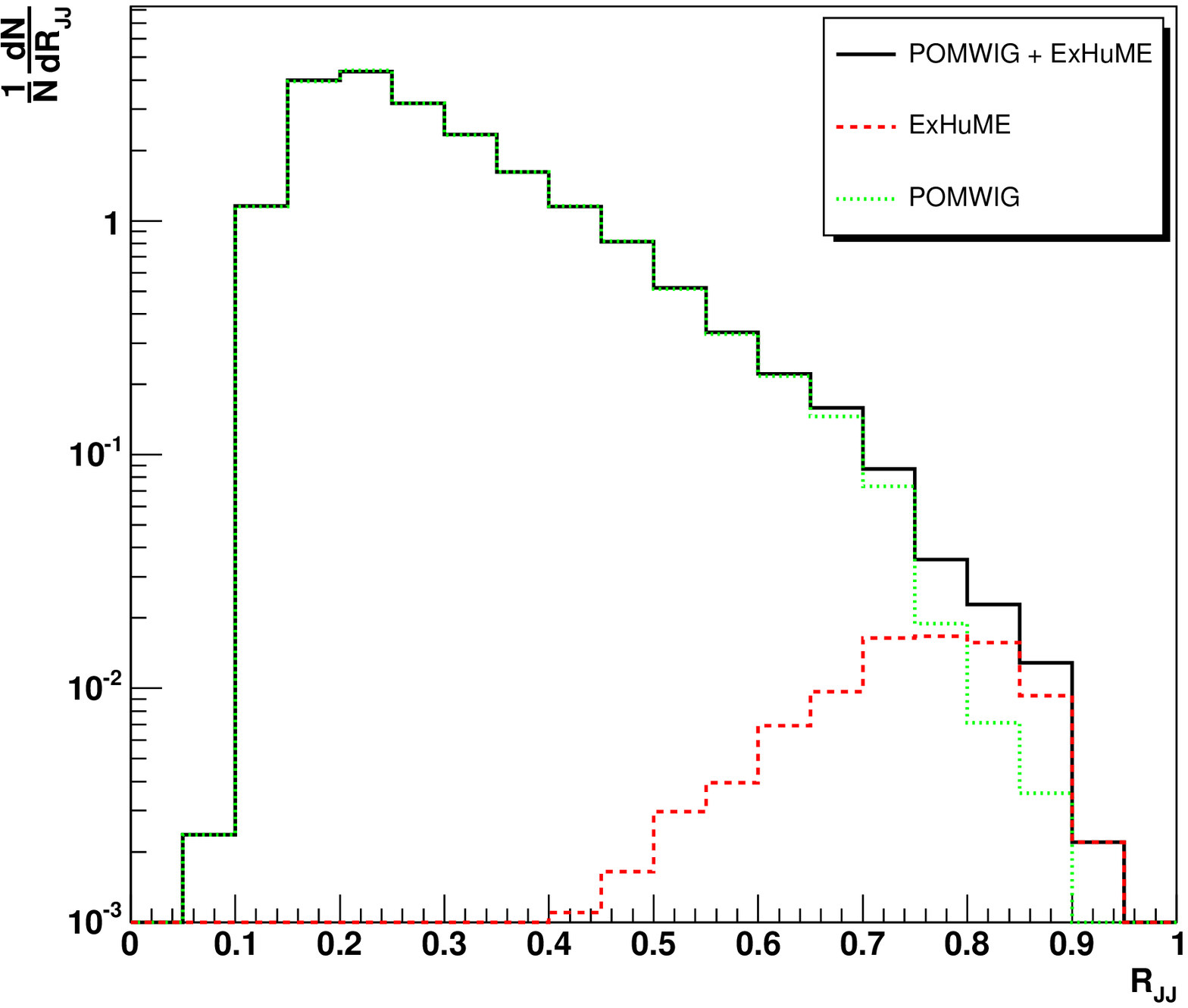}}
\subfigure[]{\includegraphics[width=.5\textwidth]{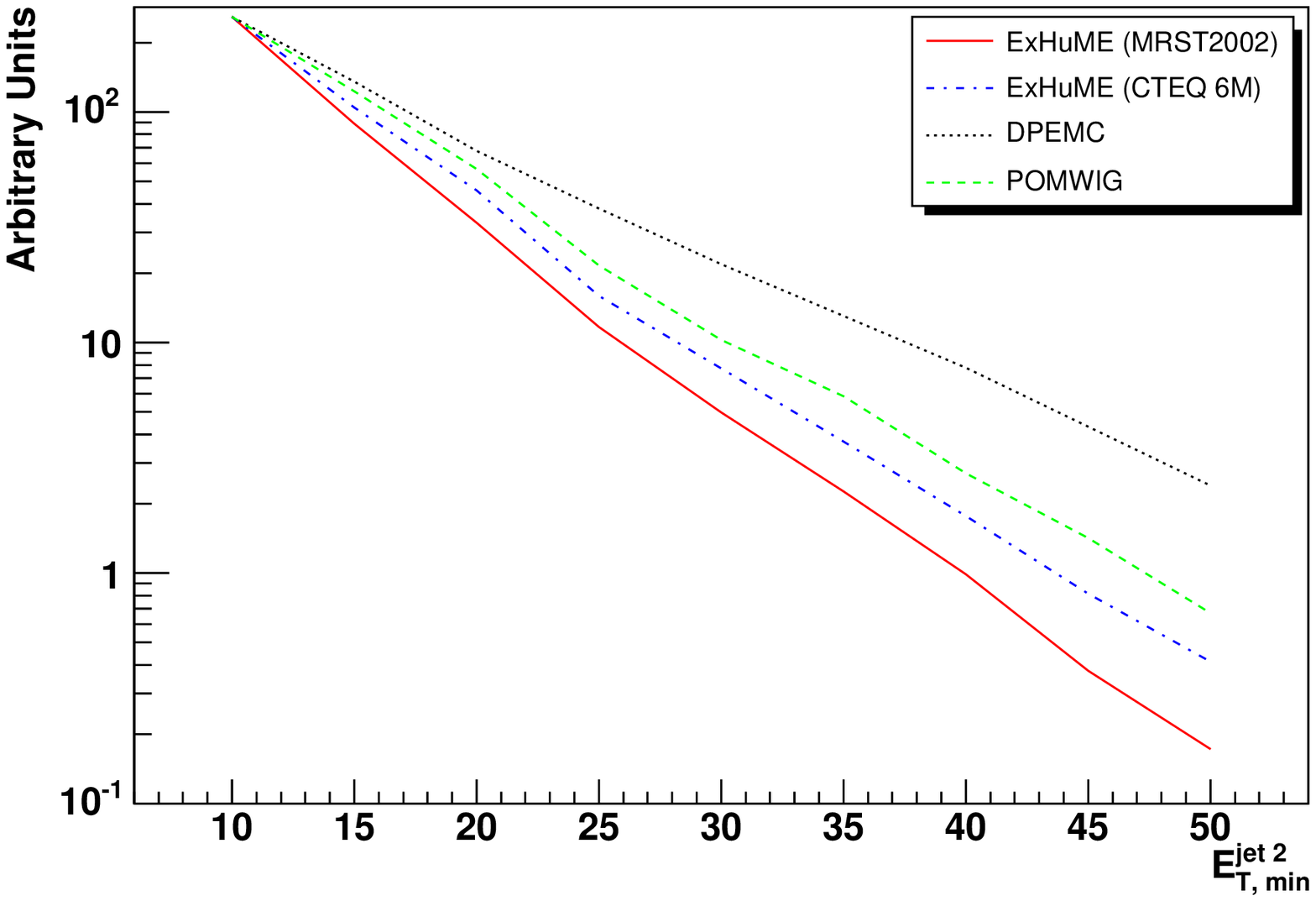}}
\caption 
  {(a)The $R_{JJ}$ distributions at the hadron level predicted by POMWIG 
+ ExHuME (left hand plot) in the CDF 
Run II kinematic range;  (b)
The $E_T$ distribution of the second highest $E_T$ jet in the region 
$R_{JJ} > 0.8$. The predictions of  ExHuME alone (with MRST2002 and 
CTEQ6M structure functions), POMWIG alone and DPEMC alone are shown with 
the curves 
normalised such that they all pass through the same point at $E_T = 10$ GeV.}
\end{figure}

A phenomenological analysis of the di-jet data of the Tevatron was 
performed in \cite{Cox:2005gr}. It was found that the CDF run-I data are 
consistent with the presence of an exclusive di-jet component, and this 
component should become visible in the data with the Run II statistics.
An important finding is that a non-perturbative model for the central 
exclusive process, as included in DPEMC~\cite{Boonekamp:2003ie}, predicts a 
different dijet $E_T$ dependence compared to ExHume, possibly allowing 
to discriminate between these two models.
Fig.~\ref{dijets} shows the prediction of the $R_{JJ} = M_{JJ}/M_X$ 
distribution, where $M_{JJ}$ is the mass of the dijet system and
$M_X$ the total mass of the centrally produced system, for a combination of 
inclusive diffractive (POMWIG) and exclusive (ExHume) central di-jet
production. The figure shows also the $E_T$ distribution of the 
second highest $E_T$ jet in the region $R_{JJ}>0.8$ for Exhume and DPEMC.
Finally it is worth noting that for such a measurement the jet finding 
algorithm needs to be optimized, a study which has not happened yet.
V.Khoze and M.Ryskin propose\cite{rjkhoze} a different variable, $R_j = 2E_{T1}.cosh(\eta)/M_X$
using only the leading jet ($\eta$ is the pseudorapidity of this jet in the $M_X$ rest frame).

Most interesting and relevant to exclusive Higgs production at the LHC is the observation of exclusive two-photon events in
CDF. We find $p\bar{p} \rightarrow p \oplus \gamma \gamma \oplus \bar{p}$ events with $p_T(\gamma) > 5$ GeV/c and $|\eta|
< 1.0$.The (QCD) diagram is $gg \rightarrow \gamma \gamma$ through quark loops (mainly $u,c$) with another
$g$-exchange, just as in the Higgs case. In fact from a QCD viewpoint the diagrams are \emph{identical} with
non-strongly interacting final states. Thus the observation (so far only 3 events .... out of about $10^{12}$
collisions!) demonstrates \emph{that the exclusive Higgs process exists} (if indeed a Higgs boson exists), and also,
because the $\gamma\gamma$ cross is in agreement with the Durham group calculation, detectable, i.e in the range
$\sigma(pp \rightarrow p \oplus H \oplus p) \approx$ 1 - 10 fb. It is very important to try, if at all possible,
to measure this exclusive $\gamma\gamma$ production at the LHC, to give a closer calibration of $p \oplus H \oplus p$. To do this
(it can only be done without any pile-up) we need enough ``single-interaction-luminosity" ($L_{eff} \approx 100 pb^{-1}$) to get a
useful sample of events (say 100 events to give a 10\% statistical uncertainty). It will (presumably) not be
possible to measure the forward protons (small-$t$ and small $\xi$ with low-$\beta$), so one must use the CDF
technique of requiring the whole detector to be consistent with noise (apart from the two photon showers). To get
enough rate one must go down to $p_T(\gamma) \approx$ 5 GeV/c, and be able to trigger on that at L1. This will need some
forward gap requirement at L1, from scintillators/calorimeters. A concern is whether an interaction in the previous bunch crossing
(with 25 ns crossing interval) leaves enough signal in the detectors to spoil the cleanliness. This needs
further study (and possibly some new counters). It is important because \emph{if} the exclusive $\gamma\gamma$ can
be measured (say to 10\%) the theoretical prediction for exclusive Higgs production can be made correspondingly 
precise, which may enable one to \emph{exclude} a SM Higgs if no exclusive signal is seen, test the SM prediction
if one is seen, and in the case of SUSY (or other BSM) make important measurements of the $Hgg$ coupling. 

Two photon (initial state) processes producing exclusive lepton pairs are also important for the LHC. These are
highly peripheral collisions (so the protons emerge with very small $p_T$) in which photons from the protons'
field collide: $\gamma \gamma \rightarrow e^+e^- (\mu^+\mu^-)$. The process is pure QED (a QCD correction from
simultaneous pomeron exchange is very small) and calculable to better than 0.1\%. Therefore if it can be measured
the luminosity for the period can be measured as well as one knows the di-lepton acceptance and efficiency. This
\emph{has} to be done in the presence of pile-up; if one had to require no other interaction in the crossing (as
for $\gamma\gamma$ final states) one would have to know precisely the inelastic cross section, which defeats the
object. We believe that this can indeed be done, even without detecting the forward protons, 
thanks to three criteria : (a) the associated charged multiplicity on the $\mu^+\mu^-$ vertex is
$n_{ass}$ = 0 (b) $\Delta\phi (\mu^+\mu^-) = 180^\circ$ (c) $p_T(\mu^+) = p_T(\mu^-)$. These events are now (belatedly) being looked for in
CDF. Exclusive $e^+e^-$ pairs have now been measured with no pile-up. There are two electrons with $p_T >$ 5 GeV/c and nothing
else in the detector (which extends to $|\eta| \approx 7.4$ ... the scattered protons are not seen). This is the first time
2-photon collisions have been seen in a hadron collider. The highest mass pair has $M(e^+e^-) \approx 38$ GeV! 
Some lepton pairs, especially more forward and higher mass, are accompanied by a (anti-)proton in the roman pots. The $\bar{p}$ momentum
is very well known from the dimuon kinematics, the main uncertainty coming from the incident beam momentum spread $\delta p$.
This provides an excellent (probably the best) calibration for the momentum (or missing mass) scale for the $p \oplus H
\oplus p$ search at the
LHC. This can be tested in CDF and D0, but this analysis is just starting. Hopefully there is enough data on tape now, as the CDF
roman pots were removed in March 2006 as the space is needed for a new collimator (and the diffractive program winds down as
the luminosity climbs; typical runs now start with $L \approx 1.5 x 10^{32} cm^{-2}s^{-1}$ with $\approx$ 6 
interactions per crossing, and end a factor $\approx$ lower.) In CDF we will retain the exclusive $\gamma\gamma$ trigger but $l_{eff}$ will be low, and continue to study $\gamma\gamma\rightarrow \mu^+\mu^-$. However there is much data to analyze, to measure
exclusive $\chi_c$, possibly $\chi_b$, $J\psi$ (photoproduction), di-jets and $b\bar{b}$ dijets, and of course $\gamma\gamma$. D0 will also have low
mass exclusive states in DPE with both protons tagged.

\subsubsection*{Forward Detectors at the LHC}
The LHC will collide protons at a centre of mass energy of 14 TeV, starting 
in 2007. Hadronic collisions thus enter a new regime, and will be mainly 
used to unveil the mystery of electro-weak symmetry breaking and search 
for new physics, such as supersymmetry and extra dimensions. Recently 
however diffractive physics was added to the physics program of the 
experiments. This followed two events: the new experimental opportunities 
and the possibility to discover new physics via exclusive production using
tagging forward protons.

The opportunities are the following. The TOTEM experiment was approved 
in 2004. This experiment uses forward detectors for total cross section,
elastic scattering and soft diffraction measurements~\cite{TotemTDR}. TOTEM
 uses the same 
interaction point as the general purpose central detector CMS. 
CMS also proposes to extend its forward 
detector capabilities, and has sent an EOI\cite{eoi} to the LHCC to express 
its interest in forward and diffractive physics.
The study of  common data taking  by the 
CMS and 
TOTEM detectors (roman pots and 
inelastic telescopes) is being addressed in a CMS/TOTEM common study 
group.

ATLAS has also submitted a letter of intent to build roman pot stations,
primordialy for measuring the total and elastic cross section\cite{atlastdr}.
A diffractive program will be addressed in a later stage.

CMS proposes to study diffractive and low-$x$ QCD phenomena, 
and to enhance its detector preformance for this physics by
extending its acceptance in the forward region
and including the TOTEM detectors as a subdetector of CMS in common mode 
of 
data 
taking.

The acceptance of CMS in pseudorapidity $\eta$
is roughly $|\eta| < 2.5$ for tracking
and $|\eta| < 5$ for calorimetry.
CMS is considering  extending its forward acceptance by adding a 
 calorimeter in the region of roughly $5.3<\eta<6.5$,
 approximately 14 m from the IP, using the 
available free space. 
Presently  CASTOR
is conceived to be a Tungsten/Quartz 
fiber
calorimeter of about $10.3\lambda_I$ long, with an electromagnetic and hadronic section.

\begin{figure}[htb]  
\centerline{\psfig{file=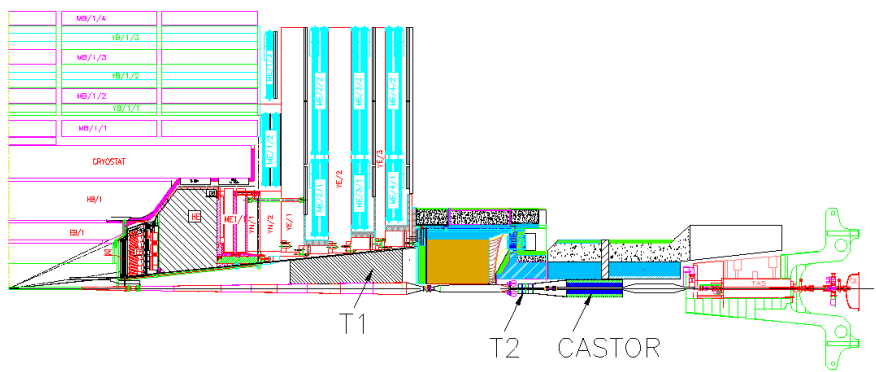,bbllx=0pt,bblly=0pt,bburx=700pt,bbury=400,height=6cm}}
\caption{Position of the T2 inelastic event tagging detector of 
TOTEM  and CASTOR integrated with CMS}
\label{fig:castor}
\end{figure}

A tracker in front of CASTOR 
is being proposed by the TOTEM collaboration, namely the 
T2 inelastic event tagger. In order to be viable for CMS, the tracker
must be usable at  a luminosity of up to $2 \cdot 
10^{33}$cm$^{-2}$s$^{-1}$
which is the  nominal CMS low luminosity operation, based 
on  a LHC optics with $\beta^* = 0.5$ m.
The position of the T2 tracker and CASTOR calorimeter, along the beamline
and integrated with CMS, is shown in Fig.~\ref{fig:castor}.

\par

Common runs are planned for CMS and TOTEM, which will include the 
TOTEM roman 
pot (RP) detectors in the CMS 
readout, in order to tag protons scattered in diffractive 
interactions. The acceptance of the RPs is large and essentially extends
over the full $\xi$ (fractional momentum loss of the proton)
range for the high $\beta^*$ LHC optics.
This will allow  tagging of protons from 
diffractive interactions independent of $\xi$ and will therefore be 
instrumental in obtaining a deeper understanding of
 rapidity gap events that will be collected.
Therefore there can be an interest in collecting some limited amount 
of data 
with such optics at a time and for a duration dictated by the overall LHC 
physics program 
priorities,  depending on the evolution of the LHC at startup.

Roman pot detectors will be also useful 
for the nominal low $\beta^*$ data taking, but the acceptance is 
 limited
to $\xi > 0.02$ with the presently planned TOTEM RPs up to 220 m.
Events with smaller $\xi$ values  can be 
tagged by rapidity gaps in the CMS detector, for
luminosities $<2 \cdot 10^{33}$cm$^{-2}$s$^{-1}$.

A zero degree calorimeter, at a distance of about 140 m from the interaction 
point, with both an electromagnetic and hadronic readout section
is  being
studied for the Heavy Ion program of CMS and can also be used 
for the forward physics program, in particular for charge exchange processes.
With these detector upgrades in the forward region
 CMS  and TOTEM will be a unique detector 
having an  almost 
 complete acceptance of the $pp$ events over the full rapidity range.

Note that also the ATLAS collaboration aims to add zero degree calorimeters and 
there is a specific experiment proposed, called LHCf, which intends to 
measure electromagnetic energy at zero degrees for studies relevant to 
cosmic rays, placed at 140 m distance of  IP1.
ATLAS also has a Cerenkov Counter proposal (LUCID) with acceptance 
over $5.4 < \eta < 6.1$, but its use for a diffractive program has not yet been 
addressed.

An R\&D study has been launched for beampipe 
 detectors at distances of 420 m from the IP,
 ie. in the 
cold section of the machine.  A collaboration called FP420 has been 
formed~\cite{fp420} which has submitted an LOI to the LHCC\cite{LOI}.
Detectors at a distance  of about 420 m would
be required to measure the 
protons from  central 
diffractive Higgs production, e.g. the exclusive channel
 $pp\rightarrow p + H +p$~\cite{DeRoeck:2002hk}. 
The technical feasibility, in 
particular w.r.t.
the LHC machine itself,  still needs to be assessed for these detectors 
options.
The FP420 studies are largely independent from the ATLAS and CMS
IP details, and will be discussed in Section 5.

Further studies include detectors between 18 m (before the TAS) 
and 60 m from the IP\cite{Alekhin:2005dy}.
With the help of the latter the region to detect
particles of $7<\eta <8.5$ could be covered. There are presently no plans 
yet to build such detectors.

\subsubsection*{Forward Physics Measurements at the LHC }
Investigations of hadronic structure at the LHC provide new
possibilities to explore important aspects of QCD.
One of the main problems of QCD is the relative role of 
perturbative QCD and non-perturbative QCD, low-$x$ phenomena
and the 
problem of 
confinement. The latter 
is often related to diffractive phenomena.
The 
common study group of CMS and TOTEM 
is preparing 
for a detailed account of the physics opportunities with such a detector.
The forward physics program presently contains the following topics
\begin{itemize}
\item {Soft and hard diffraction}
\begin{itemize}
\item Total cross section and elastic scattering
\item Gap survival dynamics, multi-gap events, soft diffraction, 
proton 
light cone studies (e.g.
 $pp \rightarrow 3 jets +p$)
\item Hard diffraction: production of jets, $W, J/\psi, b, t$ hard photons, 
structure of diffractive exchange.
\item Double pomeron exchange events as a gluon factory
\item Central exclusive Higgs production (and Radion production)
\item SUSY \& other (low mass) exotics \& exclusive processes, anomalous 
WW production.
\end{itemize}
\item Low-$x$ dynamics
\begin{itemize}
\item Parton saturation, BFKL/CCFM dynamics, proton structure, multi-parton
 scattering.
\end{itemize}
\item New forward physics phenomena
\begin{itemize}
\item New phenomena such as Disoriented Chiral Condensates, 
incoherent pion emission, Centauro's, Strangelets,...
\end{itemize}
\item Measurements for cosmic ray data analysis
\item Two-photon interactions and peripheral collisions
\item Forward physics in pA and AA collisions
\item QED processes to determine the luminosity to O(1\%) 
 e.g. $(pp\rightarrow peep, pp\rightarrow p \mu\mu p)$. 
\end{itemize}

Many of the topics on the list, except the Higgs and exotics
 can be studied best with luminosities of order
$10^{33}$cm$^{-2}$s$^{-1}$, ie. at the startup. 
Apart from Higgs production, discussed below, central exclusive 
production has been discussed as a discovery tool for other new phenomena.
For example, in a color sextet quark model~\cite{hep-ph/0510034,Phys.Rev.D72.036007}, where  these 
quarks couple strongly to the W, Z bosons and to the gluons in the 
pomerons, the exclusive WW production is expected to be many orders of 
magnitude larger that expected from SM processes and would be an easily
detectable and very spectacular signal.
Other possibilities include the production and detection of 
Radions~\cite{Phys.Lett.B630.100}. These graviscalars appear in theories of extra 
dimensions, and can mix with the Higgs boson. These particles have a large 
coupling to gluons and are therefore expected to be produced abundantly 
in central exclusive production processes.

\begin{figure}[htb] 
\subfigure[]{\includegraphics[bb=50 0 760 550,width=.5\textwidth]{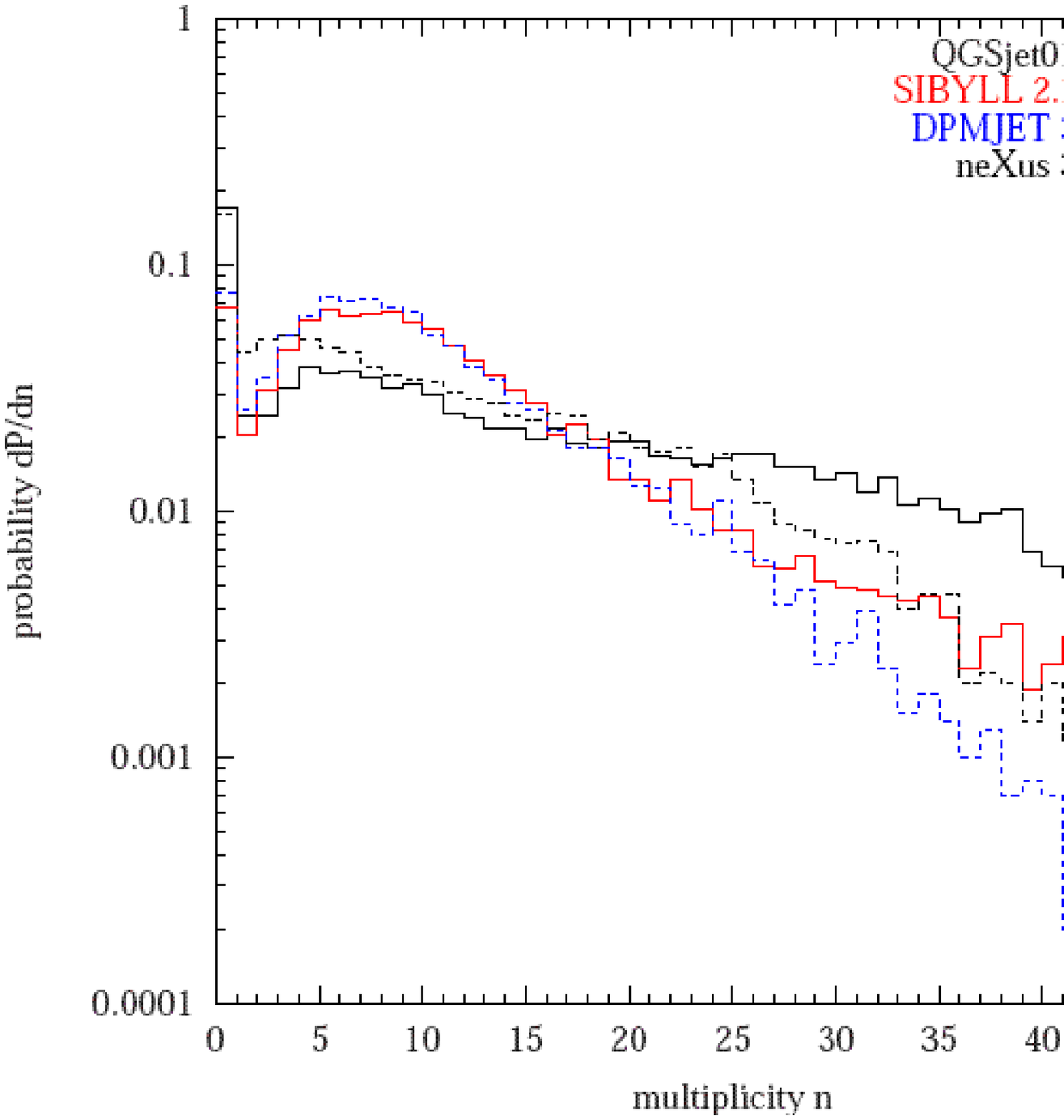}}
\subfigure[]{\includegraphics[bb=0 100 750 500,width=.5\textwidth]{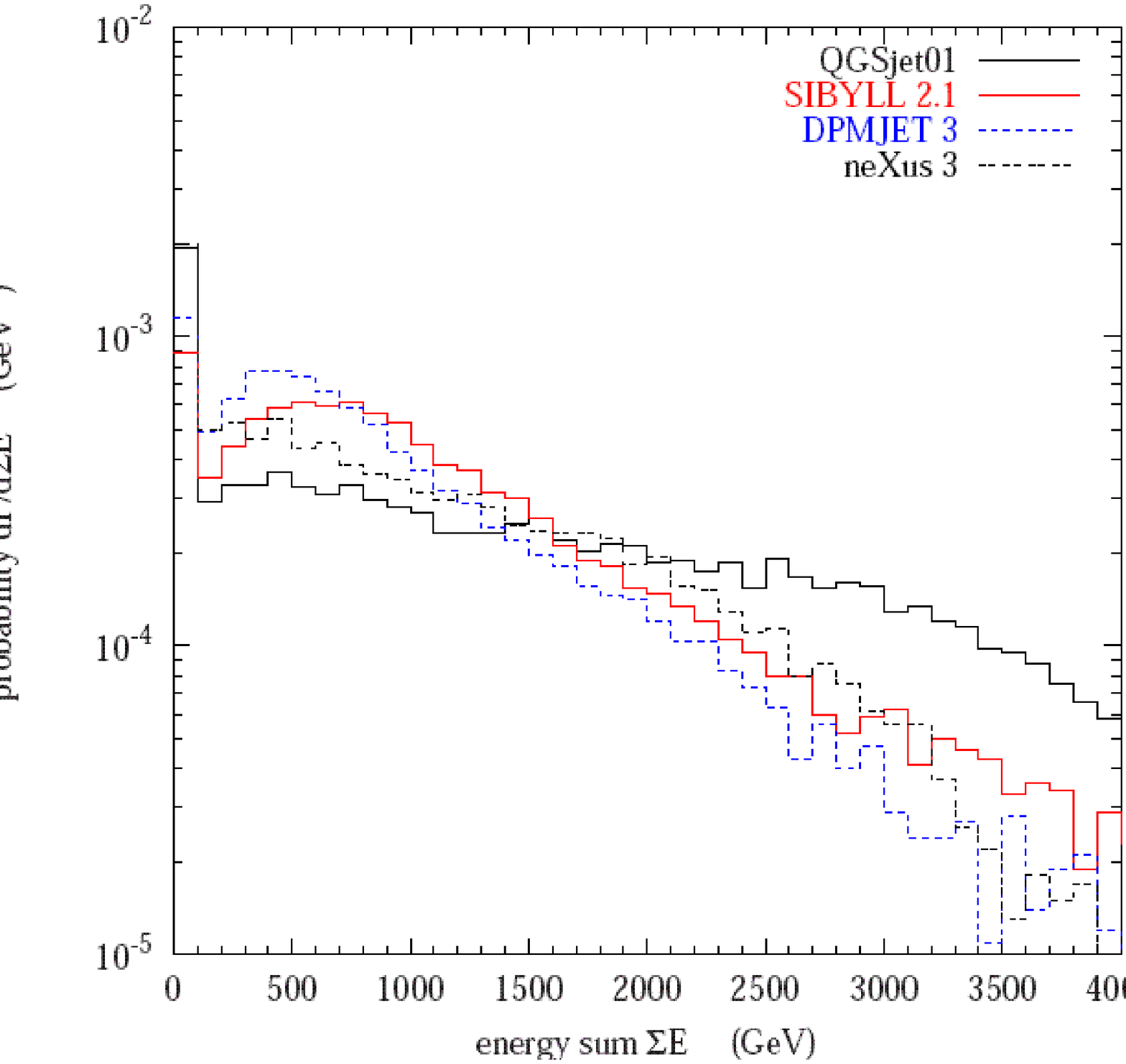}}
\label{models} 
\caption{The total particle multiplicity and total energy sum in the 
pseudorapidity range $5<\eta<7$ for different models used in cosmic ray 
studies\cite{engel,astro-ph/0210393}.}

\end{figure}

TOTEM will use a special high $\beta^*$ 
optics for the measurement of the total 
cross section. The aim is to measure the total cross section with a precision
of order of 
1\%\cite{Deile:2006tt}, but using a prediction of $\rho$. ATLAS plans to get 
information on $\rho$, trying to measure $|t|$ down to 
$6 \cdot 10^{-4}$ GeV$^2$.
Its often stressed that a measurement of $\rho$ is important for understanding 
the energy behaviour of the cross section at even higher energies than 
reachable with present machines.

The LHC data in the forward region will also help to refine 
the interpretation of  data from
ground array cosmic ray experiments. 
Correspondingly there is a considerable interest from the 
cosmic ray community in measurements
 from the LHC at large Feynman-$x$ or rapidity~\cite{engel}.
Cosmic rays interact with the gas in the atmosphere and produce 
extended air showers. Monte Carlo techniques are used to reconstruct
the original incident particle energy and type 
from e.g. the measurement of the 
muon and electromagnetic component at the earth's surface. The uncertainty
of the models for the forward particle production -- which is crucial 
for the energy reconstruction and incident particle type 
determination-- is large:
there exists too little collider data to constrain the models
in this region. In order to make considerable improvements, measurements
of the particle and energy flow at large rapidity 
          are needed at the highest $pp$ and $pA$ energies, i.e. at the 
LHC~\cite{engel}. 
Examples of model predictions for the total particle multiplicity and 
total energy flow in the forward region are shown in Fig.~\ref{models}.
Also the measurement of the total cross section and its 
diffractive components at LHC energies will be of great value and input for 
cosmic ray shower models. Hence the forward physics program at the LHC 
offers an opportunity to improve our knowledge on the leading particle 
production in comsic ray showers, and tune shower simulation programs.

\subsubsection*{The FP420 Proposal}

\begin{figure}[htb]
\centering 
\includegraphics[width=.85\textwidth]{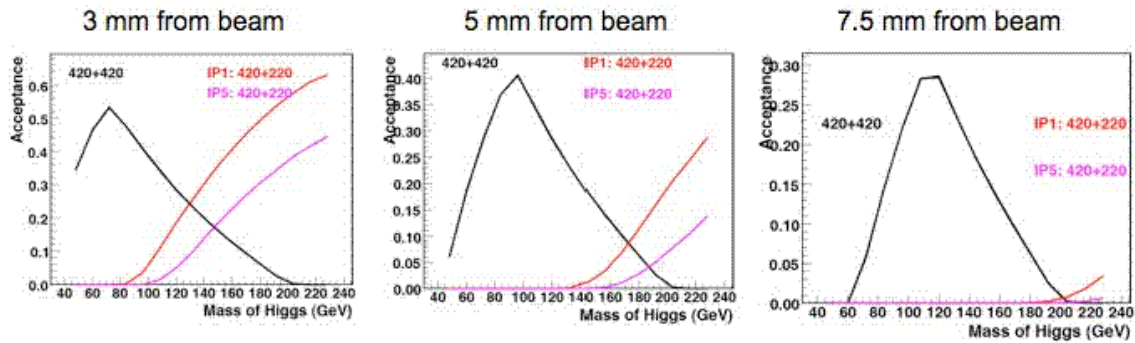}
\label{acc} 
\caption{ 
The acceptance as a function of the Higgs boson mass for the detection of both
 protons at 420 m (black lines) and one proton at 420 m and one proton at 220
m at ATLAS (red line) and CMS (purple line).}

\end{figure}

FP420 is an R\&D effort that has formed in 2005, in order to study the 
feasibility of installing 
proton tagging detectors in the 420 m region of the LHC. Presently 58 
scientists from 29 institutes have 
signed the LOI. This collaboration is still open for 
new participants.

In order to detect the protons produced in central exclusive collisions 
for masses around 120 GeV, detectors in  the region of about 420 m 
away from the interaction point are needed as shown by the   
acceptance calculations.
Fig.~\ref{acc} shows 
the acceptance as function of the particle (Higgs) mass for 
different values of detector approach to the beam. It shows that 
for low Higgs masses there is no acceptance with detectors at 220 m, and 
detectors at 420 m are needed. The differences between CMS and ATLAS for 220m 
are due to the different crossing angle plane at the IP.

The main task of the R\&D project is to redesign the region at 420 m, 
which 
presently consists of
 a connecting cryostat. This 15m long cryostat needs to be redesigned such 
that detectors at (almost) room temperature can be installed. Present proposals 
for the mechanics of such  detectors are mini-roman pots or a moving 
beampipe, as used in experiments at DESY, Hamburg.
Both have advantages and disadvantages as far as precision, stability and RF
issues are concerned. We aim for a precision of order of 10 micron for the 
position of the detectors with respect to the beam.
Beam position monitors will be integrated in the moving mechanics of these
detectors. 

The sensitive detectors will be either 3D silicon detectors or planar silicon 
with 3D edges. The position resolution should be of the order of 10 microns.
We also plan to use fast timing detectors with a resolution of 10-20 ps.
These detectors can be used to check that the protons came form the same 
interaction vertex as the central tracks, and will be extremely instrumental
to remove background or pile-up protons, not associated with the hard 
scattering in the central detector. 
Options for these detectors are Quartz Cerenkov detectors or gas Cerenkov 
detectors with microchannel PMTs.

A first report of the R\&D results, including testbeam results,  is 
planned 
to be completed by spring 2007, and will then be proposed as additional 
detectors to CMS and ATLAS.
The FP420 detectors, when approved by the experiments ATLAS and CMS, can be 
installed during a few month shutdown. The first possible occasion in the present scheme of the LHC will
be the shutdown before the year 2009 run.

\subsubsection*{Conclusion}

Forward and diffractive physics is part of the Tevatron and LHC program, 
and a plethora 
of interesting measurements has and will be made. 
The present and forthcoming Tevatron data on central exclusive production will
be vital to make firm predictions for the LHC. Several key processes are 
now delivering results.
The LHC experiments have turned their attention to the forward region to
extend detector coverage and consequently their physics program.
When equipped with detectors at 420 m, the CMS and ATLAS experiment will be 
able to measure exclusive Higgs production, with excellent signal:background and allowing to study the spin
of the centrally produced particle and its mass with a resolution of $\approx$ 2 GeV per event.
In all aspects the experience gained with operating forward detectors at 
the Tevatron is extremely useful for 
the preparation of the these 
detectors for the LHC, and a continuing collaboration and information flow
is essential.

\clearpage
\section{{Measurement Opportunities at the Tevatron}} 
\subsection*{Introduction}

In this section, we highlight some of the measurements at the Tevatron 
that can improve our understanding of the Standard Model and
optimize our chances for finding new physics at the Tevatron and
the LHC.  We consider the advantages of collecting more data for some 
of these measurements. 
An accurate description of Standard Model physics is essential to 
predict backgrounds to new physics, and the refinements to theory 
that can be obtained
from Tevatron measurements can be directly applied to interpreting the results
from the LHC. 
Run II data
are important
inputs for improving Monte Carlo modeling of the complex event structure
(this includes measurements of parton density and fragmentation functions, 
tuning the modeling of the underlying event, and the validation of production 
processes that are backgrounds to new physics searches). 
Such measurements help reduce errors on theory calculations as well as 
experimental errors, and allow a more precise comparison between theory and 
experiment. Since the Tevatron and the LHC 
will operate in very different 
kinematic regions, we can test for a consistent 
picture as predictions are extrapolated from one region to another. 
The Tevatron also contributes to our goal
of finding convincing evidence to 
support a mechanism of electroweak symmetry 
breaking, whether it be new particles or new interactions 
or something unexpected.
One should not rule out the possibility of new discoveries at the Tevatron,
perhaps based on hints from the LHC.

\begin{figure}[!h]
\centerline{
\includegraphics[width=.8\textwidth]{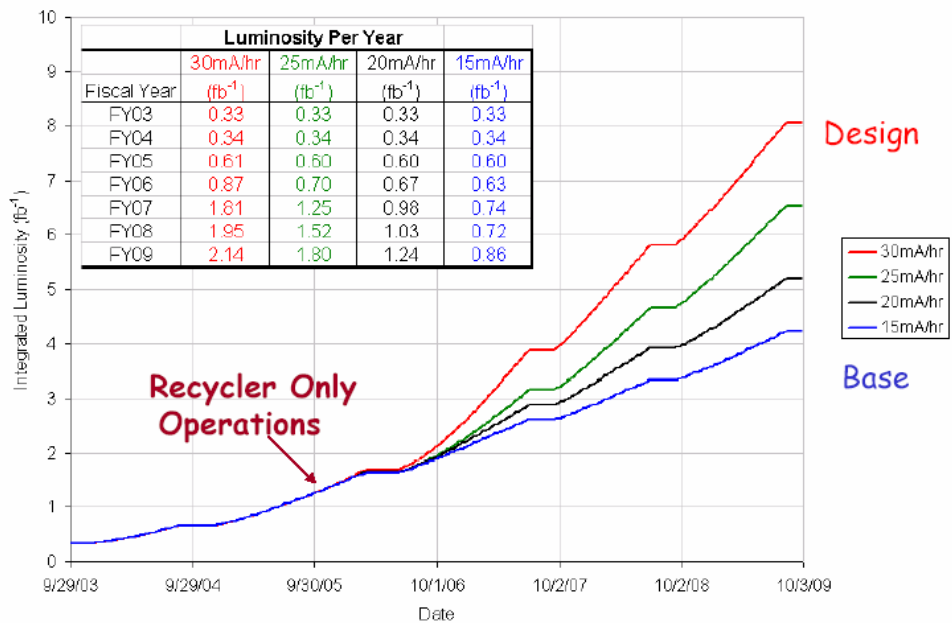}
}
\caption{\label{f:tevlumi}
Projections for the total delivered luminosity at the Tevatron.
}
\end{figure}

To date, the Tevatron has delivered nearly 2 fb$^{-1}$ of data and is 
projected to
deliver between \mbox{$4$--$8$ fb$^{-1}$} by the end of 2009
(see Figure~\ref{f:tevlumi}~\cite{b:machineStatus}).
To put this number in context, the LHC is expected to have accumulated conservatively 
from \mbox{$0.1$--$10$ fb$^{-1}$} of data by the end of operations in 2008. The precision of 
many measurements that will first be done at the Tevatron will eventually be
duplicated or surpassed at the LHC with a moderate amount of data. 
The question can therefore be asked: ``What are the advantages of running the 
Tevatron until the end of 2009 and accumulating 8 fb$^{-1}$ 
before the LHC has a comparable amount of data?''

Based on experience from Run I, it takes several years to 
commission and fully understand a new detector and to process the 
data before precision measurements can be made public. At the end of 
collider operations, 
we can expect to have mature and precise measurements available 
which approach the measurement limits of the Tevatron. 
For the next several years, the Tevatron will be operating at the energy 
frontier, providing the opportunity to: 
(1) establish physics benchmarks that will be used in the initial 
stages of data-analysis at the LHC, until they are surpassed or complemented 
by {\it in situ} calibrations, 
(2) operate on actual (not Monte Carlo) data and 
refine analysis techniques 
necessary to finish a useful physics analysis.
The continuation of this program at the Tevatron will give 
Fermilab--based scientists a head start in LHC analyses and provide 
training for the next generation of physicists and technicians.
We will need to ensure that the infrastructure remains intact to allow  
physics analyses to continue after the halt of Tevatron operations.

The conditions, such as the kinematic region, detector resolution, and background from 
additional interactions,
at the Tevatron and the LHC will be very different resulting in measurements 
with different sources of systematic errors. For some crucial measurements
such as the top or $W$ mass it will be advantageous to have precise measurements
from both facilities in order to be confident on the conclusions drawn from such
measurements. 

We are concerned that the Tevatron might be turned off prematurely,
losing unique measurement opportunities 
that will enhance the value of future LHC or ILC measurements.
A second concern is to ensure that Fermilab and the US HEP program
continues to be rich and vibrant through the period of the ILC siting
decision to maximize the chances the ILC will be in the US. 
Another concern is to have an active accelerator in the United States 
to provide the necessary training for the next generation of 
physicists who are needed to maintain a critical mass of experts in the field. 
We do not see a way to address these concerns without a program at Fermilab that is as
broad and deep as the opportunities at the Tevatron. It should go without 
saying that the case has to be made based on physics in the context
of the ILC being the highest priority for the future HEP program. We 
believe that the case to ensure that the physics potential 
at the Tevatron is fully exploited is very strong, and 
can be made convincingly.

It will be in this context that we evaluate the merits of continuing 
Run II beyond the year 2007. This paper does not attempt to cover all aspects
of the physics program at the Tevatron but highlights some of the measurements which
form a foundation for understanding the Standard Model. In addition to these 
measurements presented here, the Tevatron has rich programs in $b$ physics and 
searches for new physics which are not discussed in detail but can be found elsewhere 
in these proceedings as well as other publications.

\subsection*{The Tevatron Advantage}

The Tevatron is complementary to the LHC in a number of ways, and
there are specific measurements that can consequently be done either
better, or with entirely different systematics, at the
Tevatron. Complementary differences include:

\begin{itemize}
\setlength{\itemsep}{-0.02in}
\item The Tevatron is a $\mathrm{p\bar{p}}$ machine, and is dominated by 
  quark--antiquark collisions in the mass range up to 500 GeV (the ILC
  energy).   In particular, valence quark annihilation is likely the
  dominant contribution to the production of new heavy states with
  masses comparable to the electroweak scale.

\item The Tevatron has a larger reach in $x$ and $\displaystyle x_T={\PT\over
\sqrt{S}}$ for
  measurements of energy--dependence, evolution, and scaling.

\item The Tevatron operates at zero crossing--angle so that
  missing--transverse energy in the electroweak range ($<$100 GeV) is
  azimuthally symmetric and centered on the beamline. 

\item The number of hard ($\PT > 10$ GeV) multiple--parton interactions
per $\mathrm{p\bar{p}}$ interaction is predicted by the present best fits to Tevatron data
to be a factor of 10 lower at the Tevatron.

\item The shorter rapidity plateau at the Tevatron means that 
      missing--energy due to multiple--parton interactions and initial
      and final state radiation is better measured.

\item The longer bunch length at the Tevatron allows counting of  
      multiple vertices by separation in both space and time.

\item Tau identification, important in Higgs physics, using $\pizero$
  detection by sampling in orthogonal views in a ``shower--max''
  detector is unique to the Tevatron (CDF).

\item Triggering on displaced vertices from heavy particles
with lifetimes ($\tau,c,b,$ and exotics) in an all--purpose detector 
     is unique to the Tevatron.

\item Fewer extra collisions per beam crossing 
helps in searches for rare 
exclusive processes that veto jets, 
e.g. rapidity gaps, Vector Boson Fusion (VBF) processes, 
exclusive Extra Dimension-searches, {\it etc.}

\item Fewer extra jets and photons from multiple collisions allow
  searches for complex events with soft jets and/or photons.

\end{itemize}
Different experiments, in general, access different regions of kinematic phase space and can measure
different processes and channels. A challenge of the Standard Model is to be able to describe 
{\em all} observables. 
Predictions can be extrapolated from measured regions 
to new regions such as will be explored at the LHC, providing an important
consistency check of the Standard Model.

\subsection*{Precision Tests of the Standard Model:
the $M_{top}$--$M_W$--$M_H$ Triangle}

\begin{figure}[!h]
\centering
\includegraphics[width=.85\textwidth,trim= 50 50 0 0]{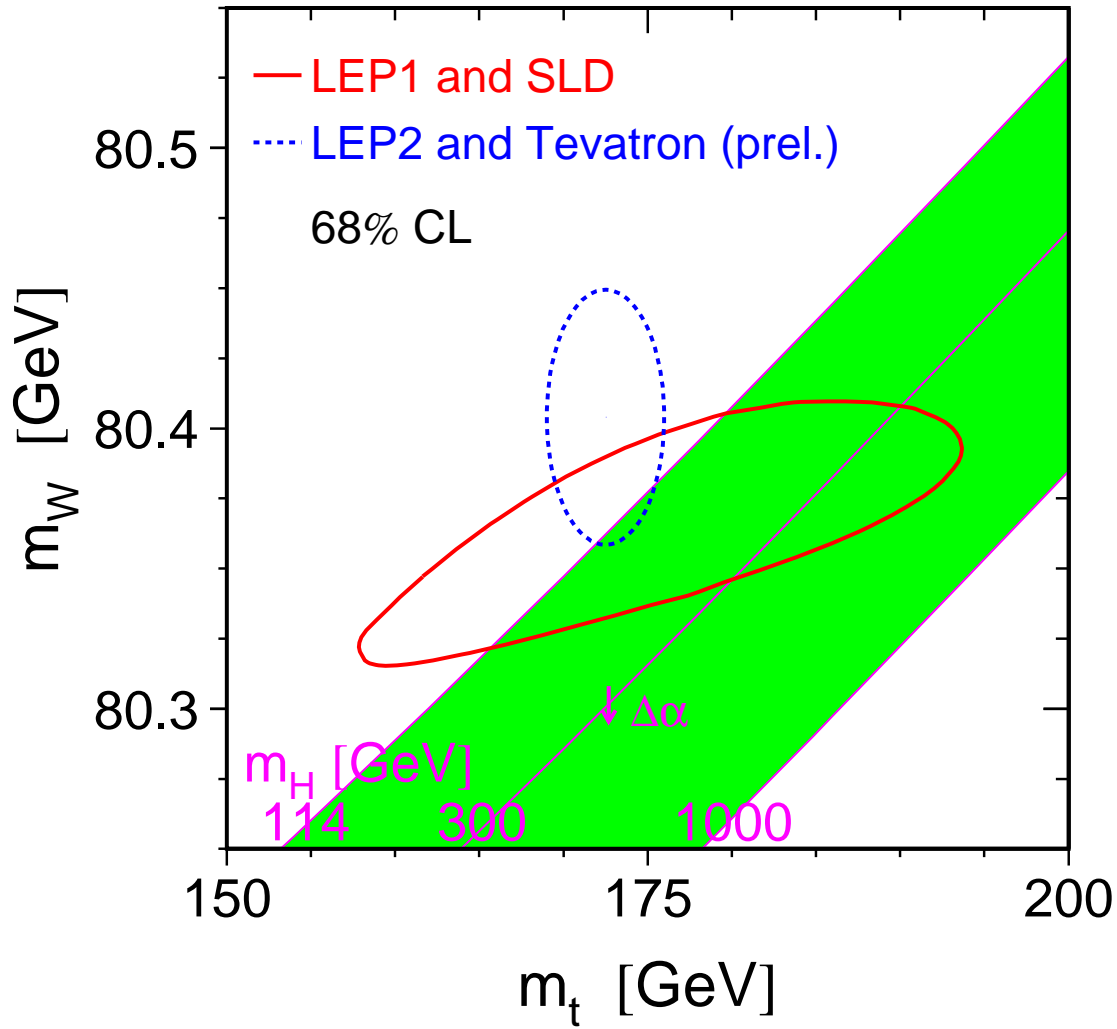}
\caption{\label{f:mtwtContours}
For a fixed Higgs boson mass, Electroweak radiative corrections imply 
a relationship between the top and $W$ boson mass.  Given a precision
of the top mass $\Delta m_{\rm top} = 1~{\rm GeV}$, the range 
of allowed $W$ boson masses is $\Delta M_{W}(m_t) = 6~{\rm MeV}$,
where the theoretical uncertainty is
$\Delta M_{W}(\rm theory) = 5~{\rm MeV}$.
}
\end{figure}

The top quark, $W$-boson, and Higgs boson masses are predicted by the SM to
have a `triangle' relationship, i.e. given two of the masses, the third
is precisely determined. 
Constraints from the $W$ boson and top mass 
measurements on the Higgs mass boson are shown in Figure~\ref{f:mtwtContours}~\cite{Group:2005di}
which include the recent combined results from CDF on the top mass based on 760 pb$^{-1}$ (new 
results for the $W$ boson mass are not yet available).
Both the top and $W$ masses can be better
measured at the LHC and Tevatron. The $W$ mass in particular can significantly 
constrain the triangle, and requires a measurement with a 
precision on the order of $1$--$2$ parts in $10^4$
(of order 10 MeV). The measurement is very sensitive to many details
of hadron collisions, including multiple parton interactions in the same
collision, multiple hadron interactions in the same crossing, initial
state radiation, quark-antiquark parton distribution functions, and
QCD backgrounds to leptons and missing transverse energy.
At the Tevatron, there is a wealth of experience in understanding the errors
associated with 
this measurement. LHC projections are in the 10 MeV range, but the
measurement is of sufficient importance to merit a second measurement
with comparable or better sensitivity and completely different
systematic uncertainties.

The top mass measurement is competitive at the Tevatron, with the
lower cross section being compensated by the benefits of valence quark
production close to threshold and hence quieter events. Like the $W$ boson
mass measurement, this measurement is already systematics
dominated. A Tevatron measurement with comparable sensitivity and
completely different systematics to the LHC one 
will give great confidence in this
number, as well as improve the precision on it.

\subsubsection*{Top Mass Measurement}

\begin{figure}[!h]
\centerline{
\psfig{figure=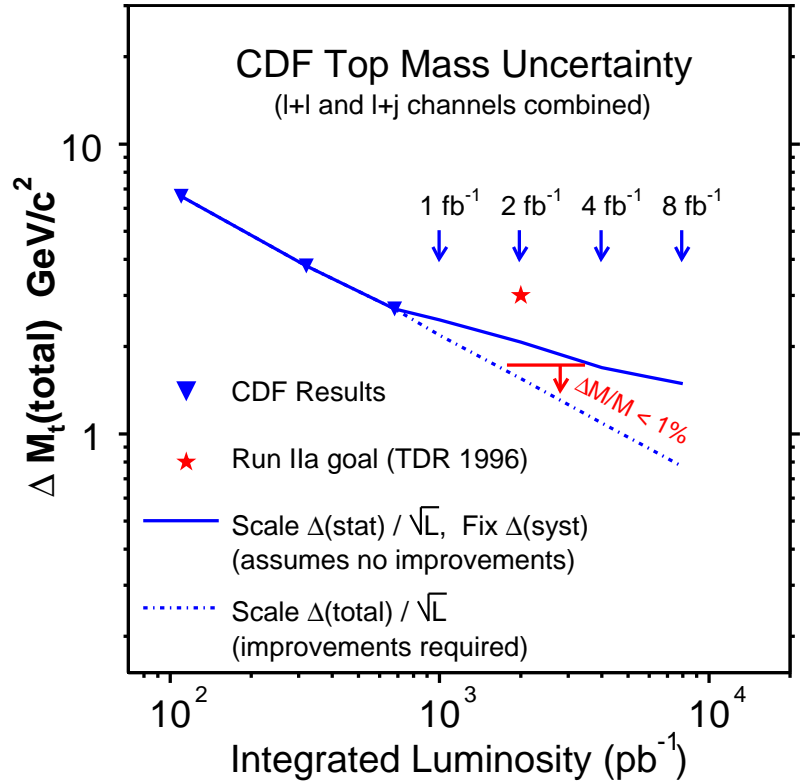,width=11cm}
}
\caption{\label{f:tMass}
The expected improvement on the error of the top mass measurement as
a function of the integrated luminosity.
}
\end{figure}

The uncertainty on the top mass measurement
is composed of a part which scales with luminosity and a part that 
does not. The expected improvement of the precision of the top mass measurement is shown 
in Figure~\ref{f:tMass} as a function of integrated luminosity. Recent results from CDF 
are already more precise than the projections made in the Run IIa Technical Design Report. 
Further improvements
can be expected as we collect more data, with between 4 to 8 ${\rm fb^{-1}}$ of recorded
data we can hope to achieve a precision of $\delta m_t  / m_t < 1\%$. 

Event selection, in particular the jet thresholds, at the LHC will be quite 
different from those at the Tevatron. Although both will be using a 
constrained fit to the $W$ mass, they are likely to be sensitive to 
different types of systematic errors. For example, at the Tevatron, unlike the 
case at the LHC, there is little or no bias from 
angular resolution and jet separation, hence it is important to determine 
the accuracy of the $W+$jet kinematics in the Monte Carlo. 
$Z+$jet data provides a way to do this test, but it provides
low statistics.
The expected precision at the LHC is estimated to be between 
1.0 GeV~\cite{Borjanovic:2004ce}   
and 1.5 GeV~\cite{Heinemeyer:2004gx} which is comparable to what can be achieved
at the Tevatron provided sufficient data is collected. The sources of systematic 
errors at the Tevatron include:
\begin{itemize}
\item Jet energy scale: derived from $W \rightarrow q q^\prime$, detector resolution

\item Background: systematic uncertainties in modeling the dominant background sources

\item $b$-jet modeling: variations in the semi-leptonic branching 
fraction, $b$ fragmentation model, differences in color flow between $b$-jets 
and light quarks.

\item ISR, FSR, UE: tuning of the different models in different Monte Carlo programs

\item Method: Fit method, Monte Carlo statistics, and $b$ tagging efficiency
 
\item Generator: Differences between \PY \ or ISAJET and \HW \  
when modeling the $t\bar{t}$ signal.
\end{itemize}

The magnitude of the errors are given in Table~\ref{t:topErrors} together with 
an estimate of their size assuming the full data set. 
Further improvements on the precision of the top mass will require more refined modeling
and more precise PDF sets.
\begin{table}[!h]
\centering
\begin{tabular}{l|l||l|l}
Source              & $\Delta m_t$ ${\rm (GeV/c^2)}$      &
Source              & $\Delta m_t$ ${\rm (GeV/c^2)}$      \\ \hline
Jet Energy Scale    & 2.5  $\rightarrow$ 0.7 &    ISR                 & 0.4  \\
BG shape            & 1.1  $\rightarrow$ 0.3 &    MC statistics       & 0.3  $\rightarrow$ 0.1 \\ 
$b$-jet modeling    & 0.6  & PDF's                & 0.3  \\                  
FSR                 & 0.6  & Generators          & 0.2  \\                   
Method              & 0.5  $\rightarrow$ 0.2 & $b$-tagging         & 0.1  \\  
\end{tabular}
\caption{\label{t:topErrors}
Sources of errors on the top cross section measurement.
}
\end{table}

\subsubsection*{$W$ Mass Measurement}

The total uncertainty on the $W$ mass measurement can also be decomposed into 
a component that scales with luminosity and a part that does not.
The uncertainties which scale with luminosity include:
statistical uncertainties and systematic uncertainties such as the energy and momentum scale 
and hadron recoil against $W$. As we collect more data we are 
able to better calibrate the energy response and reduce the associated uncertainty.

\begin{figure}[!h]
\centerline{
\includegraphics[width=\textwidth]{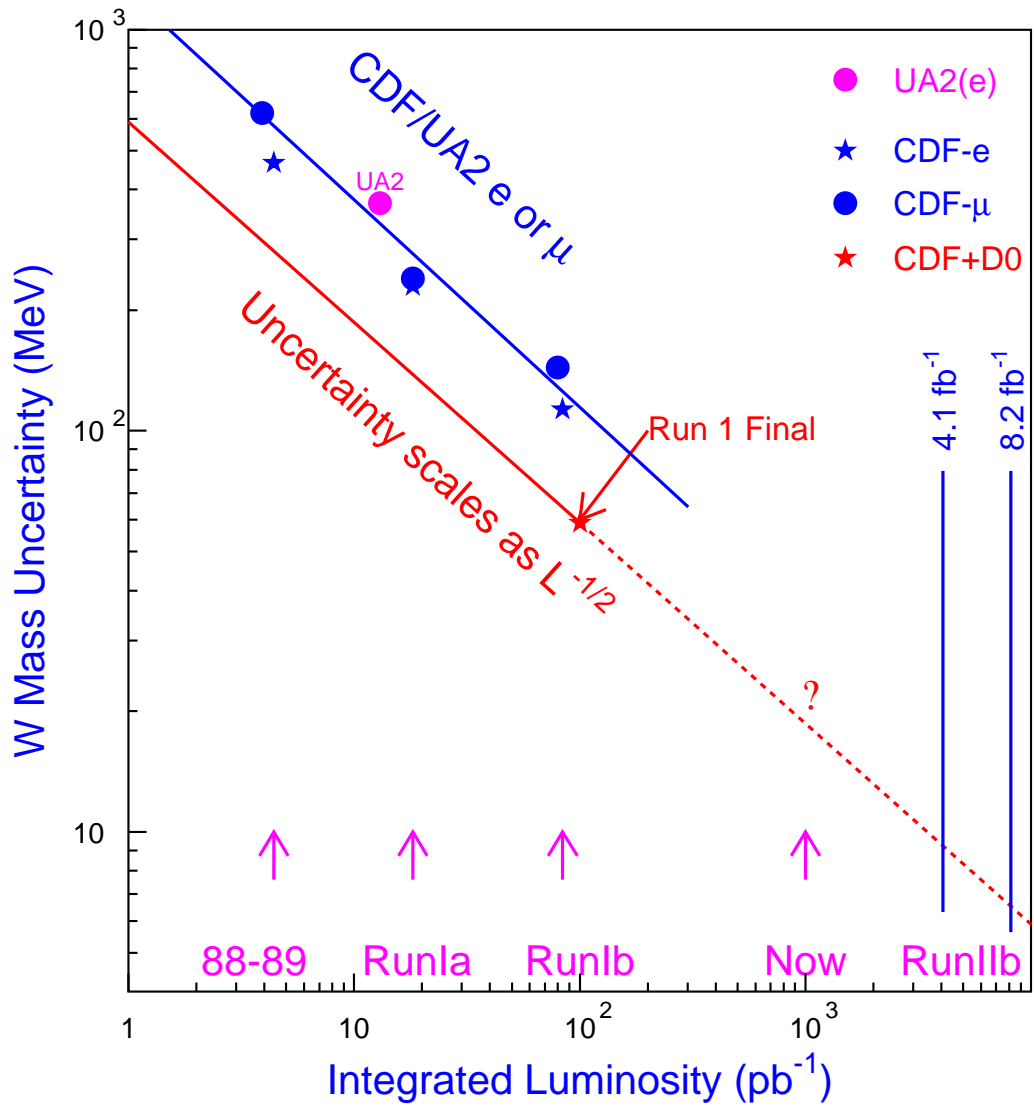}
}
\caption{\label{f:wMass}
The precision on the $W$ mass measurement is expected to improve 
with increasing luminosity. 
}
\end{figure}

Uncertainties which do not scale with luminosity include:
$W$ production and decay, PDF's, the shape of the
$W$ boson $\PT$ distribution and higher 
order QCD/QED effects, assumed to be between $20$--$30$ MeV.
Figure~\ref{f:wMass} illustrates the expected improvement in the $W$ mass precision
as a function of the integrated luminosity. With the full data set 
at the Tevatron we can expect to measure the $W$ mass to a precision of
$\delta m_W \sim$  $20$--$30$ MeV.\footnote{These error estimates are
discussed later.}
The ultimate precision estimated
for  the LHC is $\delta m_W  \sim 10$--$20$ MeV.
In order to achieve this precision at the LHC, it will require low luminosity running and an excellent understanding of the detector.

\subsection*{Validation of Standard Model Processes}

Modern experimental particle physics relies, more so than in the past,
on theoretical predictions, usually in the form of Monte Carlo
programs.  This is understandable as we evolve from a qualitative
to a quantitative understanding of the Standard Model.  
Validating these tools on high--luminosity, high--energy,
hadron--collider data is important, but can be complicated and
time--consuming.  
Current Tevatron data is helping to constrain phenomenological
models and to indicate directions for theoretical improvements, but
the validation process is limited by data and a lack of appreciation of
its importance.
By running in an energy regime that is not tainted by (potentially)
large contributions
of new physics, we can begin to build a complete description of the
important Standard Model processes.  
If our MC tools are not adequate,
then analyses at the LHC may rely on background estimates
that are imprecise and cannot be easily cross-checked with independent
samples.  Indications of new physics could potentially be absorbed into 
``background'' distributions, limiting or jeopardizing new physics
searches.

An obvious goal for Run II should be the establishment of a ``complete''
description of Standard Model backgrounds to new physics.  To expand 
beyond our current knowledge, this means obtaining a good understanding 
of diboson and single top production processes and an excellent understanding
of the $t\bar t$, $W/Z+$jets and multijet processes.

\subsubsection*{Di-Boson Production}

The study of di-boson production at the Tevatron provides complementary tests of the 
electroweak sector of the Standard Model to those made at LEP. 
Anomalous pair production of gauge bosons could be an indication
of deviations from the Standard Model gauge structure and/or
$W/Z$ substructure.  Anomalous production could manifest itself
as an increased production rate or change in the kinematic distributions, 
such as the gauge boson $\PT$.
The size of anomalous couplings can be used 
as metrics for evaluating the sensitivity to new physics and characterize 
any deviation of the $W$ and $Z$ bosons from point particles.
 
A good understanding of di-boson production is also needed to estimate 
the background for other important physics. 
The production of $WZ$ and $ZZ$ boson pairs at the Tevatron
occurs at a rate of
order 100 fb, and constitutes a significant background 
in searches for the SM Higgs boson or SUSY trilepton signatures.
In $t\bar{t}$ events 
when the $W$ bosons both decay leptonically, the signature is similar 
to $WW$ production with an ISR gluon splitting to a heavy quark pair.

\begin{figure}[!h]
\centering
\includegraphics[width=.75\textwidth]{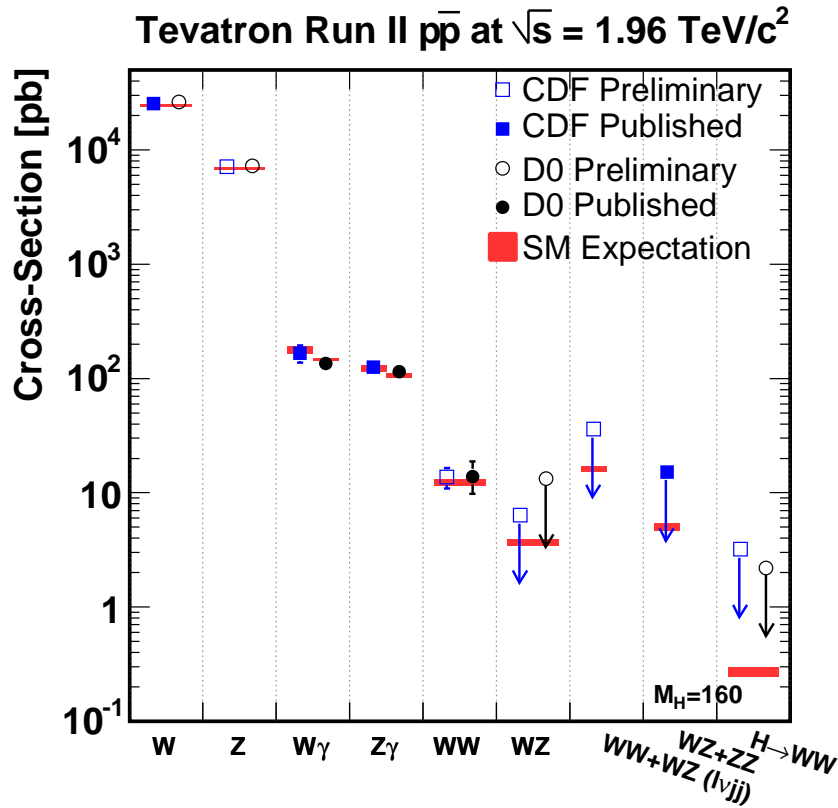}
\caption{\label{f:diboson}
Recent measurements of electroweak gauge boson production from CDF
and D\O.
}
\end{figure}

The current status of di-boson production is 
summarized in Figure~\ref{f:diboson}.
The $WW$ process is 
observed with an uncertainty on the cross section measurement that 
is $6-7\times$ the theoretical uncertainty \cite{Campbell:2000bg}.
The cross section for $ZW/ZZ$ 
processes is bounded above at roughly $3\times$ the Standard Model expectation, 
which is about the current limit on single top production.
With a comparable data set needed to discover single top production, the 
Tevatron should begin to make quantitative measurements of $ZW/ZZ$ processes.

\subsubsection*{$t\bar t$ production}

While our understanding of inclusive $t\bar t$ production is
quickly becoming systematically--limited, more exclusive topologies
involving top
are less understood. 
An important aspect
of $t\bar t H$ searches at the LHC is understanding
the mass shape of additional jets in $t\bar t$ events.  The
rate for such events 
is roughly a factor $10^{-2}$ smaller than the total rate.
Furthermore, the $t\bar t$ process at the LHC
must often be rejected (or selected) with a high efficiency, 
based on the number of jets in the event.  
A high--statistics sample of $t\bar t$
will allow detailed studies of the 
effect of additional gluon radiation.
How often do reconstruction inefficiencies
lead to {\it fewer} jets than expected?  How often does
misreconstruction occur, so a jet is formed from calorimeter
noise or from underlying event fluctuations?  How often does
a $b$-tagged jet actually correspond to a $b$ quark at the
parton level?  With thousands of clean, double-tagged top
pair events in Run II, we can address all these questions.
More support of our understanding of $t\bar t$ production
would be found in the observation of the rapidity asymmetry,
which can be observed at a proton-antiproton collider.

\subsubsection*{Single Top Production}

The single top production process has the same signature $W^\pm b\bar b+X$
as the associated production process $W^\pm H(\to b\bar b)$.  
Also, since the production rate is proportional to $|V_{tb}|^2$,
it probes the top quark width.
The SM predictions for the single top cross section are
\mbox{$0.88 \pm 0.14$ pb} for the s--channel process and 
\mbox{$1.98 \pm 0.30$ pb} for the t--channel process.

\begin{figure}[!h]
\subfigure[]{\includegraphics[width=.45\textwidth]{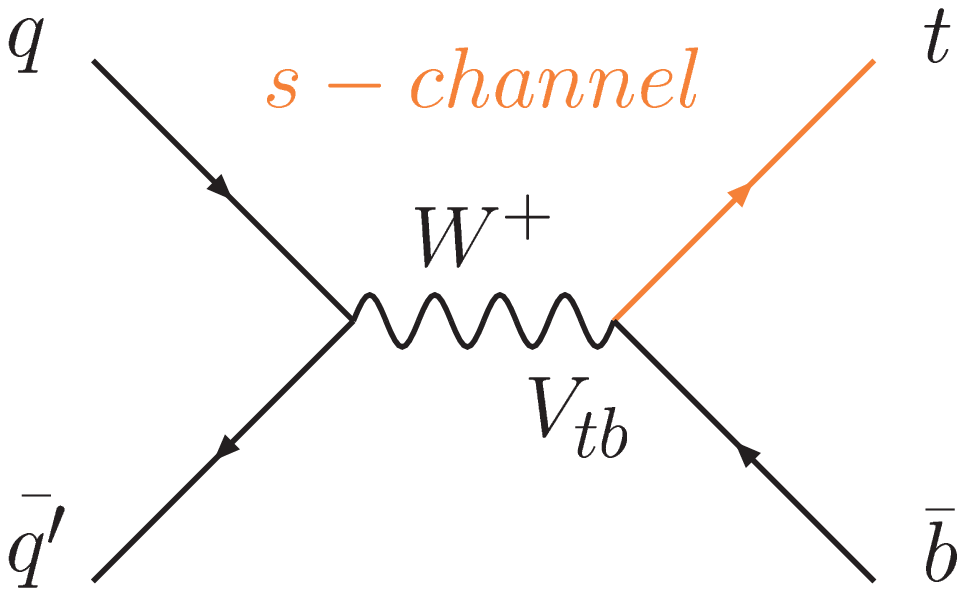}}
\subfigure[]{\includegraphics[width=.45\textwidth]{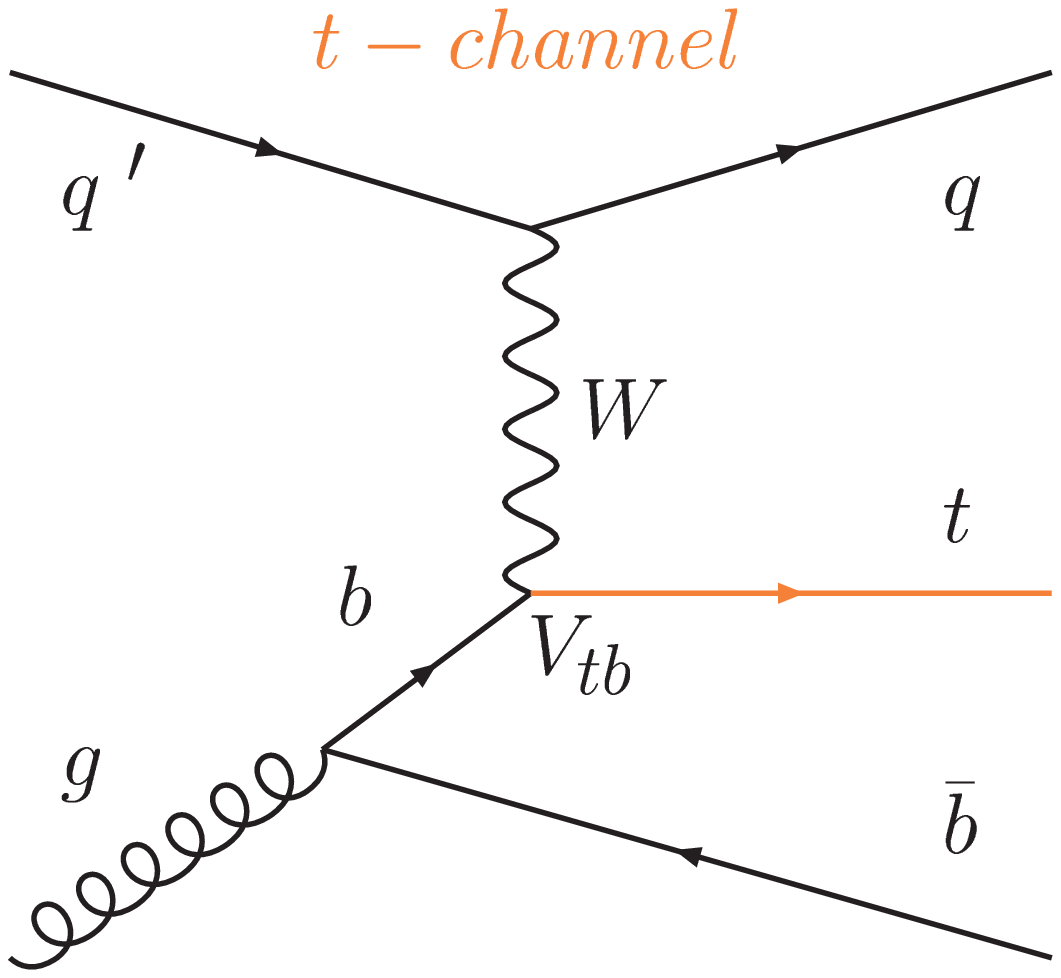}}
\caption{\label{f:singleTopF} Feynman diagrams for the two
leading single top processes:
(a) s-channel and (b) t-channel.
}
\end{figure}

\begin{figure}[!h]
\centerline{
\includegraphics[width=.95\textwidth]{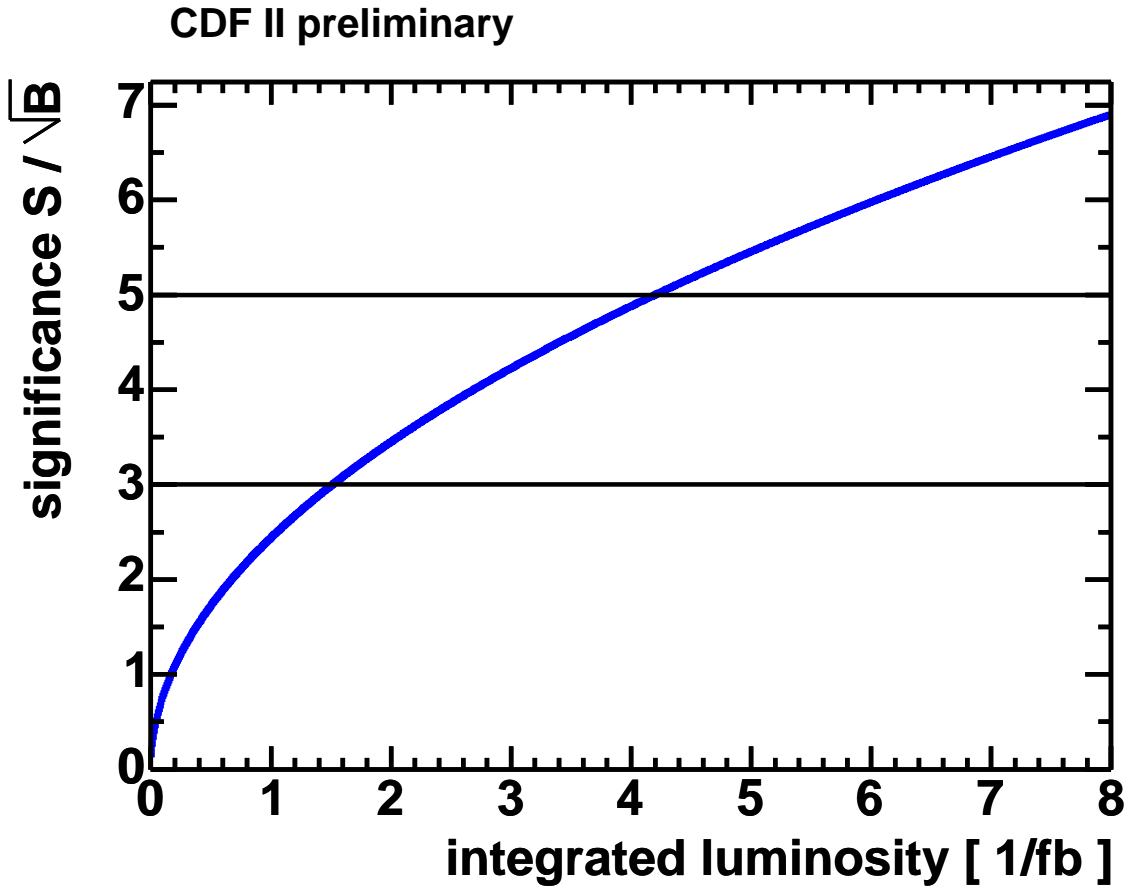}
}
\caption{\label{f:singleTopProj}
Combined channel likelihood for SM single top production.
With about 4 fb$^{-1}$ of data, CDF expects to have a $5\sigma$ signal needed
to claim a discovery.
}
\end{figure}

The combined channel likelihood for SM single top production as a function 
of the integrated luminosity is shown in Figure~\ref{f:singleTopProj}.
With $2$--$4$ fb$^{-1}$ Standard Model single top production will be established, providing 
an event sample of roughly 75 events. If the measurement is statistically
limited, then the total production cross section will be known to
roughly 10\%.  However, a detailed description of 
kinematic distributions will benefit from increased statistics.  
It is hard to quantify how well this can be done, but we can estimate
that the number of fit quantities is roughly proportional to 
$\log_{10}(N_{\rm data})$.\footnote{For a
1-dimensional distribution of say $\PT$, 10 events are typically needed to obtain a
good, Gaussian fit.  This can be generalized to many dimensions by picturing 10 slices in
each variable.}
A doubling or quadrupling of the data will easily allow for multi-variate
fits, increasing our confidence that we are observing pure Standard Model
single top production.  As a rule of thumb, we can claim an 
understanding of a process if we can predict the effect of radiating an extra, energetic jet, which
comes with a statistical penalty of 
$\alpha_s\sim 0.1.$\footnote{This is based on the fact that
many electroweak processes receive their dominant corrections
at NLO.}  Observing 15 or 30 single-top-plus-jet events versus only 7 is the difference
between Gaussian and Poisson statistics.  Further, one expects that 
cuts can be loosened {\it after} discovery, leading to larger
and more discriminating datasets.

\subsubsection*{Vector Boson plus Jet Production}

$W$ or $Z$ + jets production is a fundamental process that needs to be 
understood for new
physics searches (SUSY and Higgs) at the LHC. We will need to have sensitive tests at lower 
energy where the effects of new processes do not significantly contribute. 
Since Vector-Boson + jet is the principle background to top, we need to
understand the rate, energy spectrum and correlation between jets in this process.

Since many of the same parton topologies contribute to $W+$n jet
production at the Tevatron and LHC, Tevatron measurements could be
used to normalize the rates using the data to calculate effective
$K$--factors.  The translation from Tevatron to LHC energies, then,
would be a reweighting based on the different contributions of PDF's.
More statistics at the Tevatron could strengthen this extrapolation
by providing more $Z+$jet events. 

To further stress the importance of understanding vector boson production,
we note that this is a ``Standard Candle'' process 
that can be used to determine the 
proton-proton luminosity. This process has high statistics, can be
measured accurately, and is theoretically well understood.
At the LHC, the proton-proton luminosity will have to be known to 
better than 5\%.
Other techniques to reduce the uncertainty coming from PDF's involve using 
cross section ratios which can reduce the overall uncertainty 
on the luminosity from 5\% to $\sim$ 1\% \cite{Dittmar:1997md}.

\subsection*{Phenomenological Measurements}
Many aspects of our description of the complex structure of
hadronic events cannot be addressed from first principles.
They are by nature non--perturbative or sufficiently correlated
with other aspects of the event
to prevent a simple description.
Our lack of understanding of these event properties are often
the leading systematic uncertainties in analyses.  
The Tevatron potential for making important PDF measurements is sufficiently
rich that we will discuss it separately.

\subsection*{Improving Parton Distribution Functions}

Parton Density Functions (PDF's) are an essential input to the calculation of many 
hadron-hadron and lepton-hadron production processes. 
Uncertainties on the PDF's 
influence the measurement at several stages in the analysis. 
The inferred cross section is related to the observed
quantities and correction factors through the relation:
\begin{equation*}
\label{e:pdfMeas}
\sigma_{\rm meas} = \frac{\epsilon}{\cal{L}} (N_{\rm obs} - N_{\rm bkg}).
\end{equation*}
PDF's errors can impact the measurement through the 
calculation of acceptance ($\epsilon$), luminosity ($\cal{L}$),
event selection ($N_{\rm obs}$), and background estimate ($N_{\rm bkg}$).

The Tevatron and LHC are, borrowing a widely abused term, \emph{W/Z
factories}. The reason that their potential for contributing to the next
generation of global QCD analysis (in an analogous fashion to DIS experiments
in the last two decades) has not attracted much attention has to do with the
fact that the measured cross sections, involving convolutions 
of two PDF's, do not depend on the PDF's in as direct a way as the structure
functions of DIS scattering. Thus, it is difficult to highlight which
measurement will directly determine which particular features of PDF's. But,
since most of the open issues in current PDF analysis concern sub-dominant
effects, the more subtle role to be played by precision $W/Z$ data will be both
natural and vital. Instead of looking at LO parton formulas for motivation to
focus on particular measurements, we now need detailed phenomenological
studies of the effects of possible measurements on the remaining uncertainties
of PDF's in the global analysis context, utilizing the new tools developed in
recent years, such as the Lagrange multiplier method. Efforts along this line
are crucial for the success of the Tevatron and LHC physics programs.

\begin{figure}[!h]
\centerline{
\includegraphics[width=.75\textwidth,trim= 0 10 0 0]{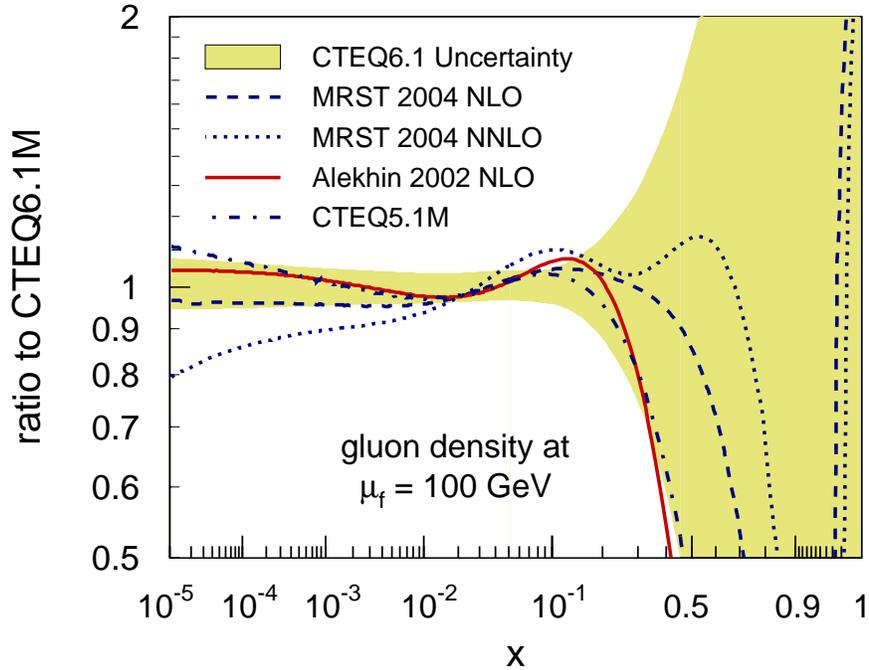}
}
\caption{\label{f:gluonUncertainty}
The gluon distribution is one of least constrained PDF's. Inclusive jet 
measurements provide important constraints on the gluon density at
high $x$. 
}
\end{figure}

Some of the PDF distributions which are not well constrained include the
gluon distribution at high $x$, strange and anti-strange quark content, strange
and anti-strange asymmetry, 
details in the $u$ and $d$ sector, the $u/d$ ratio and heavy quark distributions. 
For low $x$, the error on the gluon distribution is expected to be about 3\% and increases
dramatically for high $x$ as shown by the shaded band in Figure \ref{f:gluonUncertainty}.

The following measurements from the Tevatron can be used to help 
constrain the PDF's:
\begin{enumerate}
\item Inclusive jets at central and forward rapidity,
\item $W/Z$ total cross section,
\item $W$ Mass,
\item $W/Z$ rapidity distributions,
\item $Z$ forward/backward asymmetry,
\item $W/Z$ transverse momentum distributions,
\item $W$ charge asymmetry,
\item $W/Z/\gamma$ + jet cross sections,
\item $W/Z/\gamma$ + heavy flavor tag cross sections,
\item $\Upsilon$ transverse momentum distributions,
\item top production cross section,
\item direct $\gamma$ production cross section.
\end{enumerate}
In the following, we discuss the potential of some of these measurements.

\subsubsection*{Inclusive Jet Cross Section}

\begin{figure}[!h]
\centerline{
\includegraphics[width=.75\textwidth,trim=0 50 0 0]{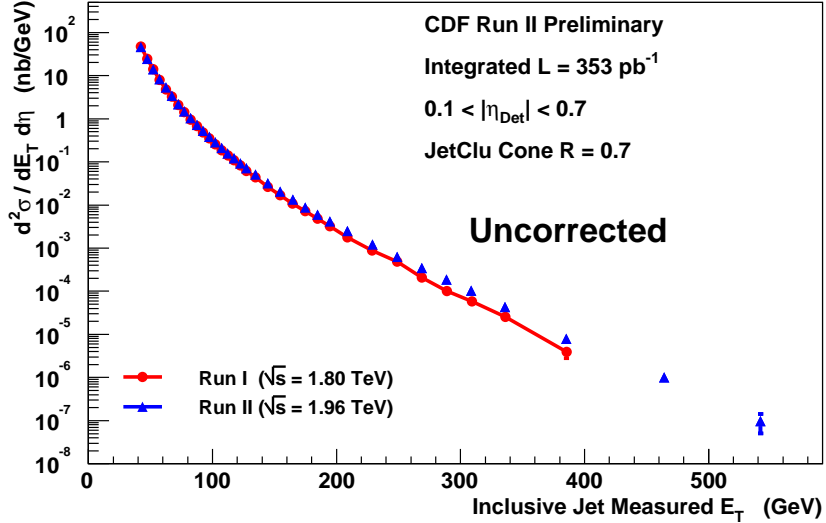}
}
\caption{\label{f:centralJets2}
The increased center--of--mass energy 
and the increased luminosity of the Tevatron
allows us to extend the Run I results by more than 150 GeV, probing the shortest distance scales.
}
\end{figure}

Measurement of the inclusive jet cross section 
is a stringent test of pQCD over many orders of magnitude
(see Figure \ref{f:centralJets2}).
New physics can show up 
as an excess of events at high $\PT$ compared with pQCD predictions. 
As is now well-known, high $\PT$ jet production also probes the high $x$ 
gluon distribution, and there is some flexibility in the PDF 
parameterization which allows it to accommodate some excess at high $\PT$. Having data at
both high $x$ and low $x$ constrains the fits through the momentum sum rules.

\begin{figure}[!h]
\centering
\includegraphics[width=\textwidth,trim= 0 10 0 0]{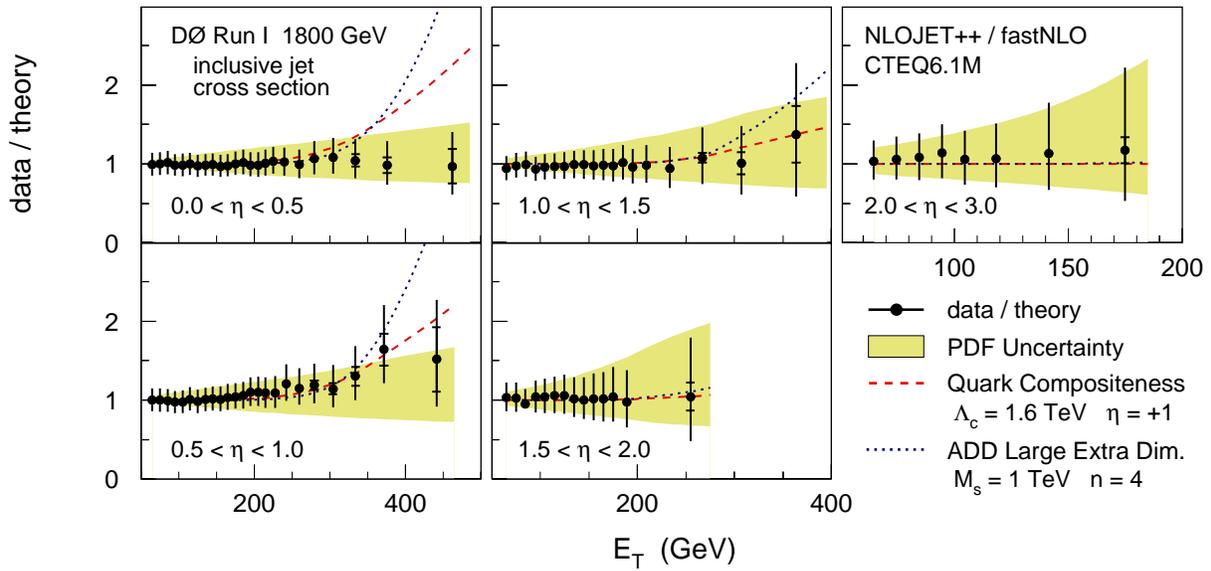}
\caption{\label{f:forwardJets}
Measurements from D\O~in Run I demonstrate 
the importance of the forward rapidity regions
for disentangling PDF effects from 
new physics.}
\end{figure}

\begin{figure}[!h]
\centerline{
\includegraphics[width=.75\textwidth]{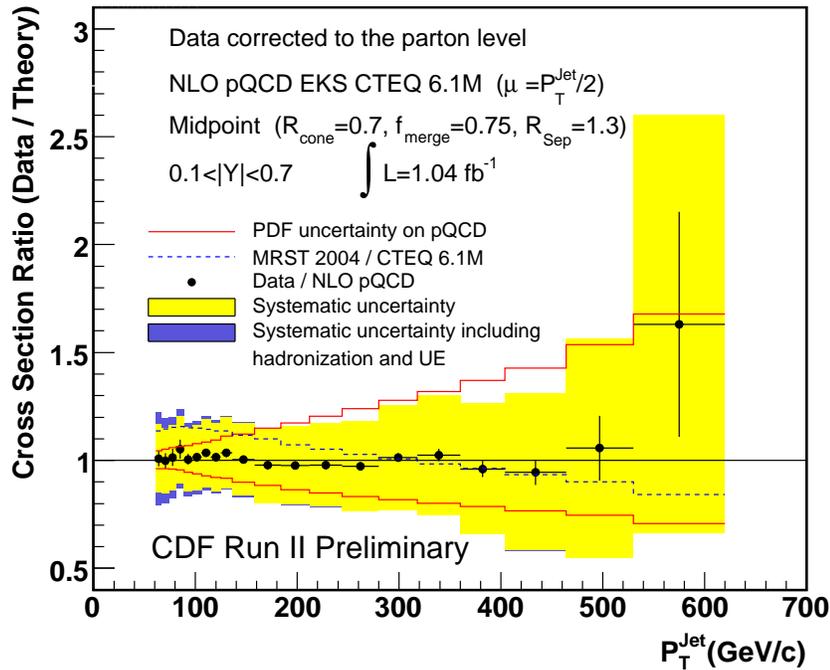}
}
\caption{\label{f:centralJets}
The inclusive jet cross section for the central rapidity region.
}
\end{figure}

\begin{figure}[!h]
\centering
\subfigure[$0.1<|y|<0.7$]{\includegraphics[width=.49\textwidth,trim= 0 20 0 0]{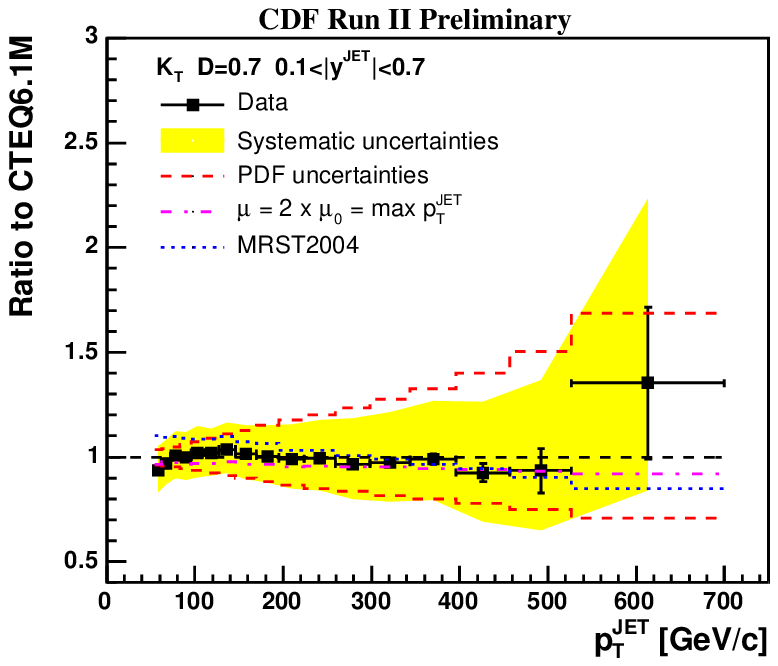}}
\subfigure[$0.7<|y|<1.1$]{\includegraphics[width=.49\textwidth,trim= 0 20 0 0]{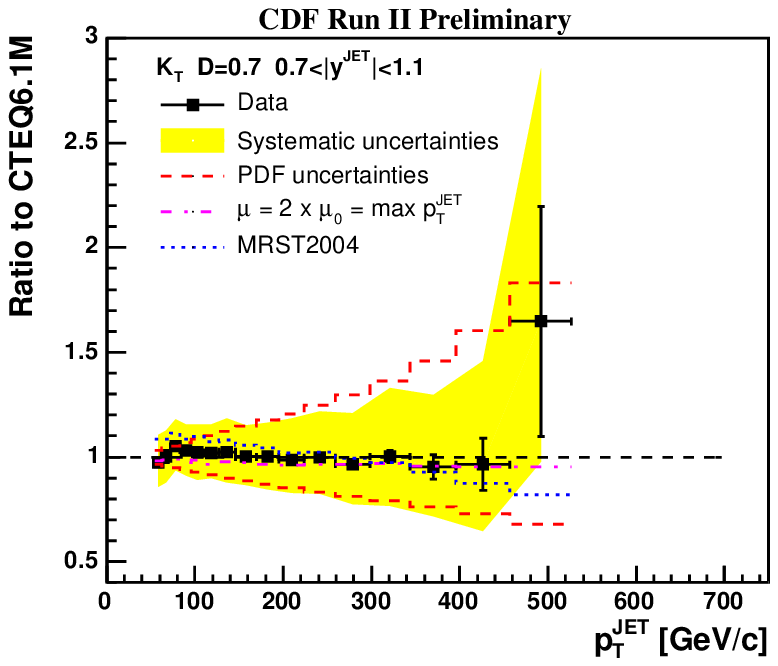}}
\subfigure[$1.1<|y|<1.6$]{\includegraphics[width=.49\textwidth,trim= 0 20 0 0]{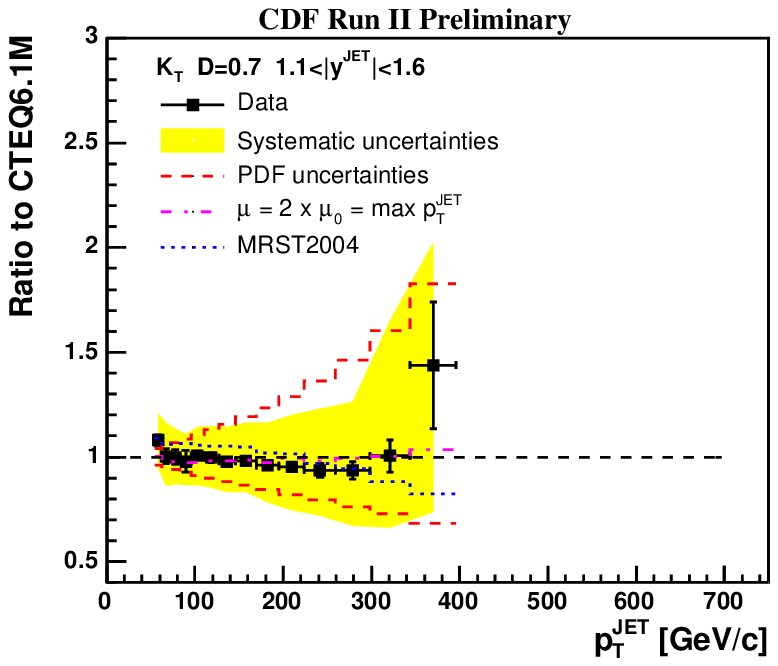}}
\subfigure[$1.6<|y|<2.1$]{\includegraphics[width=.49\textwidth,trim= 0 20 0 0]{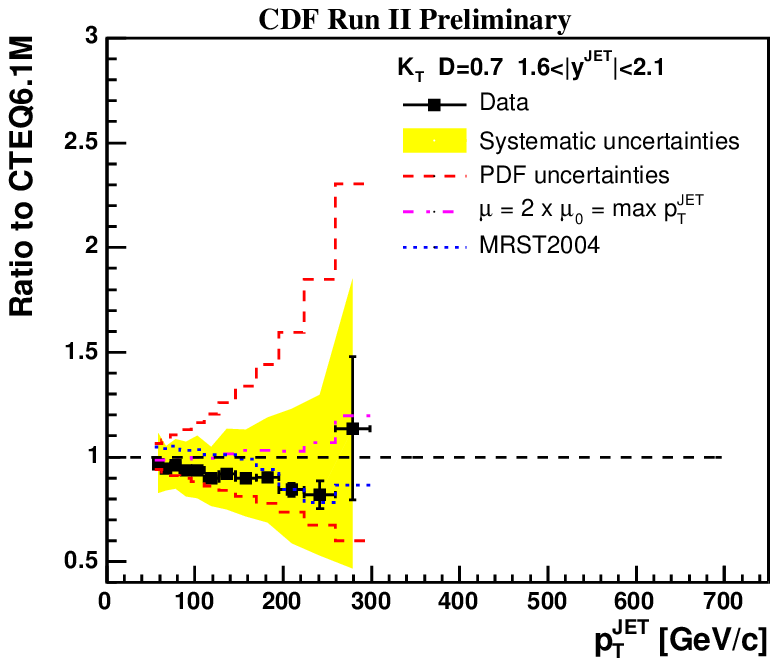}}
\caption{\label{f:forwardJets1}
The inclusive jet cross section from CDF for different
slices of rapidity.
Forward jet measurements provide additional constraints for global PDF fits and
are important when separating new physics from PDF effects. 
}
\end{figure}

In order to disentangle new physics from PDF effects, one needs to measure jets 
in the forward region. Figure~\ref{f:forwardJets} shows the increased cross section 
at high $E_T$ one expects from a quark compositeness model. If data was not available 
in the forward pseudorapidity region it would be difficult to separate new physics from
PDF effects. Inclusive jet cross section results from CDF are shown in 
Figure~\ref{f:centralJets} and Figure~\ref{f:forwardJets1}.
Note the different sensitivity to PDF's from the central region
to the forward rapidity regions.

The low transverse energy ($E_T$) region for the inclusive jet production has large errors
associated with the uncertainty of jet fragmentation and underlying event.
Data from the Tevatron has been used to help understand this region as well.
As suggested by the jet figures,
the large data set of Run II offers the opportunity for a precision comparison of the Cone (Midpoint) 
algorithm with the $k_T$ algorithm.  These algorithms have differing, and to some extent complementary, 
strengths and weaknesses.  Careful studies of the results from both algorithms applied to the Run II 
data will allow a detailed understanding of each in preparation for their employment at the LHC.

\subsubsection*{Precision $W/Z$ Measurements, Parton Distribution Uncertainties,
the W-mass and Top/Higgs Physics}

The differential cross section for $W/Z$ production $d^{2}\sigma
\,/\,dy\,dp_{T}$ (or, more practically, the cross section $d^{2}\sigma
\,/\,dy\,dp_{T}$ for one of the decay leptons in the semi-leptonic decay
channel) is sensitive to details of PDF's. Precise data on these cross sections
can play a decisive role in narrowing the uncertainties and clarifying many of
the open issues on PDF's described in the first part of this review. This is
because, first, they measure very different combinations of PDF's compared to
DIS experiments, thus provide constraints on many independent quantities not
accessible in DIS. (The leptonic asymmetry measured in Run I is a good
example.) \ In addition, the kinematic coverage of the collider cross section
data will greatly expand that of available DIS data. It is particularly
important that the $W/Z$ cross sections be measured as precisely, and in as wide
a kinematic range, as is possible at the Tevatron, in order to determine the
PDF's well enough to enable better predictions, hence improved discovery
potentials, at the LHC.
The impact that the choice of PDF set as well as the treatment of errors has on predictions
can be illustrated by the calculations of the $W$ cross section at LHC 
energies summarized in Table~\ref{t:pdfw}.
\begin{table}[!h]
\centering
\begin{tabular}{lr}
\hline
PDF Set   & $\sigma_W$ (nb) \\
\hline
MRST2002  & $204 \pm 4$ \\  
CTEQ6     & $205 \pm 8$ \\
Alekhin02 & $215 \pm 6$ \\ 
\hline
\end{tabular}
\caption{\label{t:pdfw}
NLO predictions for the $W$ cross section at the LHC using different PDF's.
}
\end{table}
The Alekhin02 fit uses a different
subset of data than the MRST and CTEQ PDF's. This will in general lead to different 
extrapolations out side of the kinematic region covered by the data used in the fit.
The choice of the $\Delta \chi^2$ definition 
leads to the different error estimate between the calculation 
using the MRST and CTEQ PDF's.
The data from the 
Tevatron can help to discriminate between choices of PDF sets.

The transverse momentum distribution of $W$ and $Z$ bosons at the colliders has been
the focus of much study, both experimentally and theoretically. The main
impetus for this effort has been the desire to achieve the most precise
measurement possible of the $W$ mass, $M_{W}$---a key parameter in precision
SM electroweak phenomenology, and hence a potentially powerful indication for
new physics. For this purpose, it is critical to reliably quantify the
uncertainty of the mass measurement, $\Delta M_{W}$. But the uncertainty
associated with the parton distributions, one of the main contributing
factors, is far from well determined. \ There is no assurance that
current estimates (previously mentioned)
of $20$--$30$ MeV at the Tevatron and $10$--$20$ MeV at the LHC are
indeed trustworthy.\footnote{For reasons described below, these uncertainties
are most likely underestimates.}

Historically, estimates of the $\Delta M_W$ uncertainty relied heavily on an assumption
that correlates it with that of the measured rapidity distribution. More
recent studies make use of the uncertainty estimates based on the Hessian
basis eigenvector PDF sets, e.g.~from CTEQ6. Unfortunately, neither of these
approaches contain reliable information on the uncertainties of PDF's
associated with the degrees of freedom in parton parameter space that are most
relevant to the $W$ mass measurement---the $p_{T}$ distribution of the vector
bosons (or their lepton decay product). In fact, there has been no systematic
study so far of the interplay between the $p_{T}$ distribution of the vector
bosons in colliders and the undetermined PDF degrees of freedom.

A fundamental unanswered
question is: what degrees of freedom in the parton distribution parameter
space are important in determining the $\PT$ distribution of the vector
boson and its decay lepton? Of particular interest is the question: are
there degrees of freedom that are, so to speak, orthogonal to those
that are already well--determined from DIS and $W$-rapidity measurements? It
would be remarkable indeed if the degrees of freedom relevant to the $\PT$ 
distributions are exhausted by those that are already well-constrained by
the $\PT$--inclusive measurements!

Detailed predictions for vector boson $\PT$ distributions are best carried out
using a formalism that includes the proper resummation of large logarithm factors
of the form $\log ^{n}(p_{T}/Q)$ (with $Q\sim M_{W/Z}$). 
Because the resummation calculation is an involved one, and the parton
parameter space is of quite high dimensionality ($\sim$ 20 or more
in conventional global analysis), \textquotedblleft
intuition\textquotedblright\ is of very limited value to reach a conclusion
on this important issue. We need to incorporate the $\PT$--resummation
calculation into the global QCD analysis, and probe the correlation between
parton parameters and measurable $p_{T}$ distributions in a fully integrated
approach. Fortunately, due to recent progress in streamlining the
resummation calculation and the global analysis tools, this goal appears to
be within reach.

The strategy, when the tools are fully developed, would be:

\begin{enumerate}
\item Use the expanded global analysis tools to perform new PDF fits,
incorporating existing data on Drell-Yan and $W/Z$ $p_{T}$ distribution data,
to explore the impact of these on the determination of parton distribution
parameters and their uncertainties (compared to currently existing results).

\item Use the Lagrange Multiplier method (cf.~CTEQ papers) to map out the
directions in parton parameter space that are particularly sensitive to the $%
p_{T}$ distributions; and compare these with the basis eigen-vector
directions in current Hessian analysis, as well as those directions closely
associated with rapidity distributions.

\item  Use the results of the Lagrange Multiplier method to quantify (much
more reliably than current methods) the uncertainty of the $W$ mass
measurement.

\item Use the same results to study the impact on the Higgs search efforts,
particularly the associated production channels $WH$ and $ZH$, and on
single-top production investigations.

\item Reversing the direction of inquiry: ask the question \textquotedblleft
How can the uncertainties (on $W$ mass and $WH$ and $ZH$ signals) be
reduced, if we can improve certain measurements at the Tevatron that can be
used as input to the expanded global analysis?" This question can be
answered with the same analysis tools by using, for instance, hypothetical
goal-oriented data sets. Such studies can provide powerful motivation for
refined experimental plans.
\end{enumerate}

The task of carrying out this program is complicated by the fact that
resummation calculations introduce certain additional \textquotedblleft
non-perturbative\textquotedblright\ parameters of their own. These parameters
have been studied before in the context of fixed PDF's. In the expanded
analysis, new efforts are needed to differentiate between these and the PDF
parameters whenever possible. The inevitable residual correlations between
them then need to be systematically taken into account in the physics
applications. The methodology is the same as the case without the $\PT$
factor.

We have highlighted the $\PT$ distribution in the above discussion. The basic
idea applies to the full range of possible precision $W/Z$ measurements
possible at Run II of the Tevatron and the LHC, such as the
rapidity and charge asymmetry discussed below. The identification of the most
productive measurements requires close collaboration between theorists and
experimentalists in an iterative mode, following the strategy outlined 
above.

\subsubsection*{$Z$ Rapidity Distributions}    

\begin{figure}[!h]
\centering
\includegraphics[width=.75\textwidth]{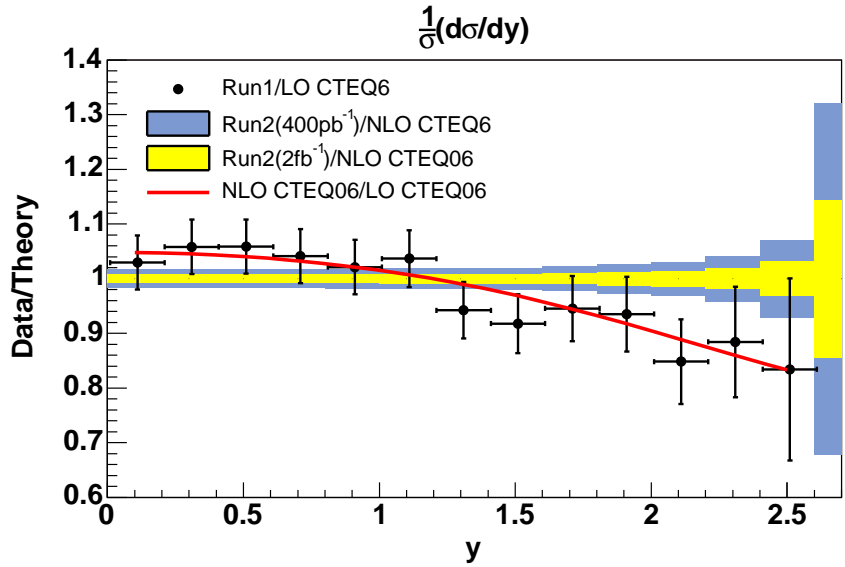}
\caption{\label{f:zrapidity}
The expected improvement in the $Z$ rapidity measurement with increasing luminosity.
}
\end{figure}

$Z$ + jets provides a different constraint on PDF's when considering semi inclusive final states.
$Z$ + jet production as a function of rapidity is sensitive to PDF's and differences between 
LO, NLO and NNLO. It is hard to quantify the luminosity required for this study as it has
not yet been attempted with present data. 
Possibly if the rapidity sensitivity becomes observable 
at 400 pb$^{-1}$ (see Figure \ref{f:zrapidity}), then the semi-inclusive
sensitivity will require at least 5$\times$ as much data.

\subsubsection*{$W$ Charge Asymmetry}    

\begin{figure}[!h]
\centerline{
\subfigure[$25<E_T^e<35$ GeV]{\includegraphics[width=.49\textwidth,trim= 0 10 0 0]{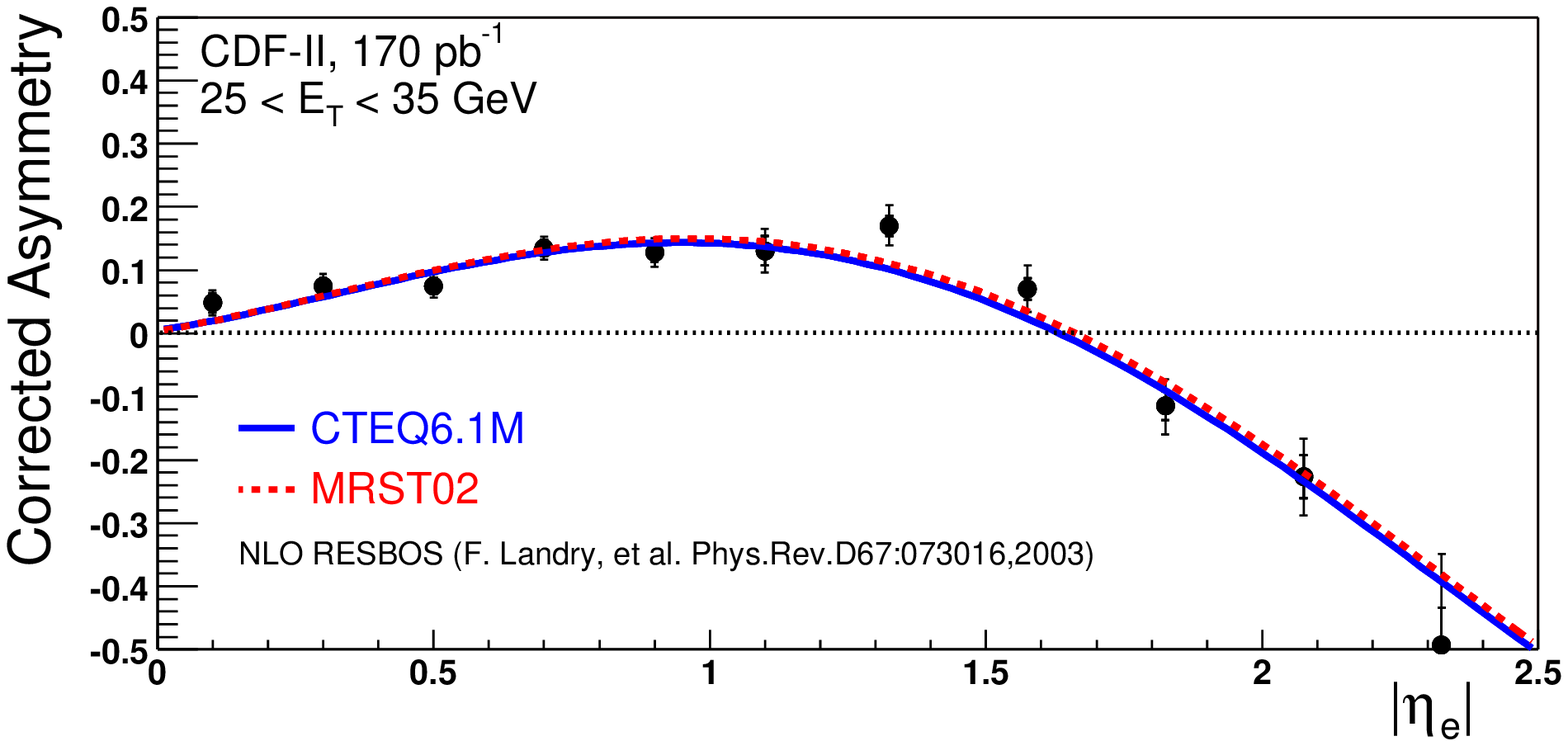}}
\subfigure[$35<E_T^e<45$ GeV]{\includegraphics[width=.49\textwidth,trim= 0 10 0 0]{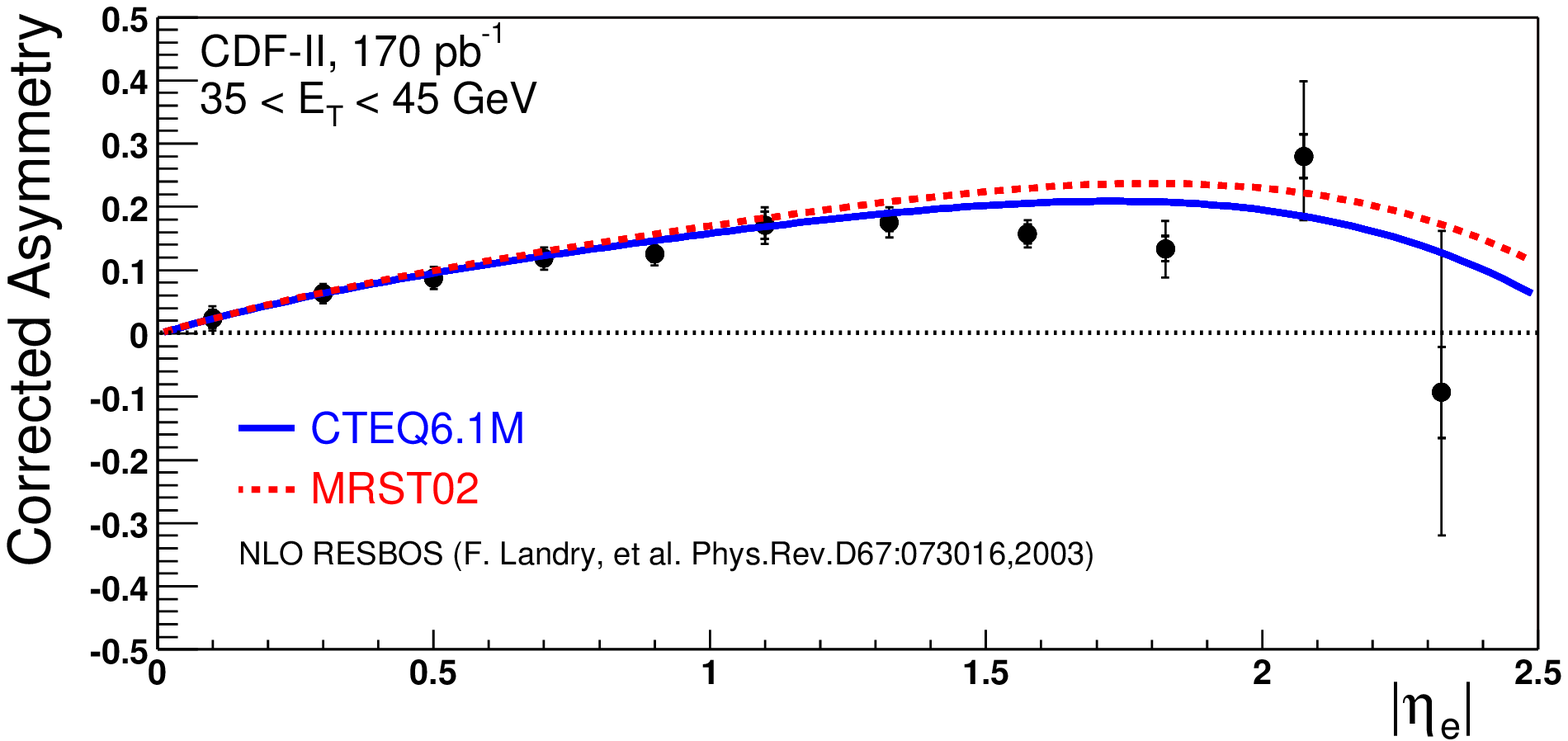}}
}
\caption{\label{f:wasym}
Electron pseudorapidity dependence of the $W$ charge asymmetry for
different slices of electron $\ET$.  More data will allow an exploration of this $E_T$ 
dependence of the asymmetry.
}
\end{figure}

\begin{figure}[!h]
\centerline{
\subfigure[]{\includegraphics[width=.49\textwidth]{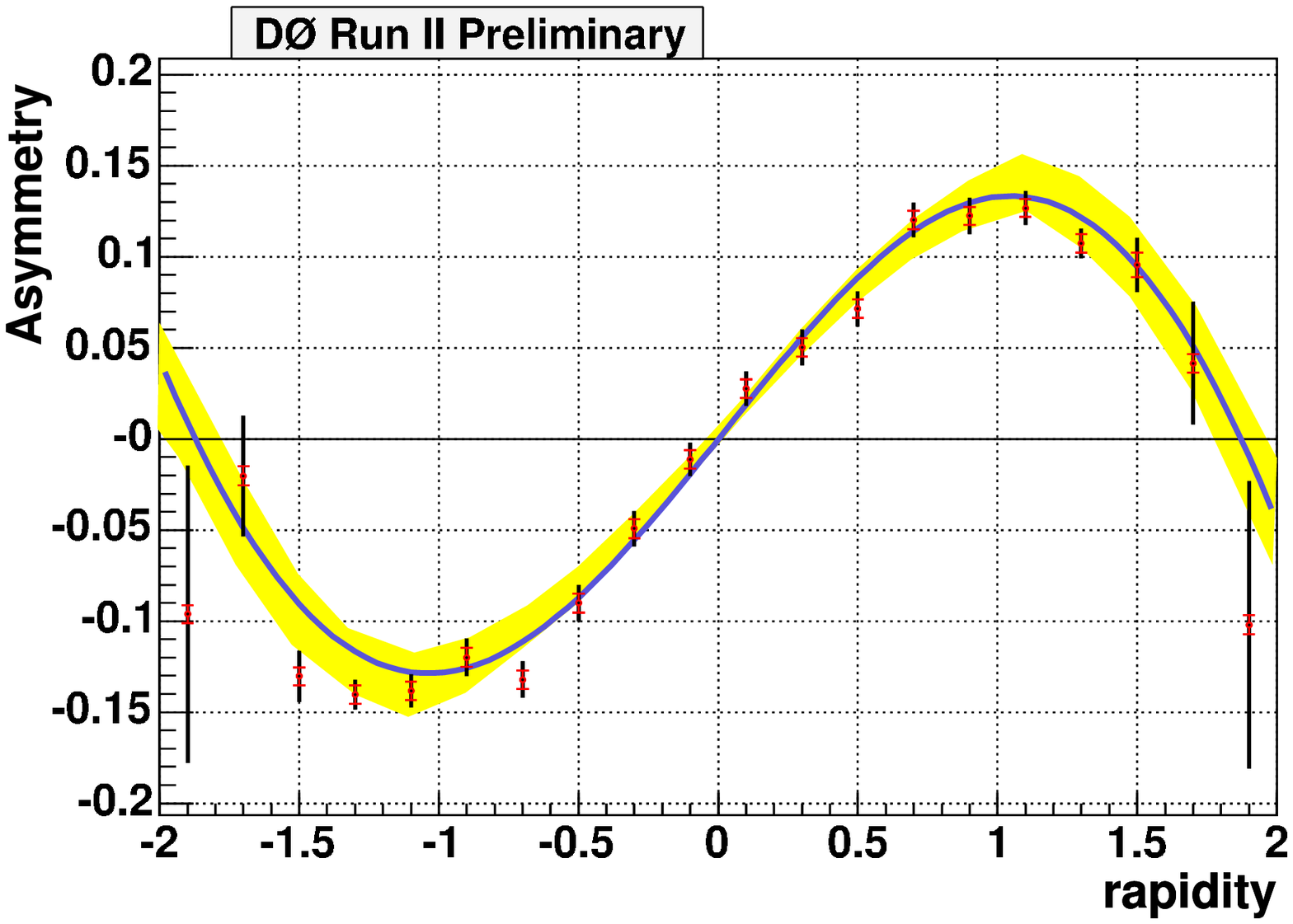}}
\subfigure[]{\includegraphics[width=.49\textwidth]{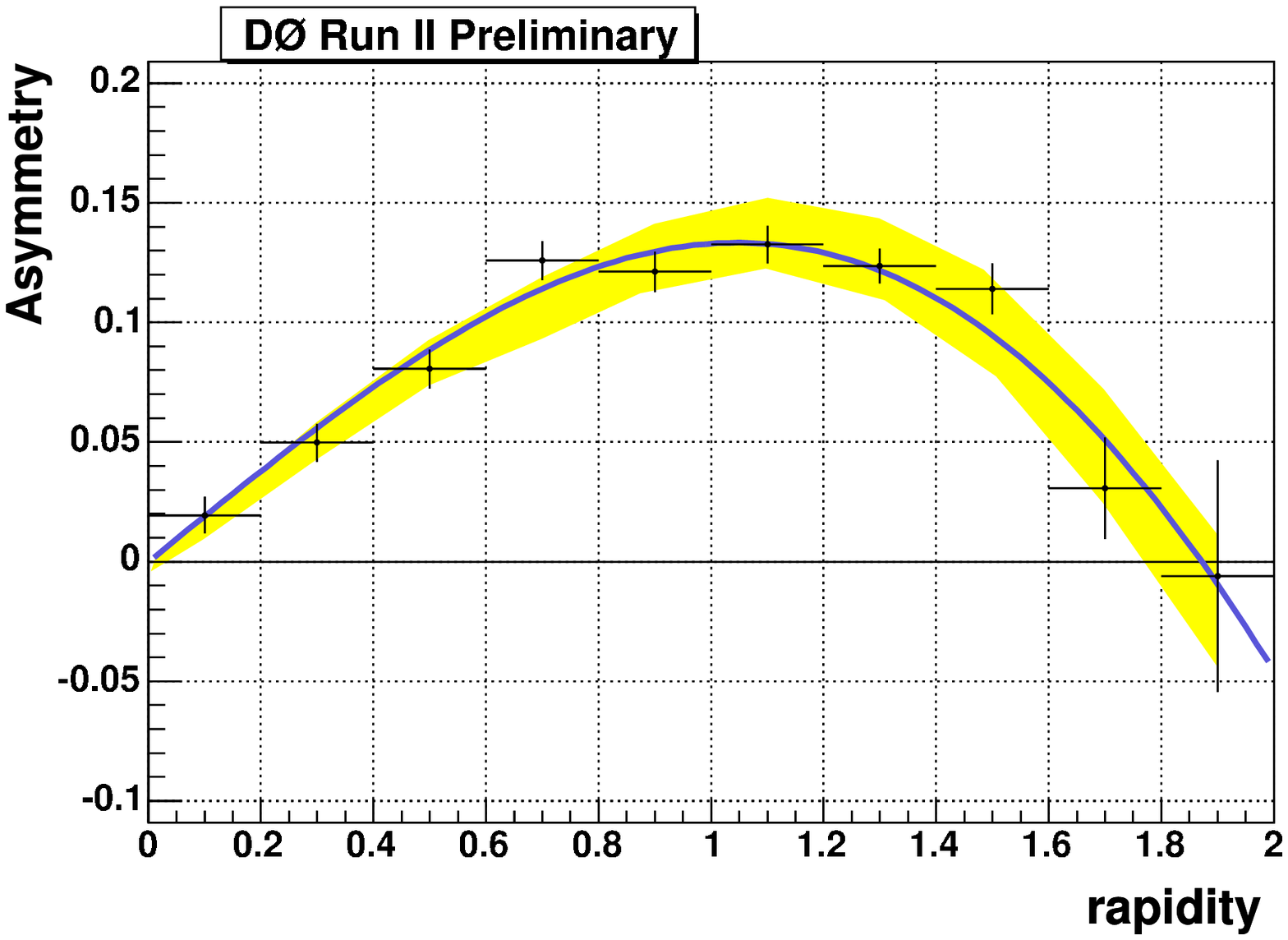}}
}
\caption{\label{f:D0wasym}
(a) the corrected muon charge asymmetry distribution with the
statistical (inner) and systematic (outer) error bars. The shaded band is the 
uncertainty determined using the 40 CTEQ6.1 PDF error sets. The solid line
shows the prediction obtained when using the MRST02 PDF set;
(b) the CP folded muon asymmetry with the total measurement error.
}
\end{figure}

A measurement of the $W$ charge asymmetry constrains the ratio: $d(x, M_W)/u(x, M_W)$ as 
$x \rightarrow 1$.
Having more data allows us to explore the lepton $\PT$ dependence of the $W$ charge
asymmetry. Recent results from CDF are shown in Figure~\ref{f:wasym} while data
from D\O \ is shown in Figure~\ref{f:D0wasym} with the error associated with the PDF 
uncertainty shown as the shaded band.
The $W$ charge asymmetry is an important input to global QCD fits and can be used
to refine PDF's.

\subsubsection*{PDF Error Estimates}    

\begin{figure}[!h]
\centerline{
\includegraphics[width=.75\textwidth]{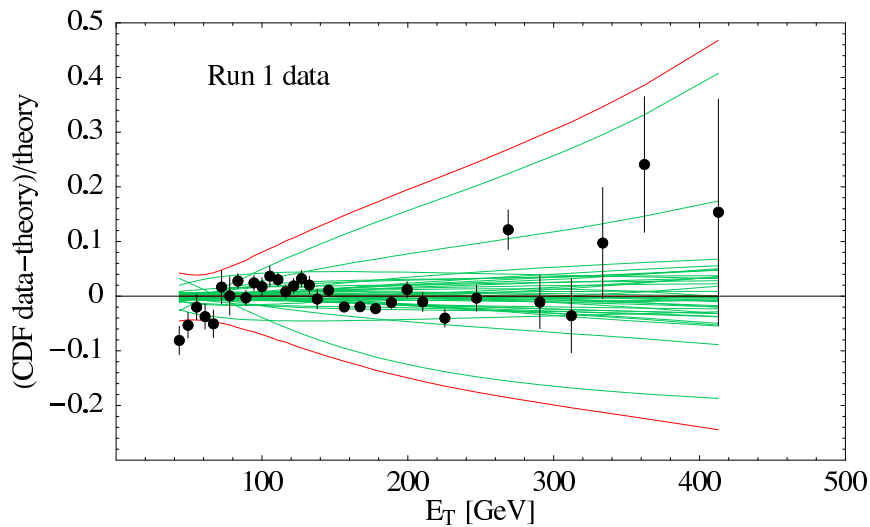}
}
\caption{\label{f:centralJets3}
An application of CTEQ PDF's with error estimates to the Run I inclusive
jet measurement.
}
\end{figure}

A significant advance in quantitatively understanding the impact of PDF errors on measurements
was the development of new techniques to estimate errors. In the past, 
an error associated 
with PDF's was determined by running Monte Carlo using two different sets
and taking the difference. This is clearly not rigorous, since
different PDF sets are usually based on different
assumptions, include different data sets in the fits, and parameterize the PDF's differently.
However, the practice was carried out for lack of a practical alternative.
At the Tevatron, PDF errors can be estimated more quantitatively
(see Figure~\ref{f:centralJets3}).
Consistency between different sets tests the universality of the PDF's.
This is an important cross check of our methodology.

\subsubsection*{Heavy Flavor PDF's} 

\begin{figure}[!h]
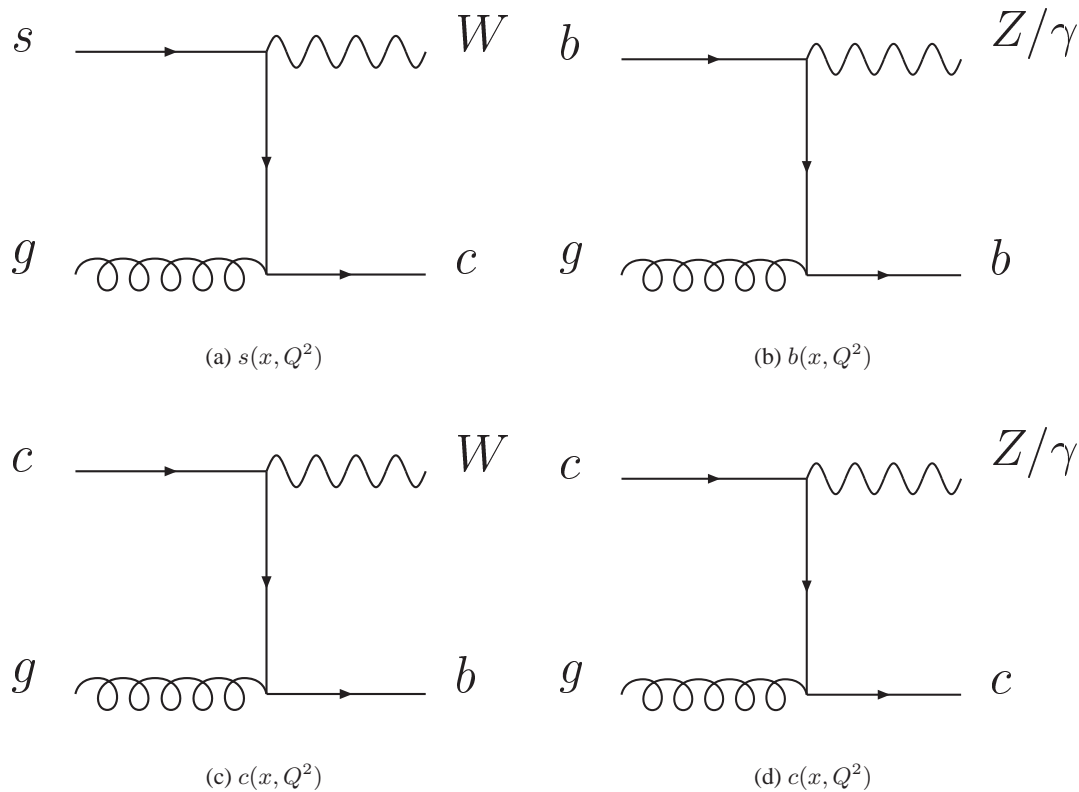

\centering
\subfigure[$s(x,Q^2)$]{\includegraphics[width=.45\textwidth]{qcd/qcdp5/V4/sea_sx.eps}}
\subfigure[$b(x,Q^2)$]{\includegraphics[width=.45\textwidth]{qcd/qcdp5/V4/sea_bx.eps}}
\subfigure[$c(x,Q^2)$]{\includegraphics[width=.45\textwidth]{qcd/qcdp5/V4/sea_cbx.eps}}
\subfigure[$c(x,Q^2)$]{\includegraphics[width=.45\textwidth]{qcd/qcdp5/V4/sea_cx.eps}}
\caption{\label{f:heavyFlavor}
Processes which can be used to probe the heavy flavor content of the proton.
}
\end{figure}

\begin{figure}[!h]
\centering
\subfigure[$\gamma+b$]{\includegraphics[width=.49\textwidth]{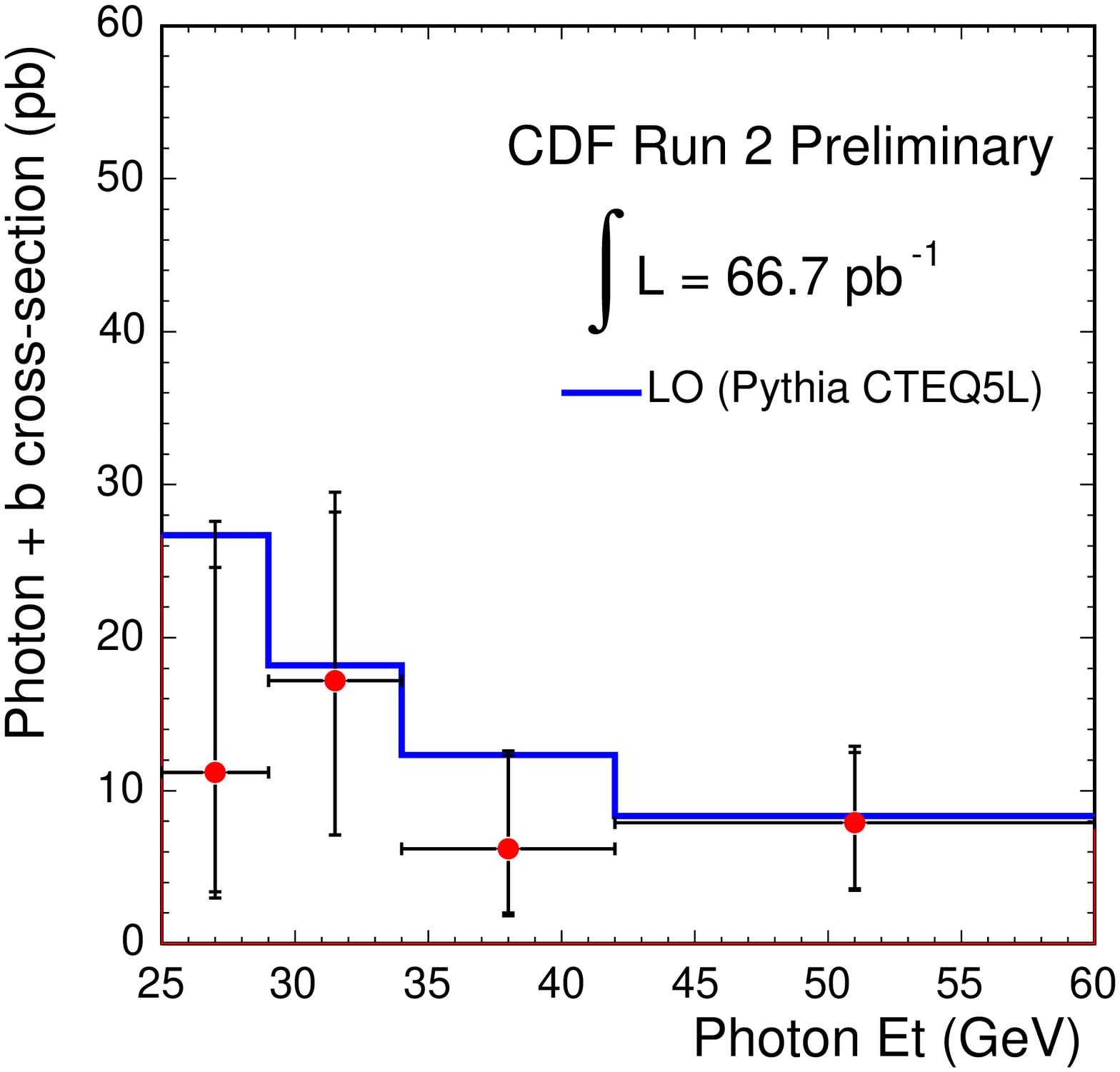}}
\subfigure[$\gamma+c$]{\includegraphics[width=.49\textwidth]{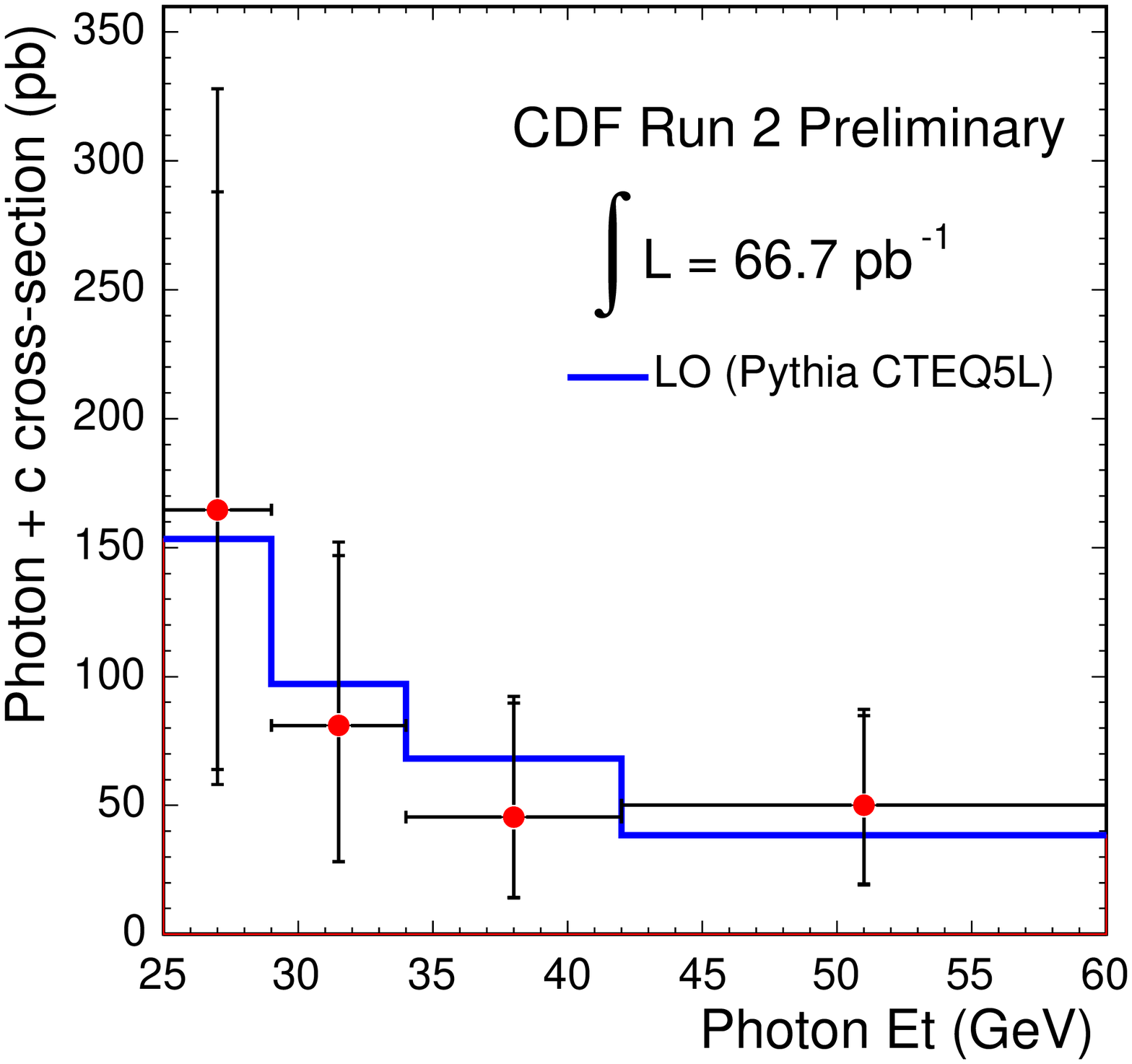}}
\caption{\label{f:gammcb}
Run II measurements of $\gamma$ plus tagged heavy flavor.}
\end{figure}

There is very little direct experimental input on intrinsic heavy flavor of the proton; all 
$c$ and $b$ distributions in existing PDF sets are radiatively generated from the gluons.
The heavy flavor content of the proton can be probed through measurements of
$c\gamma$, $b\gamma$ and $c$+jet, $b$+jet production via the processes shown in 
Figure~\ref{f:heavyFlavor}.
An understanding of the heavy flavor PDF's
is necessary for precise predictions of Higgs boson production rates.
Run II measurements of $\gamma$ plus tagged heavy flavor distributions are shown in Figure~\ref{f:gammcb}.
Currently, the results are  dominated by statistical errors. 
The largest sources of systematic
errors arise from: energy scale, tagging efficiency and the trigger.
Single top production in the t-channel process is also sensative to the 
$b$ PDF at high $x$.

\section*{Other Important Phenomenological Measurements}

\subsubsection*{Fragmentation of Quarks and Gluons and the Structure of Jets}
Analogous to our exploration of the structure of
the proton, the fragmentation of quarks and gluons into hadrons is
fundamental science. The large $z$ region ($z_i$ is the fraction of the
parton momentum carried by the $i$th hadron) is accessible at the
Tevatron, and must be understood to determine how often a ``jet''
fluctuates into only one observable charged track or photon.
This is critical for understanding backgrounds to $\tau$ leptons and
photons in Higgs boson final states.
Some of the interesting properties of fragmentation that can 
be studied at the Tevatron are:
quark versus gluon jet fragmentation;
heavy quark jet fragmentation ($c$, $b$, and even $s$);
the high $z$ limit of jet fragmentation for different
species of particles;
and fragmentation distributions $dN/dz$.
The Tevatron gives complementary measurements of these
quantities for different kinematic slices
of $x_T$ and $\PT$.
A particularly interesting and phenomenologically important question is the 
fraction of gluon and light quark initiated jets that fragment into heavy quarks.  
A precise determination of this fraction can likely be obtained from a study of the large Run 2 
sample of $W + 1 b$-tagged jet events.  At LO this sample has only a negligible contribution 
from short--distance $Wb$ states and is dominated by b quarks produced in the fragmentation process.  
Knowledge of this fragmentation process and its associated rates will lead directly to better control 
of the dominant background for single top production, the backgrounds for measurements of heavy flavor 
PDF's and a variety of backgrounds to beyond the Standard Model processes at the LHC.

\subsubsection*{Details of the Underlying Event}

The underlying event (UE) is an unavoidable background to many measurements at the 
Tevatron and the LHC.  There is also interesting QCD physics in the UE since, in general, 
it contains particles that originated from initial and final state radiation, beam-beam 
remnants, and multiple parton interactions.  CDF has studied the UE in high transverse 
momentum jet production, but there is still much to be done.  In particular, one would 
like to measure the cross-section for multiple-parton collisions and establish precisely 
how much it contributes to the UE in various processes.  Also, one would like to study the 
UE in color singlet (e.g. $\gamma^*/Z$) production, 
and compare to the UE in high $\PT$ jet production.  CDF 
can utilize the miniplug and the CLC to extend measurements to large rapidity.   Multiplicity 
distributions in $W$, $Z$, Drell Yan, $WW$, $ZZ$, and $WZ$ would be very 
interesting.  In the first 200 pb$^{-1}$, 
CDF had a clean $ZZ$ event (the only one) with 70 associated tracks, 34 in
$\PT>0.4$, $|\eta| < 1$ region, while it had a clean $WW$ event with zero
tracks in that fiducial region (out of 17 events) and almost nothing forward.
Such effects are worthy of more study.
Certainly the tails of the distribution are sensitive to the UE and possible
anomalies.  Large fluctuations are presumably due to differences in the impact parameter
(an interesting variable). In addition, we should try and establish the rate of vector boson 
fusion (VBF) and study the probability of rapidity gaps.  The following is a list of some of 
the UE related measurements that need to be completed:
\begin{enumerate}
\item The UE in color singlet production ($W$, $Z$, photon, Drell Yan, VV, di-photon). 
\item The rate of multiple parton collisions.
\item Distributions in the UE (multiplicity, $dN/d\eta$, $dN/d\PT$).
\item Correlations in the UE.
\item VBF and rapidity gaps.
\end{enumerate}

\subsubsection*{Heavy Flavor Fragmentation}

\begin{figure}[!h]
\centering
\includegraphics[width=.8\textwidth]{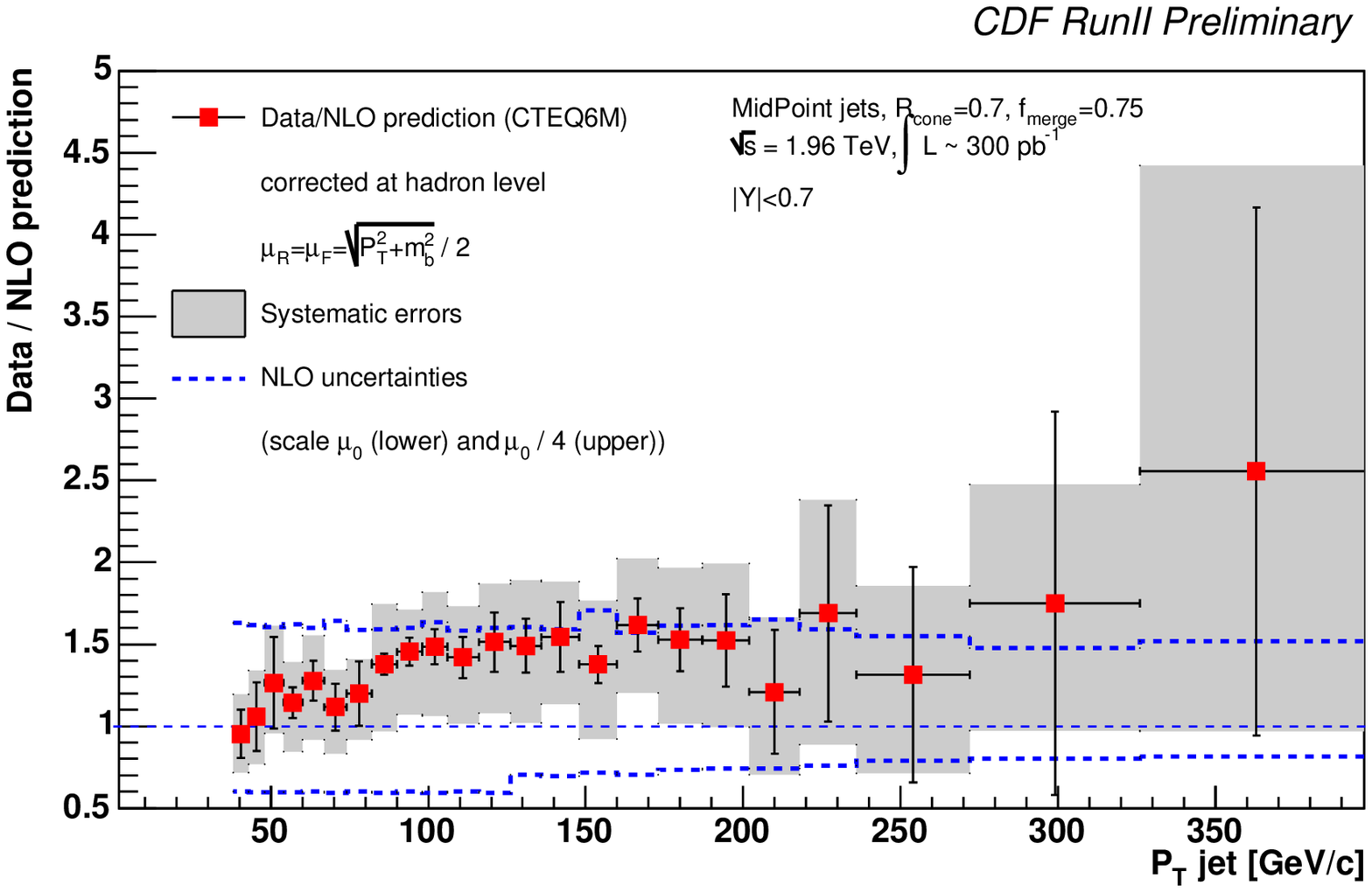}
\caption{\label{f:RunIIBJetRate} Ratio Data/Theory as a function of $\mathrm{\PT^{jet}}$
for $b$-jets from Ref.~\cite{DOnofrio:2006}.}
\end{figure}

$B$ production and backgrounds to Higgs production have never been satisfactorily 
understood. 
In Run I the rate of $B$ jet production was larger than expected from theory calculations.
A more careful theory calculation was performed using up-to-date information on the $B$ 
fragmentation function and resulted in better agreement \cite{Cacciari:2002pa}.
Recent results from CDF are shown in Figure~\ref{f:RunIIBJetRate}.
Data from the Tevatron will enable us to
reduce the systematic uncertainties associated with $B$ hadron production.
This is important, given the importance of understanding the details
of top production at the LHC and the special role of $b$ quarks in Higgs
boson physics.

\subsection*{Diffractive Physics and Central Exclusive Production}
Another section of this report deals more completely with diffraction, as a
class of interactions containing large rapidity gaps (typically $>$ 4 units)
with no hadrons. This implies color singlet exchange, requiring two or more
gluons with a (minor) contribution of $q\bar{q}$. This is a frontier of
QCD, not fully understood but where much progress has been made through
experiments at the Tevatron and HERA. Here we give a brief summary of the two main areas,
\emph{diffraction} and \emph{central exclusive production}. The latter has become very topical
as a possible window on the Higgs sector at the LHC.

\subsubsection*{Diffraction}
 Elastic scattering $\frac{d\sigma}{dt}$ and the total cross section $\sigma_T$ are basic
 properties of $p\bar{p}/pp$ interactions, which will be measured at the LHC by the TOTEM
 experiment. Unfortunately, they are not very well known at the Tevatron, with three inconsistent
 ($>3\sigma$) measurements of $\sigma_T$, and only one measurement of $\frac{d\sigma}{dt}$ into
 the Coulomb region and none into the large $|t|$ region beyond 2 GeV$^2$ (interesting
 from a perturbative point of view). There are no measurements at $\sqrt{s}$ = 1960 GeV, although the
 LHC could run at that $\sqrt{s}$ to make a comparison of $pp$ and $p\bar{p}$. The Forward
 Proton Detectors (FPD) of D\O, in a special planned high-$\beta$ run, may make a competitive
 measurement of $\sigma_T$ and a new measurement of $\frac{d\sigma}{dt}$ through the dip region
 $|t| \approx 0.6$ GeV$^2$. They should also make some measurements of low mass double pomeron
 exchange, $p\bar{p} \rightarrow p\oplus X \oplus \bar{p}$ where $\oplus$ means a rapidity gap
 (\emph{no} hadrons) and $X$ is a completely measured
 system, e.g. $\pi^+\pi^-$ or $\phi \phi$. This is a potentially rich field, both for
 studying diffractive mechanisms and for spectroscopy ($X$ is rich in glueball and hybrid states).
 Single diffractive excitation of low mass and high mass (di-jets, $W$, $Z$, heavy flavors) has been measured, but there
 is a case for a more complete systematic study, e.g. $\frac{d\sigma}{dtdM^2}$ conditional on such massive final states,
 at different $\sqrt{s}$ values. From the $s$-dependence at fixed ($t,M^2$) one could derive a ``hard pomeron" trajectory
 to extrapolate to the LHC. Monte Carlo event generators which have $p\bar{p}$ interactions and
 include diffraction, such as \HW~\cite{Corcella:2000bw,Corcella:2002jc} and \PY~\cite{Sjostrand:2003wg} could then be tested and
 tuned, to improve predictions for the LHC.

\subsubsection*{Central Exclusive Production}

  The above mentioned process, $p\bar{p} \rightarrow p\oplus X \oplus
  \bar{p}$, with $X$ a simple completely measured state, is called
  central exclusive production. The possibility that $X$ can be a
  Higgs boson $H$ has generated much interest in this process at the
  LHC. Precise measurements of the scattered primary protons
  ($\frac{dp}{p} \approx 10^{-4}$) allow one to measure the Higgs mass
  with $\sigma(M_H) \approx$ 2 GeV per event, independent of decay
  mode (e.g. $b\bar{b}, W^+W^-, ZZ$).  The ratio of signal to
  background can be $\approx$ 1:1, and possibly considerably larger
  for a MSSM Higgs (in the MSSM the Higgs cross section can be an
  order of magnitude higher than in the SM). The Higgs quantum numbers
  can be determined from the azimuthal $pp$ correlations: proving that
  is a scalar and has CP= {\tt ++} is essential to establishing its
  identity.

  The key question for exclusive central production is what is the
  cross section? 
  been proposed~\cite{Albrow:2005fw} that $p\bar{p} \rightarrow p
  \oplus \gamma\gamma \oplus \bar{p}$ has an identical QCD structure,
  might be measurable at the Tevatron and, if seen, would confirm that
  $pp \rightarrow p\oplus H \oplus p$ must occur and ``calibrate" the
  theory.  The Durham group (see
  e.g. Ref~\cite{hep-ph/0311023,hep-ph/0409037,hep-ph/0508274})
  calculated the cross sections and they have been incorporated into
  the ExHume~\cite{Monk:2005ji} generator.  The observation of the
  $\gamma\gamma$ process in CDF confirms that the exclusive cross
  section for (SM) $M(H) \approx$ 130 GeV is $\approx$ 3 fb or perhaps
  a factor $\approx$ $2-3$ higher, which is very encouraging. Other
  exclusive processes which can be related to exclusive $H$ production
  are $p\bar{p} \rightarrow p \oplus \chi_{c(b)} \oplus \bar{p}$ and
  $p\bar{p} \rightarrow p \oplus jet-jet \oplus \bar{p}$

The FP420 R\&D collaboration aims to add high precision
  forward proton detectors to CMS and/or ATLAS. In addition to $H$
  observations, exclusive central $W^+W^-$ produced by 2-photon
  exchange should be seen, $\sigma(pp \rightarrow p \oplus W^+W^-
  \oplus p) \approx 100$ fb, and final state interactions between the
  $W$'s can be studied. Other important 2-photon processes are central
  $\mu^+\mu^-$ and $e^+e^-$. These have recently been observed in CDF,
  the first time $\gamma\gamma \rightarrow X$ processes have been seen
  in hadron-hadron collisions.

\subsection*{Tevatron Experience}
Our field is full of new ideas.  However, the practicality
of those ideas can often only be judged {\it after} they have
been applied to real data.  The Tevatron serves as
a proving ground for ideas developed ``in shop'' and those
originating from the LHC perspective.

\subsubsection*{Measurement Techniques}

Systematic uncertainties are difficult to estimate without
data in hand.  ``Rare'' effects, such as a jet fragmenting to
mostly one leading particle, are nonetheless important when
convoluted with the enormous jet cross section.  Dedicated studies
at the Tevatron continue to improve our understanding of several
outstanding experimental issues.

\begin{itemize}
\item Rejection rates:
The rejection rate for the copious and hard--to--simulate
background to photons in hadronic collisions.  These are backgrounds 
to the signal of Higgs boson decay to photon pairs.

\item $b$-tagging efficiencies:
Determination of $b$-tagging efficiencies in hadronic collisions with many
background tracks from other interactions.

\item $\tau$ reconstruction efficiencies
\end{itemize}

\subsubsection*{Search Strategies}

The Tevatron Run II data can be used to
validate new and powerful analysis methods, particularly with
many of the complications of the LHC environment, at least during
the early running. 
Examples of these methods are:
\begin{itemize}
\item{\bf Matrix element weighting:}  The mapping of observed
objects back to the ``theoretical objects'', with are then
weighted according to the fully differential theoretical 
predictions; 
\item{\bf Neural Network analysis:}  The disentanglement
of (supposedly) complicated correlations between observables
based on theoretical training sets of signal and background; and 
\item{\bf Quaero/Sleuth:} An algorithm and automated procedure
to find deviations from Standard Model predictions and quantify their
significance based on the
observed data and without the bias of specific new physics scenarios\cite{Knuteson:2003dn}.
\end{itemize}
\noindent
Other examples are:
\begin{itemize}

\item Development of $b$-charge tagging techniques -- a useful application for
top--mass and $W$--helicity measurements, 
but an enormous effect on reducing combinatorics in $t\bar tH$.

\item Application of $b$--jet--likelihood methods to separate
signal from background.

\item Studies of lepton isolation, jet reconstruction, and 
missing $E_T$ in a hadron collider with many interactions per bunch crossing.

\end{itemize}

\subsubsection*{Early or Post-- Discovery of New Physics}

The design of the LHC provides significant partonic luminosity in
the energy range near $\sqrt{\hat{s}}=1$ TeV, and thus the LHC is
positioned to
discover almost any new phenomena associated with electroweak symmetry
breaking.  The Tevatron was not designed with this goal in mind, but
still has the potential to probe new phenomena up to several hundred GeV.
If the last piece of the particle puzzle is a Standard Model Higgs boson, then
the Tevatron can probably only provide evidence for its existence in a narrow
mass range.  However, theoretical arguments suggest this is an unlikely scenario.
Almost all alternatives suggest a broad spectrum of new particles and possibly
new interactions.  
The increase in energy from
the Tevatron to the LHC is so great the one may be quickly swamped by 
a full spectrum of new particles.  The Tevatron may only be sensitive to
the lighter particles of this spectrum, and could provide measurements
that are free from other sources of new physics.  
The Tevatron experiments have proven their capabilities for discovering and studying a heavy
new particle (the top), and stand as the center of expertise on the subject.
The Tevatron reach for a supersymmetric partner of the top quark with similar mass is
documented elsewhere.   Certain supersymmetric processes are sensitive to the nature of
the lightest superpartner, and there is no compelling study that the LHC can discern
a Higgsino from a Wino LSP.  The non--observation of certain associated processes 
(squark or gluino + LSP) at the Tevatron would immediately identify the LSP as a Higgsino. 
Furthermore, the Tevatron to LHC transition from
valence-- to gluon--sea dominated partons means that the rates for
QCD backgrounds have increased faster than those typical of 
quark--annihilation processes.  The signal--to--background ratio for new
light states may be more favorable at the Tevatron, and the 
systematic errors for searches will be different.

\section*{The Health of the Field in the Pre--ILC Data Era}
A key consideration in a fore-front field of science is making sure
the door is open and inviting to the best and the brightest young
scientists starting their careers. If US graduate students stop going
into a field due to lack of opportunity for individual initiative and
discovery, and the satisfaction of developing and realizing fruits of
their ideas, the US strength in many areas will wither before the time
to bid on the ILC site.

\subsubsection*{Responding to the Unexpected}

The concurrent running of a mature Tevatron and a developing
LHC opens many possibilities.
One can easily imagine scenarios in which a discovery (or a
non--discovery!)
at the LHC would point to complementary measurements at the Tevatron
to explore the space of possible theoretical explanations. 
To study some phenomena, it may be advantageous to have data at a 
lower--energy, valence--quark--
dominated collider {\it in conjunction} with a higher--energy, 
gluon--and--sea--dominated one. 
One doesn't
know what is unexpected, but some possible examples for which the
Tevatron would be critical are:
\begin{itemize}
\setlength{\itemsep}{-0.02in}
\item The Higgs is discovered, and the LHC measurements of the $M_{top}$-$M_W$-$M_H$ Triangle doesn't close. Nothing else is
  seen. Are the LHC top and $W$ mass measurements correct?
\item An ``invisible'' particle is observed at the LHC.  Is this
particle stable, or does it have a small mass splitting with a 
lighter ``invisible'' particle that produces soft, but observable
decay products?
\item A heavy, exotic particle (fractional charge?  or a heavy SUSY hadron) is observed at the LHC.  Are there lighter exotics that
were missed, possibly with lifetimes on the order of the $\tau$ lifetime?  
\end{itemize}

Re--analysis of Tevatron data can be performed after the shutdown of the Tevatron 
accelerator, provided we are prepared to do so and there is enough data to make
it worthwhile.
Even if we do not have enough
information to find new phenomena now, feedback from the LHC could
indicate which channels
to study, or how to reduce certain systematic errors.
We note that the JADE collaboration has recently performed several
interesting QCD analyses despite the fact that the data is over 20
years old!

\subsubsection*{The Health of the Field: Physicists and  Engineers}
Jack Steinberger once asked ``Why is it that the US produces such
wonderful graduate students and builds such lousy (not the actual
word)  detectors?''.
The Tevatron provides opportunities that are very attractive to the
best experimental students- playing a central role in 
world-class measurements, being responsible for entire subsystems,
and the opportunities for improvements at the scale of small but
first-class groups. The lack of manpower is in many ways an
opportunity if handled well- with appropriate support from the Lab and
an investment in stream-lining the Tevatron detectors provide
positions of responsibility for students and postdocs 
that will otherwise largely disappear.

A similar situation applies to accelerator physicists and accelerator
engineers. Fermilab has a wonderful record of surpassing accelerator
goals, dating back to the 400 GeV Main Ring complex, and now happening
again with the upgraded Tevatron. \footnote{The original CDF trigger was designed for a peak luminosity
  of $10^{30}$; the Tevatron is now running at more than 200 times that.}
An operating accelerator offers opportunities that are complementary
to those of designing a new facility; the LHC upgrade will be very
attractive to young accelerator physicists, and having opportunities
in the US we believe to be essential to attracting young physicists
and engineers into the field.

\subsubsection*{Additional Capabilities}

We should not rule out the prospects of adding new capabilities to the
detectors or running at a different center-of-mass energy in order to
leverage the experience and facilities at Fermilab. As one example,
improving particle identification so that every particle up to a Pt of
20 GeV or so is identified would open up new capabilities unavailable
to the LHC detectors.  would allow distinguishing the charm-strange
quark final state from the up-down quarks in W decay, and could allow
distinguishing the b from the b-bar in top pair events, eliminating
combinatoric smearing of the top mass. While this could also be done
at the LHC, upgrades may be much easier to implement and will provide
much more opportunity to youth than at the LHC.

The online event selection is extremely flexible. As we approach the 
systematic limits of some of the current priority measurements, such 
as top and $W$ mass, one
could change the trigger in order to collect more data for measurements
which we are not systematics limited. As an example, the $b$ physics program
would benefit from such a change in priority.

\subsection*{Conclusions}

``What are the advantages of running the 
Tevatron until the end of 2009 and accumulating 8 fb$^{-1}$ 
before the LHC has a comparable amount of data?''
These include opportunities for measurements and experience
that are worthwhile on their own and valuable as
input and guidance to LHC analyses.  In determining the
future of our field, we believe that time is of the essence.
Any advantage the Tevatron can give to the LHC physics program
allows us to make strategic decisions for the future earlier.

The Tevatron will provide the most precise measurement of the top and $W$ mass 
for many years and will likely remain competitive to what can be achieved 
at the LHC, provided the Tevatron is not prematurely shut down. 
The precise measurements will 
allow for an independent consistency check (with different systematics)
of the SM if a Higgs boson is 
discovered.  If the $m_t-M_W-M_H$ triangle relation does not hold,
then the deviation might indicate the energy scale for new physics. 

Tevatron data is being used to validate our understanding of physics
processes that are important backgrounds to new physics searches, such
as $t\bar t$, $W/Z$ + jets, diboson, single top, and multijet processes.
This includes an understanding of cross sections and kinematic distributions.
A side--benefit of this measurement program is an improved 
understanding of QCD Monte-Carlo models.  Our physics description of
an event requires detailed modeling of effects such as 
parton showering, fragmentation and hadronization, and the underlying event,
which in turn relies on fitting and tuning parton distribution and
fragmentation functions.

Many of the uncertainties on the measurements of the physics program at the LHC can 
only be addressed using data, of which the Tevatron data is the most relevant source. 
The program of measurements described here draws on the unique talents and expertise 
of those scientists working on experiments at the Tevatron.  We have the opportunity 
of completing this exciting program at the Tevatron before the LHC has accumulated large 
amounts of data.  The more we learn at the Tevatron now, the more successful the 
LHC will be. 

The decision of when is the right time to stop Tevatron operations needs to take
into consideration more than the possibility of discovering new physics.
It should also ensure that we utilize fully the Tevatron to strengthen the foundation of our 
understanding of the Standard Model. There is much interesting and important physics 
still to be explored.  Some measurements would benefit from running longer while others can be done
after Tevatron operations stop. It is essential to maintain Fermilab as an interesting 
place to do physics in order to attract the best graduate students into our field
and provide the 
necessary training for them so as to ensure that the expertise does not dissipate.




\bibliographystyle{tev4lhc}
\bibliography{tev4lhc}

\end{document}